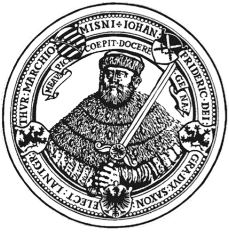

Friedrich-Schiller-Universität Jena
Physikalisch-Astronomische Fakultät
Thüringer Landessternwarte Tautenburg

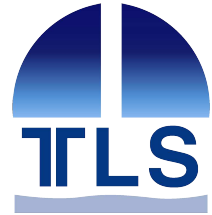

# Radiative transfer modeling of outbursts of massive young stellar objects

**Dissertation**

To fulfill the requirements for the Degree of
*doctor rerum naturalium* (Dr. rer. nat.)

Submitted to
The **Council of Faculty of Physics & Astronomy**
of the **Friedrich-Schiller-Universität Jena**

by **M. Sc. Verena Wolf**
born on **02.03.1990** in **Jena, Germany**

**Referees**

1. Prof. Dr. Artie Hatzes
   *Thüringer Landessternwarte Tautenburg, Sternwarte 5, 07778 Tautenburg*

2. Prof. Dr. Sebastian Wolf
   *Universität zu Kiel, Institut für Theoretische Physik und Astrophysik, Leibnizstr. 15, 24118 Kiel*

3. Dr. Agnes Kospal
   *Konkoly Observatory, Konkoly-Thege Miklos ut 15-17, 1121 Budapest, Hungary*

**Date of defense: 8. February 2024**

# Zusammenfassung

Junge Stellare Objekte (kurz: YSOs) sammeln bis zur Hälfte ihrer Materie in kurzen Perioden mit einem stark erhöhtem Massefluss auf. Das führt zur Aufheizung des protostellaren Kerns auf allen Skalen, von den innersten wenigen 10 au (Sublimationsradius), über wenige 100 bis 1000 au (protostellare Scheibe und Geburtsstätte der Planeten) bis hin zum ausgedehnten Reservoir des kalten Staubes. Die erhöhte thermische Abstrahlung des Staubes ist bis zu Jahren nach dem Ausbruch im IR sichtbar. Für massereiche YSOs (MYSOs mit $M_* \geq 8M_\odot$) haben solche Ausbrüche eine besondere Bedeutung, da sie Einblick in sonst Verborgenes ermöglichen (MYSOs sind selten, weit entfernt und hinter Staub verborgen). Inzwischen wurden sechs Akkretionsausbrüche bei MYSOs gefunden. Wir haben zwei davon näher untersucht und wichtige Burst-Parameter bestimmt. Es handelt sich dabei um G358.93-0.03 MM1 (junge Quelle) und G323.46-0.08 (weiter entwickelt). Dafür haben wir insgesamt drei Burst-/Post-Beobachtungen zwischen 50 und 200 $\mu m$ mit der fliegenden Sternwarte SOFIA durchgeführt. Die Burst-Parameter werden benötigt, um zu verstehen, wie massereiche Protosterne wachsen. Der Ausbruch von G323 ist der energiereichste, der jemals für ein MYSO beobachtet wurde (mit einer freigesetzten Energie von rund $10^{47}$ $erg$). Der Ausbruch von G358 war etwa zwei Größenordnungen schwächer und deutlich kürzer (2 Monate statt 8 Jahre). Wir vermuten, dass der Ausbruch von G358 durch die Akkretion eines Spiralfragments (oder eines kleinen Planeten) verursacht wurde, während G323 ein schweres Objekt (einen Planeten oder sogar einen potenziellen Begleiter) 'verschlang'.

Wir modellieren den Energietransfer in diesen Quellen durch Streuung und Absorption/Re-Emission der Strahlung an Staubteilchen (Radiative Transfer - RT), sowohl statisch, als auch zeitabhängig. Das ist die erste Arbeit, in der zeitabhängiger RT (TDRT) des Staubkontinuums auf astrophysikalische Objekte angewendet wird. Wir ermitteln realistische Nachleuchtzeit-skalen in verschiedenen Wellenlängenbereichen und untersuchen, wodurch diese maßgeblich beeinflusst werden. Da die Dauer des Nachleuchtens sehr verschieden sein kann (abhängig vom Burst und der protostellaren Umgebung), entwickeln wir ein Python Programm (den TFitter), mit dem sich Modelsätze mit zeitabhängigen Beobachtungen simultan an eine große Modeldatenbank anfitten lassen. Dies ermöglicht die systematische Eingrenzung der Burst-Parameter, sowie der lokalen Staubverteilung.

# Abstract

Young stellar objects (YSOs) accrete up to half of their material in short periods of enhanced mass accretion. These episodic accretion outbursts imply afterglows, which are observable at IR wavelengths (sometimes even years after the end of the burst). They are caused by the reprocessing of the burst energy within the dense natal environments of the YSOs. Bursts can impact the surrounding cloud core on all scales, from the innermost tens of au (dust sublimation), up to a few hundreds and 1000 au (disk, the site of possible planet formation), and the more extended cloud core (reservoir of cold dust). For massive YSOs (MYSOs with $M_\star \geq 8M_\odot$), accretion outbursts are of special importance, as they serve as diagnostics in highly obscured regions (MYSOs are rare, distant, and deeply embedded). Only since 2015 six MYSO bursts have been discovered.

Within this work, two outbursting MYSO's within different evolutionary stages, the young source G358.93-0.03 MM1 and the more evolved one G323.46-0.08, are investigated, and the major burst parameters are derived. For both sources, follow-up observations with the airborne SOFIA observatory were performed to detect the FIR afterglows. All together, we took three burst-/post-observations in between $50...200\mu m$. The burst parameters are needed to understand the accretion physics and to conclude on the possible triggering mechanisms behind it. Up to today, G323's burst is the most energetic one ever observed for a MYSO (with a released energy of $\approx 10^{47} erg$). G358's burst was about two orders of magnitude weaker and shorter (2 months instead of 8 years). We suggest that G358's burst was caused by the accretion of a spiral fragment (or a small planet), where G323 accreted a heavy object (a planet or even a potential companion).

To model those sources, we use radiative transfer (RT) simulations (static and time-dependent). This is the first work to apply time-dependent RT (TDRT) to astrophysical objects. We determine realistic afterglow timescales in different wavelength regimes and investigate the influences on the afterglow appearance (parameter study). Additionally, we develop (and benchmark) a Python routine (TFitter), that works similar to the sedfitter [122] but incorporates fluxes from various epochs. With the TFitter, we can fit large sets of time-dependent models to the data, thereby constraining the burst and the local environment in great detail. The TFitter will be used in the near future to analyze the SOFIA-measurements of four MYSOs (including a periodic burster).

# Acknowledgements

**"Das Weltall ist voller magischer Dinge, die geduldig darauf warten, dass unsere Sinne schärfer werden."**

Eden Phillpotts, 1862-1960

Many people have been part of the past five fruitful and enjoyable years of thesis work. First, I thank Dr. Bringfried Stecklum for being my scientific advisor. You did a great job supervising me and I am lucky to benefit from your experience. Without our numerous discussions, I would probably know less than half of what I know today. You always provided me with good literature and never complained that I asked that many questions. You have always looked out for my future. I can stay at the TLS for three more years only thanks to you, Dr. Jochen Eislöffel, and the DLR. This is the best that could happen, and I am very grateful for that. It makes me lucky that I don't have to stop researching when there is so much more to explore.

I also thank Prof. Tim Harries, who provided the TORUS code and modified it according to our needs. Without your help, I could not have run one single time-dependent simulation. During my visit, you gave me a warm welcome and a good introduction to TORUS. If something was not working with my simulations, I could always ask you for your help. You included features such as certain YSO-configurations, aperture-dependent SEDs or dust sublimation, just for us. Thanks a lot for that.

The best data one could get is the one obtained with the help of an instrument scientist. Therefore, I thank the entire SOFIA team and especially Christian Fischer. You really did a lot to get us valuable data. I also thank you for your continuing work on the FIFI-LS pipeline, which will also be useful in the forthcoming episode.

I also thank Prof. Artie Hatzes, my official supervisor. Even if the director's job is a quite busy one, you always had time for your students and I always felt welcome at the institute. I appreciate that you had a smile and a good word, always at the right time. Hope you enjoy reading my thesis. I know it will be a lot of work to evaluate it. Therefore, I am very happy that we found experienced referees who agreed to do so. I really appreciate your time and knowledge. In this context, I also thank the scientific community for their mutual support and scientific input. I want to mention Dr. Jochen Eislöffel, Dr. Alessio Caratti o Garatti, Dr. Thomas Sperling and Dr. Hendrik Linz.

It is important to feel valued. And I always had that, thanks to my colleges here at the TLS. This includes not only the star formation group but also all other scientists and non-scientists. You have always been there for me, with your advice, your company, and kindness.

I thank the Administration of the TLS (and the university) and our IT. I never had to worry, no matter what happened, whether my laptop died three weeks before submission, I got pregnant, the contract ran out, or I just needed a certain document. There were always options and always help. My special thanks go to Miss Schmidt, who made it possible to shift my funding when my little daughter was born.

I still cannot believe that I was the first asking for that.

Even with the help of the administration, it is possible to mess things up. Therefore, I thank Prof. Mundhenk for letting me do the exam twice, when I forgot to invite Prof. Hatzes the first time. I also thank you for the content of your lecture (advanced programming with Python), which was very useful for the programming of the TFitter.

As a PhD student, you also have teaching duties. Mine have been less nasty than I thought they would be. And for that, I thank my students. Your respect and curiosity made me really enjoy the time together in the laboratory.

Of course, this work is not possible without the right environment. Therefore, I thank my friends and my family for always being there. I could always count on you. You are my distraction and source of motivation.

Special thanks are given to my husband and my mum (who supported us a lot during the pandemic), but also to my parents-in-law, all my many siblings, my dad, and grandparents. You all provided the background, I need to focus on my thesis.

I also thank my kids for always reminding me not to give up. You have been very patient with me, especially towards the end, when things got tense. You are great kids.

Finally, I thank everyone I may have forgotten. Please, feel acknowledged anyway.

# Contents

**Conclusion** 

**Bibliography** 

# List of Figures

# List of Tables

# Acronyms

**SOFIA**     *Stratospheric Observatory for Infrared Astronomy*

**FIFI-LS**     *Field-Imaging Far-Infrared Line Spectrometer*

**HAWC+**     *High-resolution Airborne Wideband Camera Plus*

**FORCAST**  *Faint Object infraRed CAmera for the SOFIA Telescope*

**ALMA**     *Atacama Large Millimeter Array*

**SED**     *Spectral Energy Distribution*

**LTE**     *Local thermodynamical equilibrium*

**(M)YSO**     *Massive young stellar object*

**maser**     *6.7 GHz Class II methanol maser*

**M2O**     *Maser monitoring Organisation*

**G358**     *G358.93-0.03 MM1*

**G323**     *G323.46-0.08 MM1*

**NIRS3**     *S255-NIRS3*

**(TD)RT**     *(Time-dependent) radiative transfer*

**LE**     *Light echo*

**ZAMS**     *Zero-age main sequence*

**GI**     *Gravitational Instabilities*

**TORUS**     *Transport Of Radiation Using Stokes (Intensities), name of the TDRT code*

**TFitter**     *Time-dependent (SED) Fitter, name of our Python program*

**Suffix 'T'**     *Time-dependent + x (e.g., TSED, Tmodel, Tfit)*

# Aims and structure of this work

---

**Aims**  A prime objective of this thesis is to qualitatively characterize the thermal afterglow of two dedicated MYSO (massive young stellar object) bursters. We also aim to infer the major parameters of their bursts by means of radiative transfer (RT) simulations. With this, we can contribute to a better understanding of episodic accretion around MYSOs and the possible burst-triggering mechanisms behind them. Furthermore, information on the local dust distribution shall be inferred. We use time-dependent RT (TDRT) to model one of these MYSOs. As this is the first time that TDRT has been applied to astrophysical objects, we include a general parameter study. The wavelength-dependent nature of the thermal afterglow has been the subject of previous work (see, e.g., [70, 33]). We want to complement these studies with our TDRT models. We also aim for a more general approach, where we consider a variety of models to analyze the afterglow of a MYSO. Similarly to the (static) sedfitter [122], we want to develop a time-dependent fitting tool. This makes this work a foundation for future (quantitative) time-dependent analysis, which will complement the analysis of the (yet small number of) observed (MYSO) accretion events.

**Structure**  In order to achieve our goals, we perform several steps. The structure of the work is as follows:

Sect. 2 contains some **basics** helping to understand the current picture of **massive star formation** (Sect. 2.1.2) in the context of **episodic accretion** (Sect. 2.2), including the discussion of the impact of the burst on its environment (Sect. 2.2.5), and an overview over all known MYSO bursters (Sect. 2.2.6). We introduce our objects (Sect. 2.3) and science instruments (Sect. 2.4). Furthermore, the basics of **RT** (static and time-dependent) are introduced, and the 'typical' simulation setting is presented (Sect. 2.5). This is necessary to understand the applied modeling.

The **main part** of the radiative transfer analysis is presented in Sect. 3, which is divided into the following parts: We start with a summary of the different **modeling strategies** applied (Sect. 3.1). The **results** are presented in detail in Sect. 3.2 (G358) and Sect. 3.3 (G323). Although the modeling strategies differ, our objective is to maintain a similar substructure within both sections (i.e., G358 and G323). First, we give a brief *overview of the observations* used (Sect. 3.2.1 and 3.3.1).

Second, we introduce the *modeling steps* (Sect. 3.2.3.1 and 3.3.2). We start with a list (which is generally sufficient to get the overall concept of the modeling). After that, a detailed description of each step follows (this helps to understand the modeling, including its challenges).

Finally, the results are presented (Sect. 3.2.4 and 3.3.3) corresponding to the applied modeling. The structure is as follows: The *dust configuration* of the 'final' model is presented at the beginning (Sect. 3.2.3.2 and 3.3.3.1 respectively), after which the *burst parameters* are derived (starting with Sect. 3.2.4.1 and 3.3.4.2). Finally, the results *are summarized in* (Sect. 3.2.4.3 and 3.3.5).

Since G323 is the first source to be analyzed with a TDRT code, we include a **parameter study** (Sect. 3.3.3.3), showing the influence of local parameters (such as dust density), viewing geometry, and burst energy. G358 is probably the best studied MYSO burster with Class II methanol **masers**, and we provide a comparison with our model in Sect. 3.2.4.2.

In Sect. 4.1 we take a step from a qualitative to a more quantitative method by developing (and benchmarking) a generic fitting method, which incorporates the time dependency, the **TFitter**. The TFitter is benchmarked (Sect. 4.2.1) and applied to the G358 data (Sect. 4.3).

At the end of this thesis, we put our **work in context** (Sect. 5), which includes the burst parameters derived and the time scales of the afterglow. Furthermore, we discuss the reliability of the derived parameters and the applied method (challenges and caveats). Finally, we outline **future steps** and possibilities. The conclusions are summarized at the end.

This work was founded by the DLR (Grant: 50OR1718). This work was carried out under the supervision of Dr. Bringfried Stecklum.

The main results have been published in [130, G358] and Wolf+ in preparation (G323).

# Basics

## 2.1 The current picture of star formation in the context of MYSOs

### 2.1.1 The role of massive stars for the universe

Massive stars (i.e., stars with masses exceeding $8\,M_\odot$) account only for $\approx 1\%$ of the total Galactic stellar population, whereas they input more energy and momentum into the interstellar medium than all other stars together [89, 159]. Massive stars set the initial conditions for planetary formation on small scales and regulate Galactic evolution on large scales [89]. They are generally born in multiple systems, located in dense centers of stellar clusters [13], and evolve much faster than low-mass stars. Feedback via jets, disk winds, radiation pressure, ionizing radiation, photoevaporation, and supernova explosions impacts the parental clouds up to parsec scales [159]. They interact not only with their own parental cloud but also with nearby star-forming regions. The interaction of the protostellar jet (or the disk wind) with the ambient medium can introduce large-scale turbulence, which influences the star formation efficiency. Massive stars alter ISM chemistry by producing heavy elements and may form complex organic molecules by repeated outbursts when they are just forming [67]. Massive stars are progenitors of neutron stars and black holes. When 'dying' in supernova explosions, they create giant bubbles in the ISM where matter is swept up, which may become the seeds for the next generation of stars, e.g., [60].

### 2.1.2 Low-mass vs. high-mass star formation

Low-mass YSOs have been observed in the optical for almost a century (which is, in general, not possible for deeply embedded MYSOs) and their formation has been a subject of investigation for quite some time. In the following, we briefly summarize the current picture of low-mass star formation, where we follow the classification scheme introduced in 1987 by [83] and [2]. After that, we move on to high-mass star formation, discussing differences, challenges, and limits.

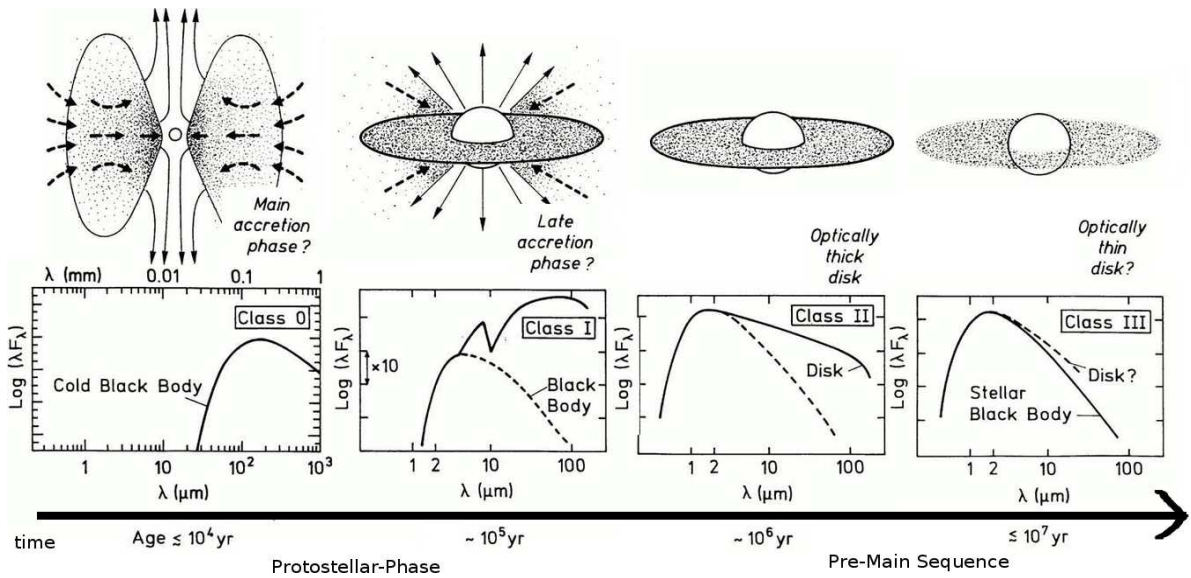

Figure 2.1: Classical phases of protostellar evolution for isolated low-mass YSOs. Modified from [119].

When a single *isolated low-mass protostar* forms, it passes through different stages, before reaching the zero-age main sequence (ZAMS). A cold cloud core collapses under its own gravity once the thermal gas pressure is no longer sufficient to prevent collapse. This is known as the Jeans criterion with a critical mass of $M_{Jeans} \approx \sqrt{\frac{1}{\rho}(\frac{kT}{G\mu})^3}$, with k as Boltzmann constant, G as gravitational constant, $\mu$ as mean molecular weight, T as gas temperature and $\rho$ as gas density. Initially, the collapse happens isothermal. With increasing density, the core becomes optically thick and its temperature gradually rises. The first hydrostatic core has formed [84, first Larson core]). It slowly continues to contract, whereby it heats up (adiabatic process). Once the core temperature exceeds the $H_2$-dissociation temperature ($\approx 2000\,K$, endothermic process), the core becomes unstable and the second collapse phase begins. When most of the $H_2$ is dissociated, the second hydrostatic core forms. This occurs within the first $10^4\,yrs$ [84]. In the low-mass regime, this time span is essentially the one of the first core (the lifetime of the second core is much shorter). In the high-mass regime, the first core phase is short and the second core is formed almost immediately [14, hydrodynamic simulation]. After the formation of the second hydrostatic core, the 'evolution' proceeds as shown in Fig. 2.1 (upper row). A disk forms (surrounding the protostar), and jets build up and start to remove angular momentum. Material falls from the envelope onto the disk, is transported inward, and accreted by the protostar (disk-mediated accretion). Most of the accretion occurs during the Class 0 and I phase. The envelope becomes thinner until it is completely removed, and only the disk is left (Class II or classical T Tauri star). Until the YSO reaches the ZAMS, the disk loses more and more material and may become optically thin (Class III). Within the disk, the planets are formed. Possibly, this process starts already as early as in the Class 0/I stage (see [140] and references therein). The cavity widens during protostellar evolution as jets and winds entrain material and clear it. The lower row shows the spectral energy distributions (SEDs), which is the flux density over the wavelength. During its evolution, the peak of the SED shifts from the far infra-red (FIR) towards the near infra-red (NIR), while the slope steepens. Essentially, this is the result of the decreasing optical depth due to dust clearing because of accretion and dust removal. A widely used method to distinguish the classes introduced above is given by means of the spectral index [83, $\alpha = dlog(\lambda F_\lambda)/dlog(\lambda)$, for $\lambda > 2\,\mu m$], which describes essentially the slope of the SED.

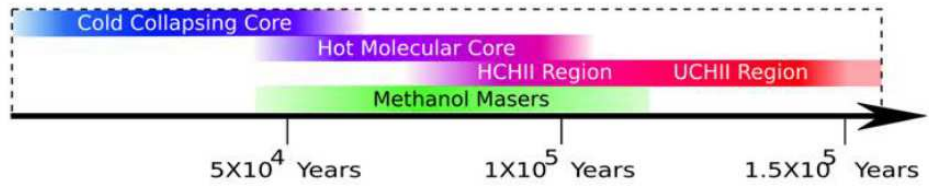

Figure 2.2: Phases of protostellar evolution for a MYSO. Taken from [67].

For *high-mass protostars* the situation is slightly different, because they are much more deeply embedded, evolve much faster (order $10^4$ – $10^5$ instead of $10^7$ years), accretion rates are much higher ($10^{-4}$ – $10^{-3}$ instead of ($10^{-7}$ – $10^{-6})\,M_\odot/yr$), and much higher forces are at play [159]. Massive stars can still be deeply embedded when reaching the ZAMS, and thus they are invisible in the optical and NIR (most of the time). On top of that, MYSOs are rare[1] and distant.

The observational stages of massive star formation are visualized in Fig. 2.2 (cut-out) and Fig. A.1. At first, a clumpy molecular cloud forms a massive prestellar core that evolves into a hot molecular core. Hot molecular cores are compact (0.1 $pc$), dense ($10^6\,cm^{-3}$), and have temperatures of about 100 $K$ [80]. Later, the protostar starts to ionize its surroundings. However, high column densities lead to absorption of most UV photons, which quenches ionization. A hypercompact HII region forms. With a further increase of the ionized region, the hypercompact HII region becomes an ultracompact HII region (UCHII), and later a (more extended) HII reg/OB association. During the molecular core and hypercompact HII region states, most of the accretion occurs. Fortunately, this is also the time when the densities and temperatures are in the range required for the Class II methanol maser stimulation. As a result of the heating of the dust grains, their ice mantles sublimate, thereby releasing molecules into the gas phase. In addition, the IR radiation of the grains is in the appropriate wavelength range for maser excitation. The connection between maser flares and accretion events enabled a follow-up program that deepened our understanding of episodic accretion processes in massive star formation.

---

[1] The distribution of stars (i.e., the observed number $\Delta N_*$ of stars in a volume and time-interval per logarithmic mass bin $\Delta log(M_*)$) can be described by the initial mass function (IMF), which was introduced already by [125] in 1955. Toward the high-mass end, the ratio $\Delta N_*/(\Delta log(M_*))$ decreases linear with the stellar mass.

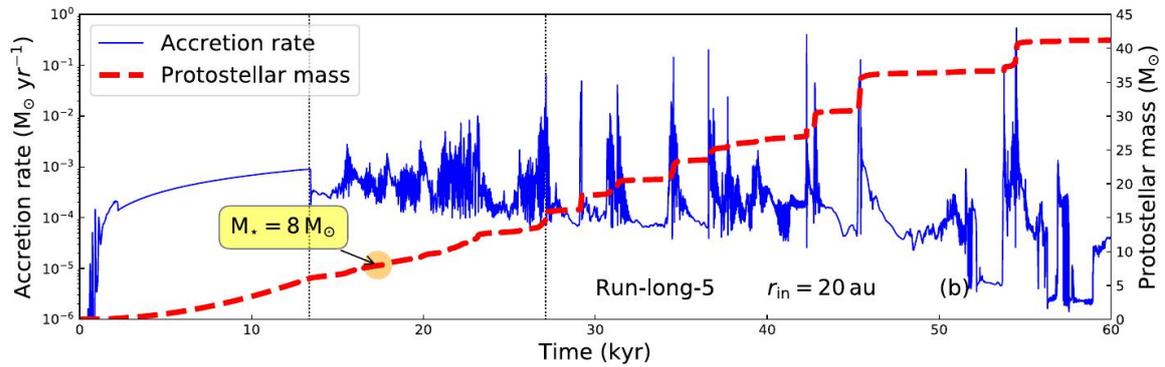

Figure 2.3: The protostellar accretion is expected to be episodic in the high-mass regime (during most of the time). This is visible in both, the partly 'stepwise' increase in protostellar mass (red) and the corresponding peaks in the accretion rate (blue). A huge fraction of the final stellar mass is actually 'delivered' within (powerful) outbursts. The Figure is taken from [100, Hydrodynamical models for MYSOs, Fig. 2].

Obviously, stars cannot become infinitely heavy. Theoretically, the Eddington limit sets the upper limit. It is reached once the radiation pressure overwhelms the self-gravity of the star and starts to rip it apart. It depends on opacity and is around $\approx 300 M_\odot$ [35, for low metallicity]. The most massive stars reported are as heavy as $\approx 250 M_\odot$ [153, Westerhout 49].

## 2.2   Accretion outbursts

Accretion is not a steady process; instead, the protostar shows variability on all time scales for most of its formation time. The topic has been reviewed in 'Protostars and Planets' (PPVI and PPVII) [45, 5, both mostly focussed on low-mass YSO's]. In the following, these reviews are summarized to highlight the importance of episodic (and powerful) accretion outbursts, which are the subject of the present work. A wide range of bursts with different strengths and durations have been observed and modeled for low-mass YSOs (see Fig. A.2). Episodic accretion outbursts are also expected for MYSOs (see Fig. 2.3).

### 2.2.1 Outbursts in the context of protostellar variability

On the *small* scales there is 'normal' variability implying changes in the order of a few $1 - 2\,mag$ for hours to days ($\approx$ stellar rotation period), which are related to the protostar or the innermost disk. Examples are short-duration accretion flares and extinction dips [45]. On the *intermediate* timescale (between one week and less than one year), modest variability (factor of $\leq 10$) occurs, likely driven by instabilities in the flow from the innermost disk onto the forming star.

On top of that, *huge outbursts* happen with durations from months to decades, and luminosity increases by tens to a few hundreds (or $2.5 - 6\,mag$) at optical and mid infra-red (MIR) wavelengths. Although these outbursts are rare (1 event per 100 to a few $10^4\,yrs$ from indirect tracers [15, 73]), they deliver a significant fraction of the final stellar mass ([46, $\approx$25%, observations in Orion], [101, up to 50%, hydrodynamical simulations for massive stars]). Additionally, they play an enormous role in stellar growth and disk evolution, impacting their surroundings in various ways (see Sect. 2.2.5) and can lead to long-lasting consequences for disk (and envelope) chemistry and for the evolution of the stellar radius [45, 114]. Huge bursts have been observed for low-mass stars at all stages from Class 0 [124, 72, among others] to the pre-main sequence (Class II / III) [5, 45, and references therein], although their frequency and characteristics may change with evolutionary status [64]. These powerful outbursts have been mentioned as a solution to the luminosity problem[2] or to explain the wide spread of observed protostellar luminosities, although both are still a matter of discussion [45]. There has been a lot of direct and indirect evidence for accretion outbursts, most of them for low-mass YSOs. The story started in 1939, when a rapid brightness increase in FUori had been found [61, missclassified as Nova], which was recognized as an accretion outburst 27 years later [58].

### 2.2.2 Historic classification

Accretion outbursts show a wide range of characteristics (see Fig. A.2). Historically, two main classes of outbursts have been established [5] (based on optical light curves of low-mass YSOs). These classes are called the FUori (most powerful, decade(s) long) and EXori (a few tens increase, months to years) outbursts. Many observed outbursts obey characteristics from both classes. A third class, the V1647ori outburst (basically all bursts in between), is established [45, 5]. With the increasing number of sources and the access to multiwavelength observations (revealing more and more deeply embedded objects), this classification needs to be 'improved' in order to cover the variety of events, which display a surprising range of

---

[2]Protostars tend to be less luminous on average than they should be according to the 'final' stellar mass distribution. This was first noted in 1977 by [128] and is called the luminosity problem.

luminosities, light curves, and spectra [78, 45].  Nevertheless, the 'historic' classes shall be
summarized briefly in the following:

- The prototype of an *EXori*-bursts was EX Lup.  Bursts of this kind, show an optical
  brightening by 2.5 – 5 mag, last for a few months to years, with a significant variety of
  the light curves (see [45] and references therein).  EXori bursts may repeat, but due
  to their extended dormant periods, the duty cycle has not yet been established [45].
  During quiescence (optical and NIR) absorption lines (metallic, such as Na, Ca, K, Fe,
  Ti, and Si, CO bandhead) are present, whereas during the burst many of these lines
  appear in emission (the heated disk surface "outshines" the protostellar photosphere)
  and existing emission lines get stronger [88].  The EXori bursts show indications of
  ongoing magnetospheric accretion from the disk onto the star [45].

- *FUori*-bursts are the most luminous ($\approx 5 \, mag$ increase in optical brightness) and longest
  ($\approx$ decade) bursts.  The FUor light curves show a wide variety (see [45] and the references
  therein).  The 'classical FUor' shows a strong initial brightening, followed by a longer
  plateau phase, with a relatively long decay timescale [5].  They can show minor short-
  time (hours to days) variability on top of the flare [76].  In the optical range, the spectrum
  shows broad blue-shifted absorption lines caused by strong (variable) winds, an $H_\alpha$-line
  with a P Cygni (or pure absorption) profile, and strong lithium absorption.  Common
  spectroscopic characteristics are widely interpreted as signs of a dominant viscous
  heated accretion disk [45], with extinction in the upper layers of the disk.  In the IR, they
  feature characteristics such as strong CO bandhead and weak metal absorption, as
  well as water and TiO or VO absorption bands and blue-shifted HeI-absorption.  The line
  width decreases with increasing wavelength in some FUors [75, e.g. in FUOri [57] and
  V1057 Cyg], which is possibly a sign of disk rotation (longer wavelengths trace colder,
  more distant, and slowly rotating disk regions) [82].

### 2.2.3   Evidence

In the following, we will list direct and indirect proofs of the accretion variability among different
wavelength regimes.  This will give an overview of the observables and help place our work in
context.

   For *low-extinction* cases (as face on, more evolved low-mass YSOs) the most direct
observation is possible via **UV and optical emission lines** and the **UV continuum**.  These
observations trace the heated shocked gas close to the photosphere.  For more embedded
sources, **proxy lines in the NIR** can be used.  Especially, FUors show a typical line spectrum

[57], which might prove the burst state (even if the rise itself lacked detection and the source is not yet/slowly fading).

**IR and (sub)mm continuum** observations give access to deeply *embedded* objects, in particular MYSOs or young and inclined (edge-on) sources, which may not be visible in the NIR (and sometimes not even in the MIR). IR observations (mainly) trace the **thermal dust re-emission** (plus possibly the viscous heated disk in the MIR).

The (MIR/) FIR observations are of special importance, as they cover the SED peak. At (sub)mm the envelope emission is optically thin, but there are some challenges. The emission comes from colder and more distant regions (the outer envelope). Therefore, the thermal response is delayed and short variations (shorter than months) may not be visible due to smoothing [45]. On top of that, even for strong bursts, the increase in the (sub)mm flux can be small, as it corresponds to the Rayleigh-Jeans tail of the SED. This means that the change is proportional to the temperature (whereas at shorter wavelengths it is proportional to the increase in luminosity) [93]. In this context, we regret that access to the FIR was lost (at least for the next decade) with the shutdown of SOFIA in 2022.

For *massive protostars*, further evidence comes from **maser** emission in the microwave range. Masers can be triggered by collisions (water masers[3] in wind shocks) or (enhanced) MIR pumping (Class II methanol masers). For low-mass YSOs, no Class II methanol masers have been found so far.

Indirect evidence comes from **chemical changes**[4] within the disk and envelope [73, Review Paper] and **signals in outflows** [5] [48, 39, PPVI chapters]. Both methods imply outstanding long records of the burst history (the most recent past ≈millennia), which allows for estimates on the burst frequency (although with a large portion of uncertainty). Values in the literature vary between a $100\,yr$ timescale [15, dynamical, from $^{12}CO$ knots in the outflow] up to $(2,-5) \cdot 10^4\,yrs$ [73, $C^{18}O$ gas phase abundance]. It was suggested, that the time intervals between bursts increase with the evolutionary sequence [64, statistical survey of 39 protostars in Perseus yield 1 per 2.4 (Class 0) and $8 \cdot 10^3\,yrs$ (Class 1)], which may reflect that fragmentation (as a cause for the burst) occurs more likely for younger YSOs. With more observations, the statistics will get better, and the burst occurrence rate (and its dependence on the chemical tracer or jet model) will be much better determined. For comparison, the modeled burst rate for MYSOs strongly decreases with the burst amplitude (>4 mag bursts are much rarer, than 1-2 mag ones). MYSOs spend only about ≈ 2% of their formation time in outbursts [100].

---

[3]Water maser observations of the MYSO S255-NIRS3 reveal episodic ejection events (possibly caused by other accretion events) prior to its recent outburst [20].

[4]The chemical changes include the dissociation of molecules, destruction of ices, dust sublimation and evaporation of CO and complex organic molecules. The freeze-out time scales depend on the density and species (it happens inside out and is faster in the disk than in the envelope). Typical freeze-out time scales are much longer than the burst duration (on the order of $10^4\,yrs$ for $C^{18}O$ at the densities at which it is found in the gas phase for these protostars [73]). Abundances (and or ratios) of different molecules in the gas (and ice phase) can be used to estimate the burst occurrence rate.

[5]The accretion-ejection connection leads to structures as knots and clumps within the jet (see Sect. 2.2.5). Their position (and velocity) can be used to infer the dates of past outbursts. Uncertainties are: projection effects (on position and velocity), assumptions on the origin (locally entrained or directly ejected from the driving source), and assumptions related to the history of the outflow velocity. Possibly, this method is not sensitive to long bursts (FUori) [146].

The vast majority of detected outbursts are for low-mass YSOs. Only a few examples are known in the high-mass regime (Sect. 2.2.6), mainly due to their highly embedded nature.

### 2.2.4  Triggering mechanisms

Instabilities are the rule rather than the exception; therefore, bursts are expected. Just to give an example: the envelope infall rate of V346-Nor's (onto the disk) is a factor of a few higher than its quiescent accretion rate (from the disk onto the star) [81], which will inevitably lead to mass build-up in the disk (which might develop an instability). Observed outbursts span a wide range of variability in terms of duration, strength, and occurrence. This opens the question of a uniform triggering mechanism. Are there different mechanisms at play, and (if so) on which scales do they act? Does it change during stellar evolution? What are the differences between low-mass and massive YSO bursts? Several mechanisms have been suggested in the literature. In the following, we give a brief overview of the most commonly accepted mechanisms. This is essential to compare the characteristics of the resulting bursts with the burst parameters derived in this work.

The thermal (disk) instability (**TI**, [10, considered as cause for the FUori outburst]) is a runaway process that 'occurs when the mass flux through the disk exceeds a critical value'. Once the disk reaches the hydrogen ionization temperature ($T ≈ 10.000\,K$), the opacity increases dramatically, causing the disk to heat up even faster (until it is fully ionized). High viscosity (proportional to temperature) can rapidly increase the accretion rate. This process happens inside-out [85] or, if it is induced by a planet, it can also occur outside-in (see [5] and references therein). For a $15 M_\odot$ protostar, such a burst would last ≈ 200 $yrs$, with a rise time of 60 $yrs$, a peak accretion rate of the order of $10^{-4} M_\odot / yr$ and an accreted mass of ≈ $5 M_J$ [41]. Note that it can also come in combination with the (extreme) evaporation of a heavy planet, which feeds the inner disk [106], leading to bursts very similar to FUori.

Magneto-rotational instabilities (**MRI**) are caused by 'connected' ionized elements (string-like connections lead to the exchange of angular momentum, such that both particles are pulled apart from each other, which causes the infall) in a magnetic field (the Lorentz force is responsible for the 'connection'). An illustration can be found in the Appendix. The MRI can only be activated in the innermost disk, the upper layers, and the outer parts (thermal ionization and cosmic ray ionization). For a $15 M_\odot$ protostar, such a burst would last a few $1000\,yrs$, with a rise time of $50\,yrs$, a peak accretion rate of the order of $10^{-4} M_\odot/yr$ and an accreted mass of $\approx 100 M_J$ [41].

If the thermal motion (gas pressure) is too small to compensate for the gravitational forces, parts of the disk could fragment, which is called gravitational instability (**GI**). For a differentially rotating disk, the stability is given by the Toomre criterion [138]:

$$Q = \frac{c_S \kappa}{\pi G \sum} \qquad with\ Q > 1\ stable \qquad (2.1)$$

Where $c_S$ is the local sound speed, $\kappa$ the epicyclic frequency, G the gravitational constant, and $\sum$ is the surface density. Fragmentation preferably occurs in the cold outer regions of the disk (little gas motion). With a few simplifications, the criterion can be transformed into a relation of disk mass $m_d$ and disk scale height h plus radius $R_d$: $m_d/M_\star > h/R_d$, which means that for typical parameters, the disk becomes unstable if its mass exceeds $\approx 0.1 M_\star$. Clumps can spiral inward, where they become accreted (possibly they are tidally disrupted). GI can be triggered externally, as e.g. by a fly-by (perturbation by an external body) or infall (from the envelope or accretion streams).

Another possibility is the accretion of a **'small body'**, such as a pebble or planet(esimal) accretion. Such a scenario had been considered as soon as 1977 [59, in the context of FUori], who estimated the energy released by a Jupiter falling onto the sun. The resulting burst would be sudden and short with an energy of $\approx 5 \cdot 10^5\,erg$ and could not explain the FUor outburst. But the idea of 'small body' accretion remains valid until today. These bodies form in the disk and spiral inward (similarly to what is described above). For a $15 M_\odot$ protostar and a $5 M_J$ planet (which gets disrupted), the burst would last a few years, with a rise time of half a year, a peak accretion rate on the order of $10^{-3} M_\odot/yr$ and an accreted mass of $\approx 2 M_J$ [41], which is more similar to an EXori outburst. Strong outbursts can also be explained by the **accretion of a companion** (as MYSOs have a higher stellar multiplicity).

## 2.2.5  Effects on the (M)YSO environment

Accretion bursts impact their natal protostellar environment, which leads to a couple of phenomena. This section is supposed to give an overview of all the phenomena detected for MYSO bursters so far. We emphasize that this section already includes some of our results (for completeness). For a brief overview, see Fig. 2.4, as well as our poster presentations [152, 6].

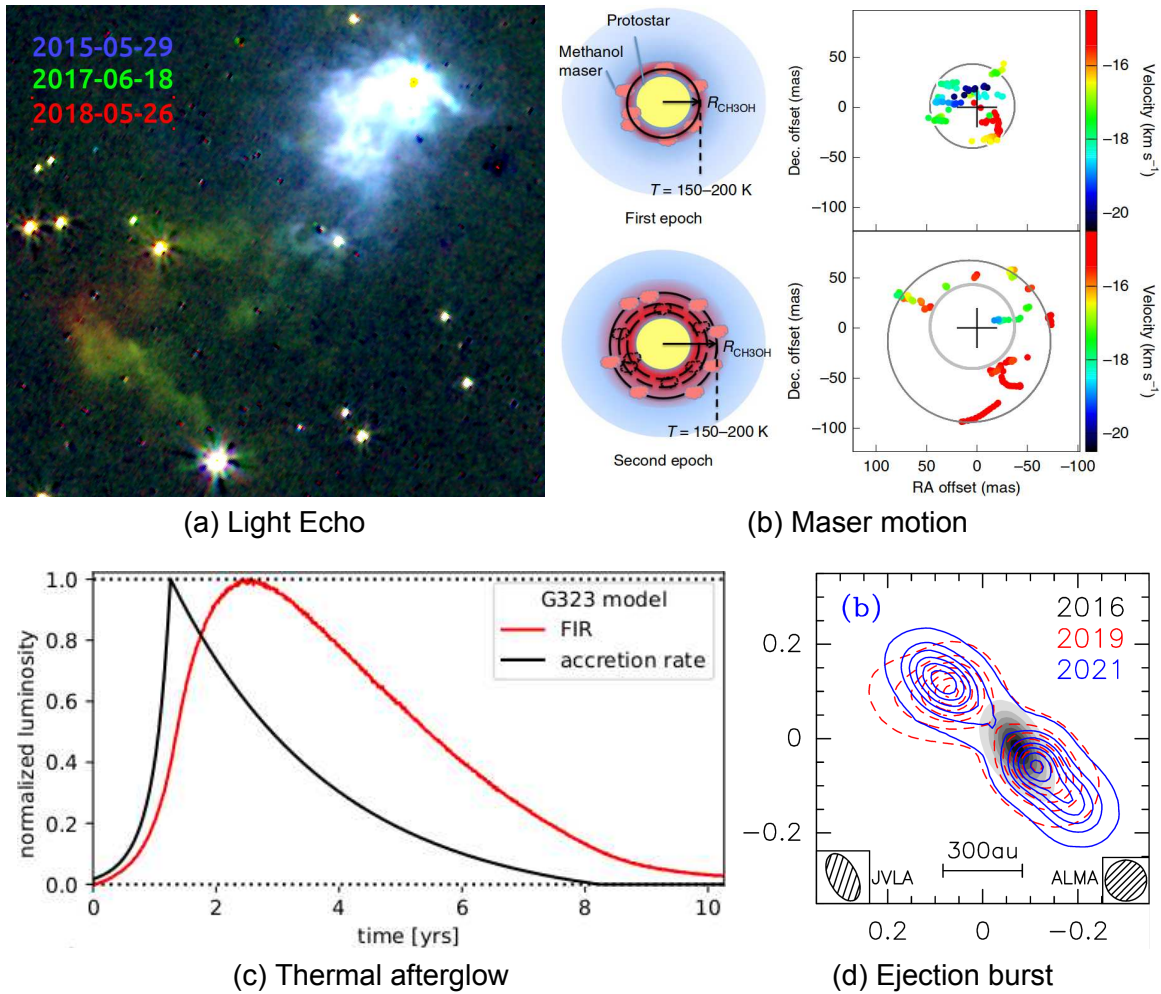

(a) Light Echo

(b) Maser motion

(c) Thermal afterglow

(d) Ejection burst

Figure 2.4: Examples of the impact of an outburst on the MYSO environment. **Upper left:** Ks-band images at three epochs (color-coded) show G323's light echo motion. **Upper right:** Expansion of G358's maser-ring due to the burst. **Lower left:** G323 models show that the FIR-luminosity (red) is elevated even after the 'end' of the accretion outburst. **Lower right:** ALMA 3 mm (colors) and JVLA 7 mm observations show the expansion of NIRS3's ionized jet-component, that appeared 1yr after the burst's onset.
Image b is taken from [21]. Images a and d are taken from [6]. Image c is a result of this work and has been presented at the TLS 'Advisory Board Meeting' in 2020.

**Light echos (LEs)**   Dusty environments scatter photons into the line of sight, causing the additional path to delay the arrival time. This phenomenon has been known for supernovae for many years (e.g., [132]) and appears for (M)YSO bursts as well (although the appearance differs due to the complex morphology[6]). Fig. 2.4 shows Ks images at 3 epochs in the G323 example (upper left). The LE reveals extended structures. The apparent outward motion of the echo can be clearly seen. The first MYSO LE observed due to an accretion burst was that of NIRS3, which had a biconical shape [25].

Interestingly for dense regions the echo appears slower, than what's expected from single scattering (due to multiple scatterings, that effectively prolongate the photon paths).[7] This effect however will be only relevant at the 'smallest' scales ($\approx$ dense core).

**Thermal Afterglow**   Our modeling focuses primarily on the thermal afterglow. The thermal afterglow covers a wide range of scales (spatial and time-wise) and is likely not sensitive to small-scale structures (such as spiral arms, clumps, etc.). In the following, we are going to explain it in more detail.

The released burst energy heats the surrounding dust, and its thermal re-emission is visible (in the IR) up to years after the end of the burst. The longevity (and the increase) depends on the wavelength. In general, the timescale increases with wavelength and the maximum flux decreases towards the (sub)mm (while it is still high in the FIR). These dependencies are known/expected [70, 33, for low-mass YSOs] and are further explored by us in this work (first TDRT models, extension to the high-mass regime). An example simulation is given in Fig. 2.4 (bottom left). It underlines the longevity of the FIR afterglow, which peaks about 2 years after the burst.

**Maser flare and -relocation**   Maser stands for microwave amplification by stimulated emission of radiation. They work similar to optical lasers, but amplify microwave-radiation. The population-inversion can be created via collisions (Class I) or via radiative pumping in the MIR or FIR (Class II). Different maser molecules and transitions exist. The most common ones are methanol (with the strongest lines at 6.7 and 12.2 GHz), water (22.2 GHz) and hydroxyl (1.035 and 1.665 GHz).

Masers require certain excitation conditions (see Sect. 3.2.4.2 for Class II methanol masers). During the burst, these local conditions (e.g. temperature, pumping rate) change rapidly, which naturally shapes the maser landscape. Four out of six MYSO bursts have been accompanied by Class II methanol maser flares[8]. Maser monitoring has proven to be successful in finding MYSO bursts. We want to highlight G358 in this context, where some interesting features have been observed in great detail. These are the appearance of new maser transitions [96] and the ring-like propagation of the maser spots (methanol Class II). The latter is really remarkable, as it allows us to trace the propagation of the heatwave through the disk [21] and revealed the spiral arm structures within [22]. A visualization of the propagation of the heatwave can be found in Fig. 2.4 (upper left), which shows the positions of the maser spots (and the corresponding temperature scheme) at two epochs. The maser ring expands with an apparent speed of (4 – 8) % the speed of light, which is much faster than the gas motion. This means that new masers are stimulated (in the new locations), while the old ones are no longer excited [21].

---

[6]Usually SN LEs show a ring like structure, which is not the case for MYSO LEs.

[7]This result has been obtained by us via simulations with the RT code Mol3D [107, extended by S. Heese to feature scattering LEs]. It is unpublished, but has been presented in 2 talks: 2019 in Exeter (astrophysics department) and 2020 in the 'Tautolloquium'.

**Ejection burst**   A fraction of the infalling matter can be ejected by jets/outflows (instead of falling on the protostar). Fig. 2.4 (lower right) shows NIRS3 radio observations at 3 different epochs. About a year after the burst onset, the radio continuum began to rise. This is probably caused by the launch of an ionized jet [27], which is expanding (Cesaroni et al., in prep.). Knots and clumps in the jet can serve as indirect evidence of outbursts for as long as a few kilo years (Sect. 2.2.3). Only recently, Fedriani+ (subm.) found new knots in the NIRS3 jet, related to its 2015 outburst.

**X-ray detection**   Chandra detected X-ray emission from NIRS3 during its burst, which was not previously present. Most likely, the emission comes from wind shocks (related to radio jet activity), rather than from the protostellar photosphere (bloating prevents high effective temperatures) [6].

## 2.2.6  Summary of known MYSO accretion bursts

This work aims to characterize particular MYSO bursters in order to answer questions on the physics of their growth process. Due to the highly embedded nature along with their scarcity and distance, no accretion outbursts have been detected for MYSO until recently. This changed in 2015, when two accretion-outburst happened almost simultaneously [25, S255 NIRS3, $M_* = 20\,M_\odot$] and [66, NGC 6334I, $M_* \approx 7\,M_\odot$]. Since then (and thanks to the maser monitoring) a handful of accretion bursts have been discovered for MYSOs: [25, S255

---

[8]M17MIR [29] and V723Car [137, 136] do not have Class II methanol maser detection entries in the maser database [9, `maserdb.net`].

NIRS3 (**NIRS3**)], [66, **NGC 6334I**], [136, **V723Car**] and [29, **M17MIR** ('only' $5.4 M_\odot$)], as well as **G358** and **G323** (subject of this work).

Despite their small number, the bursts (and MYSOs) already indicate a wide range of properties (e.g. NIR-dark and bright bursters, short and long bursts, and more). This immediately leads to the question of whether there is a common burst-triggering mechanism behind this. The burst durations range from a couple of months to more than a decade, with burst energies spanning 2 orders of magnitude (see also the results of this work). Surprisingly, the mean accretion rates $\dot{M}$ are quite comparable.

## 2.3   Two objects in the focus of this work

In this work, we performed RT simulations for two MYSO bursters, which we will introduce in more detail in the following. This is needed to put our results in the context of current observations.

### 2.3.1   G358.93-0.03

G358.93-0.03 (G358 region for short) is a star forming region located in the Southern Hemisphere ($\alpha = 17^h 43^m 10\overset{s}{.}02$, $\delta = -29°51'45.8''^9$) at $6.75 \pm^{0.37}_{0.68}$ kpc [19] (kinematic distance). This implies a distance to the galactic center of $1.6\,kpc$. The G358 region hosts eight sources (MM1-MM8 in Fig. 2.5), including 2 hot cores (MM1 and MM3, with CH3CN as tracer) [19, ALMA observations].

The region received little attention until 2018[10], but this changed with the burst [19, 96, 17, 21, 28, 144]. Its outburst was the **first** one alerted by the M2O[11] (from the monitoring of Class II methanol masers) in January 2019 [133] (see Fig. A.12).

   The brightest continuum source G358-MM1 (G358 for short) was identified as the outbursting source [19]. Follow-up maser observations show a wealth of maser species, including new transitions [19, 96, 17]. The relocation of the maser allows us to trace the heatwave that moves through the disk at a subluminal speed [21]. The disk shows spiral arm structures, which might be prone to GI [21, 22, 28]. The appearance of the spirals confirms the face-on view. Furthermore, it allows estimating the enclosed mass, which amounts to $\approx 10\,M_\odot$ [22, 28].

---

[9]J2000-coordinates
[10]Only eight entries in the astronomical database SIMBAD [148].
[11]The M2O is the Maser monitoring organization, which was founded in 2017. For further information, see `https://www.masermonitoring.com/` or [23].

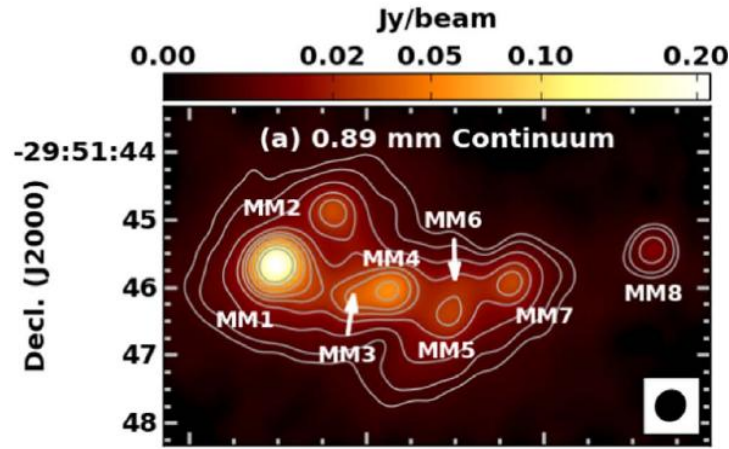

Figure 2.5: The G358 star forming region as seen with ALMA. Taken from [19]

A quick and weak sub-mm rise occurred as well. There are no resolved preburst observations (making it hard to estimate the increase in the MM1 flux between the ATLASGAL preburst observation in 2007 and the first ALMA burst observation in April 2019), but a decline of 15% between April and October 2019 has been observed with ALMA (ALMA Proposal Cycle 8, subm. by Brogan).

The FIR afterglow was detected with SOFIA (2 observations in May 2019 and August 2020). SOFIA data are used in this work and were published in [130]. G358 is MIR dark, i.e. not seen at wavelengths up to 24$\mu m$.

### 2.3.2   G323.46-0.08

The massive star-forming region G323.46-0.08 (G323) is also known as IRAS15254-5621, it is located at $\alpha = 15^h29^m19\overset{s}{.}02\overset{s}{.}4$, $\delta = -56°31'23''$ [12]. It has a local standard of rest velocity of $v_{LSR} = -67.2\ km/s$ [113]. This implies a distance of $4.08^{+0.40}_{-0.38}$ kpc (with the kinematic model A5 from [116]), which is consistent with the highest GAIA-DR2 stellar distances [7] of up to 3.7 kpc for stars in the foreground of the nebulosity associated with G323.

G323 is accompanied by a compact ATLASGAL clump [142], which is most likely a massive ($\approx$600$M_\odot$) cluster progenitor [36] hosting the MYSO. The bolometric luminosity amounts $L\approx$(1 – 1.3)$\times10^5\ L_\odot$ [91, integration of the SED]. Broad radio recombination lines in this region indicate the presence of a hypercompact HII region [4], further supported by studies of [105, 77]. G323 is less deeply embedded than G358 (it is visible in the NIR and MIR) and hence is likely much more evolved. This assumption is supported by the existence of an (NIR) light echo and signatures of a molecular outflow (see [4, red and blue CS(3–2) as well as $^{13}$CO(2–1) line wings] and [155, similar observations for the $^{18}$CO(2–1) line]).

---

[12]J2000 coordinates

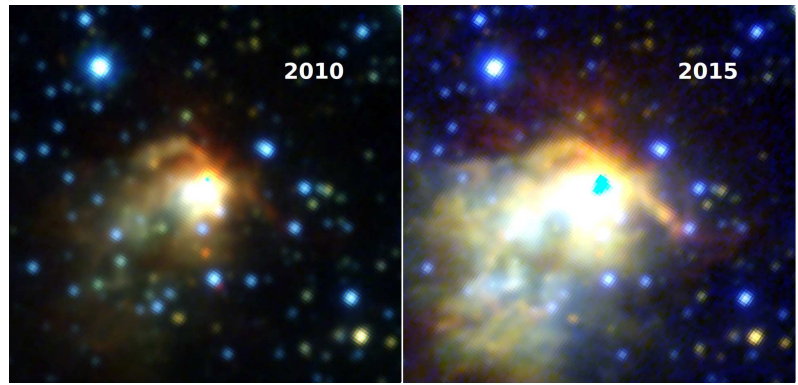

Figure 2.6: VVV $JHK_s$ preburst (left) and burst (right) color composites (FoV $44'' \times 44''$, north up, east left) of G323.46-0.08. The central cyan-colored areas are due to detector saturation. The burst is clearly visible in all 3 bands. This Fig. is part of Wolf+ (in prep.).

The region is associated with masers (methanol, water, and hydroxyl). The total flux of the methanol maser increased from $\approx 20\,Jy$ [50] to $\approx 7500\,Jy$ in between 2011 and 2015 (the latter date is two years after the flare peak)[13] [113]. Interestingly, the decay of the methanol maser flare is overlaid by a short-term periodic variation with a period of $\approx 93\,d$ [95].

Throughout the whole burst, the source had been monitored by the VISTA Variables in the Via Lactea Survey (VVV, 103) and its extension VVVX (from 2016 to 2019) in the NIR (Ks). NIR color composites before and during the outburst are provided in Fig. 2.6. The burst was much longer and stronger than the one in G358 (this work). We detected its thermal FIR afterglow about 1.5 years after the burst ended.

## 2.4 Science instruments

We performed observations with the **S**tratospheric **O**bservatory **F**or **I**nfrared **A**stronomy (called SOFIA), which was the largest and most potent IR facility at its time. SOFIA was observing the IR sky between 2010 and 2022. It was a joint project of NASA (National Aeronautics and Space Administration) and DLR (Deutsches Luft- und Raumfahrtinstitut). The observatory was a modified 'Boeing 747 SP' harboring a telescope with an effective mirror-size of 2.5 m and several science instruments (operating in between 0.36 and 612$\mu m$). Originally, the mission was designed for 20 years, but it was shut down earlier by NASA to save costs.

---

[13]During the peak of the burst, no maser observations have been performed.

In total, SOFIA performed 732 science flights from destinations around the world. The results achieved with SOFIA data are fundamental for our understanding of the (F)IR universe. It allowed for important discoveries in numerous research fields[14], and made some really important discoveries[15]. Without it, this thesis would not have been possible.

A pointing stability of 0.5" was achieved, which is technically amazing (taking into account the weight of the telescope[16]). A chop-nod observing strategy allowed for a correction of the background and thermal telescope emission[17]. For more information on SOFIA and its instruments, we refer the reader to the quick guide[18], the observer handbook[19], and the following publications: [87, 44, 32, 54] (among others). We also provide some further information in the Appendix.

### 2.4.1   FIFI-LS

We observed G358 with the **F**ield **I**maging **FI**R **L**ine **S**pectrometer (called FIFI-LS) aboard SOFIA ([87, 44, 32]). FIFI-LS was commissioned in 2014 and was used almost until SOFIA's shutdown. It has two channels, red (115-203$\mu m$) and blue (51-125 $\mu m$), which were operated simultaneously. FIFI-LS is an integral field spectrograph that features one wavelength and two spatial dimensions (3 dimensions in total) for both channels. To fit in the 2D detector, the pixels are rearranged (according to the scheme given in Fig. A.9). The FoV amounts to 30"×30" (blue) and 1'×1' (red), with a pixelsize of 6"×6" and 12"×12" respectively.
FIFI-LS is suited to detect several bright FIR fine structure lines (such as [OI] at 63 and 145$\mu m$, nitrogen lines, CII at 158$\mu m$) and molecular lines (such as CO lines). An overview is given in [44, Tab. 1.]. FIFI-LS played a major role in the investigation of (massive) star formation, as well as for the interstellar medium (in our and external galaxies).
For more detailed information, we refer the reader to the papers mentioned in the beginning of this Sect. (and references therein). We provide the schematic light path (including a brief description) in the Appendix, Fig. A.8.

For G358 we chose 6 bands (centered between 52.0 and 186.4 $\mu m$)[20]. These bands cover the entire spectral range of the detector and the SED peak. The spectral scan length per subband ranged from 0.3 to 1.0 $\mu m$ for our observations. Unluckily, the postburst could only be observed with the red channel (due to problems with the blue filter wheel). More information on our FIR data (used for G358) can be found in the corresponding section and our publication [130].

---

[14]This includes research fields as galaxies, stars and star formation, planets, molecular clouds, interstellar medium and more.

[15]One example is the first discovery of the helium hydrideion (HeH$^+$), which was probably one of the first atoms formed in the early universe[53, an observation with GREAT].

[16]The primary mirror weights about ≈ 850 $kg$.

[17]A detailed description of the chop and nod can be found here (on the example of the FORCAST detector): `https://www.sofia.usra.edu/proposing-observing/proposal-calls/past-proposal-calls/cycle-4/cycle-4-phase-ii/why-chop-and-nod`

[18]`https://www.sofia.usra.edu/sites/default/files/Other/Documents/quick_guide.pdf`

[19]`https://www.sofia.usra.edu/proposing-observing/proposal-documents`

[20]We used only the continuum, but those bands are chosen to include CO lines, which might be analyzed in the future.

### 2.4.2 HAWC+

G323 postburst observations have been performed with the **H**igh-resolution **A**irborne **W**ide-band **C**amera (called HAWC+), see [54] (and references therein). This instrument was commissioned in 2016 and became a facility instrument in early 2018. It combines a high-angular resolution imaging photometer and a polarimeter with five broad bands centered between 53 and 214 $\mu m$. With HAWC+ SOFIA's capabilities have been extended to study interstellar magnetic fields. The spectral range is similar to that of FIFI-LS. The 5 filters (centered at 53, 62, 89, 154 and 214 $\mu m$) provide a wavelength spacing similar to that we applied for G358. The bandwidths range from 8.7 to 44 $\mu m$ (increasing with wavelength). The instrument provides a 64x60 pixel array (for imaging). The size of the pixels ranges from 2.55" to 9.37" (with an FoV of 2.8'×1.7' to 8.4'×6.2'). For an overview of the basic optical specifications, see [54, Tab. 1]. A sketch of the principal optical components is provided in Fig. A.10. For a more detailed description, see [54].

## 2.5 Basics of radiative transfer and simulation setting

In this section, we will give a brief overview of the applied fitting method. At first, we introduce the principle of radiative transfer simulations, static (used for G358) and time-dependent (used for G323). After that, we introduce the underlying dust configurations (density grid) used for our simulations. For further reading, we refer the reader to [47, general introduction], [120, static RT with HYPERION] and [56, 64, time-dependent RT with TORUS], as well as [121, YSO grid (dust distribution)].

### 2.5.1 Static radiative transfer (RT)

Radiative transfer describes how radiation is transported through a medium. RT codes attempt to solve the radiation transfer equation, which connects the specific intensity $I_\nu$ at the frequency $\nu$ with the optical depth of the medium $\tau_\nu$:

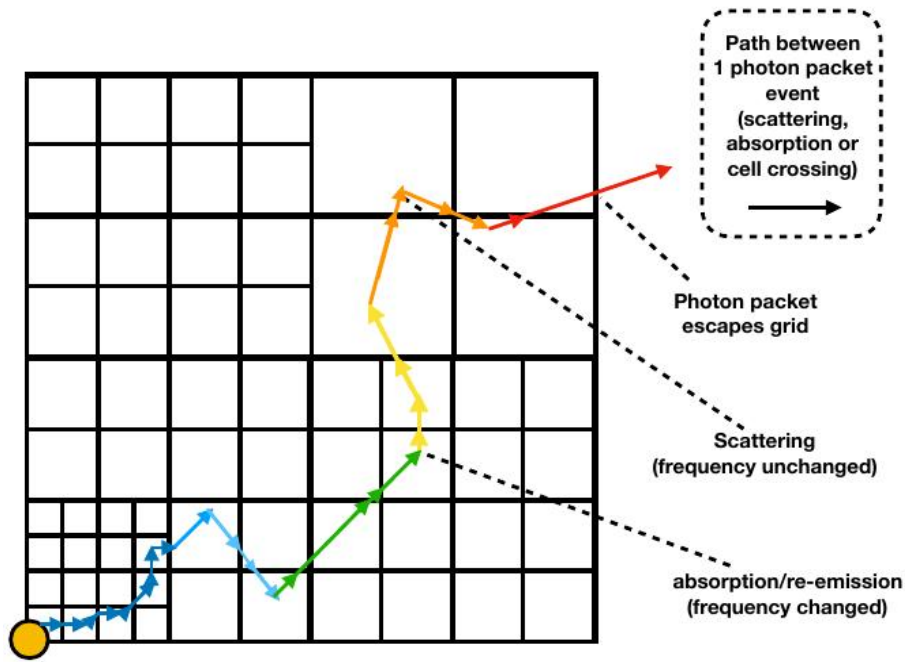

Figure 2.7: The principle of radiative transfer. Taken from [55, Fig. 6].

$$\frac{dI_\nu}{d\tau_\nu} = S_\nu - I_\nu \qquad (2.2)$$

$S_\nu$ is the source function, which is the ratio of opacity and emissivity ($S_\nu = j_\nu / \kappa_\nu$).

The radiative transfer equation is usually a complex function of the radiation field. It can be solved analytically only for a few cases (e.g., spherical symmetries and homogeneous radiation fields). Numerical codes are therefore widely used.
There are a variety of codes. In the following, we explain the basic principle for the example of TORUS[21]. We assume that we have one source that sits in the center of a given density grid (for multiple sources and/or different source positions, it works similarly). At the source, photon packets[22] (or photons) are started, whereby their energies re-sample the source spectra. The photon packets are processed through a defined density grid, where the paths are random walk-like (Monte Carlo method). Fig. 2.7 visualizes this principle for TORUS, which features adaptive mesh refinement, allowing smaller grid cells in the densest regions. The photon packet starts at the left corner and follows the colored way until it leaves the grid. On its way through the grid, the packet can be scattered (with the frequency remaining the same) or absorbed. In the latter, it gets immediately re-emitted, with a new frequency. All physical parameters, such as scattering-/absorption-efficiency, scattering angle, and frequency of the photons by (re)/emission, are sampled from realistic probability distributions. The properties of the grid cells (such as local density/optical depth[23], local temperature, and the properties of the dust/gas[24]) along the path are taken into account.

---

[21]TORUS stands for "Transport Of Radiation Using Stokes (Intensities)". It was originally a static code that was modified for time-dependent treatment by [56].

[22]In TORUS photon packets are used. Each packet has the same energy; therefore, the number of photons within one packet varies with its frequency (and is higher at lower frequencies). As a consequence, the number of photon packets decreases towards longer wavelengths (as more photons are included in one photon packet).

In the static case, the absorption and emission rates within each cell (and time step) are balanced (radiative equilibrium, $\dot{A} = \dot{E}$). The photon packets are processed through the grid until the final temperature is reached within each cell. Once the temperature calculation has been completed, synthetic observables, such as SEDs and images, can be computed.

## 2.5.2  Time-dependent RT (TDRT)

In the time-dependent case the assumption of radiative equilibrium (i.e. $\dot{E} = \dot{A}$) is dropped, and grid cells can gain (they heat, $\dot{A} > \dot{E}$) or lose (they cool, $\dot{E} > \dot{A}$) energy during each time step (the net change of the energy density of the gas is then $\dot{u}_g = \dot{A} - \dot{E}$). Assuming that the (cell) temperature over one time step $\delta t$ is constant leads to an updated energy density of $u_g^{n+1} = u_g^n + (\dot{A} - \dot{E})\delta t$. The calculation loops over all photon packets (with the different frequencies). The procedure is repeated for each time step. The photons are stored according to their respective travel times, which are continuously updated. The local optical depth is taken into account. For a more detailed description, we refer to [56].

The choice of *time step* is crucial. If it is too small, the probability of an interaction within one cell and a time step decreases and the computation becomes intractable. If it is too long, stability may not be guaranteed [56]. In particular, $\dot{A}$ and $\dot{E}$ are assumed to be constant for one time step, which may not be fulfilled for large steps. Clearly, the time step should be less than the speed of the luminosity variation and much shorter than the grid crossing time[25]. For our simulations, time steps on the order of days have been proven to be a good choice. Note that for a smaller time step, a higher number of photons is needed (to achieve a good S/N), since an SED is stored at each time step. This rapidly increases the simulation time and storage.

Today, 2 time-dependent codes exist: TORUS [56, 55] (which we use) and a new code, which was developed by [11] in 2022 as an extension to the POLARIS-RT code [118]. Both codes will be compared in the near future.

---

[23]The optical depth sets the likelihood of an absorption or scattering event to occur (a photon packet can pass a grid cell without interaction).

[24]The size of a dust grain e.g. determines its emissivity at a certain frequency.

[25]I.e., the time for one photon to travel from the source to the outer edge of the grid.

### 2.5.3   The density grid - components of a 'typical' YSO grid

RT codes in principle, allow for the usage of arbitrary density grids. Within this work, we restrict ourselves to (adaptions of) the spubsmi-configuration from [121] according to the structure of an YSO in a very early evolutionary stage. This is probably the most simple approach, but nonetheless includes the most important components, such as the following.

The **s**tar, a **p**assive circumstellar disk (with its inner radius governed by the dust **s**ublimation radius), a **b**ipolar cavity, an **U**lrich-type envelope and an ambient **m**edium. A schematic visualization can be found in Fig. A.11. In the following, we briefly describe each of the components and name the free parameters (at the end of each paragraph).

The settings are rotation-symmetric and cannot reproduce small-scale 3D structures, such as spiral arms. Although these structures exist, they will likely not impact the SED (which is dominated by the thermal emission of the "smooth" extended envelope).

**Protostar**   The protostellar luminosity of a YSO is the sum of its intrinsic luminosity (contraction, deuterium burning) and its accretion luminosity $L_{acc}$. For embedded sources, the exact shape of the spectral energy distribution of the central source is not important, since the (proto)stellar radiation is reprocessed [121]. Therefore, we use a black-body approximation[26] The free parameters are protostellar radius $R_*$ and temperature $T_*$ (which give the **luminosity** $L_* \propto R_*^2 \cdot T_*^4$ using the Stefan-Boltzmann law).

**Passive disk**   The disk is assumed to be a standard flared accretion disk as described in [127], with this density distribution:

$$\rho_{disk}(r) = \rho_0^{disk} \left(\frac{R_0}{r}\right)^{\beta-p} \cdot e^{-\frac{1}{2}\left(\frac{z}{h(r)}\right)^2} \qquad (2.3)$$

Where $r$ is the radius in the disk midplane, p is the surface density power, and $h(r) = h_0 \left(\frac{r}{R_0}\right)^{\beta}$ is the scale height, with $R_0 = 100\,au$, $h_0$ as the scale height in $r = 100\,au$, $\rho_0$ as mid-plane density at 100 au, and $\beta$ as the scale height exponent (or flaring power).

We assume a passive disk (i.e., no viscous self-heating), which is probably inappropriate. However, for embedded YSOs, the effect is likely to be minimal [121]. The viscosity of the disk is strongly dependent on the radius [112], thus active disks differ from passive ones only in the innermost regions, where most of the dissipative energy is released. Although the MIR might rise earlier (because of viscous self-heating), the FIR/sub(mm) will not be influenced. Free parameters are **mass** $m_{disk}$ **(or density** $\rho_0^{disk}$**), outer radius** $r_{max}^{disk}$**, surface density** power $p$ or $\alpha$ (with $\alpha = p - \beta$), **flaring power** $\beta$, **scale height** $h_0$; the inner radius $r_{min}$ is set to the sublimation radius of the dust.

---

[26]In the YSO grid of [121] the stellar temperature $T_*$ is used to select the 'correct' photospheric model a [26] (for the sources with $T_* > 4000\,K$). This is more appropriate but will lead to the same results.

**Ulrich envelope**    The envelope is a rotationally flattened envelope as described in [141]. Envelope and disk densities are added, ensuring a smooth density distribution. The envelope density is given as:

$$\rho_{env} = \rho_0^{env} \left( \frac{r}{R_c} \right)^{-1.5} \left( 1 + \frac{\mu}{\mu_0} \right)^{-0.5} \cdot \left( \frac{\mu}{\mu_0} + \frac{2\mu_0^2 R_c}{r} \right)^{-1} \tag{2.4}$$

where $\rho_0^{env}$ is related to the mass infall rate $\dot{M}_{env}$ via

$$\rho_0^{env} = \frac{\dot{M}_{env}}{4\pi} \cdot \frac{1}{\sqrt{GM_\star R_c^3}} \tag{2.5}$$

With G as the gravitational constant, $M_\star$ as stellar mass, $r$ as radius in spherical coordinates, $R_c$ as centrifugal radius (set to the disk outer radius), $\mu = cos(\Theta)$ and $\mu_0$ as the cosine polar angle of a streamline of infalling particles for $r$ to infinity. The angles $\mu, \mu_0$ are obtained by solving the streamline equation:

$$\mu_0^3 + \mu_0 \left( \frac{r}{R_c} - 1 \right) - \mu \frac{r}{R_c} = 0 \tag{2.6}$$

The free parameters are **density** $\rho_0^{env}$ or $\dot{M}_{env}$ and **the outer radius** $r_{max}^{env}$ (only for TORUS models, in the [121] YSO pool, the grid extends until temperature and density reach ambient level).

**Bipolar cavities**    During star formation, powerful jets and winds entrain material and carve out bipolar cavities. The shape of the cavity can be described by (following [121]):

$$z(r) = r_0 \cdot cos(\Theta) \left( \frac{r}{r_0 \cdot sin(\Theta)} \right)^c \tag{2.7}$$

with $r_0 = 10.000\,au$, $\Theta$ as the opening angle of the cavity and c as the power-law exponent of the opening of the cavity. Inside the cavity ($|z| > z(r)$) the density $\rho_0^{cav}$ is assumed to be uniform. This is a simplification and is likely not true in reality. The disk and cavity densities are added, in order to ensure that the disk is not 'removed' by the cavity.

The free parameters are the **density** $\rho_0^{cav}$, **opening angle** $\Theta$ and the **power-law exponent** $c$ (which is fixed to $c = 1.2$ in all of our TORUS settings).

**Ambient medium**   The static YSO model grid [121] includes an ambient medium, with a minimum density of $\rho_{amb} = 10^{-23} g/cc$ and a minimum temperature of $T_{amb} = 10\,K$. We did not include an ambient medium in the TORUS models. However, the minimum values (density and temperature) within our grids are above those just mentioned (for all our settings). Note that no external heating is included.

**Dust**   We use the same dust for all components and all regions. This is a simplification. In reality, the size and composition will depend on the region (dust sublimation, dust settling, etc.). We use MRN dust for the G323 settings (TORUS) as described in [99]. With a size distribution that can be described by a power law with $n \propto a^{-3.5}$ with grain sizes ranging from $a_{min} = 5\,nm$ to $a_{max} = 250\,nm$. The dust consists of compact, homogeneous, and spherical grains with a composition of 62.5% silicate and 37.5% graphite (with optical properties of [147]). For G358, Milky Way dust with Rv = 5.5 is used according to the underlying YSO grid of [121]. For the TDRT database, pure silicate was used. This is sufficient for a first test, but probably not for realistic YSOs. As dust sublimation temperature we assume $T_{sub} = 1600\,K$ in all cases (in accordance with other authors, as e.g. [123]).

# Radiative Transfer Analysis

## 3.1 The modeling strategy

### 3.1.1 Why the FIR/thermal afterglow?

We used *FIR data to model the thermal afterglow*. This wavelength region is suited to serve as a 'measure', aka a 'calorimeter', of the burst energy for three reasons. It covers **most of the emission** (SED peak). It **lasts considerably longer** than the burst itself (i.e., months to years). And finally: it is '**stable**' as the (FIR) emission is produced in the colder (outer) regions of the envelope, which will not be affected (other than heated) by the burst. (All photons that reach there are re-processed many times; therefore neither stellar bloating nor dust sublimation etc. can influence the FIR afterglow.)

In other words: FIR observations are particularly suited to detect the thermal afterglow, whereas the thermal afterglow is suited to infer the main burst parameters.

### 3.1.2 Two different approaches

This section is supposed to *guide the reader* through the main part of this thesis. This is done, since it is easy to lose track. Essentially, the radiative transfer analysis consists of two parts: A semi-static approach (used for G358, Sect. 3.2) and a time-dependent approach (used for G323, Sect. 3.3). This section gives a rough idea of the modeling strategy; the detailed methods are described in Sect. 3.2.3.1 (G358) and 3.3.2 (G323). The TFitter is presented in Sect. 4.1. An overview of the strategy is not necessary at this point.

**G358**   When G358's accretion burst was discovered, TORUS (i.e., the TDRT option) was still under improvement. Therefore, we use a **semi-static approach** to obtain G358's burst parameters. We call it 'semi-static' because all models are computed with a static RT code, but the burst epochs are included in a more 'self-consistent manner'. We adapt the sedfitter database of [121], where we use the static RT code HYPERION to obtain more realistic burst-/post- models (which are essentially the best preburst configurations, but with higher source luminosities). Out of the best-fitting models for all three epochs (pre-, burst-, and post-), we deduce a (epoch-) combined **mean**-model. We used the **burst/post-luminosity** (together with the mean model) to estimate the main burst parameters. The results have been published in [130] and are described in detail in Sect. 3.2.

**G323** Until our discovery of the accretion outburst in **G323**, the time-dependent modeling had been substantially improved. Therefore, we could use TORUS to predict the afterglow time scales, which was crucial for the acceptance of our ToO (Target of Opportunity)-SOFIA proposal. We compute a **TORUS database of static models**, from which we obtain a **mean preburst-** model (using the best models). For this model, we then applied TDRT using the **Ks light curve as a proxy for the source luminosity variation**. We use the $1\sigma$ confidence intervals (of the mean preburst model) to **predict the minimum and maximum afterglow timescales**. Furthermore, we run a set of TDRT simulations, where we varied the input energy in order to constrain the burst energy. The method is described in more detail in Sect. 3.3. With this, we are able to constrain ranges for the main burst parameters (by using only a few models), but we cannot extract the full information encoded in the post-SED.

## 3.2 G358.93-0.03

### 3.2.1 Overview over the data

In this section, we briefly show the data on which our results are based. A more detailed summary is given in our publication [130].
The **preburst SED** (as given in Tab. A.2) is based on archival data from these surveys:

- **NIR**: 2MASS and VVV

- **MIR**: ISOGAL, GLIMPSE, (NEO)WISE and MIPSGAL

- **FIR**: MIPSGAL, FIS (AKARI) and HI-GAL

- **(sub)mm**: ATLASGAL

Note that G358-MM1 (the out-bursting source) is MIR dark, and the flux at the short wave-lengths actually corresponds to MM3 (another source in the field). The MM3 SED is given in the Tab. A.1. A visualization is provided in the left panel of Fig. 3.1. We use the MM1 SED (blue) for the preburst SED (where we decompose the total SED, as described in Sect. 3.2.2). Our modeling is essentially based on the **burst and post-FIR** observations, which have been taken with **SOFIA** in 2019 May 1 and 2020 August 28 respectively. Due to problems with the filter wheel onboard SOFIA, no data could be obtained in the blue channel at the second epoch. The data is summarized in the Tab. A.3 (burst) and Tab. A.4 (post), see also Fig. 3.1 (right). Again, the SEDs had to be decomposed (to obtain the MM1 SED), before applying the RT models (Sect. 3.2.2).

### 3.2.2 SED Decomposition

YSOs in massive star-forming regions are, in general, not isolated single objects but rather accompanied by other sources close by. The same holds for G358, which consists of eight sources detected from ALMA observations (see Fig. 2.5, or the original in [19]). As the YSO-grid includes only single sources (and no SEDs resulting from the superposition of multiple objects), we had to remove the contribution of the other sources prior to fitting MM1 (which showed the burst). **MM2** and **MM4-8** are presumably in an early evolutionary stage, as neither of them has NIR counterparts. Therefore, the best way to remove their contribution is by applying **graybody-**fits. We used the temperature and emissivity index derived from the total pre-SED and fit the individual solid angles of emission (see also our publication [130]).

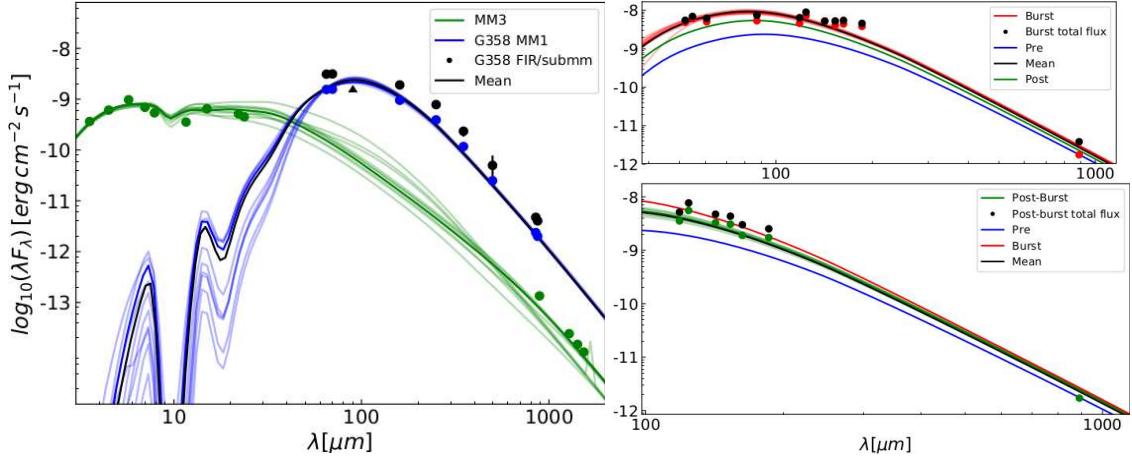

Figure 3.1: **Left:** Contribution of MM1 (blue) and MM3 (green) to the total flux densities (black). Solid lines show the models, triangles mark lower limits. **Right:** Burst (**upper**) and post- (**lower row**) decomposed measurements. Note the different scales.

MM1 accounts for ≈ 50% of the (sub)mm flux (Tab. A.2).

**MM3** is more evolved, which means that for this source, a simple gray body is not sufficient to describe it. However, there exist NIR/MIR data, which allows for SED fitting. We remove the contribution from the best model (dark green line in Fig. 3.1), where we use the **sedfitter** 'spubhmi'-setting [121] (same as 'spubsmi' but with an inner gap, which is more suited for more evolved sources). For further details on the MM3 fit, see [130]. The contribution of MM1 (blue) and MM3 (green) to the total preburst flux densities is visualized in the left panel of Fig. 3.1. Below ≈ 40 $\mu m$ MM3 is the dominant source. The black dots indicate the total FIR and (sub)mm values; blue dots are the MM1 values after all other sources have been removed.

The decomposition is necessary for **all epochs**. It is reasonable to assume that MM1's outburst changes only MM1's SED, while it does not affect the SEDs of the other sources. MM1's burst-/post- SED thus is the total SED of the respective epoch, diminished by the preburst SEDs of all other sources in the field (as described above). A visualization is given in Fig. 3.1 (right panel), where the total values are given in black and MM1's-contribution is indicated in red (burst, upper) or green (post, lower panel) respectively.

A similar approach is used by Hunter 2021 and is shown in [67]

### 3.2.3   Semi-static mean model

In this section, we present our final model. At first, we develop the fitting method (in Sect. 3.2.3.1). After that, we present the components of the final epoch-combined mean model (Sect. 3.2.3.2) and derive major burst parameters in Sec. 3.2.4.1.

### 3.2.3.1 Method

**Overview of the modeling steps**    In order to obtain our results, we perform the following steps, which we will explain in more detail below (Steps 1 to 5):

1. The MM1 preburst-SED was fitted with the sedfitter. (**see Step 1**)

2. RT-simulations have been performed, providing suitable burst/post models. (**Step 2**)

3. A new database was created from these models and fitted to the SEDs (burst and postburst). (**Step 3**)

4. The ten best models of each epoch have been combined into a weighted mean model. (**Step 4**)

5. RT-simulations of the mean model have been performed for each epoch. (**Step 5**)

6. Major burst parameters have been derived. (**Sect. 3.2.4.1**)

7. Possible maser-sites have been identified. (**Sect. 3.2.4.2**)

**The steps in Detail**    In the following, we explain the steps (Steps 1 to 5), as listed above, in more detail. We focus on the most important points. Additional information can be found in our publication [130].

**Step 1.**    We use the RT sedfitter [122] with the 2017 'spubsmi' data set, which includes 40.000 models at nine inclinations each (360.000 SEDs in total) to fit the MM1 **preburst** SED. The underlying density grid is introduced briefly in Sect. 2.5.3. All parameters vary within typical ranges, as described in more detail in [121]. We use a distance of d=$(6.75 \pm ^{0.37}_{0.68})$ kpc (see Sec. 2.3.1) and a foreground extinction of $A_V = (30 - 70)\,mag$ (as discussed in our publication [130]). A radial aperture size of $3''$ has been applied for the fits. This choice minimizes the $X^2$-value of the fit (see also [130]). The result of the preburst fit is given in the Appendix. The SED's of the ten best fits are shown in Fig. 3.1 (blue lines, plot on the left), where the best model is darkest.

**Step 2.**    In principle, we could have fitted the burst-/postburst SEDs with the 'spubsmi' data set (used in Step 1) as well. The problem is that this can result in (totally) different dust

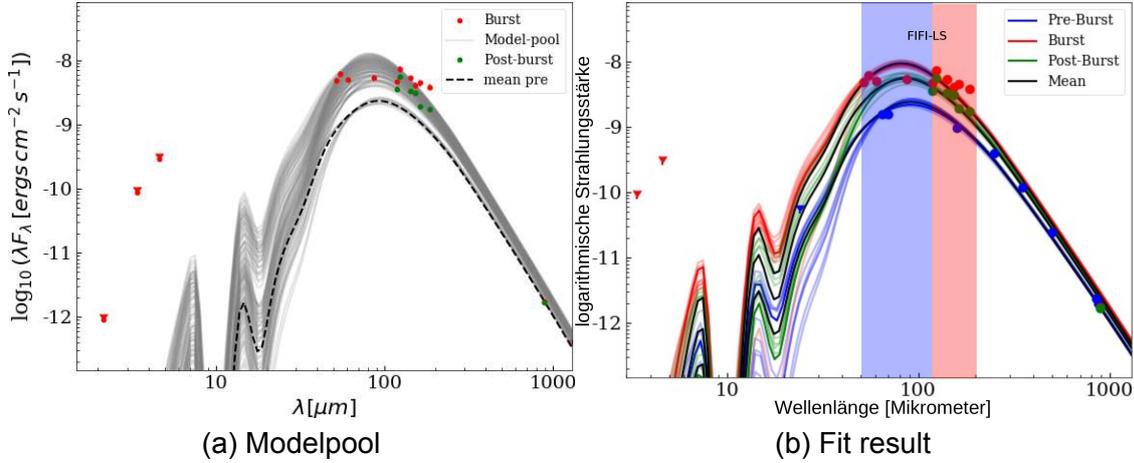

(a) Modelpool                                      (b) Fit result

Figure 3.2: **Left:** The SEDs in the 'burst-/post-'Pool (gray lines) cover the observations (color) pretty well. Triangles are upper limits, the dashed black line indicates the mean preburst model. **Right:** Epoch SEDs (color-coded) of the 10 best models (the best one is darkest), together with the mean model (black). The Figs. have been published in [130, and the corresponding press-release]

configurations or viewing geometries.[1] But clearly, the outburst is not expected to change the YSO configuration entirely. Therefore, we choose a more sophisticated approach, where we design a **new model pool based on the preburst** fit result. We call this a **semi-static** approach. In order to get more realistic burst-/post- models, we use the 10 best pre-models but increase the source luminosity of each of these models. We used nine linearly spaced steps between 2 and 6 $L_{pre}$, in agreement with the observed increase in luminosity (in both epochs). All parameters are kept constant, despite the inner radius being shifted to the new sublimation radius.[2] Simulations were carried out with the HYPERION RT code [120], which was also used for the 2017 YSO grid [121]. The new data set covers the measured burst/post-SED's pretty well, as shown in Fig. 3.2. The SED's are computed by means of static RT, which is not ideal to describe non-equilibrium cases (such as outbursts). However, the new pool allows for a more consistent, although still semi-static treatment of the distinct epochs.

**Step 3.**   For the burst/postfit, we use the sedfitter [122]. The HYPERION output (i.e., the burst/post files produced in Step 2) does not have the same file structure as the sedfitter input. Therefore, we had to change their format prior to the (burst-/post-) fit. These are basically technical details, but the main point here is that we fixed the foreground extinction and the distance to the values obtained from the preburst fit. In other words: We make sure that the burst does not change the distance and/or the foreground extinction of the MYSO.

---

[1]For illustration, the SED fits of NIRS3 (another MYSO burster) with the 'spubhmi' dataset, lead the funny 'fact', that 'obviously' during the burst the inclination is lowest. This is not true. But the burst SED mimics an MYSO with a lower line-of-sight extinction.

[2]We adapt $T_{sub}$ = 1600 K (the same as in [121]). For simplicity, we do not perform an extra run to determine the corresponding radius from the simulation; instead, we use the empirical relation $r_{sub} = R_* \cdot (T_{sub}/T_*)^{-2.085}$ [149, Eq. 1].

We exclude the (sub)mm observations at $\lambda > 890\,\mu m$ from the SED fit of the burst, since their deviation from the stationary models is biggest as the timescales at those wavelengths are longest. The static models are therefore expected to overpredict the (sub)mm flux. For a more detailed discussion of that, we refer to Sect. 9.1 ('Misfit of the (sub)mm fluxes') in our paper [130].

**Step 4.** In this paragraph, we elaborate how the weighted mean model was obtained. The model itself is presented in Sect. 3.2.3.2. The mean model combines the fit result of all three epochs. The analysis was carried out from the ten best models for each epoch, where we weight the parameters according to their respective $X^2$-values. For the log-sampled values, we use the geometric weighted mean, where for the lin-sampled values we use the arithmetic one. Since the quality of the fit is very different for each epoch, we decide to normalize the sum of the respective weighting factors to unity and split up to 0.5 for the preburst and 0.25 for the other two epochs. With this choice, 'stationary' and 'nonstationary' contributions are equalized. The fit result can be found in the Appendix. An important outcome of the fit is the luminosity for each epoch (used to derive the burst parameters). Of course, the respective epoch luminosity is determined for each epoch separately.

| epoch | date | $n_{data}$ | $L[10^3 L_\odot]$ | $L[L_{pre}]$ | instruments |
|---|---|---|---|---|---|
| pre | ≤ 2016 | 8 | $5.0^{+1.1}_{-0.9}$ | 1 | |
| burst | 1. May 2019 | 11 | $23.4^{+4.4}_{-3.7}$ | $4.7^{+2.1}_{-1.5}$ | FIFI-LS (red+blue), ALMA |
| post | 28. Aug. 2020 | 7 | $12.4^{+2.0}_{-1.7}$ | $2.5^{+1.1}_{-0.8}$ | FIFI-LS red, ALMA |

Table 3.1: Luminosity at the epochs as obtained with our fit. The number of data points does not include limits.

**Step 5.** RT simulations of the mean model (with parameters presented in Sect. 3.2.3.2) have been performed with HYPERION for the three epochs (similar to Step 2). The resulting SED's are shown as black curves in Fig. 3.1 in the respective panels. We utilize these (epoch-combined) mean models to obtain further results (Sec. 3.2.4 and [130]).

|      | view |      |       | star        | disk            |                  |      | env              | cav              |      |
|------|------|------|-------|-------------|-----------------|------------------|------|------------------|------------------|------|
|      | i    | d    | $A_V$ | $L_{pre}$   | $m_{disk}$      | $r_{max}^{disk}$ | $\beta$ | $\rho_0^{env}$ | $\rho_0^{cav}$   | $\Theta$ |
|      | °    | kpc  | mag   | $L_\odot$   | $10^{-5}\,M_\odot$ | au            |      | $10^{-18}\,g/cc$ | $10^{-22}\,g/cc$ | °    |
| samp | lin  | log  | lin   | log         | log             | log              | lin  | log              | log              | lin  |
| mean | 21.8 | 6.77 | 60.5  | 4984        | 8.4             | 952              | 1.16 | 2.98             | 6.37             | 33.7 |
| $\sigma$ | 10.1 | 1.05 | 9.7 | 1.2         | 72.2            | 2.5              | 0.06 | 4.14             | 7.42             | 10.2 |

Table 3.2: Selected parameters of the mean model, the full fit-result is given in the Appendix and our publication [130]. All masses and densities refer to the dust.

### 3.2.3.2  G358's epoch-combined mean model

In this section, we (briefly) discuss the parameters of the resulting (epoch-averaged, semi-static) mean model. Selected parameters of the mean model are summarized in the Tab. 3.2. The corresponding SEDs are given in Fig. 3.1 as black lines for all epochs. Additionally, we include scatter plots (showing the scatter of selected preburst models/parameters together with the mean model, as deduced from all epochs) in the Appendix (Fig. A.13 and A.14). The main features can be summarized as follows.

- **Protostar:** The preburst luminosity amounts $(5.0^{+1.1}_{-0.9})\,10^3 L_\odot$, where $R_*$ and $T_*$ are not very well constrained. That is, the source can be anything in between 'small' and hot or bloated and 'cool'. This is what is expected for high optical depths [121]. During burst/post-epochs, the luminosity is increased by a factor of $4.7^{+2.1}_{-1.5}$ (burst)/ $2.5^{+1.1}_{-0.8}$ (post), as summarized in Tab. 3.1.

- **Disk:** The disks parameters are poorly constrained. This is expected for face-on systems, where the (protostellar) radiation does not pass through the disk. This is supported by the fact that most of the FIR radiation originates from the extended envelope (only a minor fraction is produced within the disk itself). There may be a slight anticorrelation between the flaring power and the disk mass (see Fig. A.13, right panel). The mean mass of the disk is $m_{disk}$ = 8.4 $10^{-5}M_\odot$ in unit dust mass. Note that the ten best models cover almost the whole mass range.

- **Envelope:** The protostar is surrounded by a dense envelope with $\rho_0^{env} = (3.0^{+9.0}_{-2.3}) \cdot 10^{-18}\,g/cc$ (in units of dust). For comparison, this can be translated to $\dot{M}_{env} \approx 1.5 \cdot 10^{-3}M_\odot/yr$ using Eq. 2.5 (with $M_* = 12M_\odot$ derived from [28] and a gas-to-dust ratio of 38, which is appropriate for the G358 region [49, Eq. 2]). Note that all densities of the best models are in the upper range. Its density decreases with the corotation radius $r_c$ (see Fig. A.13, left panel) as $\rho_0$ is defined at that location.

- **Cavity:** Neither cavity density (NIR dark, high foreground extinction) nor the opening

angle Θ (which probably is not at the higher/lower end of the range) are well-defined (see Fig. A.14). The system is likely seen through the cavity as the inclination $i$ is always smaller[3] than Θ (while i increases with Θ).

- **View:** The system is probably seen face-on (no model has $i \geq 60°$). This is in agreement with [21, 22, 28, observations of maser spiral patterns].

### 3.2.4 Results

#### 3.2.4.1 Burst parameters

In this section, the (epoch-combined) **mean** model is used to obtain the main burst parameters. The aim is to put G358's accretion burst in context and evaluate possible conclusions about the triggering mechanism behind. The main burst parameters are: decay time $\Delta t$, released energy $E_{rel}$, accreted mass $M_{acc}$ and mass accretion rate $\dot{M}$.

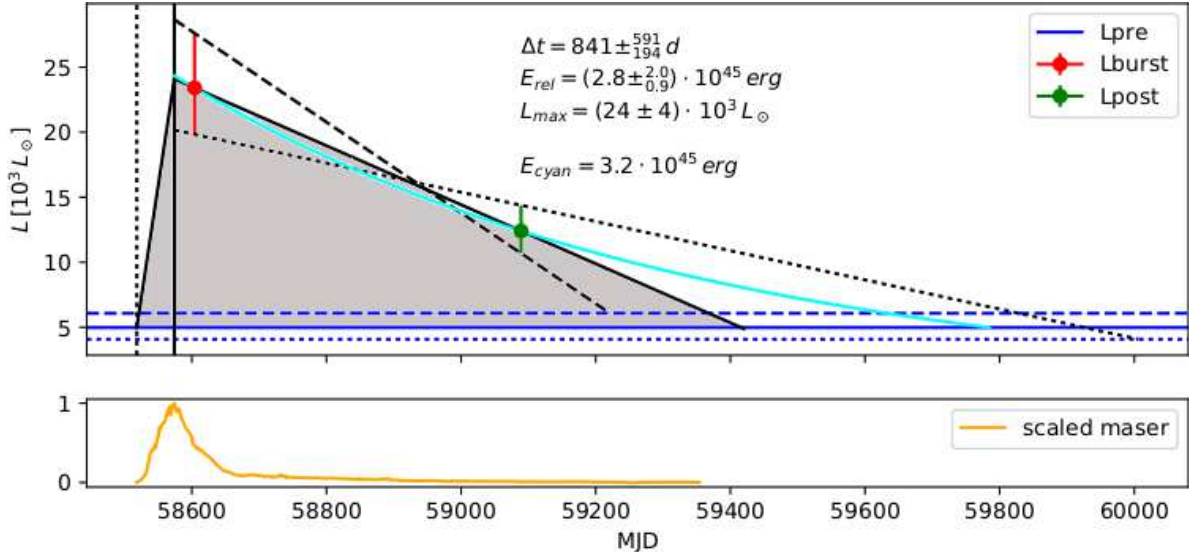

Figure 3.3: **Upper panel:** Visualization of the burst-energy estimate: If the luminosity L decays linear in between maser-peak (solid vertical line) and return to the pre-level, the burst energy equals the gray area. Dashed/dotted lines indicate minimized/maximized decay times. The cyan curve shows the same, if the decay is not linear in L but in log(L) instead. For a detailed explanation, see text. **Lower panel:** scaled light-curve of the total 6.7 GHz maser flux.

---

[3]The cavity opening angle is defined at $r_0$ = 10.000 $au$ and not at the outermost radius, therefore this comparison has some weaknesses. However, the envelope density decreases with $r^{-1.5}$ outside the co-rotation radius, and its density is already quite low outside $r_0$.

**Decay time and energy-release:**   The two epochs can be used to obtain an estimate of
the energy release of the burst. The most simple approach is to assume that the luminosity
decays linearly until it finally returns to the preburst level. This is illustrated in Fig. 3.3 with the
3 black lines (dashed, dotted, and solid for lower, higher limits, and estimated decay time,
respectively). We further assume that the peak time is the same as for the maser peak and that
the luminosity increases linearly in between the onset of the maser-flare and flare-peak. With
this we obtain an estimate of **the burst energy of** $E_{rel} = (2.8^{+2.0}_{-0.9}) \cdot 10^{45} \, erg$ as corresponding
to the gray area in Fig. 3.3. With $E_{rel} = \frac{L_{peak} - L_{pre}}{2} \cdot \Delta t'$, where $\Delta t'$ is the time between the onset
of the maser flare and the expected return date, and $L_{peak}$ is the luminosity at the peak. The
deviation leads to the error estimate for the released energy as given above with Gaussian
error-propagation. The **decay time** amounts approximately $840^{+590}_{-190}$ days (approximately 2.3
years). It is a measure for the duration of the FIR-afterglow. During the burst, the luminosity
increases by $\Delta L = (19 \pm 5) \cdot 10^3 L_\odot$ (a factor of $\approx 4.8$).
Another possible assumption is that the luminosity decays linearly in log(L) (instead of L).
This, however, leads to similar values for $\Delta t$ and $E_{rel}$ as shown by the cyan curve in Fig. 3.3.
The lower panel of the figure shows the total maser flux (scaled). It decays much faster
(as compared to the FIR). Due to the proximity to the source and the nature of the radiative
pumping of the masers, they react to the source variation almost instantaneously. The duration
of the protostellar flare is well reflected by the maser flare.
However, the increase in the masing radiation is not necessarily a good measure of the
'total' injected burst energy since its 'efficiency' depends very much on the local conditions
(such as column density, local temperature, and coherent path length). In the FIR, the 'entire'
released energy contributes at some point to the reprocessed enhanced thermal dust emission.
Therefore, the above estimate is the best we can do, at least with static RT.

**Accreted mass and mass accretion rate**   If the released energy equals the potential energy
of the infalling material, a lower limit for the accreted mass $M_{acc}$ can be derived (assuming
infall from a distance d much bigger than the protostellar radius $d >> R_\star$):

$$M_{acc} = \frac{E_{rel} \cdot R_\star}{G \cdot M_\star} \qquad \text{with G as grav. constant} \qquad (3.1)$$

The **accreted mass** depends on the energy input $E_{rel}$ (as estimated above), the protostellar
mass $M_\star$ and its radius $R_\star$. The mass was derived as $M_\star = (12 \pm 3) M_\odot$ by [28, kinematic model
of the spiral arm accretion flow]. The radius derived from our fit is $R_\star = (8.4^{+15.7}_{-5.5}) R_\odot$. The
lower value is likely not below the ZAMS value (which amounts to $R_\star = (3.9^{+0.1}_{-0.2}) R_\odot$ for solar
metallicity [139]). This slightly reduces the uncertainty. These values together lead to:
$M_{acc} = (170^{+350}_{-120}) \cdot M_{earth}$, where the error is dominated by the uncertainty of the radius. The

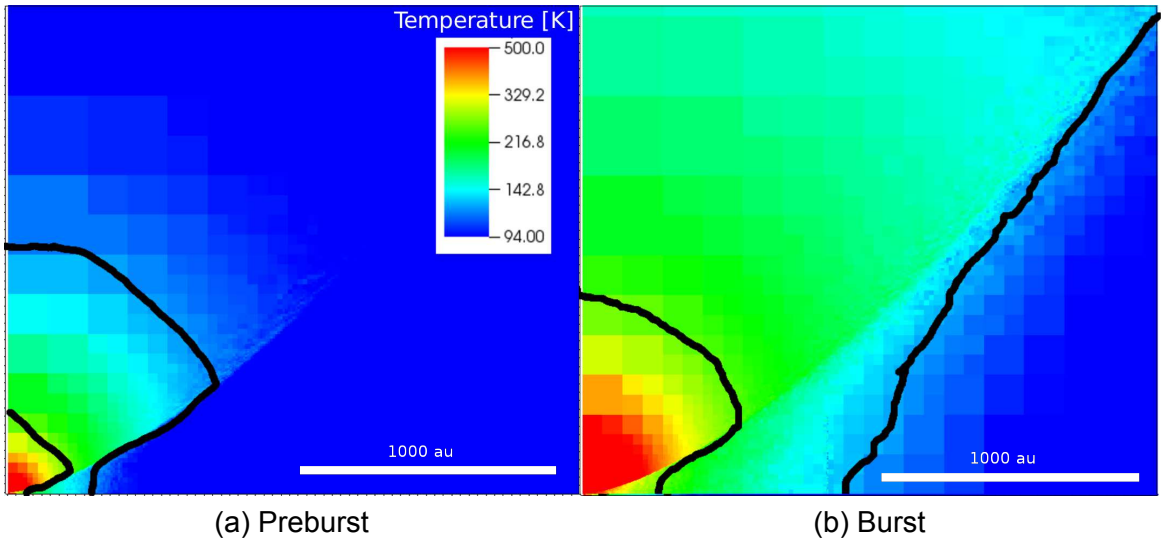

(a) Preburst                                    (b) Burst

Figure 3.4: Temperature-map during the pre- (**left**) and burst (**right**). Bold black lines enclose the temperature range suited for maser-excitation.

protostellar radius $R_*$ is not very well constrained by our simulations (see Fig. A.14, upper left). The fitting in principle allows for a (strongly) bloated protostar, which would translate into a higher value for the accreted mass. In that sense, the lower limit is stricter.

From the burst duration and the total accreted mass, the mean **accretion rate** can be inferred. As mentioned above, the maser curve is probably a good proxy for the duration of the (main) accretion phase. It lasts about $\Delta t_{acc} = (60 \pm 5)\,d$ (MacLeod, priv. com.); this is basically the 'FWHM' of the total maser-curve. The resulting mean accretion rate amounts $\dot{M}_{acc} = (3^{+6}_{-2}) \cdot 10^{-3} M_\odot / yr$.

### 3.2.4.2 Temperature-changes and (Class II methanol) maser relocation

G358 is probably the best studied MYSO burster. Changes in its maser landscape have been the subject of many publications (e.g. [28, 9, 22, 19]). It is possible to use the result of our simulations to constrain possible maser[4] sites. In our paper, we publish such a study for the static mean model (as presented in the previous sections). We will also present a similar approach here (but with a time-dependent treatment). At first, the masing conditions are introduced. After that, we show the expected shift of the maser regions compared to what is observed in [9, 22].

---

[4]With 'maser' we refer to Class II methanol masers, which are radiatively pumped.

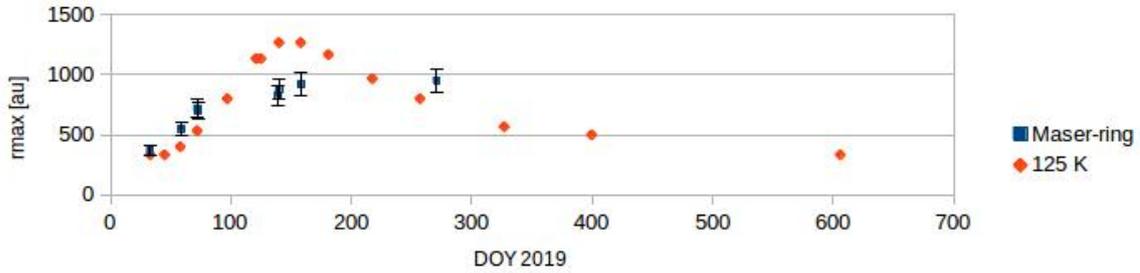

Figure 3.5: The outward motion of the maser-sites can be seen in both the observations and the simulations. Blue dots indicate the outermost radius of the masers in [21, 22], Fig. 1, where the error-bars refer to the distance uncertainty. The orange points indicate the radius (in the disk mid-plane), where the temperature equals 125 K (minimum for maser excitation). At the postburst epoch (last data point), the masers are expected to be back at the preburst location.

**The masing conditions**   During a flare, maser components close to the protostar are no longer excited, while new components arise farther out. This is related to a change in the masing conditions. Masers can only occur under certain conditions:

- **Enough methanol in the gas phase:** Only when the temperature increases above 94 K thermal desorption of methanol (main desorption mechanism in dense HII regions) becomes possible and masers can arise. Below that temperature, all methanol is bound in the ice-mantles of the grains. When the gas temperature exceeds the dissociation temperature of methanol, which is at about $377 - 477\,K$ [1] the molecule quickly gets destroyed.

- **Gas densities are in the right range:** For column densities exceeding $NH_2 \geq 10^9\,cm^{-3}$ collisional de-excitation will hinder the population inversion of the energy levels [34]. On the other hand, for column densities below $\approx 10^5\,cm^{-3}$ [129, Fig. 1a, upper left], the brightness of the masers drops rapidly.

- **MIR radiation to pump masers**: methanol masers are pumped through MIR radiation. Pumping is effective if the dust temperature is greater than 125 K [34, Fig. 3]. The minimum wavelength for pumping radiation is $15\,\mu m$ [110, Fig. 1]. At gas temperatures around 250 K the maser brightness has already decreased by an order of magnitude. Note that in dense disk regions, gas and dust are well coupled ($T_{dust} \approx T_{gas}$). Together, we adapt a range of $T = 125 - 250\,K$.

- **Sufficient coherent path length**: Since G358 is seen face-on, the velocity dispersion along the line of sight is likely small. The cause of this is mainly thermal motion, turbulence, and convection in the disk rather than differences in rotation speed. The small dispersion can be probed by broadening of molecular lines (e.g., [38]). The criterion of a sufficiently coherent path length is likely fulfilled. Nevertheless, differences in the path length may exist (especially in the innermost parts of the disk) and imprint the observable maser 'efficiency'.

**The modeled maser ring expansion** We can trace the spatial shift of possible maser sites with our models. This is illustrated in Fig. 3.4, which shows the temperature maps for pre- (left) and burst (right). This simulation is done with TORUS, using a slightly adapted mean model[5] together with the scaled[6] maser curve as a template for the luminosity variation. A similar approach has been published for static models in [130]. The range suited for maser excitation is enclosed by bold black lines. It clearly moves outward as a result of the burst. Fig. 3.5 compares the outermost possible radius (that is, $T = 125\,K$) from our simulation (orange) with the maximum radius of the maser rings (blue) as published in [21] and [22]. Here, we use the radius measured in the disk midplane. The midplane of the disk is probably too dense (masers are quenched there), but the masers are likely associated with spiral structures within the disk [21]. The principal behavior agrees well, although there are some differences. We emphasize, that the predicted speed of the maser ring expansion is in agreement with the observations [9, (4 – 8)% of c, with c as speed of light]. It amounts to ≈ 6% between flare start and maximum maser-ring radius for our model. In between burst onset and ≈ $DOY\,100$ the maser-ring quickly expands for both simulation and observation. After the initial fast expansion, the observed maser ring expands slower, contrary to the simulation, where the fast expansion continues until it peaks at ≈ $DOY\,170$ (and $1300\,au$). Since the disk ends at ≈ $1000\,au$ the maser ring may not become larger than that, even if the pumping condition is fulfilled outside the disk. However, a slow expansion is clearly observed in between DOY 140 and 270 (where at the latter the simulated radius is already smaller than the observed one). This can be an indication that the accretion phase was indeed longer (and possibly featured a lower peak accretion rate) as compared to our assumption. Clearly, a more careful analysis is needed, but this nevertheless indicates that our results are realistic.

### 3.2.4.3 Our results in the context of current observations of G358

This part is supposed to give an overview over our results and put them in the context of current knowledge about this particular MYSO and its burst.

The burst was relatively weak and short (EXor-like). We estimate a burst energy of $E_{rel} = (2.8^{+2.0}_{-0.9}) \cdot 10^{45}\,erg$, an accreted mass of $M_{acc} = (170^{+350}_{-120}) \cdot M_{earth}$ and a mean accretion rate of $\dot{M}_{acc} = (3^{+6}_{-2}) \cdot 10^{-3} M_\odot / yr$ within the mean accretion phase of 2 months.  These parameters agree well with the accretion of a planet (such as a small Jupiter) [40].  Another possibility is the accretion of a disk fragment, such as a spiral arm fragment.  Spirals within G358's disk are found via highly resolved maser observations (e.g., [28, 9]).  Such structures are expected in heavy disks.  From G358's 4-arm spiral pattern, the authors conclude that the disk-to-star mass ratio exceeds 25% [22].  Together with their value for the enclosed mass, which amounts to $(11.5 \pm 4.8) \cdot M_\odot$ (in agreement with [28]), this yields a (minimum) disk mass of $\approx 3 M_\odot$.  The total disk mass, derived with our models, is about 2 orders of magnitude smaller.[7]  However, there are several issues.

---

[5]We used astrosilicate instead of Milky-Way dust and a smaller grid (according to the extent of the disk).
[6]We scaled it, such that the energy equals our semi-static estimate.

At first, the disk is not well constrained by our models, as the system is seen face on (the uncertainty is given by a factor of $\approx 70$).  The second issue is that the SED modeling is only sensitive to the dust 'within' the 'isolated' (small) grains (responsible for the thermal emission). However, the most mass is probably located within large clumps (or spirals), which contribute only little to the thermal emission (but essentially to the total mass).  Therefore, the disk mass as derived by our models probably always underestimates the 'true' disk mass (unlike the kinematic models, as used by [28, 22]).  Furthermore, the disk and envelope mass are added (for the models in the YSO-grid [121]).  Therefore, a fraction of the mass is 'hidden' in the envelope.  Another point is that we use the same dust in all regions.  But within the midplane of the disk, the grains are likely larger due to dust growth and grain settling.  This also increases the 'true' disk mass.  We conclude that all these effects together can explain 2 orders of magnitude difference, and G358 features indeed a massive disk.

Note that, if the disk was really as light, as suggested by our models, it might even be a transient feature.  The accreted mass amounts to $\approx 16\%$ of the total disk mass[8] (as derived by our models).  Transient disks may be a feature in very early systems, but even if G358 is extremely young (MIR dark), this scenario is (likely) not true for this system.

G358's disk is likely heavy, prone to GI, but stable (no runaway fragmentation, the spirals transport angular momentum) [22] and the burst parameters fit the accretion of a spiral fragment.  G358 is another example of a disk-mediated accretion burst.  It fits well in the current picture of massive star formation.  There is recent evidence that disk fragmentation is common in MYSOs, e.g., [3, kinematic study of 20 MYSOs].

Our simulations indicate an expansion of the location of the possible maser sites at subliminal speeds ($\approx 6\%$ of c between the start and maximum radius) due to the burst.  This is in agreement with the observations [9, estimated expansion of the maser ring at $(4-8)\%$ c].  It is much faster than any physical gas motion but much slower than the speed of light.  According to [70], the timescales are set rather by the photon travel times, than by the timescale of the dust heating[9]. In this context, the 'slow' expansion of the heatwave is remarkable, and indicates the 'effective' slowdown of the photon travel speed due to multiple scatterings, as well as absorption and reemission events, within the dense geometries (random walk-like propagation paths). We emphasize that the latter is already a result of the time-dependent modeling, which is the subject of the following chapters.

---

[7]It amounts only $\approx 3 \cdot 10^{-2} M_\odot$, with a gas/dust ratio of 38 (appropriate for the location of G358, [49, Eq. 2]).

[8]Adopting a gas/dust ratio of 38 (appropriate for the location of G358, according to [49, Eq. 2])

---

[9]The heating of the dust grains at 1000 au needs approx 100–1000 s, ignoring the finite speed of light [70, pure thermodynamical considerations for low-mass YSOs].

## 3.3   G323.46-0.08

### 3.3.1   Overview over the data

This section is supposed to provide an overview of the data on which this thesis is based. Further information is provided in the Appendix. We refer also to our corresponding publication (Wolf+ in prep.).

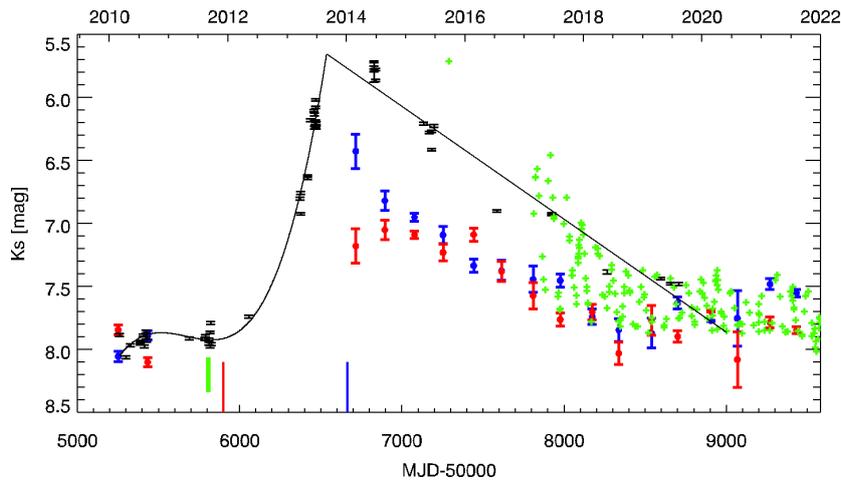

Figure 3.6: Light curves based on VVV(X) (black symbols) and (NEO)WISE photometry (W1 - blue, W2 - red) as well as 6.7 GHz total maser flux (green crosses, [95]). Vertical green, red, and blue lines mark the dates of the last pre-flare methanol measurement [50], the burst onset and first flare evidence from the 6.035 GHz exOH maser [95]. The rise of the $K_s$ light curve has been approximated by a polynomial, while its decay is roughly linear on a log-scale (black line). The (NEO)WISE magnitudes are shifted to match those of $K_s$. Similarly, the maser total flux is shown on a log-scale using a range that matches the IR variability. Its scatter is due to the short-term periodicity. This figure will be published in Wolf+ in prep. [Credits: B. Stecklum]

The source is included in various surveys; therefore, the **preburst SED** is fairly well sampled (in the whole range from NIR to (sub)mm). The pre-SED is plotted in Fig. 3.7 and Fig. 3.8 (blue dots), a table with the values is provided in Tab. A.5. Note that no further sources are detected in the ALMA field. Therefore, no source decomposition (analog G358) was necessary. The source is included in the VISTA Variables in Via Lactea Survey [103] (VVV) and its extension (VVVX, 2016-2019). G323 has been continuously observed during the burst in the Ks-band. The **Ks-light curve** is shown in Fig. 3.6 (black), together with the scaled total maser flux (green) and the (NEO)WISE observations (red and blue). We used the Ks-light curve as a template for the variation in the accretion rate during the burst (as discussed later). The essential dates are established from the Ks-light curve (that is, start, peak, and end, as given in Tab. 3.3).

| | start | begin fast rise | peak | end | HAWC+ |
|------|-------|-----------------|------|-----|-------|
| MJD | 56083 | 56100 | 56535 | 59119 | 59767 |
| date | 5. June 2012 | 22. June 2013 | 31. Aug. 2013 | 27. Sept. 2020 | 6. July 2022 |

Table 3.3: Dates of the G323 outburst (from the Ks-light curve, Wolf+ in prep.).

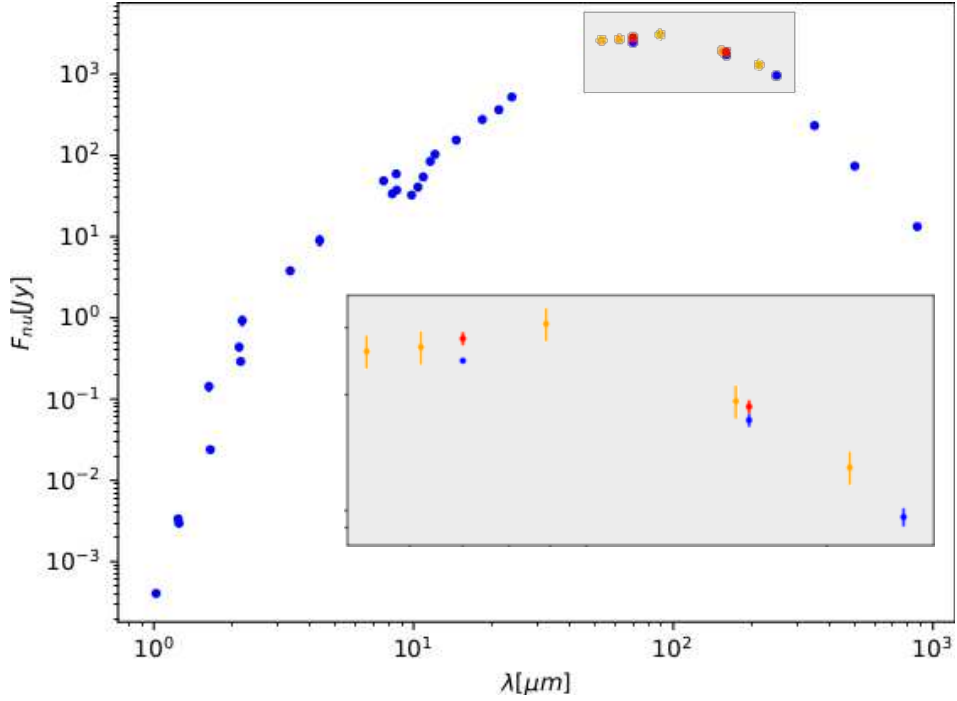

Figure 3.7: Preburst SED (blue), together with the HAWC+ postburst observations (orange). The HAWC+ observations have been interpolated to match the wavelengths of the preburst observation. The resulting data points are colored red. The inset shows a zoom-in on the region of interest. The flux excess in the postburst epoch is small (only $\approx 10\%$ at $70, 160\,\mu m$).

The Ks flux reached the preburst value around 27 September 2020. We used the results of the preburst SED fit and our burst template (based on the Ks-variation) to predict the afterglow 2 years after the end of the burst. We obtained **HAWC+ observations** (see Tab. A.6), performed in July 2022. The increase in flux in the FIR was only about 10%, making the analysis challenging.

As the increase is quite small, we highlight that the following analysis would have been suitable to obtain the upper limits of the burst energy *even in the absence* of a measurable afterglow.

The errors of each of the HAWC+ fluxes (orange dots in Fig. 3.7) are dominated by the uncertainty photometric calibration. Fortunately, in the $154\mu m$ band, there is another source in the field, which is also seen in the PACS $160\mu m$ image. It was used to refine the photometric calibration of the HAWC+ pipeline. The HAWC+ fluxes have been interpolated at the PACS wavelengths (red dots) using a polynomial approximation of the postburst FIR SED established from all HAWC+ bands (orange dots). This thesis focuses on the **interpretation of the 'red' postburst fluxes** (i.e., the data points at 70 and 160 $\mu m$). A more detailed discussion of how the values have been derived is included in Wolf+, in prep.

### 3.3.2   Method

We apply a similar approach as for G358. Again, the part of the method is split into a list of the steps (**Overview over the modeling steps**) and detailed explanations, including major assumptions (**The steps in detail**). In Step 3 we introduce three typical time scales (used to describe the afterglow). For a clearer view, these definitions are placed at the end of this section (**Definitions**).

**Overview of the modeling steps**    We use a **fully time-dependent** approach to model G323, where we use **three peculiar models** that feature mean, minimal, and maximal afterglow durations. The following steps have been performed:

1. A **static model pool** had been generated with TORUS. (**See Step 1**)

2. The preburst SED was fitted, and a **mean model** was obtained.
   (**Step 2**, with **results presented in Sect. 3.3.3.1**)

3. TDRT-simulations were performed to describe the **afterglow**.
   (**Step 3 + Definitions**, and **results in Sect. 2.2.5**)

4. A **small parameter-study** was performed. (**Step 4, Sect. 3.3.3.3**)

5. Models with **minimum and maximal afterglow timescales** have been established.
   (**Step 5, Sect. 3.3.4**)

6. The **HAWC+ (postburst) data** was used to infer the burst energy.
   (**Step 6, Sect. 3.3.4.2**)

**The steps in detail** In the following, the above steps are explained in detail.

**Step 1:** We follow [121] and create a model pool of (static) SEDs, where we sample all parameters in defined regions. The major differences are described below. Since the luminosity is relatively well known[10], a similar density in the model space can be reached with fewer simulations (2.500 compared to 42.000 for the 'spubsmi'-dataset). However, a **2-step approach** is used to reduce computational demand. In a first step, all preburst models were computed with only $10^6$ photon packets. With this number, the scatter due to synthetic noise exceeds the observational errors. Therefore, the 50 best-fitting (but noisy) models were recomputed with $10^8$ photon packets, ensuring reliable $\chi^2$ values. Contrary to [121] we vary the outer radius of the envelope. G323 is resolved (with ALMA and HAWC+) and its parent cloud core probably does not extend to distances of $\approx 10^6 \, au$ (as implied by the models in the [121] data set). We use the total fluxes only, even though TORUS in principle allows for aperture-dependent values as well. Instead of Milky Way dust (as in [121]), MRN dust is used [99] (see also Sect. 2.5.3). Fig. 3.8 shows that the models (gray) cover the observed preSED (blue dots) pretty well.

**Step 2:** We fit the preburst SED, where we vary the distance in 10 logarithmic steps between 3.70 and 4.48 $kpc$. We apply a foreground extinction of $A_V = 18 \pm 1 \, mag$ (see Sect. 2.3.2). We use our own Python routine, which basically works the same as the sedfitter [122]. We could not use the sedfitter, since our data structure is different. Furthermore, we optimize the SEDs in the first fitting step (because of the high noise). For the final preburst fit, the synthetic noise is negligible (as we use $10^8$ photons for the best models), and we assume that the simulation has no error (following [122]). Similarly to G358, we construct a **weighted mean model**, but only **for the pre-** (and not for the post-) epoch. The resulting mean SED is plotted in Fig. 3.8 (red) together with the model pool (gray) and the 10 best fits (blue), the parameters are presented in Sect. 3.3.3.1 (mean model).

**Step 3:** We perform **TDRT simulations for the mean model** (from Step 2) to describe its afterglow. This is the heart of the time-dependent analysis. Therefore, we describe this particular step in more detail.

We start with a brief summary of our assumptions. Then we present the details of the simulation and then the most important steps of our results.

**TORUS** uses a couple of **simplified assumptions**, which have implications for the simu-lation setting. In the following, we want to briefly summarize the ones we consider the most important. These are: **no hydrodynamics** (the dust distribution remains the same throughout the whole simulation, no dust is removed, moved or added), **no chemical changes due to the burst**, including **no dust sublimation** (except for the maser-curve, as described below). To account for the latter, we **slightly shift the inner radius** to $3R_{sub}$ (i.e., from 20 to 60 au for the mean model)[11] for all of our settings. This ensures that the dust does not get too hot. For the outcome (resulting SEDs) the effect is negligible (reprocessing). We kept the disk mass constant, when shifting $R_{sub}$, effectively putting the mass slightly more outward rather than removing the inner disk.

---

[10]A bolometric fit of the SED yielded $L_* = (1 - 1.3) \cdot 10^5 L_\odot$ [91], which we extend to lower luminosities, taking into account that the bolometric luminosity inferred from the SED exceeds the true one for low inclinations [151].

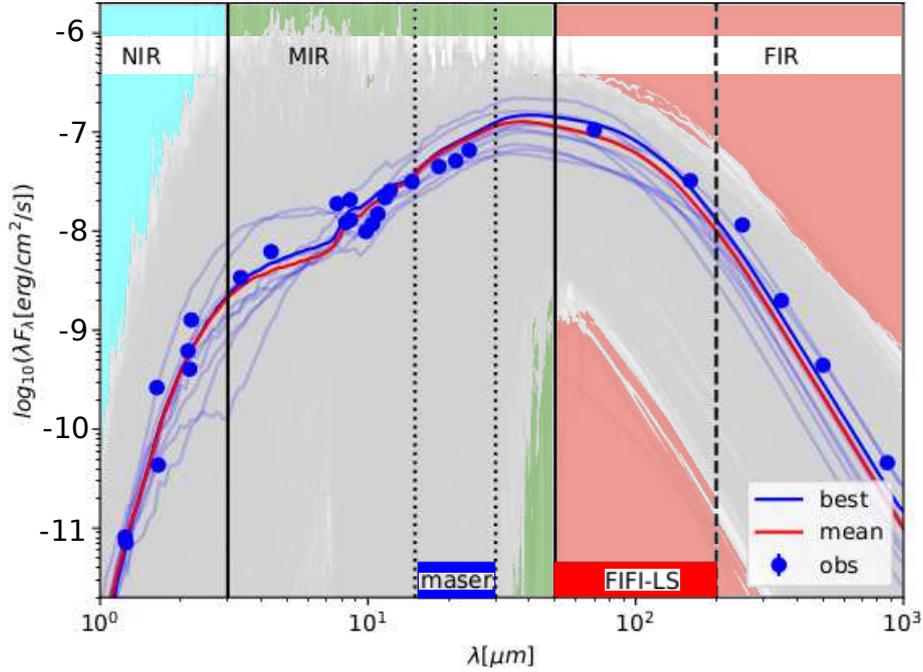

Figure 3.8: All TORUS models (gray), together with the preburst SED (blue dots), the ten best fits (blue, the best one is darkest) and the mean model (red). The wavelength-ranges are indicated by the background colors. The model-pool covers the data pretty well, and the mean-model is a good fit. HAWC+ (not shown) observes in the same wavelength-regime as FIFI-LS.

The **scaled Ks-light curve serves as a template for the source luminosity variation**. We assume that the increase in the bolometric luminosity equals the increase in Ks at all times. This leads to an accretion energy of $E_{acc} = 2.3 \cdot 10^{47}\ erg$. We emphasize, that this is only a first estimate, the value is varied to obtain a more realistic value later on (Step 6). We scaled the interpolated Ks-light curve so that the input energy is equal to the above assumption. With a preburst luminosity of $L_{pre}$ = $6 \cdot 10^4 \, L_\odot$ this implies a scaling factor of 1.2 and a peak luminosity of 13 $L_{pre}$. We note that the Ks-curve was scaled in $log(F_{K_S}(t)/F_{K_S}^{pre})$ (rather than in $F_{K_S}$). This keeps the essential dates (i.e., start, end, and peak are fixed) for all input energies (Step 6).[12]

---

[11]This value is based on an extra simulation (it is the -outermost- location, that exceeds the sublimation temperature of $T_{sub} = 1600\ K$).

For the simulation, we used a **time step of 3.65 days** (630 au at the speed of light), which is probably too large to tackle the changes in the innermost regions correctly. It is nevertheless a good compromise between computational resources (mainly computation time, but also storage) and results. For deeply embedded sources, the radiation is heavily reprocessed and the SED variation is only slightly affected by the temperature changes on a small scale. For comparison, the width of the step is equal to ≈ 5% of the extent of the grid. We emphasize that the **maser excitation curve** is obtained by invoking a much smaller time step (that is, 0.7 $d$), dust sublimation, and a grid restricted to the disk area only.

In the following, we want to give the **structure of the result** (for a better overview). The **characteristics of the static mean model** are summarized (split by components) in Sect. 3.3.3.1. From the output of the TDRT simulation (presented in Sect. 2.2.5) we create **density and temperature-maps**, as well as a **dynamic SED** and selected **light curves** (using visit [30] and Python). As the afterglow is wavelength dependent, **color changes** and a change of the **spectral index** are expected, which we also discuss.
We use specific **wavelength regimes: NIR ($\lambda \leq 3$), MIR ($3 < \lambda \leq 50$), FIR ($\lambda > 200 \mu m$) and total** (i.e., bolometric) to address the wavelength dependence of heating/cooling. To achieve comparability, we also present three **typical time scales**, which are defined at the end of this section.

At the end of the result section (that is, Sect. 2.2.5), we include a **preliminary comparison of the mean model with observations**, where we use the mean model (from Step 2) and the source template with $E_{acc}$ = $2.3 \cdot 10^{47} \, erg$ (from this step). This includes the Ks curve and the **optimized light curves at** 70 **and** 160 $\mu m$. By optimized, we mean a stepwise interpolation in time, with an interval that is smaller close to the peak and for shorter wavelengths (see also Sect. 5.3).

**Step 4:**  The mean model (Step 2) incorporates 4 components (protostar, disk, envelope, cavity) with 11 free parameters. Therefore, it is important to test which parameters influence the result (afterglow time scales) and which are more or less irrelevant. Therefore, we include a **parameter study**, where we summarize the most important results in Sect. 3.3.3.3. We emphasize that we did (most of) these tests when the G323-preburst database was not established yet. Therefore, the models are not the actual mean model (unless otherwise stated). However, the general conclusions of these tests remain. We note that all test models are similar (face-on models with similar characteristics). We divided our analysis into **three groups**: parameters that characterize the **dust distribution** (like envelope density, disk mass, cavity opening angle, etc.), **burst** parameters (shape, released energy) and **viewing geometry** (inclination, aperture size). Possible **degeneracies** are discussed briefly.

---

[12]As a consequence, the decay gets steeper with the $E_{acc}$ variation as adapted in Step 6.

**Step 5:**    In order to estimate the **range of possible afterglow time-scales**, that are consistent with the preburst fit, we **vary the parameters within the** $1\sigma-$ **confidence intervals**, so that the time scales are minimized/maximized. We did not change the source luminosity (neither the preburst luminosity nor the burst profile) and the inclination to ensure better comparability of the models. Note that even if the inclination can change the output a lot, it should not be a big concern in this case, since it is (almost certainly) smaller than the opening angle of the cavity and therefore the optical depth along the line-of-sight should not change within the confidence intervals (given the constant cavity density). Both models are included in Tab. 3.4. In summary: for **Tmin/Tmax** all densities and masses are reduced/increased, the cavity angle is maximized/minimized, respectively. Parameters that are not expected to significantly imprint the afterglow are kept constant. Furthermore, we kept the distance and foreground extinction. This is not important for the TDRT simulation, but (only) for the subsequent fitting.

**Step 6:**    The main goal of this section is to **infer the burst energy** of G323 by modeling its thermal afterglow. The **HAWC+ postburst** fluxes (measured on 6. of July 2022) only slightly exceeded the **preburst fluxes** by about 14.2 ± 4.6 at 70 $\mu m$ and 8.5 ± 6.1% at 160$\mu m$, where we use Gaussian error propagation, together with the values from Tab. A.5 and A.6. The small increase makes the following analysis rather difficult.

    Clearly, the afterglow duration increases with the amount of energy released by the burst. Therefore, we **run models with different** $E_{acc}$, where we scaled the burst profile (similar to Step 3, but for different energy inputs). We do not vary the burst shape, since we consider the Ks curve to be the most reliable proxy for the variation of the accretion rate. From the simulations, we get **'ratio-' light curves at** 70 **and** 160 $\mu m$ (for various input energies), which we compare to the ratios as estimated above. Since the increase is rather small, we **compare the relative increase (ratios)** and not the postburst SED. We emphasize that this is the best way to constrain the burst energy for such small flux excesses.

A $\chi^2$ minimization is applied to infer reliable estimates of the burst energy in the three scenarios (mean/min/max). The **Tmin/Tmax**-settings **set upper/lower limits for the burst energy** and hence the (minimum) accreted mass. The results are presented in Sect. 3.3.4.2.

**Definitions**   The following **time scales** are used to **describe the afterglow**. These time scales **depend on the wavelength** (and generally increase toward longer wavelengths).

**Définition 3.1** ($t_{peak}$)
*Peak time is the time when the light curve peaks. It is a measure for the burst rise time (face-on, NIR).*

**Définition 3.2** ($t_{80}$)
*t80 is the time when 80% of the energy is released in the respective wavelength regime (or injected in case of the source). It is a measure for the afterglow duration.*

**Définition 3.3** ($t_{25}$)
*t25 is the time when the flux excess drops to 1.25 times the respective preburst level. It is a measure of the detectability, whereby the 'exact time' depends on the instrumental setting.*

### 3.3.3   Results

#### 3.3.3.1   Configuration of the mean model

In this section, we present **G323's mean model**, which has been derived similarly to G358 but only for the **preburst** epoch (i.e., the mean model is not epoch-combined).
The mean parameters are given in the Tab. A.9 (all) and summarized in the Tab. 3.4 (selection). The temperature and density of the mean model are visualized in Fig. 3.9. In the Appendix, there is a corner plot showing the coverage of the parameter space together with the 10 best models and the mean model (Fig. A.26). In the following, we briefly summarize the resulting mean parameters for each component and discuss possible correlations. All parameters, despite the cavity and envelope density, scatter a lot. However, the mean model fits the preburst SED pretty well (Fig. 3.8).

- **Protostar:** The bolometric luminosity (integration of the SED) amounts $L_{bol} = 1.4 \cdot 10^5 L_\odot$, which is more or less consistent with the range given in [91, $L_{bol} = (1 - 1.3) \cdot 10^5 L_\odot$]. Note that this is double of the 'input'-luminosity ($L_* = 6 \cdot 10^4 L_\odot$). This is an effect of the face-on view, which maximizes bolometric luminosity [150]. G323 is 10 times more luminous than G358. $L_*$ weakly increases with the cavity angle, as evident from Fig. A.26. This correlation can be explained as follows. Through a wider cavity, the radiation escapes more easily. Furthermore, there is slightly less dust (for settings with a wide cavity) on average (because of the 'low' cavity density). As a consequence, there is slightly less dust reemission (which is then 'compensated' by a higher $L_*$ in the fit).

- **Disk:** Like for G358, the disk is not very well constrained (given the face-on view). It has a similar extent and seems to be a factor $\approx 4$ less massive. However, the errors span 2 orders of magnitude (each); therefore, no conclusions are possible. All disk parameters show no (or only weak) correlation with the other parameters. The scale height (maybe) decreases with the outer radius of the envelope. Eventually, the extent of the disk increases with increasing inclination. Both correlations are pretty weak.

- **Envelope:** The envelope density is high and fairly well constrained (since it sets the amount of cold dust). The envelope extends to $(2.4^{+1.2}_{-0.8}) \cdot 10^4 \, au$.

- **Cavity:** The cavity parameters (especially density) are well constrained, contrary to G358. This is because G323 is NIR bright (unlike G358) and has a lower, well-defined foreground extinction. The opening angle of the cavity is $\Theta_{cav} = (40 \pm 10)°$, which is consistent with a more evolved source. The cavity opening is probably weakly correlated with protostellar luminosity and inclination (see protostar/view). The cavity density is in the order of $10^{-20} \, g/cc$, which implies that it is optical thin in the MIR. This is not the case for the envelope, which gets optical thin only at wavelengths exceeding $\approx 200 \, \mu m$ (adapting a typical density in the order of $10^{-17} \, g/cc$).

- **View:** The inclination is $i = (26 \pm 17)°$. This is smaller than the cavity angle, and therefore G323 can be considered (nearly) face-on. This matches the observations (see Sect. 2.3.2). We note that some models have $i > \Theta_{cav}$, where $i$ is only slightly larger for these models. Since the cavity is curved (and $\Theta_{cav}$ is always defined at $r^{env}_{max}$ in the TORUS models), this still implies that the line-of-sight (partly) passes through the cavity. There are probably weak correlations with the extent of the disk and the outer radius of the envelope. The inclination tends to be smaller (or only slightly larger) than the cavity angle.

| | fit | | view | star | disk | | env | | cav | |
|---|---|---|---|---|---|---|---|---|---|---|
| | $\chi^2$ | d | i | $L_{pre}$ | $m_d$ | $r^d_{max}$ | $r_{max}$ | $\dot{M}_{env}$ | $\rho^{cav}_0$ | $\Theta_c$ |
| | | kpc | ° | $10^4 \, L_\odot$ | $M_\odot$ | au | $10^4 \, au$ | $M_\odot/yr$ | $10^{-20} \, g/cc$ | ° |
| samp | | log | lin | log | log | log | log | log | log | lin |
| mean | 90 | 3.9 | 26 | 6 | $2 \cdot 10^{-3}$ | 680 | 2.4 | 0.032 | 4.5 | 42 |
| $\sigma$ | | | 17 | 1.7 | 64 | 3 | 1.5 | 2.2 | 1.5 | 11 |
| Tmin | 610 | 3.9 | 26 | 6 | $3 \cdot 10^{-5}$ | 680 | 1.6 | 0.015 | 3.0 | 53 |
| Tmax | 180 | 3.9 | 26 | 6 | 0.1 | 680 | 3.6 | 0.070 | 6.8 | 31 |

Table 3.4: Selected parameters of the G323 mean model as fitted with the TORUS model-pool, the full fit-result is given in the Appendix. The lower rows show the parameters for the models with minimized/maximized time-scales. Densities/masses are total values (adapting a gas/dust ratio of 100). Only the distance of the mean model is fitted (min/max are assumed to have the same distance). However, fitting the distance for min/max leads to a similar result.

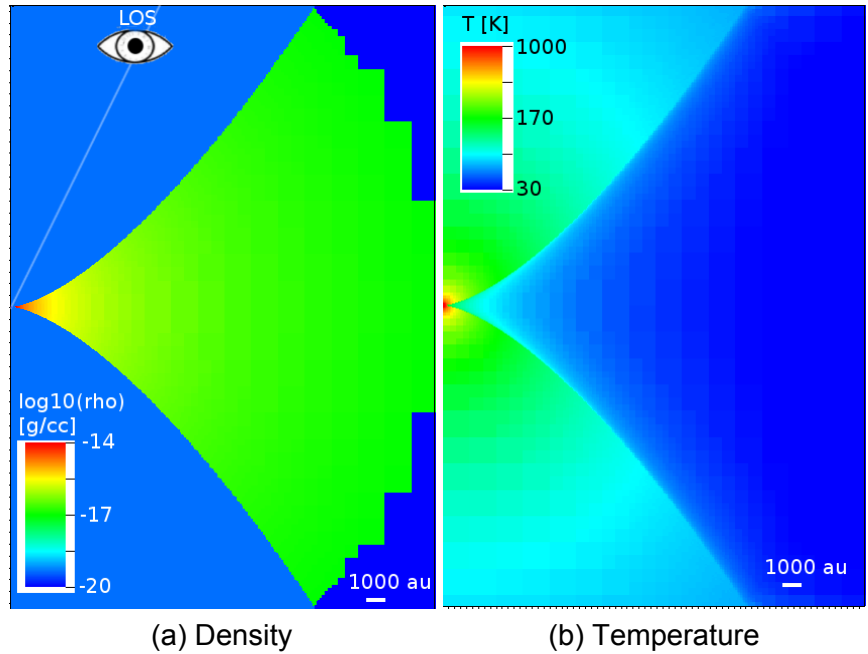

(a) Density           (b) Temperature

Figure 3.9: Density (**left**) and preburst temperature (**right**) distribution of the mean-model. The line-of-sight (LOS) is indicated by the solid line. A zoom-in is given in Fig. 3.16

### 3.3.3.2 The afterglow of the mean model

In the following, we will describe the afterglow of the mean model. We split this Sect. in the following parts: the dynamic SED, the afterglow of the different wavelength regimes, selected light curves, as well as color changes and spectral index evolution. We use the definitions (see Def. 3.1 to 3.3) and line-styles as summarized in Tab. 3.5. In this Section no fit to the HAWC+ data is presented (as follows, in Sect. 3.3.4). We nevertheless emphasize that the (preburst) mean model agrees quite well with the (postburst) data, as shown in a 'quick comparison' at the end of this section.

| NIR | MIR | FIR | maser | bolometric | source |
|---|---|---|---|---|---|
| $< 3\mu m$ | $3..50\mu m$ | $> 50\mu m$ | $15..40\mu m$ | $\int F_\lambda d\lambda$ | $\propto$ Ks |
| dotted or < | | dashed or $\circ$ | | dash-dot or + | |
| peak | | t80 | | t25 | |
| $F^\lambda(t) = F^\lambda_{max}$ | | $E^\lambda(t) = 80\% E^\lambda_{rel}$ | | $F^\lambda(t) = 1.25 F^\lambda_{pre}$ | |

Table 3.5: Overview over the wavelength-ranges (upper part) and timescales (lower part): line-styles and colors are the same for all plots.

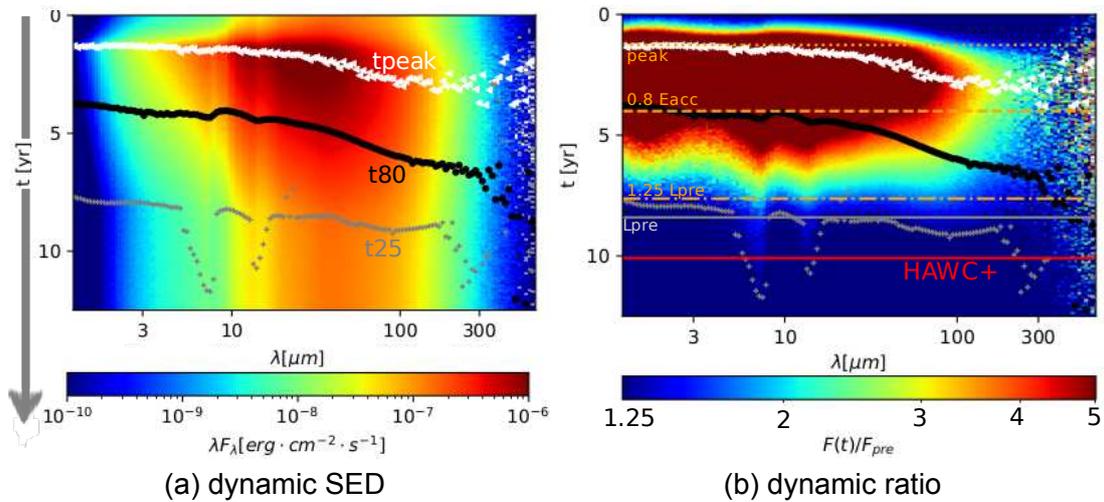

(a) dynamic SED                    (b) dynamic ratio

Figure 3.10: Dynamic SED (**left**) and ratio (**right**) based on $10^7$ synthetic photons. Dedicated times are indicated (see text). At the long wavelength end, there is some scatter due to low photon counts.

**Dynamic SED**   The dynamic SED provides probably the most condensed overview of the entire afterglow. It (the dynamic SED), that is, the flux density over wavelength and time, is shown in the **left** panel of Fig. 3.10. In the **right** panel, the dynamic ratio is given for comparison. Time zero corresponds to the burst's onset. We indicate the peak (white triangles), t80 (black dots) and t25 (gray crosses) for each wavelength.[13] For the variation in the accretion rate, the corresponding times are shown in orange (horizontal lines). Furthermore, the time when the accretion stops is indicated by the gray line. The date of the HAWC+ observation is marked by the red line. SEDs and ratios at these particular times are shown separately in the Appendix (Fig. A.18).

Obviously, the afterglow duration depends on the wavelength. Both $t_{peak}$ and $t80$ increase with wavelength. This is expected since the radiation at longer wavelengths can be attributed to regions more distant from the star (i.e., colder regions). Interestingly, the delay between NIR and FIR amounts to almost 2/3 years for $t_{peak}/t80$ (at $100\,\mu m$). For comparison, the light travel time from the source to the outer edge of the grid is only $160\,d$. The delay between source time and FIR is much more than what could be explained by geometrical/projection effects (and distinct spatial origins) alone, and it rather indicates a measurable slowdown of the energy transfer (toward the 'FIR'-emitting regions) by numerous absorption and re-emission processes due to the high optical depths in between. Toward longer wavelengths ($\lambda \geq 300\,\mu m$) both curves flatten. This is expected because at some point even the coldest and most distant regions are 'processed' (the lowest temperature within the grid is $\approx 20\,K$).

---

[13]As a reminder, the times are as defined in Def. 3.1 to Def. 3.3 and summarized in Tab. 3.5.

The *t*25– curve looks somewhat different. The time increases only slightly, is almost constant in between 30 and 80 $\mu m$ and decays for $\lambda$ exceeding 100 $\mu m$. Although the timescales increase in principle with wavelength, the peak level is much lower at higher wavelengths (at some point hindering a further increase of t25).

At 4 – 8 $\mu m$ there is a 'prominent' feature and a weaker one at ≈ 15 $\mu m$. The MIR-feature can most likely be attributed to the densest regions (disk midplane), which cannot efficiently cool. For weaker bursts/less dense environments, this will not occur (since the disk will not heat entirely/the cooling is faster). However, this requires further investigation. Toward long wavelengths, the synthetic noise increases dramatically (huge scatter due to lower photon counts). We emphasize that t80 is the timescale least sensitive to numerical scatter.

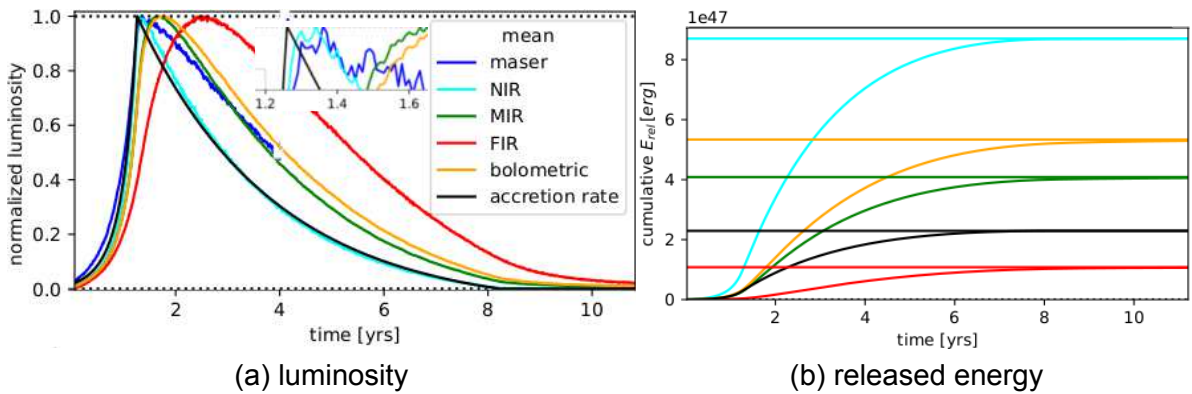

(a) luminosity  (b) released energy

Figure 3.11: Normalized luminosity-curves (**left**) and cumulative released energy (**right**, in units of $10^{47}$ $erg$) at different wavelength-regimes for the mean model. NIR is scaled by a factor of 50 for better visibility. The inset shows a zoom to the maser-peak. The maser-curve was obtained with a separate simulation (see text).

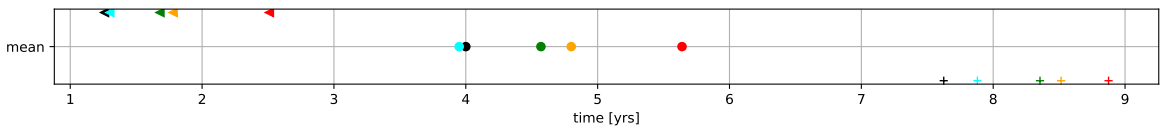

Figure 3.12: Visualization of the time-scales for the mean model, with tpeak (triangles), t80 (dots) and t25 (crosses) from left to right. The symbols and wavelength-ranges (color-coded) are given in Tab. 3.6, vertical spacing is for better visibility.

**Wavelength regimes**    To quantify the wavelength dependence of the afterglow, we split the SED into distinct regions, namely the **NIR, MIR and FIR** (Tab. 3.5). For these regimes, we deduce luminosity curves by integrating the flux (at each time step) within the defined boundaries.

| regime | $\lambda$ $\mu m$ | $t_{peak}$ | $t80$ yrs | $t25$ | $L_{pre}$ $10^4 \cdot L_\odot$ | $L_{max}$ $L_{pre}$ | $E_{rel}$ $10^{47}\,erg$ |
|---|---|---|---|---|---|---|---|
| source | input | 1.4 | 4.0 | 7.6 | 6.1 | 14 | 2.3 |
| NIR | < 3 | 1.4 | 4.0 | 7.9 | 0.2 | 21 | 0.02 |
| MIR | 3 – 50 | 1.7 | 4.5 | 8.1 | 8.3 | 14 | 4.0 |
| FIR | > 50 | 2.6 | 5.6 | 8.8 | 4.9 | 5 | 1.1 |
| bolometric | $-\infty \ldots \infty$ | 1.8 | 4.8 | 8.4 | 13 | 10 | 5.3 |

Table 3.6: Quantities of the mean model for different regimes. A visualization of the timescales is provided in Fig. 3.12.

Due to the steep decay towards longer/shorter wavelengths, a cut-off towards sub(mm)/VIS is not necessary.[14] The released energy within those regimes computes as:

$$E_{rel} \propto \int_{t_0}^{t_{end}} \int_{\lambda_{min}}^{\lambda_{max}} (F_\lambda - F_{pre})\, d\lambda\, dt \qquad (3.2)$$

Fig. 3.11 shows **normalized luminosity-curves** (**left**) for the different wavelength regimes. Only the NIR curve closely follows the source. MIR and FIR curve features **delayed afterglows**, where the delay is on the order of **months (e.g. MIR peak) to years (FIR t80)**. An overview over all time scales is given in the Tab. 3.6.
The **maser-'stimulation' curve** was obtained using a grid with the extent of the disk (this is the region where the masers should reside) and a shorter time step (allowing dust sublimation to be included). We integrate in the wavelength range suitable for radiative pumping of the masers. This yields the best approximation of the 'maser-stimulation' curve possible with our models. It peaks only slightly behind source input and NIR and therefore may serve as a template for the source luminosity variation if no optical or NIR light curve is available (e.g. G358).

The **cumulative released energy** is given in the **right** panel. The NIR curve is scaled by a factor of 50 for better visibility. Most of the flux ($\approx$ 80%) is released in the MIR (see also the Tab. 3.6), and the rest is released in the FIR. The NIR (high extinction) and (sub)mm (Rayleigh-Jeans tail of the SED) almost have no contribution. The apparent total released energy (bolometric value, output) is higher than the burst energy (source, input) as a result of the face-on view.
The dependencies described above *imply conclusions for observing strategies* as time scales and pre- and burst levels differ for different wavelengths. In particular, as the MYSO is much fainter in the NIR and the afterglow is shorter, such bursts are best observed in the MIR and FIR.

---

[14]We made a test (not shown), where we cut off the FIR at 200 $\mu m$, but this yielded the same result.

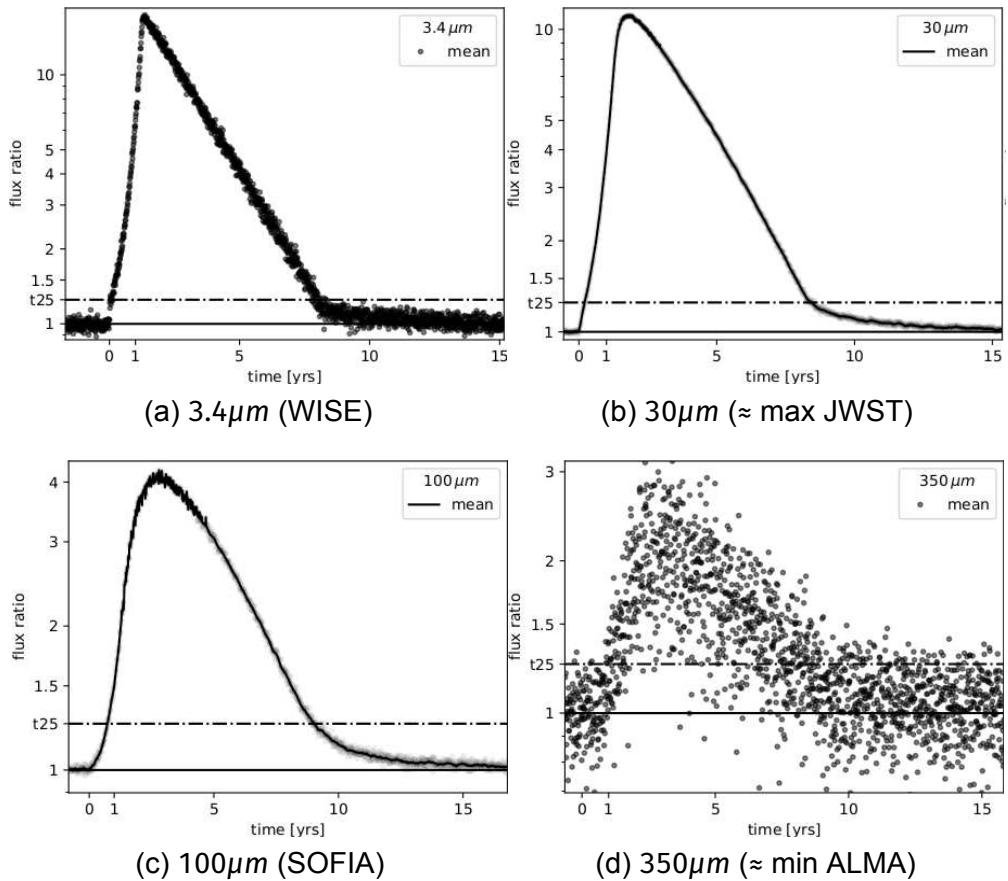

(a) $3.4\mu m$ (WISE)  (b) $30\mu m$ ($\approx$ max JWST)

(c) $100\mu m$ (SOFIA)  (d) $350\mu m$ ($\approx$ min ALMA)

Figure 3.13: Selected ratio curves (G323 mean model) reflecting the wavelength-dependent nature of the afterglow. Wavelengths are chosen according to observing windows of past and current instruments.

**Selected light curves**    In this section, we are switching from the wavelength regimes (above) to selected filters, which could be more easily compared to observations (of different facilities available). We are not showing Ks, 70 and 160 $\mu m$ here, as they are presented in Sect. 3.3.4 (fit) and at the end of this section (quick comparison). Instead, the light curves are shown at (different) dedicated MIR and FIR wavelengths (corresponding to **important past and current missions**). Our selection **reflects the principal behavior**.

In particular, we chose the light curves at $3.4, 30, 100$ and $350\,\mu m$, which are shown in Fig. 3.13. For better comparability, we divide by the respective preburst flux (ratio plot). The corresponding missions/facilities are given below the respective panels. WISE carried out full-sky surveys and therefore could allow a posterior detection of MIR bright *bursts in archival data* (similar to G323). However, not all MYSO bursts are MIR bright, and some might lack detection at short wavebands. With JWST, future bursts could be **followed up** in the NIR and MIR (even highly obscured objects). SOFIA was the best and only instrument to follow up (past) thermal afterglows in the MIR (FORCAST) and FIR (FIFI-LS, HAWC+) until 2022. In the (sub)mm, ground-based observations are possible (archival + future burst detections), e.g. with ALMA (starting at ≈ 350 $\mu m$). At these wavelengths, the numerical scatter increases dramatically (low photon counts), which makes the time-dependent analysis more challenging. The maximum ratio decreases with wavelength. In the (sub)mm, the increase becomes very small (it is proportional to the increase in temperature, e.g., [93]). However, future bursts may be observed with ALMA because of the high sensitivity. The peak shifts to later times with wavelength, while t25 increases slightly between 3.4 and 100 $\mu m$. Bursts can be caught with JWST (if immediate follow-up is guaranteed). However, without the FIR data, the SEDs will be poorer constrained (especially for deeply embedded/MIR dark objects). In this regard, we regret the end of the SOFIA mission. Light curves can only be established with a good cadence of the observations, e.g., with (NEO)WISE or with the JCMT survey in the (sub)mm [98].

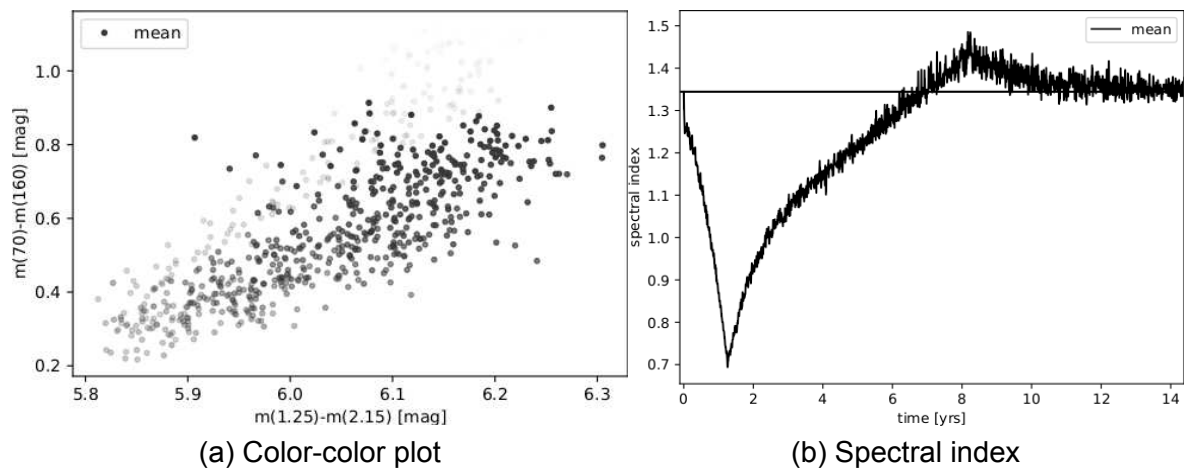

(a) Color-color plot                                      (b) Spectral index

Figure 3.14: The wavelength-dependent nature of the afterglow implies color-changes (**left**) and a change of the spectral index (**right**).

**Color change and spectral index evolution**   The wavelength-dependent timescales imply **color changes**. This has been discussed in the literature (e.g., [90]) and occurs for G323 as well (Wolf+ in prep.). Here, we present this from a more theoretical point of view. In principle, this analysis can be done for all possible combinations of colors.

An example can be found in Fig. 3.14 (left), where m(70)-m(160) is shown over J-K for 8 years after burst onset (this is almost at its end). The time is coded as the transparency level (the most transparent is the earliest). Due to the burst, both colors become bluer by 0.4/0.7 $mag$ (J-K/70-160 respectively) and then red again. The change occurs a bit earlier at J-K (the 'blueing'-curve is slightly 'above' the 'reddening'-curve). We emphasize that this color change is caused only by the difference in the timescales for the different wavelengths, and other effects (e.g. dust clearing) are not included.

Consequently, the slope of the SED changes, implying a change in the **spectral index**[15] (right panel). Although the classification according to [83] (as introduced for low-mass YSOs in Sect. 2.1.2) might not be applicable to MYSOs, the spectral index can nonetheless be determined for these sources as well. The interesting question in this context is how strong the change is and whether misclassifications due to outbursts are to be expected. During the burst, the YSOs look too evolved and slightly (too) less evolved shortly after the event. A misclassification between neighboring classes (due to an outburst) can not be excluded in general, but the time span (when a misclassification is possible) is probably short as compared to the protostellar lifetime. This might also have some implications for low-mass YSOs. Note that the idea that outbursts can lead to misclassifications is not new; it has already been discussed in the literature (e.g., [93]).

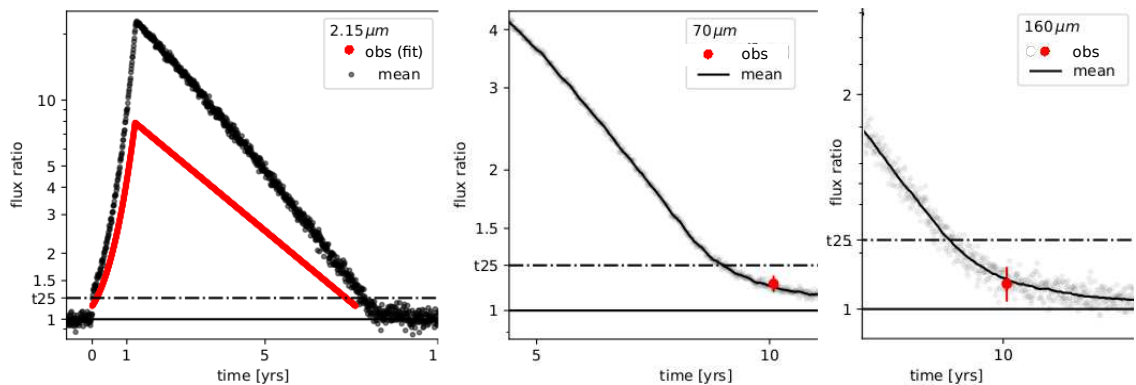

Figure 3.15: Predicted (black) FIR ratio-curves (**middle** and **right**) agree well with the data (red). The (black) dots indicate the simulation scatter. For Ks (**left**) the model overpredicts the increase (red is the fit to the observation).

**Comparison with the data**    Before we start varying selected parameters of the model, we want to show a 'quick comparison' of the mean model as it is with the observations. This will give a first impression of how good the model is already and where it could be optimized.

The modeled **preburst SED matches the observations pretty well** (as shown in Fig. 3.8 and quantified using the $X^2$-value in Tab. A.9). At wavelengths with postburst measurements (i.e., $1.65, 70$ and $160\,\mu m$), the deviation exceeds the observational errors (see Tab. A.8). For Ks, it is on the order of the scattering of the observed data points (showing that the observational error might be too small). For $\lambda \geq 70\,\mu m$ the model slightly underpredicts the observations, which indicates that the mean model has slightly too little cold dust. Nevertheless, the way in which the (static) mean model was derived is reasonable, and the model seems quite good (as it is) already.

---

[15]The spectral index is defined as $\alpha = dlog(\lambda F_\lambda)/dlog(\lambda)$, for $\lambda > 2\,\mu m$ [83].

Fig. 3.15 shows a **comparison between the observed and predicted increase** in flux density at the observing epochs (VVV/VVVX and HAWC+) for all three filters. In the FIR, the agreement is quite good, which indicates that the burst energy estimate is reasonable. For Ks, the predicted increase is about factor 2 too high. This could indicate that the burst energy was overestimated. Then the burst energy needs to remain longer within the system to explain the HAWC+ observations. Or (alternatively) something happens in the innermost regions that 'dims' the visible Ks-flux. Possible reasons for a drop in the Ks flux are that the line-of-sight extinction increases due to disruption or due to the release/entrainment of material into the jet (as expected for an ejection-shock); or a decrease in the emitting/reflecting area due to the burst. A further discussion will follow after the final modeling in Sect. 5.

### 3.3.3.3  First TDRT parameter-study

To discuss the results in a more advanced manner, it is important to verify the impact of the model parameters (burst and dust configuration) to understand which one is (how) important and on what scales (spatial and time-wise). We performed a first parameter study, which we will discuss in more detail below. A brief summary is given at the end of this section.

In the following, we split the discussion into three groups: the static model, view, and burst (as named in 'Step 4').

**Parameters of the static model**   The (standard) model configuration is given in Sect. 2.5.3. We do not vary the protostellar luminosity or the dust composition. Therefore, the free parameters (which could have an imprint) are (split by components):

- disk: $m_{disk}$, $\alpha$, $\beta$, $h_0$, $r_{max}^{disk}$

- envelope: $\rho_{env}$, $r_{max}^{env}$

- cavity: $\Theta$, $\rho_{cav}$

Some of these parameters show a strong influence on the afterglow, while others are not/less important. As these are many free variables, we start this section with a few **considerations:**

The *disk* is probably the most complex component to discuss with respect to its influence. On the one hand, the disk is **deeply embedded within the envelope**, therefore its (own) thermal radiation will be reprocessed many times (especially at short wavelengths and for inclined systems). At longer wavelengths, the envelope may dominate the emission (it is about 30 times as extended as the disk). On the other hand, there can be **large scale effects** as shielding (especially in the equatorial plane) that might alter the heat propagation within the envelope (behind the disk). The disk is flared and will cool/heat via its surface. Its temperature (and density) structure is complex and covers orders of magnitude (temperatures of a few tens of K and densities of $\approx (10^{-12} - 10^{-20})\, g/cc$). Since G323 is seen almost face-on and the disk is poorly constrained and likely lightweight (mean model), its parameters may be less crucial (at least for the FIR afterglow).

The *envelope* density sets the **amount of cold dust** (which can be heated due to the burst), therefore it should have a huge impact on the afterglow. In general, we expect that the afterglow duration increases with the amount of envelope dust. The optical depth is higher (for denser envelopes), which leads to a steeper temperature profile.

The *cavity* density is much below the envelope density, therefore, a large part of **the radiation released will escape quickly through the cavities**. This means that the afterglow duration increases with the density of the cavity and decreases with its opening angle (which increases during stellar evolution). The outflow angle is probably more important (than the density) because of the huge differences in the mean free path length between envelope and cavity density (even if the cavity density is comparably high, the light still escapes 'fast' through it).

To investigate the influence of disk, envelope and cavity on the afterglow, we use a **set of different YSO configurations, all featuring the same burst**.

We tested the influence of the **disk** with 3 settings: **mean, heavy and no disk model** as described below. The favored (mean) model has a rather lightweight disk. However, the density in the inner disk region is not small, as shown in the upper right of Fig. 3.16 (setting c). This is because of the dense envelope (in our models, both the disk and the envelope extend to the dust sublimation radius, whereby the density is the sum of both components). Additionally, (to the mean model), we include two more settings in order to test the influence of the disk: one with a massive disk (top left, $m_{disk} = 1000 \cdot m_{disk}^{mean} \approx 0.1 \cdot M_\star$)[16] and one without a disk (top middle). The envelope is cut at 684 au (i.e., its inner radius is set to the outer radius of the disk) in both cases (settings a and b). The corresponding temperature maps are shown in the lower row.

Some effects are visible in this figure. The disk **midplane heats indirectly** (from the surface and the envelope behind, setting a). The **dense disk shields the envelope behind** (which is only slightly colder for the mean model than for the heavy disk, and much warmer for the no-disk model).

---

[16]This mass is close to the limit, where the disk gets prone to fragmentation.

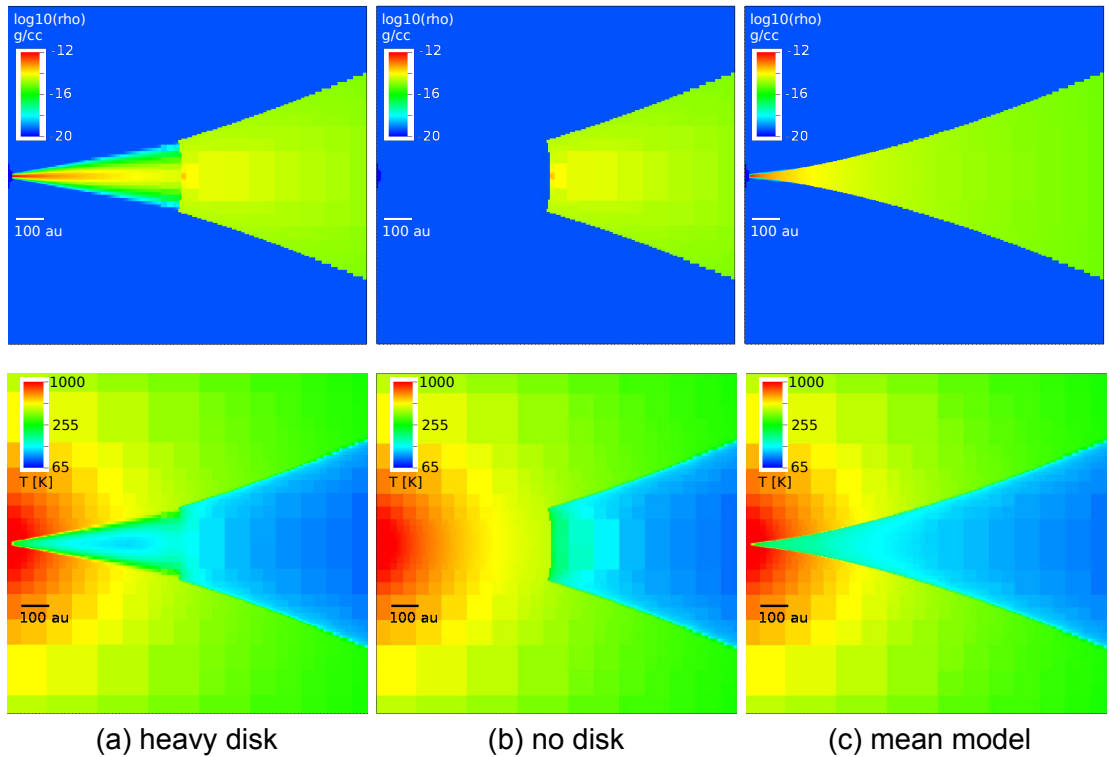

(a) heavy disk                 (b) no disk                 (c) mean model

Figure 3.16: Density (**upper**) and corresponding preburst temperature (**lower row**) of three models (left to right) used to examine the influence of the disk (see text).

Fig. 3.17 compares the preburst SED, a representative FIR ratio curve (70 $\mu m$) and a comparison of the time scales. The preburst SEDs differ in the range between (1 – 10) $\mu m$. The heavy-disk model contains the largest amount of dust at temperatures of a few hundred Kelvins. Therefore, it emits the most. The **temporal evolution** of the mean and heavy-disk model are **quite similar** except from some differences (close to the peak). The timescales (lower row) differ slightly. Even the unrealistic 'no disk '-model behaves surprisingly similar until the flux density drops below ≈ 1.5$F_{pre}$ of the respective preburst value. Then the decay (in the MIR and FIR) is much slower. This is due to the absence of dust in the inner region; the envelope 'behind the disk region' (at ≈ 1000 $au$) can heat to *higher temperatures*. The heated envelope dust acts as heat storage. For what concerns the behavior of **the $K_S$-ratio**: while the heavy-disk and mean models are almost similar (which is not necessarily to expect), the no-disk model features a maximum increase, which is a factor of 2 higher, while the rise and decay are prompt for all three settings (see Fig. A.19).

Together, the imprint of the disk is (surprisingly) small. This holds especially at the time of the SOFIA flight (as long as the dust is not completely removed).

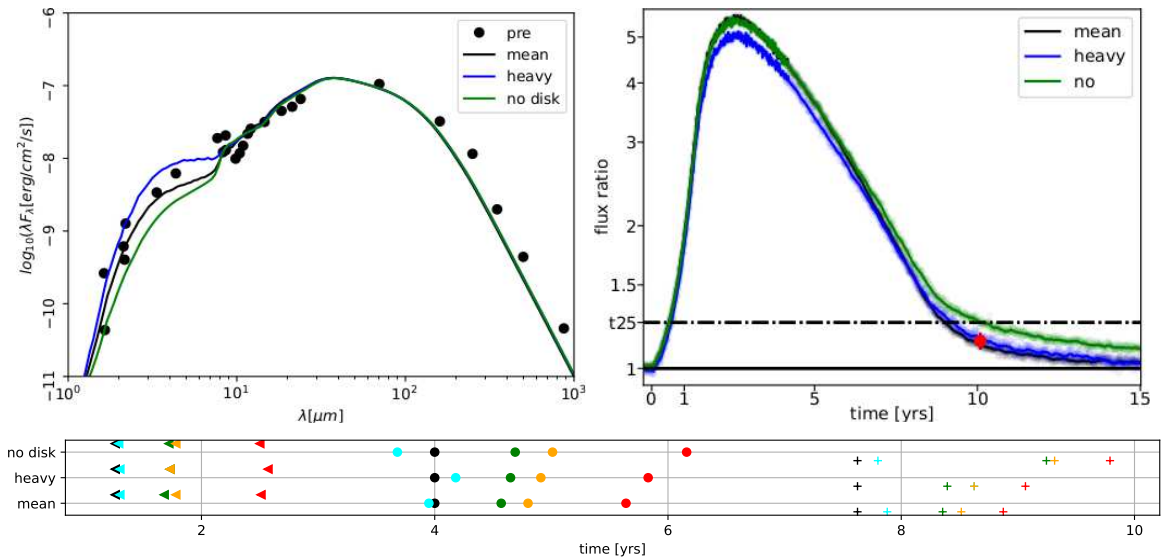

Figure 3.17: Pre SED (**upper left**), representative FIR-light curve at $70\,\mu m$ (**upper right**) and timescales (**lower row**) for the settings in Fig. 3.16. The lower row is similar to Fig. 3.12, but now different settings are compared.

We use two settings to test the influence of the **envelope density**, the **mean** model and a corresponding model, with an envelope density, that is **higher by a factor of 1.5** (slightly smaller than the $1\sigma$-range).The **afterglow-duration increases with the envelope density** (as expected). The preburst level is higher for $\lambda >100\,\mu m$ and in between ()$1 – 10)\,\mu m$ (overlap with the disk). The comparison of both models is given in Fig. 3.18. The time-shifts are comparable with the cavity angle variation (below), despite the comparably 'small' increase (in density), meaning that its imprint is actually strong. No effect of the steeper temperature profile is evident. In fact, the midplane temperature is only slightly steeper inside $\approx 120\,au$.

We tested the influence of the **cavity** with 3 models: the **reference** model, the reference model with a **lower cavity density** and the reference model with a **wider** cavity opening. The low-density model features a density that is factor 7 below the reference. For comparison, this is about 10 times as much as the '$1\sigma$-accuracy' of the mean model. The wide-model features a 20° wider opening ($\approx 2\sigma$) and the same density as the reference model. These considerations regarding the cavity are true, as shown in Fig. 3.19. Obviously, since the line-of-sight is within the cavity, the lower density results in less absorption in the short wave bands (up to $\approx 40\mu m$). The curve of the $K_S$ ratio is not affected. This is, of course, expected since (most of) the Ks radiation is not produced within the cavity (just the 'dimming' factors are different). For a wider angle, the heat escapes faster. Furthermore, there is less dust in the envelope, which could heat up (as indicated by the pre-SED beyond $20\mu m$). The **afterglow is shorter for both cases, and the angle is in fact more important**. Note that reducing the opening angle of the cavity (by 20°) has a stronger effect than increasing it by the same number, since the cavity is already relatively wide open ($\Theta_{cav}$ = 34°). Together, the imprint of the cavity is large (especially its opening angle). The amount of dust within the cavity may be reflected in the pre-SED (especially for face-on configurations).

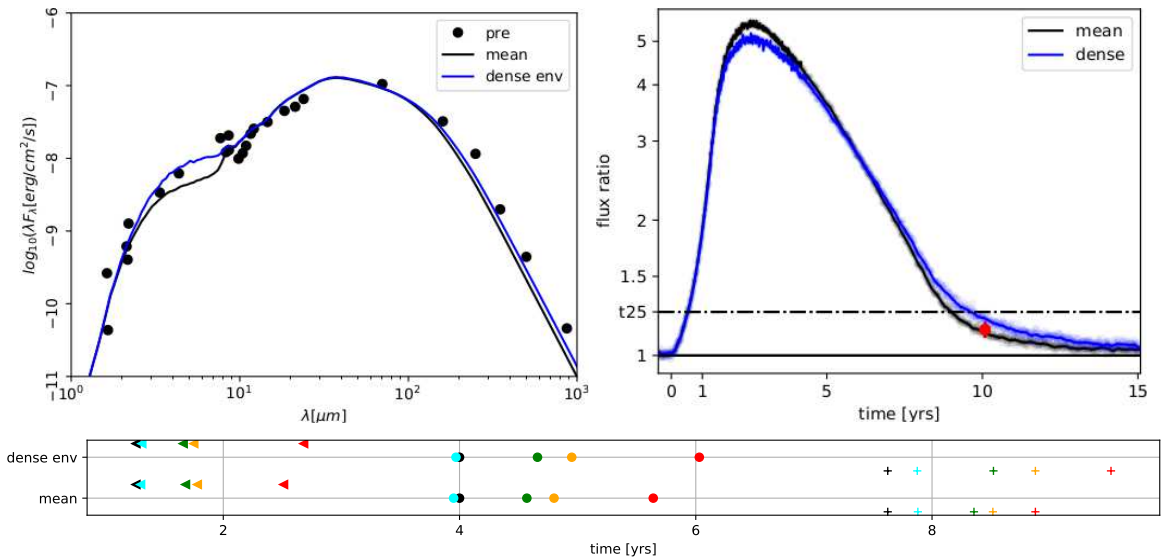

Figure 3.18: Same as Fig. 3.17, but for the mean model (black) and the same model with a higher envelope density (blue).

**Viewing geometry**    The axial-symmetric nature of the dust distribution implies a (strong) dependence of the afterglow on the viewing geometry. The following analysis is essentially divided into 2 parts: the **inclination and the aperture**. We start with the 'easy' case, which is the inclination.

The extreme cases are face-on (like G358 - and also G323) and edge-on (S255IR NIRS3) **inclination**. For edge-on sources, the line-of-sight extinction is maximized. This will especially impact shorter wave bands. However, also, for longer wavelengths, the MYSO environment may not be optically thin. Our simulations show that the flux densities become independent of the inclination only when the wavelength exceeds $\lambda \approx 200\,\mu m$ (see Fig. A.11). The effects are a **strong delay at short wavelengths** (NIR peak time) + **much lower flux densities** (the inclined MYSO mimics a less evolved -NIR dark source). The FIR is less affected, as shown in Fig. 3.20. The strong (NIR-) delay has some interesting implications, as this means that the maser flux (which is an 'intrinsic' property, that is 'independent' of the inclination) can rise prior to the NIR even for passive disks in case of highly inclined systems. TDRT simulations may provide a unique tool to measure systemic inclination (further simulations are needed to prove this).

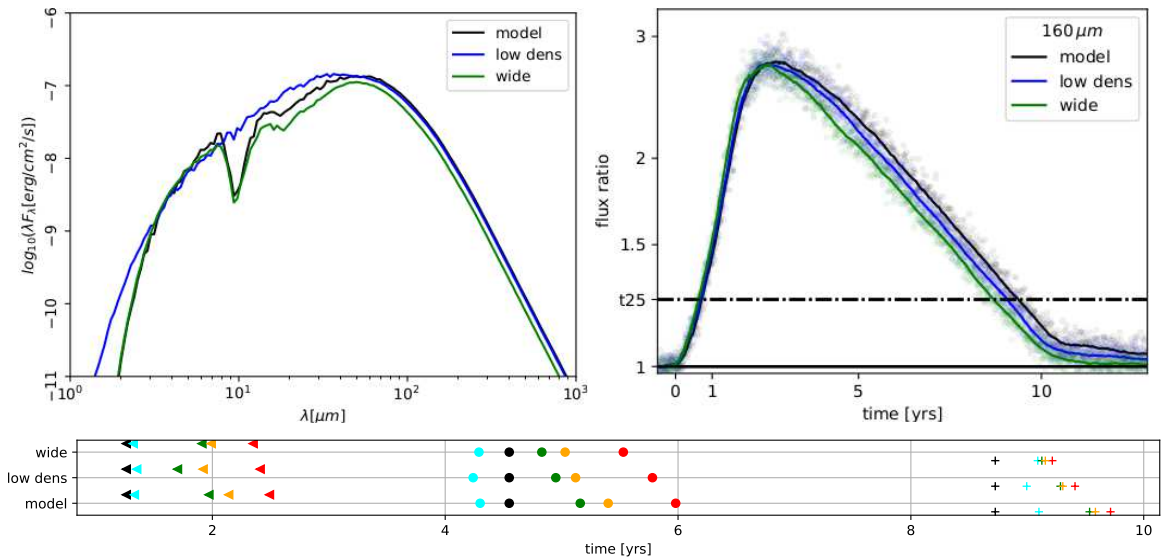

Figure 3.19: Same as Fig. 3.17, but for reference 'model' (black), the same model with a lower cavity density (blue) and the same cavity density but a wider cavity opening (green).

In addition to the inclination, we want to tackle another 'view' parameter: the **aperture**. It is illustrative to 'measure' the total released energy on different (circular) apertures, as the radiation within the different wavelength regimes is produced within different spatial regions (e.g., [70]). We did this for an extended model in a face-on and edge-on configuration on apertures with radii in between 30 and $10^5$ $au$. Interestingly, for the face-on case, all NIR radiation is already included in the smallest apertures, whereas it increases with aperture for the edge-on case. MIR and FIR fluxes increase in both cases with aperture (because they originate from the extended envelope). The shorter wavebands are heavily delayed for edge-on on all apertures. A plot with a more detailed explanation can be found in the Appendix (Fig. A.22). Our **conclusions can be summarized as follows:**

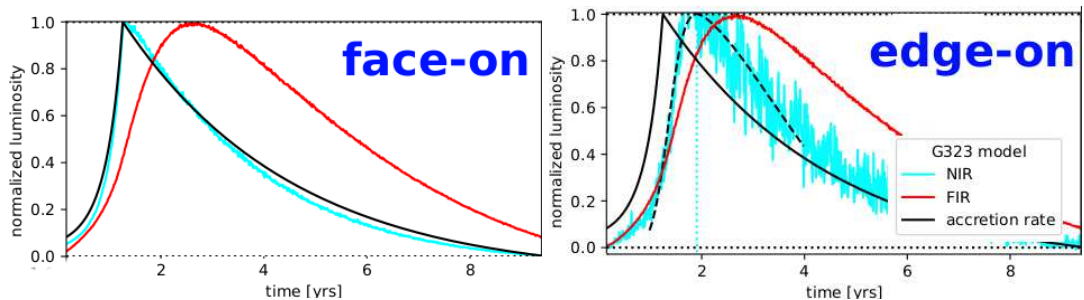

Figure 3.20: NIR and FIR luminosity increase for face (**left**) and edge-on (**right**) inclination. The NIR light curve in the right panel has been fitted (dashed black line), to obtain the peak time (vertical dashed cyan line). This Fig. was presented at the 'Advisory Board Meeting' 2023.

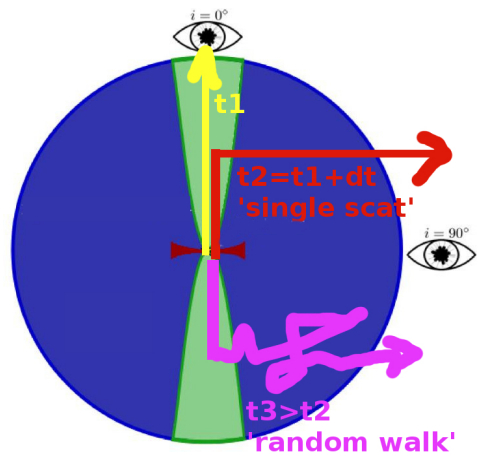

Figure 3.21: Illustration of different NIR-photon paths for face on (yellow) and edge on view (red, violet). Figure modified from [94, Fig. 1]. The NIR photons are produced in the innermost regions, they can pass through the cavity, but are subject to numerous scattering processes, when moving through the (denser) envelope regions (thereby getting 'delayed').

- Most of the radiation (80%) is released within $< 100$au (NIR), $\approx 10^4$au (MIR), $\approx 7 \cdot 10^4$ au (FIR) respectively. Therefore, **the extended outer envelope is crucial for longer wavelengths**. Note that the reference model is face-on, but less dense than the mean model (therefore the values may be slightly different).

- The time scales generally increase with aperture, as long as the apparent released energy does. This can be violated towards small apertures (with sizes matching the disk's extent), since the disk is 'over-dense' (therefore increasing the delay). However, this requires further simulations (with a better time-sampling).

- The **NIR radiation streams along the cavity**. It cannot pass through the dense disk and is heavily delayed (maybe even by years) within the envelope. This is illustrated in Fig. 3.21. The NIR photons are produced close to the star, and in the face-on case they (more or less) follow the cavity (yellow path). In the edge-on case, they start following the cavity until they get scattered at the cavity walls and pass through the envelope on random walk paths (violet). Even for a straight envelope path (red), they would be 'delayed' (but not as much). However, the fraction of NIR photons visible in the edge-on case is small (as most of the NIR-radiation is either reprocessed or escapes through the cavity).

**Burst parameters**  Despite the dust configuration and view, the characteristics of the burst (i.e., its **duration, shape, and amount of released energy**) also imprint on the afterglow. The easiest way to test the influence of each parameter is by means of rectangular bursts, which is defined by its strength and its duration. Fig. A.21 shows a model featuring a rectangular outburst with $E_{acc}$ = $2.3 \cdot 10^{47}$ *erg* and a duration of 1 yr (**left panel**). The other panels are the same model, but with an outburst with 3 times that energy, which is released within the same time (**middle**) and over 3 yrs (**right**). Obviously, more energetic outbursts (middle, right) imply longer afterglows. The afterglow appearance (even after the burst end) depends on the burst shape. The middle and right panel differ despite sharing the same $E_{acc}$. This is a proof that the afterglow implies a **record of the burst history**.

The comparison of the middle and right panel shows, that the time scales are longer for the burst with the extended duration. This is because the bursts start at the same time and the 3 yr burst (right) is ongoing, when the 1 yr burst is already over (middle). If instead the bursts would end at the same time the time scales would be longer for the higher source-luminosity increase (middle panel). This is because the longer burst (right) would have started 2 years before the short one (middle) and the injected energy could already partly leave the system. The higher the peak values, the more radiation is emitted at shorter wavelengths as a result of the higher temperature gradients involved. We emphasize that there is **no easy way to scale the results**. In principle, for each burst, a new model has to be established. This is usually not possible, and in reality it makes sense to use a proxy for the accretion-rate variation. The most simple and yet successful applied approaches are (scaled) NIR-Light curves (Ks for G323) and total maser flux (methanol Class II maser for G358). Other future possibilities are reconstructions from LE/maser regions or iterative approaches.

**Degeneracies**  After the discussion of possible imprints on the afterglow of the burst, the viewing geometry, and the dust configuration separately, possible degeneracies will be addressed below.

Even if there were not two identical afterglows, measurements always come with a certain **accuracy, wavelength, and time coverage** (they are snapshots). This is illustrated in Fig. 3.22, which shows a fictive SOFIA postburst SED (black dots) together with a set of (vastly) different models, which are all able to explain the 'fake data'. It is possible to overcome the degeneracy (at least partially), by which the following things may help: a good wavelength coverage, and sufficient time sampling, at best close to the peak, small error-bars (observational and numerical), prior knowledge about view, burst, and system, as well as the combination with other observations (for example, light echoes, maser relocation). It would be nice to judge which of those points are the most crucial. But in reality, it is difficult to weigh the points against each other since their influence is vastly different, and they 'act' on vastly different time and spatial scales. First tests indicate that knowledge about the burst is very valuable (i.e., start/end dates + at best a reasonable burst template). But the other points addressed are certainly also important, whereby some (of these points) merely help the pre-model and some the time-dependent models. A TDRT-database can be a valuable tool for testing the degeneracy in a more systematic fashion.

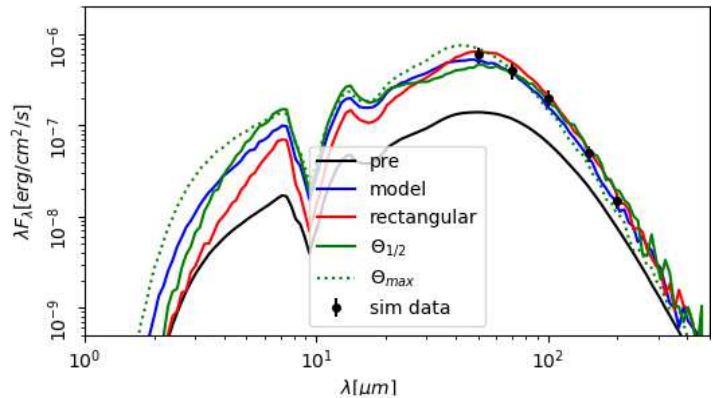

Figure 3.22: Illustration of the degeneracy among different models. The same (fictive SOFIA) postburst SED can be recovered by a bunch of vastly different bursts and YSO environments. This Fig. was presented in the 'Tautolloqium' 2022.

**Summary of the parameter-study**

- Higher densities feature longer afterglows.

- The (entire) disk (and the inner parts of the envelope) show surprisingly little imprint on the SED evolution longward 10 $\mu m$ (due to reprocessing). At the end of the relaxation process, this changes. ).

- The disk is important for the masers (possible sites and amount of pumping radiation). The innermost structures impact the afterglow between 1 and 10 $\mu m$ (Ks curve).

- The opening angle of the cavity plays a crucial role, as it determines how quickly the released energy can leave the system.

- Most of the FIR and MIR emission is produced within the extended cold outer envelope.

- The NIR/MIR afterglow can be delayed by months for highly inclined systems (and the peak fluxes are orders of magnitude lower).

- More energetic bursts imply longer afterglows. The burst shape imprints the appearance of the afterglow.

- Degeneracies between models are possible. They can be overcome with sufficient wavelength and time-sampling. Knowledge about the system (local dust distribution, burst, or/and view) further helps.

### 3.3.4 Max/Min model

We can now use the results of the parameter study to model the accretion outbursts of real MYSOs, such as G323. In this section, we present the results for models with **minimal/maximum afterglow time scales** (with parameters varied in the $1\sigma$ error ranges, as described in Step 5 of Sect. 3.3.2). We start with the timescales for **'the particular burst'** (with $E_{acc}$ = 2.3 · $10^{47}$ *erg*) and end with the **estimate of the burst energy**, whereby we vary the input accretion rate (and hence the burst profile) to match the HAWC+ data (as described in Step 6) for all three settings.

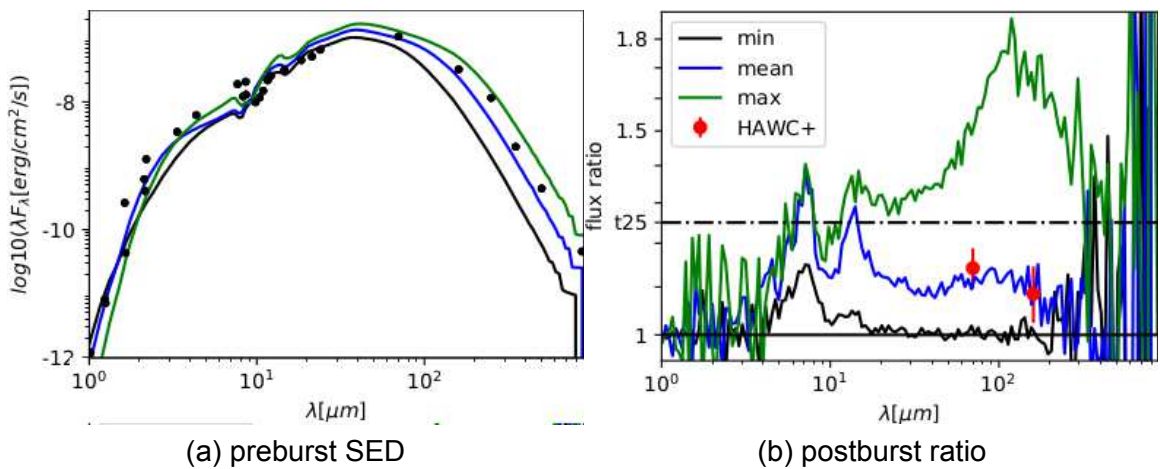

(a) preburst SED                                         (b) postburst ratio

Figure 3.23: Preburst SED (**left**) and ratio at the HAWC+ epoch (**right**) for the three models (min, mean, max) used to define minimum/most likely/maximum time-scales

The preburst SEDs and the flux ratios at the HAWC+ epoch are shown in Fig. 3.23 (left and right, respectively) for all three settings (color-coded) + the 'usual' 2.3 · $10^{47}$ *erg*-burst. The 'min' model cannot reproduce the FIR emission due to the lack of cold dust. Therefore, the upper limit of the burst energy (as derived with this particular setting) is likely much too high. The parameters of the models are given in Tab. A.9 (all) and summarized in Tab. 3.4 (selection). Out of the three models, the mean model fits best as expected.

### 3.3.4.1   Afterglow- time-scales

To compare the afterglow of the three particular settings, we use some tools already introduced in Sect. 2.2.5. The **dynamic SEDs** probably provide the best overview. They are shown in Fig. 3.24. We emphasize that all differences are caused by the different MYSO configurations (the burst and view are the same in all cases). It is surprising how much the local dust configuration actually imprints the afterglow (especially in the MIR/FIR).

The lower panel of Fig. 3.25 gives an overview of **the time scales** (Def. 3.1 to 3.3) in the different wavelength-regimes[17], which span a remarkably wide range (see also Tab. 3.7). The FIR afterglow can be observed for more than 6 $yrs$ longer for the 'max' setting (t25) and even the (FIR) peak time varies by a year.

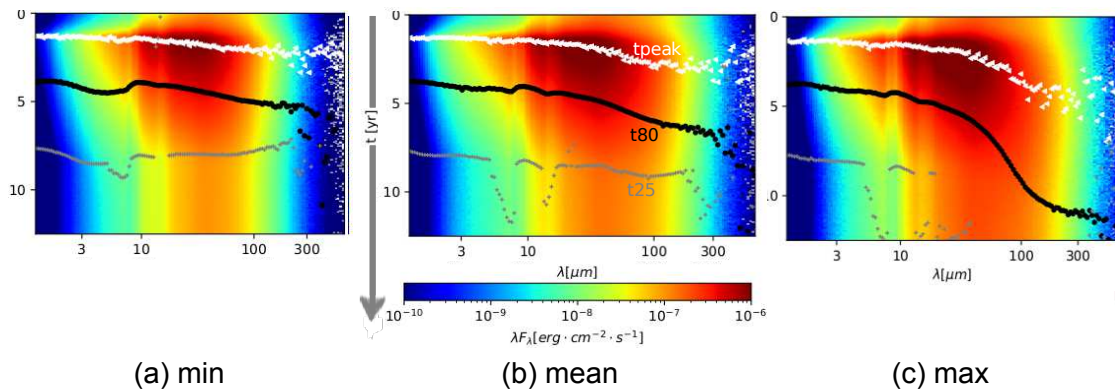

(a) min                        (b) mean                              (c) max

Figure 3.24:  Dynamic SED of min (**left**), mean (**middle**) and max (**right**) model showing possible afterglows for G323 (as they depend on the local dust-distribution). The burst is the same for all cases. The differences of the time-scales (horizontal 'lines') are remarkably large (especially in the FIR). The dynamic ratios are provided in Fig. A.24.

The **luminosity curves** within the different wavelength regimes can be found in the Appendix (Fig. A.25 upper row). They show what we just discussed. The lower row of the Fig. shows the apparent cumulative released energy. The amount released at short/long wavebands decreases/increases with the afterglow duration (from min to max). Interestingly, the total (bolometric) value is not the same (despite the fact that the energy input is the same). Instead, it increases with the duration of the afterglow. For the min/max setting, more/less radiation can escape without detection (i.e., outside the line-of-sight) at short wavelengths (lower/higher densities, wider/smaller outflow angle). Note that a similar effect also occurs if the density is fixed and the opening of the cavity is varied (see Fig. A.20).

---

[17]As summarized in Tab. 3.5.

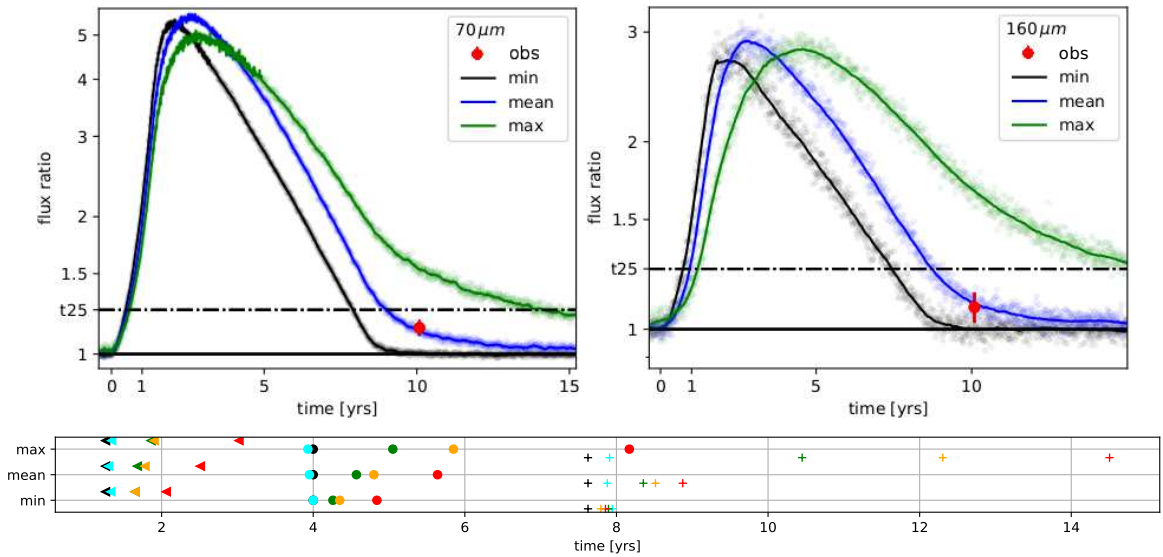

Figure 3.25: Light curves (upper) and timescales (lower row) for three models within the 1σ-confidence intervals of the preburst model, indicating the maximum range of possible afterglow time-scales for a particular burst.

| regime | λ | $t_{peak}$ | t80 | t25 | $L_{pre}$ | $L_{max}$ | $E_{rel}$ |
|---|---|---|---|---|---|---|---|
| | μm | | yrs | | $10^4 \cdot L_\odot$ | $L_{pre}$ | $10^{47}\,erg$ |
| source | input | 1.3 | 4.0 | 7.6 | 6.1 | 14 | 2.3 |
| NIR | < 3 | 1.4/1.3 | 4.0/4.0 | 7.9/7.9 | 0.1/0.1 | 21/26 | 0.02/0.01 |
| MIR | 3 – 50 | 1.7/1.8 | 4.2/5.0 | 7.9/10.5 | 6.2/10.1 | 15/12 | 3.2/4.9 |
| FIR | > 50 | 2.1/3.2 | 4.8/8.2 | 7.9/14.5 | 3.1/7.4 | 5/4 | 1.1/1.8 |
| bolometric | −∞ … ∞ | 1.7/2.0 | 4.3/5.9 | 7.8/12.3 | 9.4/17.6 | 12/7 | 4.0/6.9 |

Table 3.7: Same as Tab. 3.6, but for the min/max values.

At the **observing epoch**, the 'max' configuration predicts an increase by about a factor 1.5 (Fig. 3.23 right and Fig. 3.25 upper row), which is clearly too much. The 'min' setting reaches the prelevel (in the FIR) about one year before the observation (≈ 9 yrs after the burst onset). For the 'mean' model, the agreement is quite good (as shown already). Remarkably, the Ks ratio curve (Fig. A.23) looks almost the same for all models (it overpredicts the observed light curve, as discussed already for the mean model in Sect. 3.3.3.2). Next, we will vary the accretion rate input to estimate the burst energy for all three configurations, starting with the mean model.

### 3.3.4.2   Burst parameters

**Burst energy**   Fig. 3.26 shows the light curves for the **mean model, featuring bursts with different energies** (color-coded). The respective Ks-light curves are given in the Appendix (Fig. A.27).

The modeled ratios for each burst energy input (in both FIR filters) are plotted in the left panel of Fig. 3.27 (dots). The ratios can be fitted with a linear function in the given range, whereby the y-axis intercept equals 1 (as no energy input means no flux increase). In the right panel, we show the reduced $X^2$-values as a function of the burst energy input as obtained from the expected ratios (fit, black curve) and "directly" from the models (dots). With this, we can refine our previous estimate of the burst energy (i.e., $2.3 \cdot 10^{47}\ erg$). **The refined value equals** $E_{acc} = (2.4 \pm 1.0) \cdot 10^{47}\ erg$, which agrees very well with our initial guess. The $1\sigma$-confidence-intervals extend until the $X^2$-value becomes worse than $1 + X^2_{min}$ (in accordance with [74]).

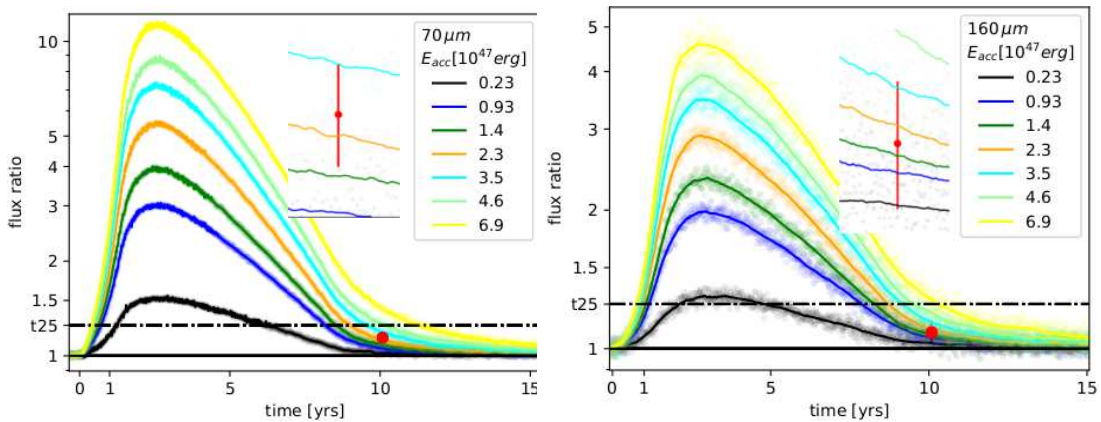

Figure 3.26: Light curves for the mean model, featuring bursts with different energies (colors). The energy is given in $10^{47}\ erg$. The inset shows a zoom in the region of interest.

The same exercise can be done for **other settings, which gives** $30 \pm 12$ **(min)** $0.41^{+0.18}_{-0.36}$ **(max) in units of** $10^{47}\ erg$ (see Fig. A.28, as well as Fig. A.29). Limits are obtained as before. The upper limit of the min setting (i.e., $42 \cdot 10^{47}\ erg$) should be taken with care, as it is an extrapolation. But we will discard it anyway, as it does not agree with the observations (see below). For comparison, no bursts (that is, ratios of 1 in both bands) lead to a $X^2$-value of $\approx 5.3$ (adopting typical errors for the simulation). This value is for the mean model; for the max model it is $\approx 2.7$ due to a larger scatter of the simulation. For the min/max-setting, the $E_{acc}$ values are about factor 13/6 above/below compared to the mean model. The higher value is not likely, as the preburst fit of the min model is much worse. Additionally, this setting overpredicts the Ks flux by far. However, it may still serve as a conservative upper limit.

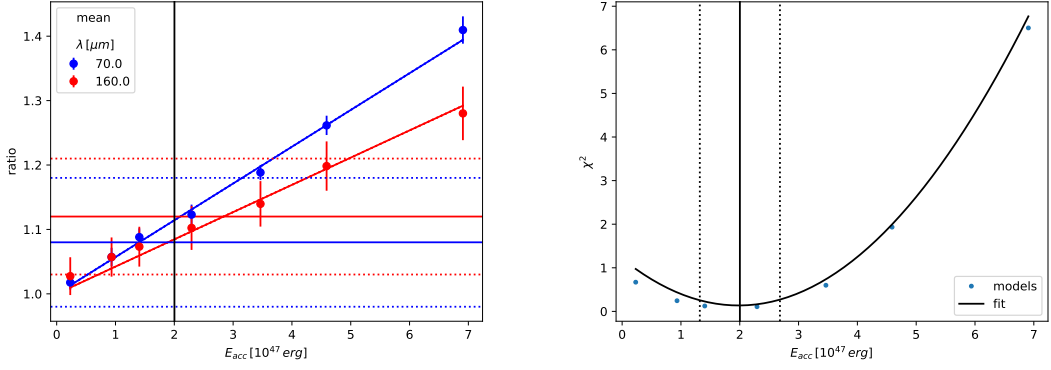

Figure 3.27: Burst energy estimate for the mean model: The flux-increase depends linearly on the burst energy at both bands (**left**). The horizontal blue/red (solid) lines are the observations (with confidence-intervals, dotted) at 70/160 $\mu m$. The vertical black line indicates the fitted $E_{acc}$. The $\chi^2$-minimization (used to determine $E_{acc}$) is shown on the **right** (see text).

The best match to the Ks curve is given for $E_{acc} = 9.3 \cdot 10^{46}\ erg$ (Fig. A.27, mean and max setting). This value lies well within the determined range and would imply a dust configuration in the range between mean and max. In principle we could do a $\chi^2$ minimization (similar to what we did before) for the Ks curve as well. However, small-scale changes might influence the appearance of the Ks curve, and the FIR is more stable.

| model | $E_{acc}$ $10^{47}\ erg$ | $\chi^2_{min}$ | $M_{acc}$ $M_{Jup}$ | $\dot{M}_{acc}$ $10^{-3} M_\odot/yr$ | $L_{peak}$ $L_{pre}$ |
|---|---|---|---|---|---|
| min | $30 \pm 12$ | $10^{-7}$ | $230 \pm 110$ | $27 \pm 12$ | $310^{+130}_{-140}$ |
| mean | $2.4 \pm 1.0$ | $0.04$ | $19 \pm 9$ | $2.1 \pm 1.0$ | $14^{+8}_{-6}$ |
| max | $0.41^{+0.18}_{-0.36}$ | $0.8$ | $3.2^{+1.6}_{-2.9}$ | $0.36^{+0.18}_{-0.33}$ | $2.7^{+0.8}_{-1.5}$ |
| 'Ks-based' | $0.9^{+2.5}_{-0.8}$ | | $7^{+20}_{-6}$ | $0.8^{+2.2}_{-0.7}$ | $5.4^{+16.6}_{-4.2}$ |

Table 3.8: Summary of the burst parameters for all three settings and $\chi^2$-values for the best $E_{acc}$. We consider the Ks-based range (highlighted) most reliable (see text).

The **best estimate of the burst energy is probably** $E_{acc} = (0.9^{+2.5}_{-0.8}) \cdot 10^{47}\ erg$, which is based on the Ks value and covers the whole **range spanned by the 'mean' and 'max' setting**. This would correspond to a peak luminosity of $L_{max} = (3.2^{+10.3}_{-2.5}) \cdot 10^5 L_\odot$, and hence an increase by factor $5.4^{+16.6}_{-4.2}$ (or $\Delta L \approx 2.6 \cdot 10^5 L_\odot$) A more detailed discussion of the derived burst energy estimate follows in Sect. 5.2. A database with models within the above range will help further refine our values.

**Accreted mass and mass accretion rate**   From the burst energy, the accreted mass and the mass accretion rate can be inferred with $M_{acc} = \frac{E_{acc} \cdot R_*}{G \cdot M_*}$ and $\dot{M}_{acc} = M_{acc}/t$ (similar to Sect. 3.2.4.1), where we adapt the ZAMS values for the protostellar mass and radius. For a star with a luminosity of $(6.1^{+3.9}_{-2.5}) \cdot 10^4 L_\odot$ (modeled), the ZAMS mass amounts to $(23^{+5}_{-4}) \cdot M_\odot$ and the corresponding radius $R_* = (6.5^{+0.9}_{-0.7}) \cdot R_\odot$ [139, Eq. 1 and 2]. Since G323 is probably in a more evolved state (NIR bright), this assumption may be justified. The resulting parameters are summarized in the Tab. 3.8. The **accreted mass is between 1 and 30 Jupiter masses** (or 300 and 9000 earth masses), which is 2 to 60 times the value of G358 and 1 to 20 times the estimate of NIRS3 [25], where the accreted mass of NIRS3 might be underestimated (since the burst was still ongoing, when $M_{acc}$ was derived. The errors are dominated by the error of $E_{acc}$.

The **mass accretion rate** for the whole burst duration ($t$ = 8.4 $yrs$) is **of the order of** $10^{-3} M_\odot/yr$ (similar to the other known outbursts). If the accretion rate is calculated for the mean accretion phase (which could be approximated by $t80 \approx 4\,yrs$, that is, the time when 80% of the total burst energy is released), the value (for the accretion rate) doubles.

### 3.3.5   Our results in the context of current observations

This section is meant to give an overview of our results on G323 and to set them in the context of current knowledge about this system in particular and the picture of massive star formation in more general.

G323s burst is the **most energetic MYSO outburst observed so far**. It is 3 to 100 times more energetic as G358's burst and more than 10 times longer. It is up to 30 times more energetic as the NIRS3 burst and 4 times longer. We derived a burst energy of $E_{acc} = (0.9^{+2.5}_{-0.8}) \cdot 10^{47}\,erg$, which would imply an accumulated mass of $(7^{+20}_{-6}) \cdot M_{Jupiter}$ and an accretion rate of the order of $10^{-3}\,M_\odot/yr$. The rise time is $\approx 1.2\,yrs$, which is about 3 times longer than for G358.
The event can possibly be explained by the same trigger scenarios (that is, planet disruption or fragment/small body accretion) as G358 (despite the different bursts). We derive a disk mass of $(3 \cdot 10^{-5} - 0.11)\,M_\odot$, which is likely too small as an 'artifact' of our modeling. Although our models (dust configurations) do not support the conclusion that G323's disk is prone to GI, this could still be the case.

There is another interesting possibility. The accreted mass was derived, assuming the ZAMS value for $R_*$ (and $M_*$). This seemed realistic, as G323 is likely more evolved. Massive protostars, however, could be extremely bloated before they contract towards the ZAMS (see e.g., [63]). Then the accreted mass would be much higher. The protostellar bloating is supported by the appearance of the observed maser pattern. Bloated protostars may be unstable to pulsations, whereby such pulsations could lead to midterm maser periodicity (in the order of $\approx 100d$) [68]. A similar periodicity has also been observed for G323 [113, with a period of 93.5 days, for Class II methanol maser].

The period of the observed midterm maser variability would correspond to a luminosity of $\approx 4 \cdot 10^4 L_\odot$ [68, Eq. 1, for a spherical accretion model] or slightly more for a thin disk model [68, Fig. 2, blue stars]. For both models (spherical and disk), the protostar is unstable to pulsations, and thus may oscillate. This agrees with our models and observations.

For the spherical accretion model, the oscillation period and the protostellar mass and radius can be related according to [68, Eq. 2 and 3]. In that case, the radius would be as large as $R_* = 336 R_\odot$ and the mass would be $17 M_\odot$ (slightly below the ZAMS value). With our estimate of the burst energy, this would give an accreted mass of roughly half the solar mass (instead of $7 M_{Jupiter}$). In reality, the spherical case is certainly not fulfilled. But it can be considered a limit.

The expected bloating depends on the accretion rate for the thin-disk model [63, Fig. 12]. The figure gives $R_*$ as a function of $M_*$ for different mass accretion rates. If G323 is heavily accreting ($\dot{M} \approx 4 \cdot 10^{-3} M_\odot / yr$) it could be bloated to a few 100 solar radii ($\approx 0.3$ times the value of the spherical accretion model), but if the protostellar accretion rate is $\approx 10^{-4} M_\odot / yr$ its radius could be close to the ZAMS value (as assumed in Sect. 3.3.4.2). The accreted mass is probably between a few Jupiter masses and a few tenths of a solar mass. Therefore, we suggest that the accreted object was either a heavy planet or a small companion (which was possibly just forming). The accretion of a companion could not be excluded with the observations (no radial velocity measurements, too tight to be resolved at any wavelength). On the other hand, planet accretion is expected to occur frequently [143]. High-resolution hydrodynamical simulations of collapsing massive cloud cores [108] show that primary formation is accompanied by the formation of a disk with spiral arms (which could be globally stable and locally unstable). Fragmentation occurs frequently, whereby fragments heavily interact, and some migrate inward. Such fragments can weigh $\approx 1 M_\odot$ and may evolve into hydrostatic cores (which may become accreted) or even secondary cores (which likely form spectroscopic companions). Although those simulations have computed outbursts with longer durations (about a few decades) and smaller primary masses, they might still be explained by similar mechanisms.

# The TFitter

## 4.1 The TDRT model pool

As a **last step**, we developed a method to generate and fit **model pools of TDRT datasets**, based on G358 models[1].

The aim of a TDRT model pool is to obtain more reliable burst parameters, as well as constraints on the environment and viewing geometry, by self-consistently **fitting all epochs simultaneously with a large sample of models**. Hopefully, this will also enable smaller confidence intervals and help establish connections between different parameters.

What we present here can be considered the alpha version, with a 'small' set of (nonperfect) G358 models. Thanks to the grant provided by the DLR, the method will be applied to G323, G358, S255IR-NIRS3 and G24.33+0.14 (G24) soon.

The structure is as follows. We start with an introduction of the method (from the creation of the data set to the final result, Sect. 4.1.1). A more detailed description of the **TFitter** is included in the Appendix (Sect. A.4.1). We present a benchmark case, where we use the TFitter to fit a set of test SEDs (Sect. 4.2.1). Finally, the test pool is fitted to the real G358 data (Sect. 4.3). We do not consider the result of this fit as better values for the burst parameters/dust configuration, as there are several issues, which will be subject to our next project.

In the following, we use the suffix 'T' in order to indicate that it refers to a TDRT (not to static RT) property (e.g., TSED, Tdir, Tmodel, and TFitter).

---

[1]As the G358 dataset includes burst- and postburst measurements (both clearly exceeding the respective preburst values) it is (better) suited to establish the Tpool method. Therefore, we use this dataset, even if (most of) the other time-dependent models that have been presented are for G323.

### 4.1.1   Method

**Summary of the steps**   These are the basic steps that we performed to fit sets of observing epochs with a pool of TDRT models. We describe it in the example of G358, but the principle remains the same for other sources.

1. We established a preburst data set. (**see Step 1**)

2. We fit the preburst SED. (**Step 2**)

3. TDRT simulations have been performed for selected models and bursts. (**Step 3**)

4. Test-data was created (only necessary for the benchmark). (**Step 4**)

5. We fit the epoch data (all real/benchmark pre, burst- and post-SEDs) with the TFitter. (**Step 5, Sect. 4.2.1 to 4.3**)

**The steps in detail**

**Step 1**   The **TORUS preburst database** is generated similarly to what is described in Sect. 3.3.2 (for G323) but with extended models and ap = 3" (as discussed in [130]). For simplification, we use pure silicate models with a size of $0.12 \mu m$, which is not appropriate for real systems, but sufficient for testing.

**Step 2**   We **fit the preburst SED** with d, Av as given in Sect. 3.2.3 (Step 1) and [130]. We flag the data points at $\lambda > 500 \, \mu m$ because the simulation is too noisy. We use a two-step approach to reduce the computational demand (similar to G323, see Sect. 3.3.2, Step 2).

**Step 3**   The selection of Tmodels (time-dependent models) is probably the most critical step. In principle, we could run a subset of burst models for each configuration (i.e. each preburst setting). However, this will be very demanding in terms of time and storage. To reduce computational demand, we will perform **TDRT simulations only for the 100 best preburst models**. This leads to unevenly sampled parameter-spaces, which is of course a drawback of this method, but the models that do not fit the preburst SED will not perform well in the fitting (of all epochs) anyway.

We use **10 Bursts with different luminosity**, whereby the total maser flux serves as a template for the source luminosity variation. This is the best guess, since G358 is MIR dark.

To simulate the different burst energies, we start simply by scaling the template with 10 scaling factors linearly spaced between 1 and 5.5. Factor 1.5 equals the static estimate of $2.8 \cdot 10^{45}$ *erg* (for the mean luminosity $L_*$ derived in [130]), factor 1 equals its lower limit. We note that the end of the burst is not fixed but shifts towards later times for more energetic bursts. The source templates are shown in the Appendix (Fig. A.33).

We used a time step of a week, which is slightly worse than what we used for G323 (i.e., 3.65 *d*).

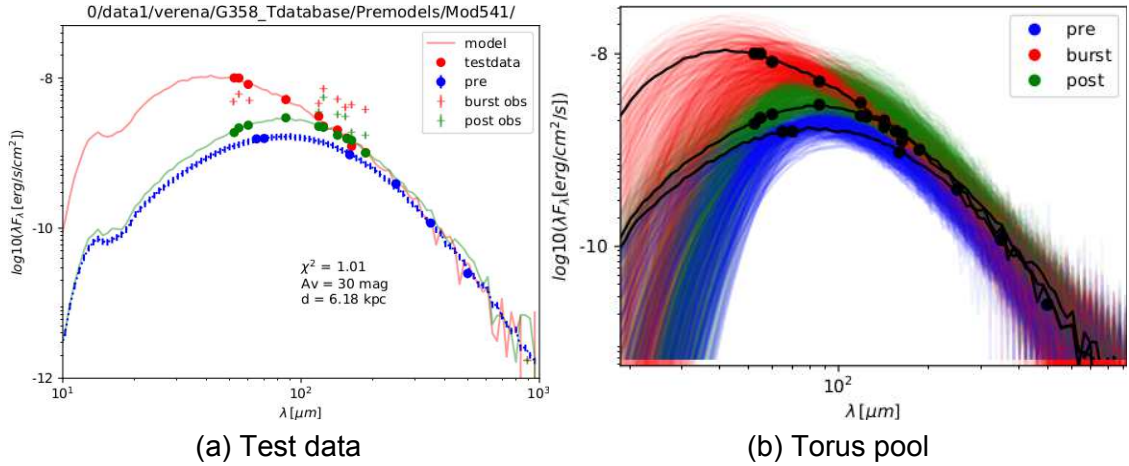

(a) Test data                    (b) Torus pool

Figure 4.1: Benchmark of the TFitter. **Left:** Test data (dots) and real data (crosses). The observations obviously do not match the test data, which shows the expectation from the best preburst model for a burst with $1.7 \cdot 10^{45}$ *erg* (60% of the static estimate). **Right:** all Tmodels at the given epochs (color-coded). The test model (black) is covered well at all epochs.

**Step 4**   The method has not yet been benchmarked. Before applying it to real data, we use test data to evaluate it. This will show whether the method works and will help to judge how powerful it is to constrain the burst energy and the system parameters. In the following, we describe the benchmark in more detail. A fit was also performed with real data. This is described in more detail in Sect. 4.3.

As test data, we use **'fake' data at the respective observing epochs**[2], based on the best TORUS preburst model. We used a burst template that corresponds to a scaling factor of 1.5 (which implies a burst energy that is equal to the static estimate for sources with the protostellar luminosity of the mean model). The energy of the test burst is 60% of the static estimate given in [130] (as the preburst luminosity of the best model is 60% of the one of the mean model). We use the observed preburst SED (instead of fake preburst data) to determine $\chi^2_{pre}$ (see Step 2). The wavelength coverage is better than that for the other epochs, which is usually valid for the observations as well.

---

[2]We use the closest SEDs, which can be off by a few days (i.e., we do no interpolation in time).

We include a 5% random error on the synthetic data points for both (burst-/post-) epochs (small random noise). We used the spectral filter set from the burst (which includes 5 bands in between 50 and $200\,\mu m$) for both epochs (even if the postburst observations lack the blue channel). With this choice, we avoid biases due to different wavelength-sampling. The test-data is plotted in the left panel of Fig. 4.1 with color-coded epochs. The observations are given as crosses for comparison. Obviously, the best preburst model (test data) does not match the observations. The right panel of the Fig. shows the test set (that is, pre, burst, and post SED), which is well covered by the Tpool.

For the ratio fit, we replace the test data for burst/post by the respective ratio with the prediction of the underlying pre-model (best simulated pre-SED).

**Step 5**  Here, we briefly summarize the Tfitting (TFitter + setup) for the benchmark case (it works analogously for the real data). A more detailed description of the TFitter is provided in the Appendix. The TFitter (which works similarly to the sedfitter [122], but incorporates all epochs) allows for **value (standard) and ratio fits (small flux excesses; see G323)** of the epochs. In the latter, instead of the SEDs, the ratios (at each epoch) are fitted. We benchmark both, as this allows us to compare both methods and find possible biases. The most important points are the following. The burst and post-light curves are optimized to reduce synthetic noise. This basically means that they are fitted in the time domain (see Sect. 5.3 and 4.2). Only after that, the burst/post SEDs are extracted from the Tdata sets. Distance and foreground extinction are adapted from the preburst fit alone (similar to what is done for G358), as the pre-SED usually features the best wavelength coverage.

For the **value fit**, we weigh the epochs equally to account for all epochs. The preburst fit is repeated, and all available pre-SEDs (for each configuration) are included (all Tdirs include at least one additional pre-SED). This slightly reduces the synthetic noise.

For the **ratio fit**, we read the previous pre-fit results. This is done because the observed ratio at the preburst epoch is (by definition) always one. It is possible to exclude the pre-fit from the results. However, we weigh all epochs equally (including the preburst). The other epochs are repeated with the modified test data (see above).

## 4.2   The TFitter

The TFitter works similarly to the sedfitter [122], but for multiple observing epochs, as mentioned already. We include a description of the basic steps in the Appendix (Sect. A.4.1). The results of the benchmark fits are shown here.

### 4.2.1 TFitter benchmark

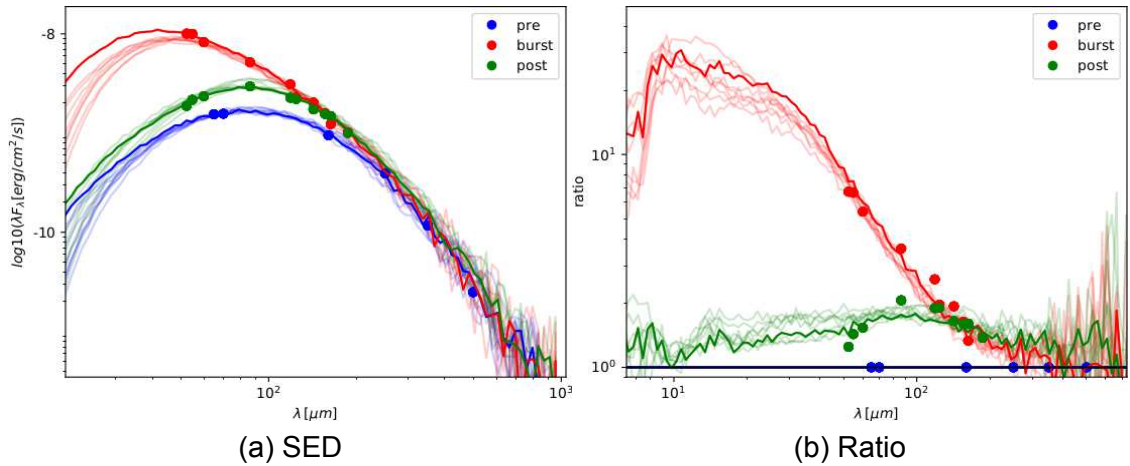

(a) SED                                        (b) Ratio

Figure 4.2: Test-data and the 10 best SEDs (left) and ratios (right) for the benchmark fits (the best one is darkest), the epochs are color-coded.

#### 4.2.1.1 Standard configuration (value fit)

In this section, we present the results of the TFitter for a standard scenario (2 epochs, both clearly exceeding the FIR preburst fluxes). This is the same setting as should be used for G358 and NIRS3.

The 10 best fitting models are listed in Tab. A.10, a visualization is provided in Fig. 4.2 (best SEDs and ratios). These 10 models fit all epochs pretty well. The SEDs match the test data better than the ratios. This is expected because the deviation of the respective epoch- and pre-values add up in the case of the ratios. The best model (darkest) is the same as the test-model (input test-data), which proves that the TFitter works correctly. Note that the test model appears twice in the list of the best models (best and second best). This is because it appears twice in the input list (at the start and in the middle). We could have removed it from the list automatically, but having it twice is a further check that the TFitter works correctly.

Example light curves are given in Fig. A.41 for the best model (blue) compared to the pool (100 best models in gray). The test values (horizontal) and dates (vertical lines) are indicated. The best model matches all test data points pretty well (for all wavelengths and epochs) as expected (see also Fig. A.36).

The TFitter returns fit information on each epoch and each data point separately. This is quite useful to check whether the $\chi^2$-value is dominated by a certain wavelength or/and epoch. Furthermore, it could help identify systematic deviations. This could, e.g., hint at an inappropriate burst template or also calibration issues. The $\chi^2$ values separated by epoch (color coded) are shown in Fig. A.38 for the 10 best models. Fig. A.37 shows the $\chi^2$ values further split in the filter bands for the best model. The result looks reasonable. The $\chi^2$ values of the best models are dominated by the burst.

To show **whether the test parameters could be recovered** by a mean model (as previously used), we provide corner plots in Fig. A.39 (premodels) and Fig. A.40 (Tmodels). The values of the best model are colored red. The 10 best models have been used to determine the weighted mean parameters (similar to the mean models for G358 and G323) shown with the vertical green lines. All the values of the mean T model agree quite well with the values of the test model within the confidence intervals (only $\alpha_{disk}$ is not recovered).

In the following, we want to use the result of the TFitter benchmark to **quantify whether there is really an advantage of the time-dependent approach compared to the pure static (or semi-static) treatment**.
For that, we compare **mean pre-** ('best static approach') and **mean Tmodel** ('best approach') with each other and the **test model** ('test reality'). Furthermore, we compare the precision of the parameters and the burst energy with the results of the semi-static approach from Sect. 3.2. A better match of mean Tmodel and test model (as compared to the mean pre-model and test model) is an indicator that the TFitter can indeed help to put better constraints on the YSO dust configuration. The mean Tmodel (Fig. A.40) seems to agree slightly better with the test model, than the static mean model (Fig. A.39) but further tests with a larger test database are needed.
The **size of the confidence intervals is mostly only slightly smaller** (or similar) to what we could achieve with the static prefit of our test set (green lines in Fig. A.39) and the semi-static approach (values published in [130] for G358 real data). We had hoped for significantly smaller error bars with the TDRT approach. However, this is not the case, but the Tmodel test sample is quite small (only ≈ 130 different dust configurations). Hopefully, with more models, this will improve.
Even if the refinement of the dust configuration is only small, **the burst recovery is quite striking**. The burst energy can be recovered with a precision of ±20%, which is much better than the precision achieved in the semi-static approach (exceeding ±50%). On top of that: all best models share the same source template. This strongly indicates that the shape of the burst-luminosity variation may be recovered with the TFitter, which would be a great success.

All together, the benchmark can be summarized by the following: the TFitter **works correctly** (all results look quite reasonable) and the first test already indicates the **potential of the TFitter** (in particular for the reconstruction of the burst history).

#### 4.2.1.2 Small flux excesses (ratio fit)

For small flux excesses (as f.e. $L_{post} \approx 1.1 \, L_{pre}$ in case of G323) it might be difficult to recover the correct burst energy with the 'standard' (value-) Tfits (as done above). The issue is that the scatter of the best preburst models (i.e., the range spanned by the best preburst models) might be larger than the actual flux increase (in a certain wavelength regime). Therefore, the TFitter includes the possibility to fit the flux ratios (ratio-fit). This approach should be used for G323 (see also Sect. 3.3.4.2).

In the following, we present the result of the fit of the ratio for the test data (as described in Step 5 in Sect. 4.1.1). We emphasize that we tested the ratio fit with the same test data as above (used for large excesses) to compare the results. The ratio-fit needs to be further tested (with small excesses) before (systematically) applying it to real data. This will be done in the future, but is beyond the scope of this work.

We provide similar figures as before in the Appendix, which are SEDs and ratios for the 10 best models (Fig. A.44), example light-curves for the best model (Fig. A.46), detailed $X^2$-values for the test model (Fig. A.43) and for the 10 best models (Fig. A.45). The parameters of the 10 best models are provided in Tab. A.11, the corner plot is shown in Fig. A.47. The results can be summarized as follows.

- The test model can be recovered, but it is 'only' second best.

- The **results agree with the value fit** (but huge confidence-intervals).

- The mean parameters **are less constrained** (including the burst energy).

- The S/N of the simulation is higher for the ratios (since the preburst adds additional noise), which leads to smaller and more similar (within the 10 best models) $X^2$-values of the fit.

We conclude that both the value and the ratio-fit work. In general (for moderate flux excesses), the value fit should be used. For small excesses, the ratio fit might be better. But it needs further testing (with an appropriate test data set). When the ratio fit is used, the simulations should be performed with a higher number of photons (especially the preburst) to achieve a sufficient S/N.

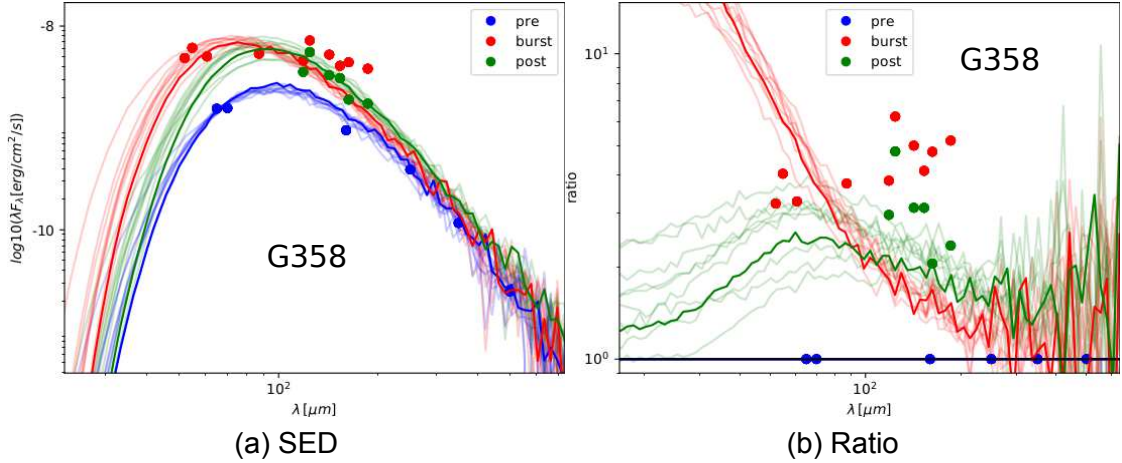

(a) SED                                        (b) Ratio

Figure 4.3: Same as Fig. 4.2 but for the fit with the real data. All best models do not fit the burst SED long-ward ≈ $120\mu m$. The observed burst-ratios increase with wavelength, which is clearly not the case for the models.

## 4.3   Fit with the real data

After the successful benchmark, we apply the TFitter (with the test dataset) to the real G358 observations.

The total maser flux serves as a template for the accretion rate variation. A close correlation between both is established in Wolf+ in prep. for G323. Time-dependent simulations show that there is a delay of a few days between the maser peak and the adapted source-luminosity variation (see Fig. 3.11). We use the adapted semi-static mean model (from Sect. 3.2.4.2) to infer the most likely onset of the burst in the case of G358. It amounts to 9 ± 7 days. We rerun the TFitter for the real data (and include the time shift as estimated above). The TFitter setup is the same as for the benchmark (except for the time shift). The results are presented in the following.

The best models are summarized in the Tab. A.12. We emphasize that the adapted mean model (from Sect. 3.2.4.2) is within these models, which indicates that the time-dependent and semi-static results are in good agreement. But the best Tfits are not good. They are much worse than the benchmark fits (especially for the burst). The corresponding $\chi^2$ values are given in Fig. A.50. All $\chi^2$-values are pretty high, whereby the burst values exceed the pre- (and post-) burst ones by far. The model pool is visualized in Fig. A.52. It is clearly visible that the burst SED at $\lambda \geq 124\mu m$ (i.e. in the FIFI-LS red filter set) is not covered. The poor burstfit is also evident in Fig. 4.3, which shows the SEDs (**left**) and ratios (**right panel**) of the ten best models at all 3 epochs (color-coded). The pre- and post-SED are fitted well by the best models, contrary to the burst-SED. Interestingly, this does not hold for the ratio (right panel), where the post-ratio is below the observed one, as a result of the slightly over-predicted FIR preburst SED. This means that a ratio-fit possibly would deliver a higher burst energy. Therefore, both methods should be tested more rigorously (even for strongly elevated burst/post fluxes, as given in the case of G358).

Clearly, more work is needed to refine the parameters from our semi-static estimate by means of TDRT. This is beyond the scope of this thesis but will be done in the future. There are a couple of issues (and open questions) that are summarized briefly below.

- The test set is not appropriate. The number of models (especially Tmodels) is small, we use pure silicate models (despite the modified mean model), the model set is not properly sampled and biased towards lightweight disks (disk masses as derived by [22, 28] are not taken into account). As the cavity imprints the afterglow a lot, we should also aim for more realistic models (the cavity has a constant low density in all our settings).

- Furthermore, the burst templates may not be appropriate; in particular, we lack short and energetic bursts (as the scaling is such that the burst duration increases with its energy). A further problem of the applied scaling is that the burst energy $\propto L_{pre}$, this introduces biases (it implies that less luminous sources feature less energetic bursts on average). The maser light curve serves as a template; deviations between the maser curve and the source luminosity variation are not taken into account (as an example for G323, the FWHM of the modeled maser curve is larger than the one from the source input). Therefore, iterative approaches should be tested. We take into account that the source likely rises before the maser flare, but only for one dedicated model (and it might not be the same for all configurations).

- As the heat wave expands radially, we could try to fit SEDs measured on different apertures (thereby taking into account spatial information). However, this may be difficult as it is impossible to include a PSF on the Tmodels. For dedicated models, we should also create images with TORUS. This will help to draw a more realistic picture.

- The data needs to be re-calibrated. A recalibrated data set will be released in July 2023, which is expected to lead to less scatter at the observing epochs.

# Discussion

The discussion is divided into **these parts**. We start with a discussion of our **results**. These are the results on **G358 and G323 in the context of current observations** (comparison with other known MYSO bursters and conclusions on the accretion physics behind), as well as our conclusions on the **afterglow timescales**.

Then we discuss the **reliability of the results**. After that, the **challenges and possible issues of the applied methods** follow (this is more technical, but is nonetheless important to evaluate the results). And finally, the possible **prospects** (of TORUS and the IR astronomy past-SOFIA) are outlined.

## 5.1 The results in context

### 5.1.1 Episodic accretion of MYSOs

| System | $M_*$ | $L_*^{pre}$ | $L_{peak}$ | $\Delta L$ | $t_{rise}$ | $\Delta t$ | $\dot{M}$ | $E_{acc}$ | $M_{acc}$ |
| --- | --- | --- | --- | --- | --- | --- | --- | --- | --- |
| | $M_\odot$ | $10^3 \cdot L_\odot$ | $L_{pre}$ | $10^3 L_\odot$ | yr | yr | $10^{-3}\frac{M_\odot}{yr}$ | $10^{45} \cdot erg$ | $M_{Jup}$ |
| NIRS3* | 20 | 30 | 5.5 | 130 | 0.4 | 2.5 | 5 | 12 | 2 |
| G358* | 12 | **5.0** | **4.8** | **19** | **0.14** | **0.5** | **1.8** | **2.8** | **0.5** |
| G323* | **23** | 60 | **5.4** | **260** | **1.4** | **8.4** | **0.8** | **90** | **7** |
| NGC* | 6.7 | 3 | 16 | 44 | 0.6 | >8 | 2.3 | >40 | >0.4 |
| V723 Car | 10? | ≈ 4 | | | 4 | ≈ 15 | | | |
| M17 MIR | 5.4 | 1.4 | 6.4 | 7.6 | | 9-20 | ≈ 2 | | |

NIRS3 (S255IR NIRS3) [25, 134, 86], G358 (G358.93-0.03-MM1) [130, 19, 28, 21, 8, 22], G323 (G323.46-0.08) [113], NGC (NGC 6334I MM1) [66, 65, 18, 16], V723 Car [135, 137, 136], M17 MIR [29]

Table 5.1: Accretion outbursts observed so far around MYSOs. Two of them (highlighted) are investigated in this work in more detail. The bold values are based on this work. The star * symbol indicates the accompanying methanol Class II maser flare. The full names are given below the table.

Although the number of known MYSO bursters is quite small, they span a considerable range of burst characteristics. This raises the question whether different trigger mechanisms are responsible. An overview of all known bursts is provided in the Tab. 5.1. We emphasize that the bold values have been derived in this work.

The G323 outburst is, with an energy of $\approx 10^{47}\, erg$, the most energetic observed so far. It lasted about 8 years, which is at the longer end. The increase in luminosity is probably also the highest ($\approx 3 \cdot 10^5 L_\odot$, which is twice as much as for NIRS3 [25]). On the other hand, the G358 outburst is rather short ($\approx$ months) and weak ($\approx 10^{45}\, erg$). The increase in luminosity is at the lower end ($\approx 2 \cdot 10^4 L_\odot$). The accreted mass differs by an order of magnitude, where both values are still in the range of a (heavy) planet. Interestingly, the accretion rates during the burst are quite similar for all known objects (on the order of $10^{-3} M_\odot / yr$).
G323 is more massive than G358 and is also more luminous. However, G358 is probably younger and may become heavier by the time it reaches the ZAMS. Both sources are likely to be seen face-on (contrary to NIRS3, which is highly inclined). G358 is deeply embedded (MIR dark), pointing to an early evolutionary stage. Its burst parameters support disk-mediated accretion (as its disk is prone to GI), possibly of a spiral fragment.
G323 is likely more evolved (NIR bright, LE, hints for an outflow), which would in principle support the ZAMS assumption. Then the accreted mass is in agreement with a heavy planet (or a disk fragment), just as for G358 (but scaled-up). However, the Inayoshi pulsation [68] points to a bloated star. Therefore, we suggest another possibility, which is the accretion of a potential companion. This is expected to occur from hydrodynamical simulations [108], where fragmentation leads to the formation of clumps (which may evolve into cores). These large fragments (potential companions) might spiral inward and get accreted, thereby leading to outbursts. Observational evidence for close substructures in high-mass star forming regions exists, e.g., [12, 12 distinct sub-sources within $\approx 10^5\, au$ in G351.77-0.54, which are consistent with thermal Jeans fragmentation].
The timescale is still short enough to be explained by the accretion of a compact object, which may be disrupted by tidal forces only within the accretion event. Longer burst durations (similar to V723 Car and M17 MIR) would point to a more diffuse object [100, in the order of $\approx 4/27\, yrs$ for compact/diffuse objects].
None of the bursts agree with MRI or TI, where the (peak) accretion rate is much lower ($\approx 10^{-4} M_\odot / yr$) and the rise time ($\approx 50$ years) and duration ($\approx 100$ for TI and $\approx 1000$ years for MRI) are much longer [41].

When all this is taken together, the range of properties is large. Disk-mediated accretion caused by gravitational instability or planet migration [143] is supported as a (dominant) mechanism by our analysis in agreement with others (e.g., [25, disk mediated accretion outburst in S255], [3, observational evidence for 13 disk candidates]). Companion accretion may also occur (and we suggest this as an explanation for G323). Clearly, more work is needed to further narrow down the possibilities.

### 5.1.2 Afterglow timescales for MYSOs

From both of our objects, we got FIR afterglow timescales for MYSOs, which clearly exceed the duration of the enhanced accretion phase. For G358, the FIR decay time is about $2.3^{+1.6}_{-0.5}\,yrs$ (semi-static estimate), compared to a burst duration of a few months (maser flare). Note that the exact value for G358 will be refined with time-dependent modeling, but the order of magnitude is correct. For G323, the values (that characterize the length of the FIR-afterglow) are $5.6/8.8\,yrs$ (t80/t25, respectively, for the mean model and a burst with $E_{acc} = 2.3 \cdot 10^{47}\,erg$). This is about $2.6/1.2\,yrs$ after the source (TDRT input). For comparison, the FIR peak is reached $1.2\,yrs$ later. The exact values depend greatly on the burst and the local dust configuration. For dense environments, the timespans in the FIR can easily be extended by years. According to [70], the timescales are set by the photon travel time, rather than the timescale of the grain heating (as the dust quickly reaches equilibrium). They conclude that the afterglow duration is set by the finite speed of light, which leads to timescales of hours to days (MIR to FIR) and days to months (in the (sub)mm). These numbers are much smaller than ours. But our sources are much more luminous, more deeply embedded, and the cloud cores are more extended. The extended envelope is crucial for the FIR afterglow [70] and the amount of energy released in the FIR increases to distances up to the edge of the grid (our simulations). A more realistic comparison is given by the grid crossing time, which is $\approx 160\,d$ (at the speed of light). The modeled afterglow duration is much longer. This points to an effective slow-down of the energy transfer due to numerous scattering and absorption + re-emission events (similar to what is suggested [24]). This effect is not included in [70], who used the light travel time even for the optically thick case. One could argue that not all photons are expected to take the direct path (even in the optical thin case). A rather conservative upper limit (for the optically thin photon travel time) is $1.3\,yrs$, which is 3 times the grid crossing time (and certainly enough time for all photons to leave the system). Both the mean and max configurations have longer timescales (t80 and $t_{peak}$, $t80$, $t25$ respectively). The strong dependence of the duration of the afterglow on the dust density (in particular, the envelope density) is a further proof that **the duration of the afterglow is determined by the effective travel speed** (which is much lower for high optical depths), rather than the finite speed of light. The implication is that **the burst energy remains much longer in younger, more deeply embedded sources**. Interestingly, a slow-down effect for dense environments also occurs for scattered light echos, which do not include absorption and thermal reemission (based on RT simulations with Mol3D [107, extended by S. Heese to feature scattering LEs] in the JHK-band for somewhat lower densities).

We emphasize that the **afterglow is strongly wavelength-dependent** (e.g., [33, 70]), which is also the case in our simulations.

The characteristics of the afterglow manifest in some *interesting effects*, which we will briefly discuss below.

One effect is the **delay in the NIR afterglow for strongly inclined sources**, which could explain the rise of the maser flare before the increase in the NIR. For active disks, this is to be expected, because disk self-heating manifests itself in an increase of MIR radiation (before the outburst actually occurs), e.g., [157] or a recent work from [31, showing that larger (FUori) outbursts can exhibit infrared precursors decades before optical bursts, while outbursts in inner discs can show time delays of a few years]. It is remarkable that a similar behavior is expected for passive disks as well (although certainly not as extreme as a decade), if the inclination is high enough, as in the case of NIRS3. Other effects are **color changes**, or a change in the spectral index. Such effects have been discussed, for example, in [92, using different classification methods]. They found that a correct classification (as Class 0) is usually achieved when the bolometric temperature is used as a proxy for the evolutionary phase, whereas the YSOs tend to be more evolved (misclassified as Class I instead of Class 0), when using the ratio of (sub)mm to bolometric luminosity during the outbursts. We did a similar study, but with only one source (for the G323 burst) and the classification scheme according to [83]. We found that the total time in which (M)YSOs appear redder/bluer is short, as this occurs predominantly during and shortly after the burst (MYSOs spend only about 2% of their total formation time in the burst state [100, hydrodynamical simulations]). Therefore, misclassification due to bursts can occur. But they are probably the exception.

Another interesting thing is the **strong imprint of the cavity opening** on the duration of the afterglow. The cavity widens during stellar formation (due to the presence of jets and stellar wind). Our cavity model is oversimplified. It features a constant density with a sharp cutoff. In reality, the structure is more onion-like, where the density increases toward the envelope. Furthermore, there might be a radial gradient. The density profile itself may not be that relevant (see Sect. 3.3.3.3), the question arises whether the sharp cutoff could be problematic. This could be addressed in future tests.

Furthermore, we predict a **subluminal motion of the heatwave** (with an initial expansion speed of about 6% of c) from the shift of the $T = 125\,K$ isoline (maser spot motion in Sect. 3.2.4.2). This compares to the value of [9]. Our disk model is not realistic (and the density was not taken into account in the final analysis). The agreement is nevertheless remarkable. Further modeling is needed and may allow us to constrain the disk much better.

## 5.2   Reliability of the results

Although the results seem quite reasonable, it is nonetheless important to evaluate their credibility. We will do this by briefly evaluating the main issues (step by step).

**Uncertainty of the accreted mass and the mass accretion rate (for G358)**   As discussed for G323 already (end of Sect. 3.3.3), the accreted mass (and hence the mass accretion rate) depends on the stellar properties (i.e., $M_*$, $R_*$). For bloated protostars, the accreted mass can be much larger (as suggested for G323). We also want to briefly evaluate our G358 result here. We used the radius of our models, which is about twice the ZAMS value. We consider the ZAMS radius as a lower limit. The upper limit is as large as $24R_\odot$. However, the radius is not well constrained by our models (only the luminosity is). The protostar could even be more bloated. According to [63, cold disk accretion models], G358 is in the right range for bloating. If the radius is $\approx 80R_\odot$, then the estimated mass (and accretion rate) are ten times too small. Although much bigger, this is still a typical planet mass. Unfortunately, because of the high extinction of deeply embedded protostars, it is hard to derive their surface gravity from photospheric absorption lines because of strong veiling.

**Uncertainty of the burst energy**   The reliability of the estimated burst energy is probably the biggest point here. There are several sub-points. We start with **G358**. We use static models, but the system is out of equilibrium. This raises the question whether the luminosity at the burst/post-epoch is realistic? We can utilize the TFitter plots to get a rough idea of the expected deviation from the static case. At the time of the burst observation, the FIR not yet reached its maximum at all observed (FIR) wavelengths, and the source is back at its prelevel. However, most of the emission is produced in the wavelength range covered by FIFI-LS. Therefore, the value for the burst luminosity (as derived from the static fit) is probably right, while the assumption of linear decay is false. A better assumption is that the luminosity stays at its peak level (or increases slightly) for at least a couple of months.

In the postburst epoch, the situation is a little different. All FIR wavelengths are clearly decaying, but the decay is much faster at the 'short' (FIR) wavelengths. The static model thus likely overestimates the total post luminosity. Both effects (in part) compensate for each other. For illustration: the principle behavior, as just described, is visible e.g., in Fig. A.49, which shows the predicted light curves of the adapted static mean model at each band as compared to the observation (albeit the predicted fluxes at the observing epochs generally do not match the ones from the respective semi-static mean model).

First fits with the TFitter suggest that the order of magnitude is correct (and that the burst might be about twice as bright as in our estimate). However, there are several issues (see

Sect. 4.3) especially for the burst epoch. The analysis shall be repeated with a proper, fully time-dependent treatment (and the recalibrated dataset).

For **G323**, the situation is different, as we use a different modeling strategy. The afterglow is much longer for denser environments, and hence the burst energy needed to explain the observation is much smaller. Moreover, the increase at the postburst epoch is quite small. The max model features very small increases, and the $\chi^2$-values of this model scatter a lot, especially toward low burst energies (Fig. A.29). The lower limit is chosen, such that all possible models are included. However, a linear fit to the ratios at 70 and 160$\mu m$ would lead to a minimum energy of $\approx 0.23 \cdot 10^{47} \, erg$. With more time and more photons the lower limit will likely shift towards that value. But this needs more simulations. A return to the preburst level is not likely, but may not be excluded. In this case, the derived values can be considered as upper limits. The range estimated by us is fairly large (more than 1 order of magnitude between the mean and the max setting, and another one between the mean and the min setting). These 3 settings were meant to give limits for the accretion energy (while they will be refined with a bigger model set in the future). However, we could ask whether these configurations are realistic. The min-setting (highest energy input) obviously has too little cold dust. Therefore, we conclude that the upper limit of the mean model is more realistic. However, there might exist settings that include enough cold dust and have short afterglow timescales (such as the mean setting with a widened cavity).

Another proxy for the burst energy comes from the Ks curve. The assumption that the increase in Ks is $\propto L_{bol}$ leads to $2.3 \cdot 10^{47} \, erg$. A more iterative approach (using our models) gives $0.9 \cdot 10^{47} \, erg$ (Sect. 3.3.4.2). We base our final result on that, although we consider our modeling not suited to correctly reproduce the Ks-flux. In the following, we will elaborate on possible influences on the appearance of the Ks curve. This also has an additional meaning, as Ks serves as a burst template. There are two factors that are not taken into account (by our models). These are line-of-sight extinction variations and changes in the emitting surface area. Extinction variations can be caused by dust removal/sublimation (flux increases after/during the burst) or disruption of the accreted object (flux decreases before, during, or after the burst). FUor/EXor outbursts are accompanied by winds, which become obvious from the P-Cygni absorption profiles. These will blow out dust grains and thus reduce extinction; see, e.g., [117, 104]. However, the Ks preflux is constant prior to the burst and returns (almost) to the prelevel at the last observation (Fig. 3.6). Therefore, the variations in LOS extinction should not be a concern here.

What about the emitting surface area? Due to dust sublimation, the area A, which radiates at $T_{sub}$, will increase during the burst. This effect is huge, since A is roughly $\propto R_{sub}^2$. Note that this is also reflected by our models, where the Ks-magnitude depends strongly on the local properties inside the innermost au (while the ratio is rather unaffected). To obtain the maser curve (blue curve in Fig. 3.11), we performed a simulation that includes dust sublimation (which is not the case for the other simulations). So, the best we can do is to compare that simulation with the corresponding mean model. The maximum Ks ratio is about a factor of 1.5 smaller for the more realistic sublimation model (see Fig. A.30). This might be surprising because the emitting surface area should increase with sublimation. But in reality, the opposite is the case, as we shifted the inner radius to $3R_{sub}$ (standard configuration) to avoid unrealistic high temperatures. This comparison implies that the most probable value (for $E_{acc}$) is slightly higher than the Ks-based one (i.e., $0.9 \cdot 10^{47}\, erg$). We emphasize that this case is covered by our results (within the given errors) and is in line with the HAWC+ data.

The most accurate estimate of the burst energy will be possible with a combined reanalysis of the Ks data together with the FIR data. Therefore, the FIR can serve as a validation and informant on the local dust distribution.

Another point that we have not addressed so far is the assumption that the dust and gas have the same temperatures. Gas timescales are longer (both heating and cooling), and the difference is greater in optically thin regions [70]. The gas is heated by the dust mainly via collisions (and IR-absorption of molecules such as water and oxygen) [62, Tab. 1, right column]. UV-heating of the gas (by the dust) can be neglected at typical core densities. In the (dense) innermost regions, where the Ks-flux is produced, the assumption that gas and dust are well coupled is justified. Therefore, our final estimate (based on Ks) of the burst energy is not affected. However, most of the FIR radiation is produced in the extended (outer) envelope. There, the coupling may not be given. If the gas and dust are poorly coupled, then the dust will heat much faster than the gas. Possibly a huge fraction of the thermal emission is released by the dust before the gas gains a significant increase in energy. If the energy is 'stored' in the gas, it will be 'hidden' (we consider only the thermal dust emission). In this case, the burst energy might be underestimated. Note that the total energy is released as thermal dust emission, even if the gas gains a significant amount of energy (its mass and heat capacity are higher). But the heat transfers back quickly (at least if both are well coupled). For static models, the problem is minor, since gas and dust (usually) have enough time to exchange their energy even if they are not well coupled.

Once improved distances based on maser parallaxes become available, the values (for both sources) could be further constrained.

## 5.3   Challenges and caveats

**Caveats of the (static) SED fitting in general**   Multiwavelength studies are really valuable as they include information about all spatial scales (from the inner disk to the outer envelope). However, they face some challenges. The data points are usually taken with different instruments (different sensitivities), the wavelength coverage can be poor (it is fairly good for our sources), the sampling is usually not even, and there are degeneracies (the same SED can be explained by different YSO-configurations). In addition, our geometries are simplified (not all settings are realistic, and small-scale structures are not included). A more general discussion of the caveats of SED fitting can be found in [121, for the 2017 YSO-grid]. To ensure more stability, we use the mean model approach (where we include the best 10 models). A quick way to test whether the SED fitting is still biased is to artificially remove (or add) data points to the SED and compare the results. We did this for G323, where we artificially removed all data points in between 30 and 350 $\mu$ m (which will no longer be accessible in the future). The corresponding corner plot is provided in Fig. A.31. The best-fitting models differ, but the mean values are stable. This may not hold in general (especially for sources with comparably lower MIR flux densities), but supports the approach (of the mean model) used.

**Noisy pre models**   For the TORUS (preburst) database, we had an additional problem. We could only use a few photons (because of the high computational demand for all models together). Therefore, the SEDs were partly quite noisy. Furthermore, the noise level was quite different for the models (despite the same number of photons). This introduced a bias towards models with poorly constrained SEDs. Our solution is a 2-step approach, where we rerun the (poor step 1, with $10^6$ photon packets) simulations of the best models with a sufficient number of photons (in step 2, with $10^8$ photon packets). This is a good compromise that ensures a reliable result at reduced costs. If useful, we also apply fits in the wavelength domain (to reduce scatter)[1]. Note that the quality of the pre-SED also has some relevance for the ratio fit (TFitter). Here, we use a preburst duration (where the flux remains at its preburst level). The average over this epoch returns a reliable pre-SED, including realistic errors.

**Noisy Tmodels**   As for the pre-models, we also have a noise problem for the TModels (especially towards long wavelengths, i.e., FIR and (sub)mm). This was probably the biggest challenge for the design of the TFitter. It is crucial for the result, especially if the flux increase is small (e.g., G323). We used an interpolation in time with a stepwise adaption of the interval size (which is shorter for shorter wavelengths and close to the peak). We could also have increased the number of photons (but much higher computational demand) or decreased the time step (worse time resolution), but the above solution seems appropriate.

---

[1]This was helpful e.g. in the step 1 fit.

## 5.4 Outlook

There are a couple of things that can be done with the existing data. The next thing is to do a fully time-dependent reanalysis of the G358 (recalibrated) data set with the TFitter. Thanks to the acceptance of our DLR grant, we will also be able to model other MYSO bursters, such as NIRS3, G24.33+0.14 (periodic) (and G323) with a suitable TDRT database for each object. NIRS3 is especially promising, as the data set is fairly large. Furthermore, the object is seen edge-on (contrary to G358, G323), which implies some interesting effects. With the release of another TDRT code [11], a comparison of TORUS with another software became possible, allowing further proof of the reliability of our (current and forthcoming) results. For what concerns new data, the future is less bright.

The SOFIA shutdown left a gap that cannot be filled by JWST (NIR/MIR) or ALMA ((sub)mm). SOFIA was the only access to the FIR, where most of the photons in the cosmos are emitted. The shutdown is difficult to understand for the entire FIR community, especially since there will be no replacement in the near future [158]. The only alternative in the FIR is perhaps balloons (if safe landing technologies can be provided). However, it is questionable whether follow-up burst observations with balloons are feasible. Therefore, the best strategy will be to hunt for future bursts with JWST (up to $28\mu m$) or/and ground-based facilities such as ALMA (longward of $350\mu m$) or the VLT (up to $24\mu m$). As these observations usually do not cover the peak of the SED (only for the less embedded sources it is in the MIR), they are less suited to model the epoch SEDs. Hence, our modeling strategy may suffer from larger uncertainties (and it will become harder to derive the burst parameters). An alternative way to derive the burst energy may be provided by dedicated lightcurves (as applied for Ks in the case of G323). This requires a sufficient cadence, which is only provided by surveys, such as (NEO)WISE (for less embedded sources) or the JCMT in the (sub)mm (but small increases).

Nevertheless, even if JWST cannot remove the need for an FIR facility, it still offers the possibility to get astonishing insights into massive star formation, also at very early stages (due to its high sensitivity and spatial resolution). Its spectral range covers various atomic and molecular lines of both gas and ice that trace different mechanisms, such as accretion, ejection, and disk heating. An example of the wealth of lines visible with JWST (MIRI) is published in [79, EXLup in quiescence], which shows that this facility provides the possibility to put more stringent constraints on molecular abundances and the thermal structure of EXLupis disks than ever before.

TORUS is a versatile code that can be applied to very different astrophysical cases that invoke variability. This includes low-mass YSOs, sources with periodic luminosity variations, and SN/Nova explosions. It might even be used on galactic scales to investigate AGN variability. Recent improvements of TORUS include useful features that have been used in this work and/or might be used in the near future (e.g., aperture-dependent SEDs, dust sublimation, or imaging capability). In the future, it might be extended further. Features such as active disks, spectral lines, electron scattering, or a time-dependent hydrodynamic would be nice. A coupling with chemical burst models would be tempting as well.

# Conclusions

Our knowledge about massive star formation has rapidly deepened since the discovery of the first MYSO accretion bursts ([25, NIRS3] and [66, NGC]). Due to the connection of accretion bursts with methanol Class II maser flares, it became possible to hunt those outbursts via maser monitoring. The M2O (maser monitoring organization, e.g., [23]) had been established, leading to the discovery of new outbursts (beginning with G358). The growing sample of MYSO bursts shows a wide range in properties. We derived the main burst parameters for two MYSO bursters, which are at the higher and lower end in terms of burst duration (spanning between months and decades) and released energy (spanning between $10^{45}$ and $10^{47}$ $erg$). With TORUS [56] it has become possible to tackle the different timescales using time-dependent RT for the first time. The thermal afterglow is highly wavelength dependent and acts on different spatial and time scales, e.g. [70, 33]. With our simulations, we confirm these dependencies and extend them to the regime of the deeply embedded MYSOs. We show that for MYSO the thermal afterglow is visible for up to a few years after the bursts end, which is much longer than what is derived in [70, for low-mass YSOs]. We confirm the importance of the extended envelope for the FIR emission and show that the timescales can differ by years (depending on the local dust distribution). We also investigated the influence of the view and the burst's appearance. We conclude that, in principle, it is possible to reconstruct the burst history from the appearance of its afterglow (and we develop a Python routine, the TFitter, that can do that).

There are a few sources that could be tackled with our approach, and we will do that in the near future. Only recently has another time-dependent code been developed [11], allowing comparison with TORUS. But also other approaches are making great progress, (hydrodynamical) simulations are getting better, and an increasing number of observations helps to put more and more constraints on the models, and not to forget the maser monitoring, which serves as systematic burst-alert system for MYSOs. There is one serious drawback, which is the shut-down of SOFIA (when suddenly the access to the FIR was lost completely) and one great opportunity, which is the launch of JWST. Although JWST will never compensate for the loss of SOFIA, it will nonetheless provide a wealth of insights in star formation (especially because of its high sensitivity and resolving power). The first observations are already there, indicating its power, e.g., [79, spectrum of EXLup in quiescence]. Episodic accretion in (massive) star formation is a research field, still full of mysteries and yet constantly filling with new insights. We are proud to contribute to the understanding of episodic accretion and curious about the upcoming discoveries that will surely be made.

Appendix A

# Additional Files

## A.1 Additional figures

### A.1.1 Basics

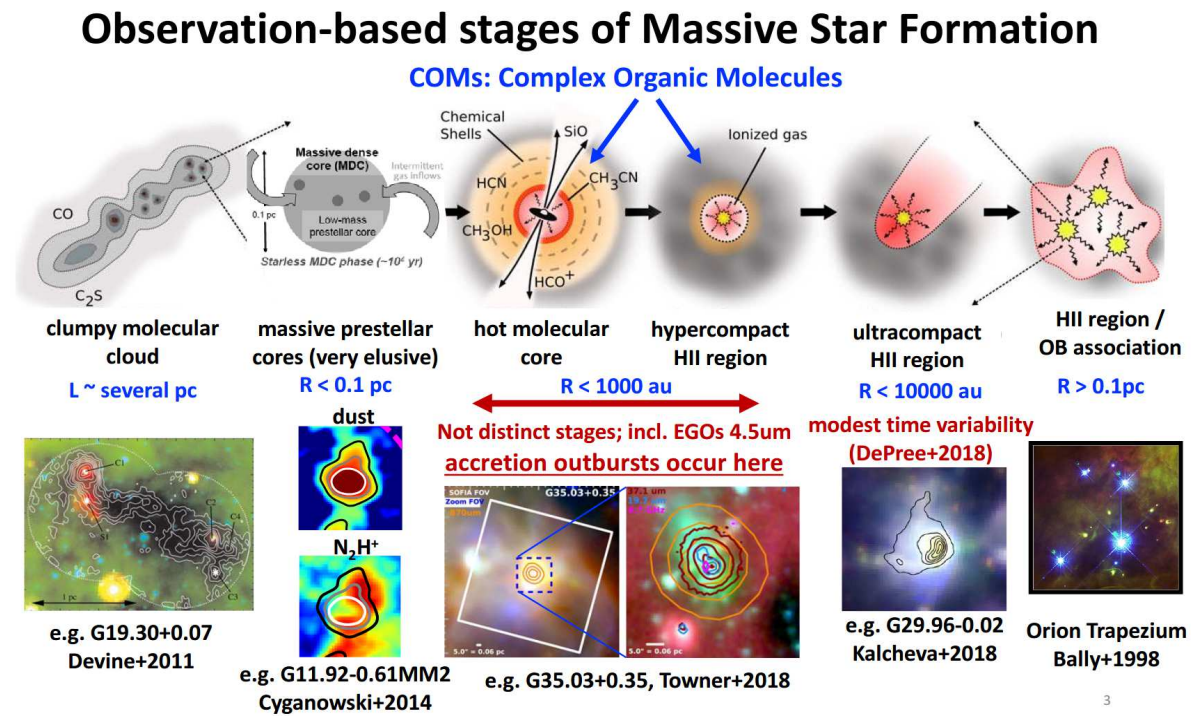

Figure A.1: Phases of protostellar evolution for a MYSO. Taken from [67].

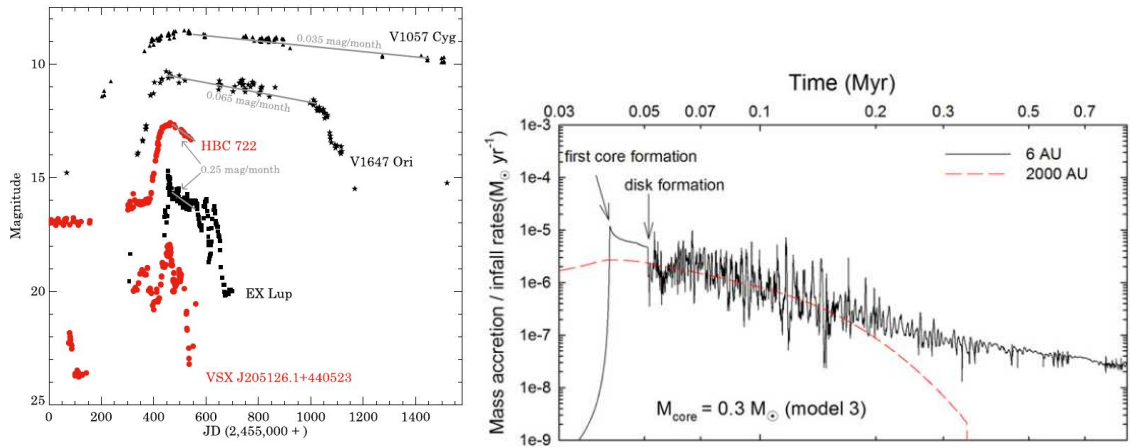

Figure A.2: **Left:** Optical light curves of different low-mass YSO outbursts, indicating the wide range of burst characteristics. **Right:** Simulations show the unsteady growth of the protostellar luminosity as a function of time. Black is the mass accretion rate and red the envelope infall rate given at the given distances. The Figs. are taken from [78, Fig. 3] and [145, Fig. 3]

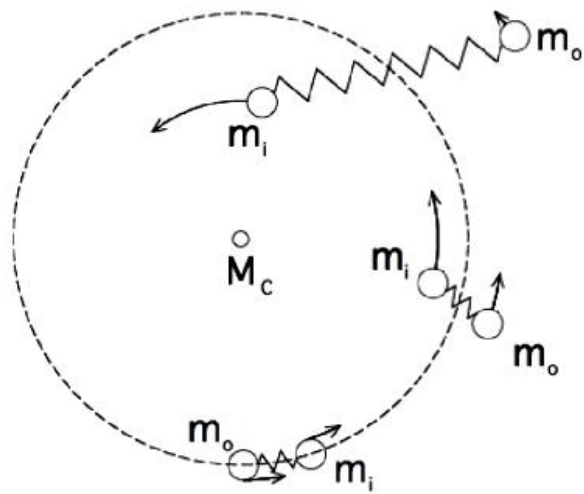

Figure A.3: The principle of the MRI (a simple mechanical analogue). Taken from `https://mri.pppl.gov/physics.html`. MRI is caused by 'connected' ionized elements m, where string-like connections (caused by the Lorentz force) lead to the exchange of angular momentum ($m_o$ speeds up/gains angular momentum, while for $m_i$ the opposite is the case), such that both particles are pulled apart from each other. The spring strength symbolizes the magnetic field strength.

### A.1.2  The SOFIA observatory

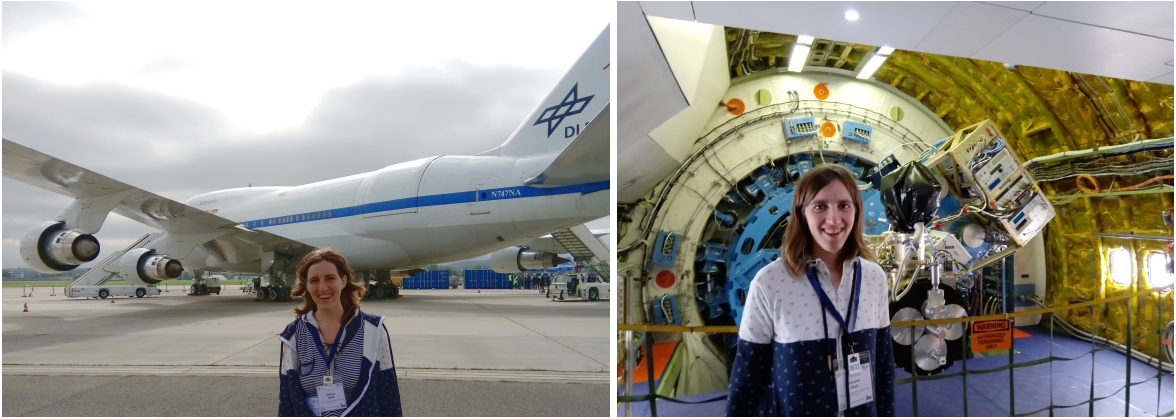

Figure A.4: SOFIA together with the author from out- (**left**) and inside (**right**). The HAWC+ detector is mounted (right panel). The telescope sits -airtight- behind two port-side cavity doors inside the airplane fuselage (just behind the science instrument). The telescope door remains open during the flight. The images are taken in Stuttgart in September 2019.

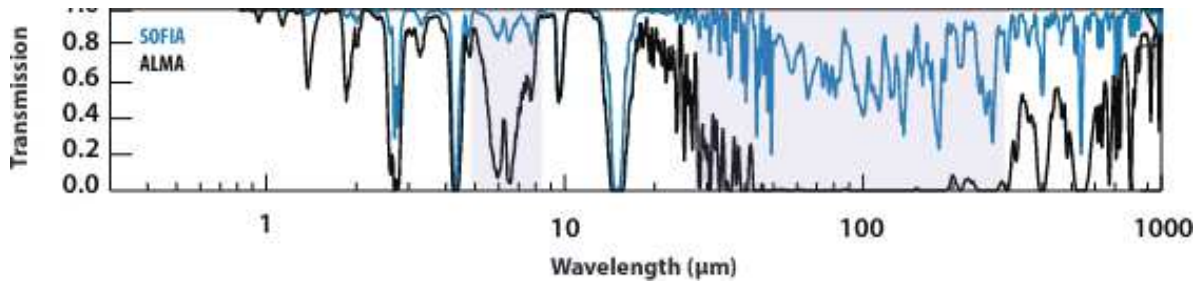

Figure A.5: Atmospheric transmission with SOFIA as compared to ALMA. In between 30 and $300\mu m$ there are no observations feasible with ground based observatories. SOFIA was operating at an altitude in between 12 and 14 km (stratosphere), which is above ≈ 99% of the earth's water vapor (responsible for the low MIR/FIR transmission). Image taken from [43].

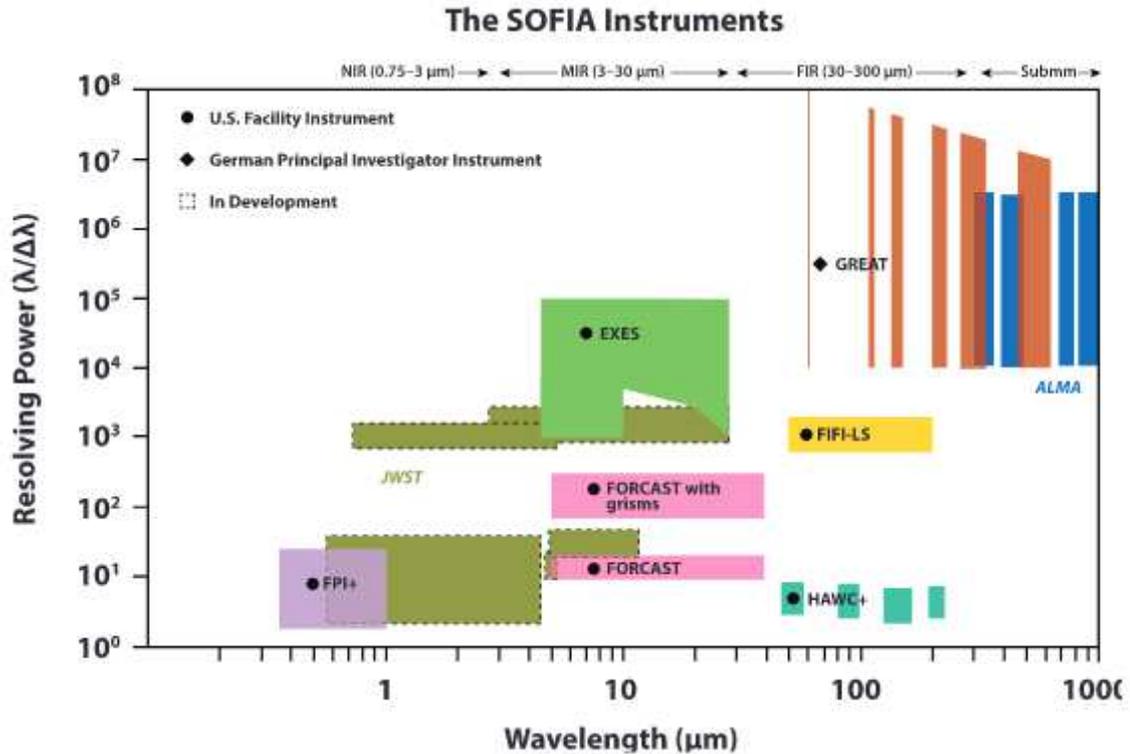

Figure A.6: Resolving power and wavelength-ranges of SOFIA's instruments (suited for imaging, spectroscopy and polarimetry). Taken from the quickguide `https://www.sofia.usra.edu/sites/default/files/Other/Documents/quick_guide.pdf`. The easy access to the instruments and the telescope for maintenance and improvement was one of the biggest advantages of SOFIA as compared to space-based observatories.

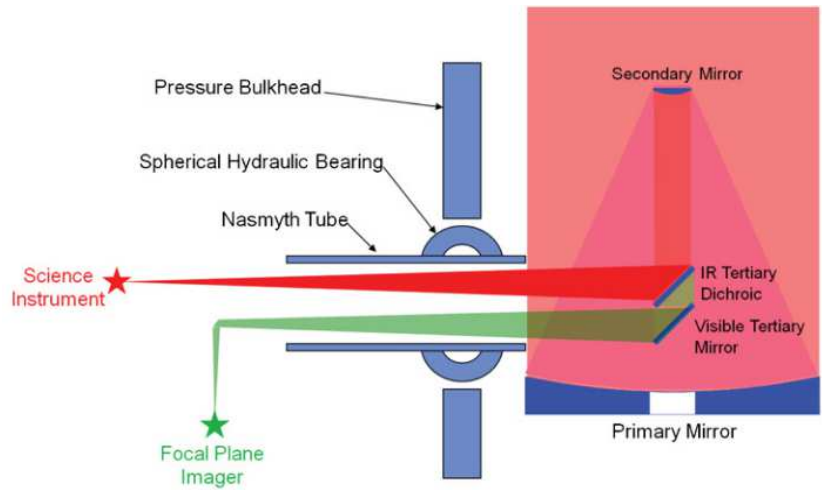

Figure A.7: Light path within SOFIA: after the parabolic primary mirror, the primary beam gets reflected at the hyperbolic secondary mirror. The tertiary mirror serves as a beam splitter: it is transparent for the optical light (green), which goes to the guiding camera (the focal plane imager 'FPI') and reflects the IR radiation (red), which enters the science instrument. The image is taken from [156, Fig. 3].

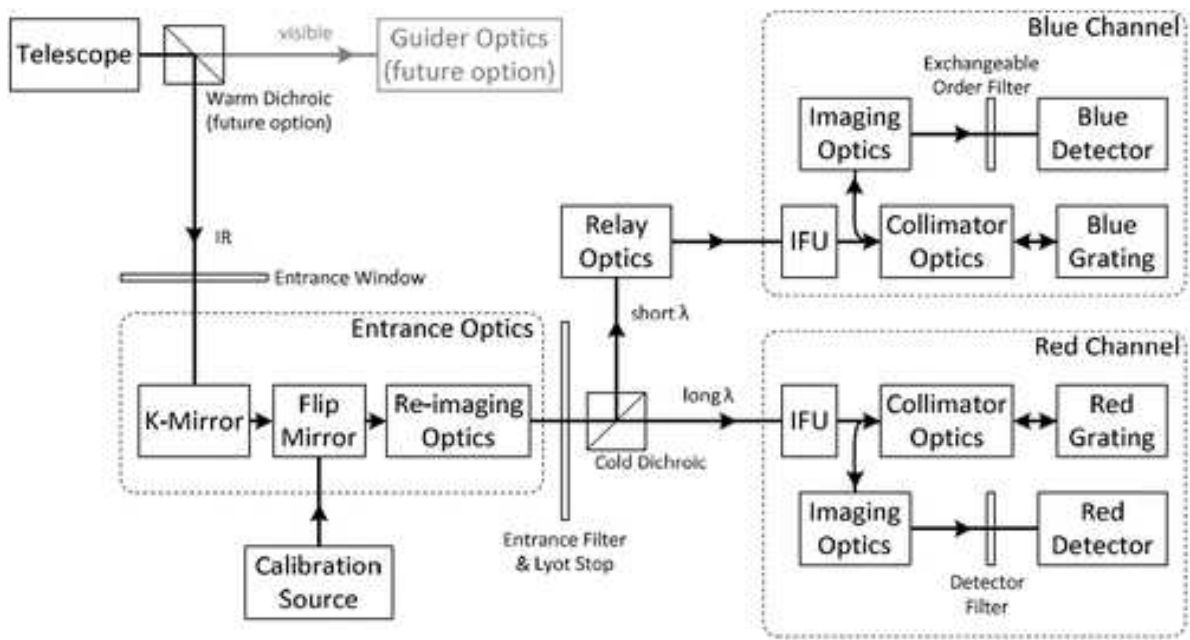

Figure A.8: Schematics of FIFI-LS taken from [44, Fig. 4]. The K-mirror is needed to keep the field orientation fixed (for long integrations) and optimize the coverage for sources. The flip mirror switches the optical path between sky and calibration source (which is used to generate spectral and spatial flat fields). The blue and red channel can be observed simultaneously (thanks to the cold dichroic beam splitter). After the splitter, the pixels have to be rearranged in order to store the 3D information on the 2D detectors (see Fig. A.9). For a more detailed description, see [44].

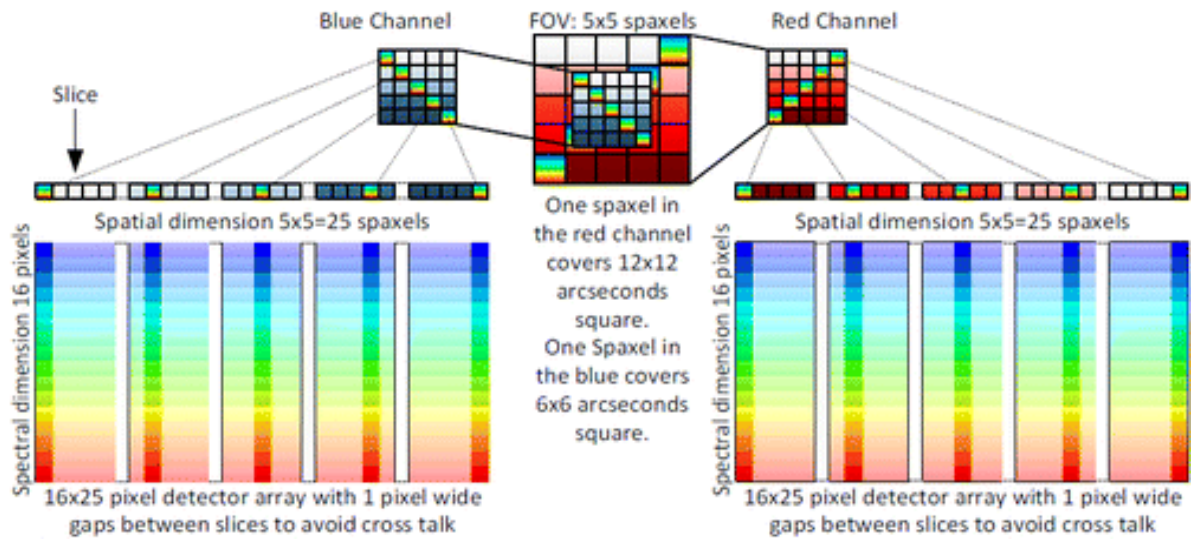

Figure A.9: Schematics of the concept of FIFI-LS taken from [44, Fig. 2]. Each detector (blue and red) features 16x25 pixels. In order to include spectral and spatial dimensions, the pixels are rearranged with 16 wavelength- and 5x5 spatial pixels each. Due to the different plate scale (i.e. 6"/px in the blue and 12"/px in the red channel) the FOV (field of view) is different.

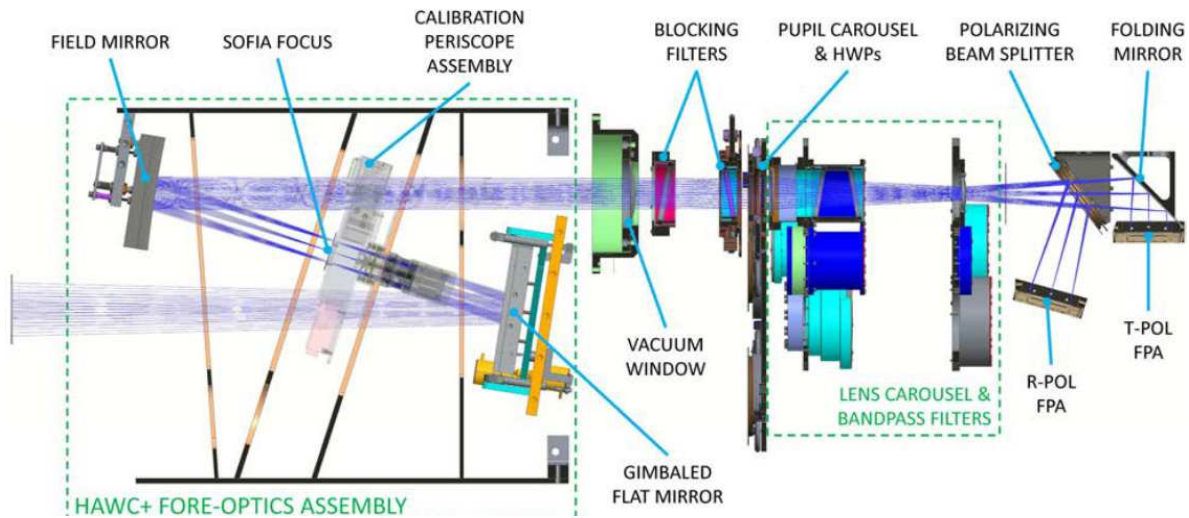

Figure A.10: Sketch with the principal optical components of HAWC+. Image taken from [54, Fig. 13].

### A.1.3 Simulation setting

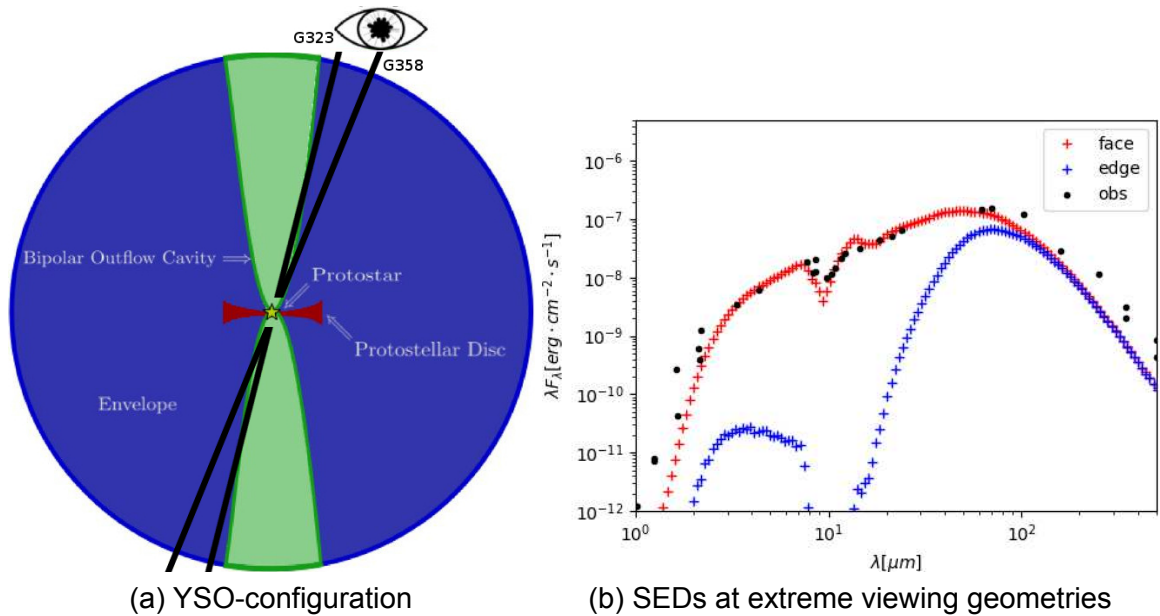

(a) YSO-configuration

(b) SEDs at extreme viewing geometries

Figure A.11: **Left**: Schematics of a prototypical YSO-environment with a star-disk-system in the center of an extended envelope featuring a bipolar outflow cavity. The lines-of-sight of G358 and G323 are indicated, for both objects, the viewing geometry is such that the optical depth is close to the minimal value (face-on view). The Figure is modified from [100]. **Right**: Example SEDs for the extreme viewing geometries. This Fig. has been presented in the 'Tautolloquium 2022'.

## A.1.4   G358

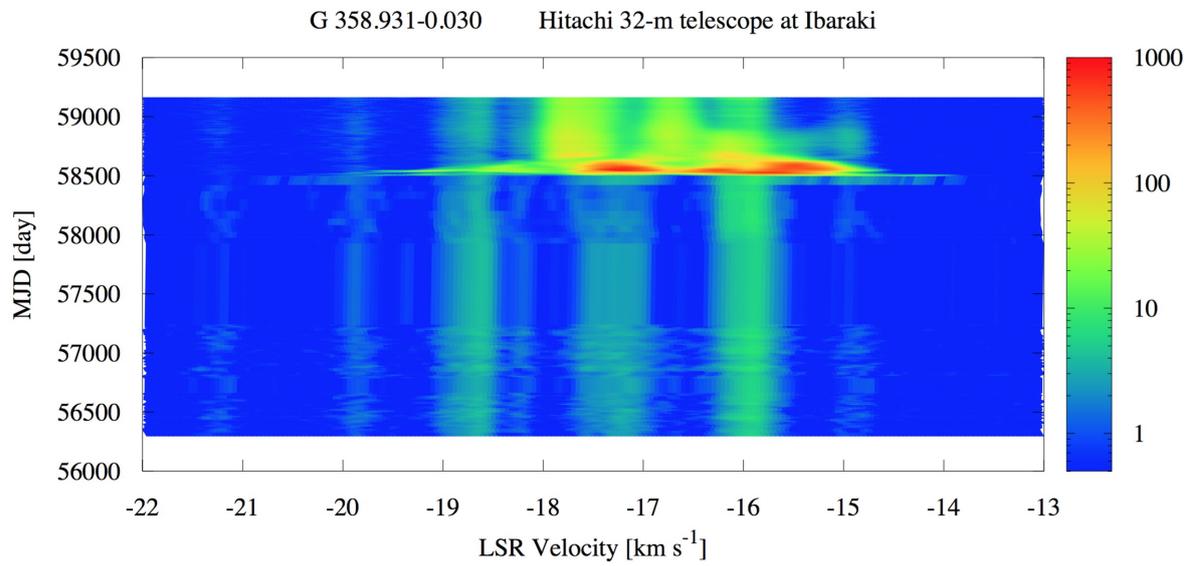

Figure A.12: The dynamic maser spectrum of G358, showing a sudden increase of all velocity components at the beginning of the flare. Taken from [133].

### A.1.4.1 Modeling

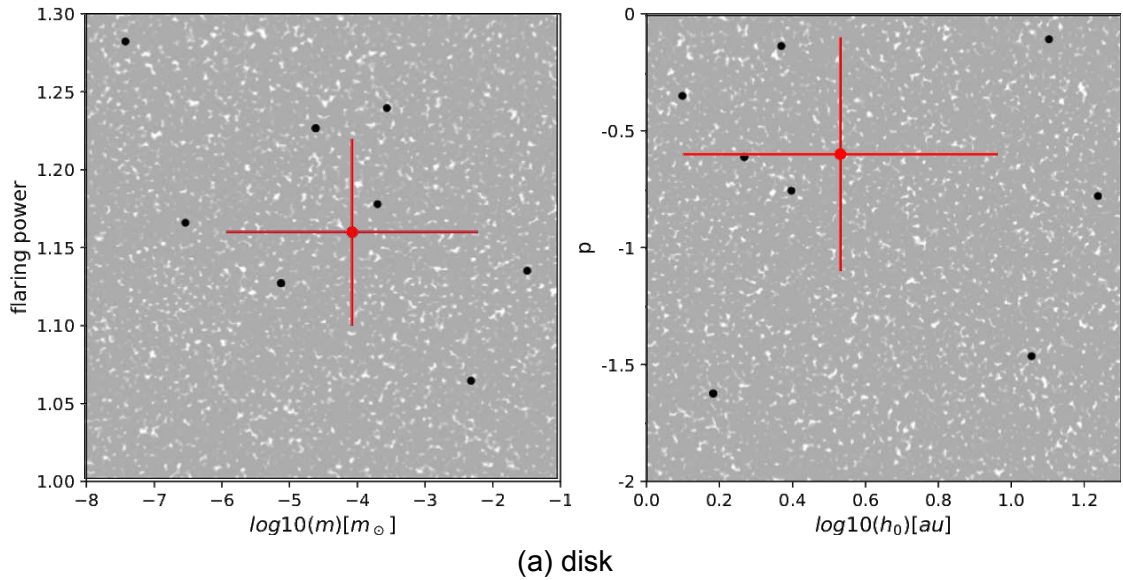

(a) disk

Figure A.13: Example for result of the static preburst fit. In each plot, two specific disk parameters are shown. Grey dots indicate all models in the pool, black dots are only the ten best models (some share parameters, therefore the number of black dots might be less than ten). The plots are generated with the sedfitter software [122]. The values of the mean model are added in red, errors correspond to the 1$\sigma$-confidence intervals. The disk values show a strong scatter. **Left:** Mass and scale-height as an example. There might be a weak anti-correlation between both. **Right:** Disk surface density power p against the scale height $h_0$.

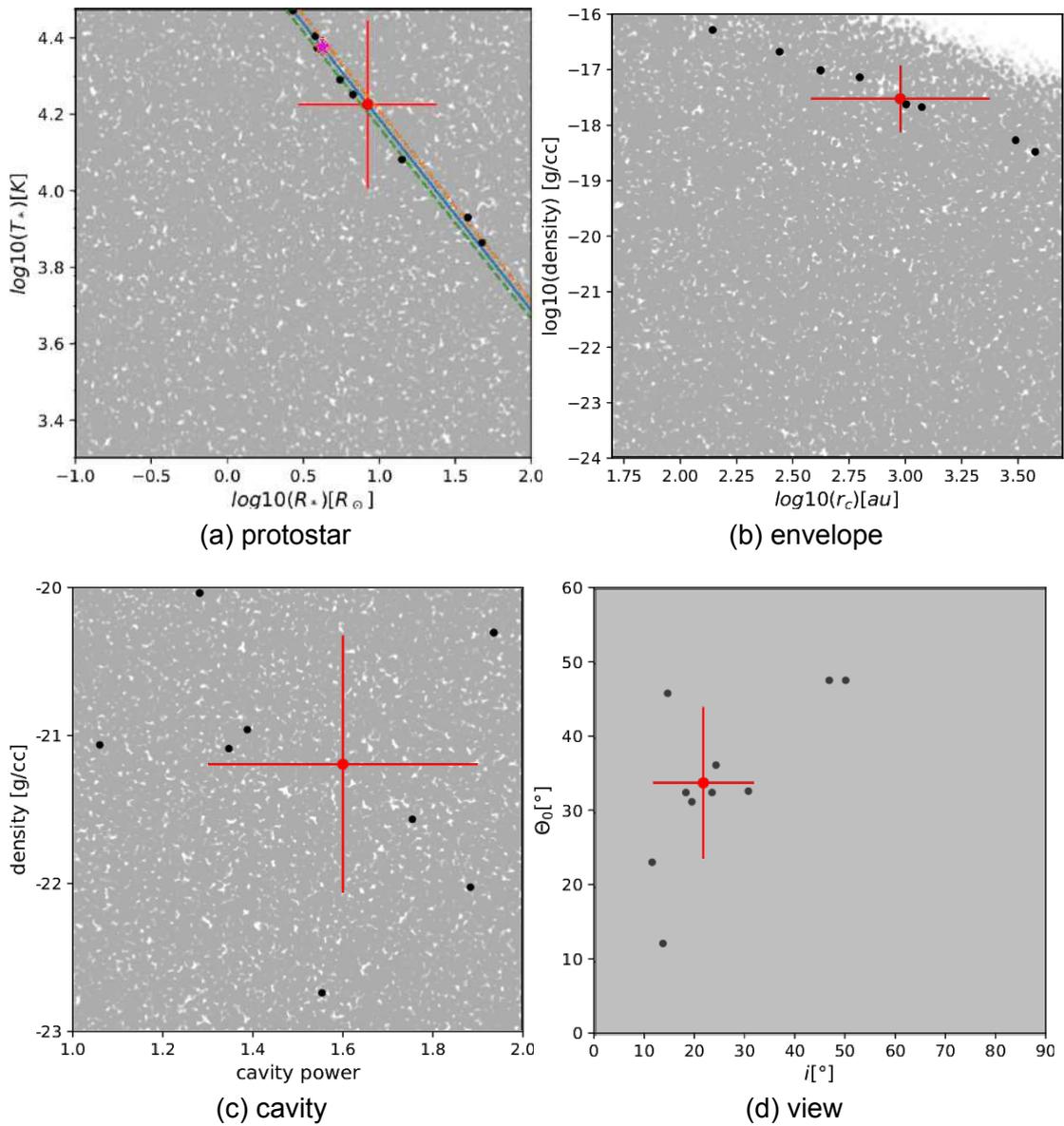

(a) protostar

(b) envelope

(c) cavity

(d) view

Figure A.14: Further examples for the result of the preburst fit. The Fig. is similar to Fig. A.13. In each plot, only two specific parameters are shown. We tried for different combinations, however no correlations have been found except for the ones discussed in Sect. 3.2.3.2. Therefore, each of the free parameters appears only once. **Upper row:** The protostellar luminosity is well-defined by the fit (straight line), whereas its temperature and radius are not very well constrained (**left** plot). The purple star marks the corresponding ZAMS radius [139]. **Right:** The envelope density over the co-rotation radius. The density is in the higher range, it decreases with the co-rotation radius (set to the disk's outer radius). **Lower row:** On the **left** side the cavity power is plotted against its density, neither of both is well constrained. In the **right** panel the cavity opening angle $\Theta_0$ is plotted against the inclination, showing that the protostar is likely seen through the cavity.

### A.1.5 G323

#### A.1.5.1 Observations

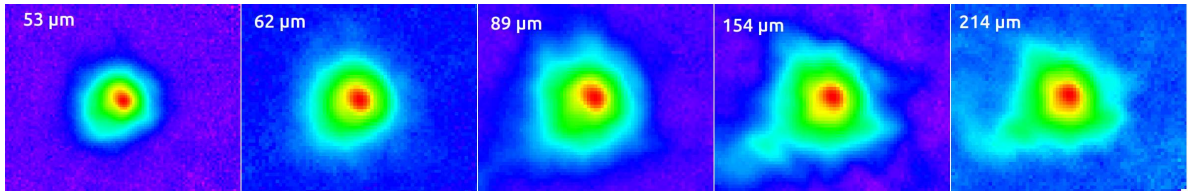

Figure A.15: HAWC+ log-scaled image cut-outs, centered on G323 and spatially scaled to the PSF size. The absence of Airy rings indicates that the source is resolved at all wavelengths. [Wolf+ in prep., Credits: B. Stecklum]

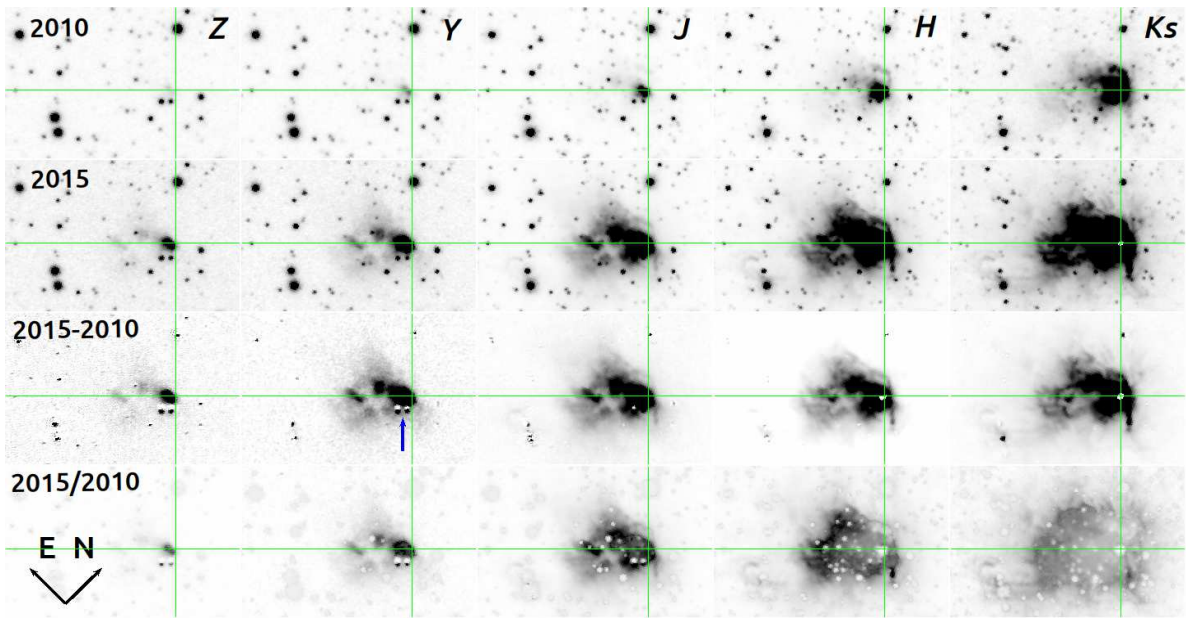

Figure A.16: G323s burst caused a scattered light echo in several NIR bands. From left to right: $Z, Y, J, H$, and $K_s$ filters. Rows from top to bottom show epoch 2010 (preburst), epoch 2015 (burst), difference image (e.g. 2015-2010) and ratio image (e.g. 2015/2010). The cross-hair marks the MYSO position. For the upper two rows, pixel values comprise 98 percentiles, displayed using a linear stretch. The two lower rows adapt a range from 0 – 17.5. The blue arrow points to a common foreground proper motion binary. [Wolf+ in prep., Credits: B. Stecklum]

### A.1.5.2  Modeling

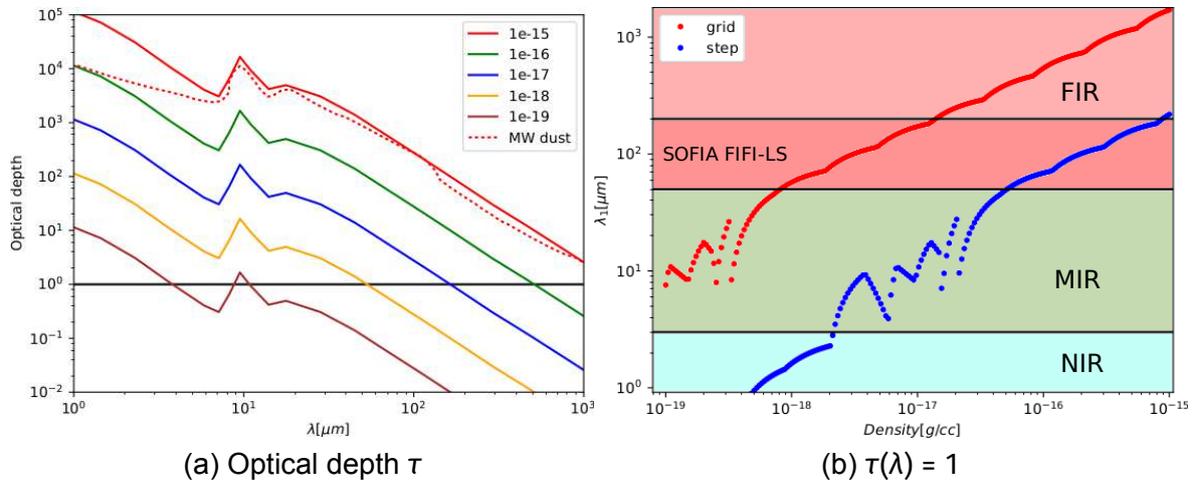

(a) Optical depth $\tau$      (b) $\tau(\lambda) = 1$

Figure A.17: **Left:** Optical depth for MRN-dust with different (constant) gas densities for a distance matching the grid extent of the mean model for G323 (i.e., $2.4 \cdot 10^4 \, au$). The lowest density is in the order of the cavity density. The Milky Way dust is given for comparison (MW dust). $\tau = 1$ is indicated by the horizontal black line. Obviously, the denser, the higher the wavelength where it gets optical thin. **Right:** The wavelength, at which $\tau = 1$ as a function of the density, for the grid (left panel) and for 1 time step (i.e., 630au or 3.65 d at the speed of light). The scatter is due to problems with the interpolation at the silicate feature. The overall trend is nevertheless visible. At the very densest regions, it gets optical thick, even at FIR wavelengths (and for one time step).

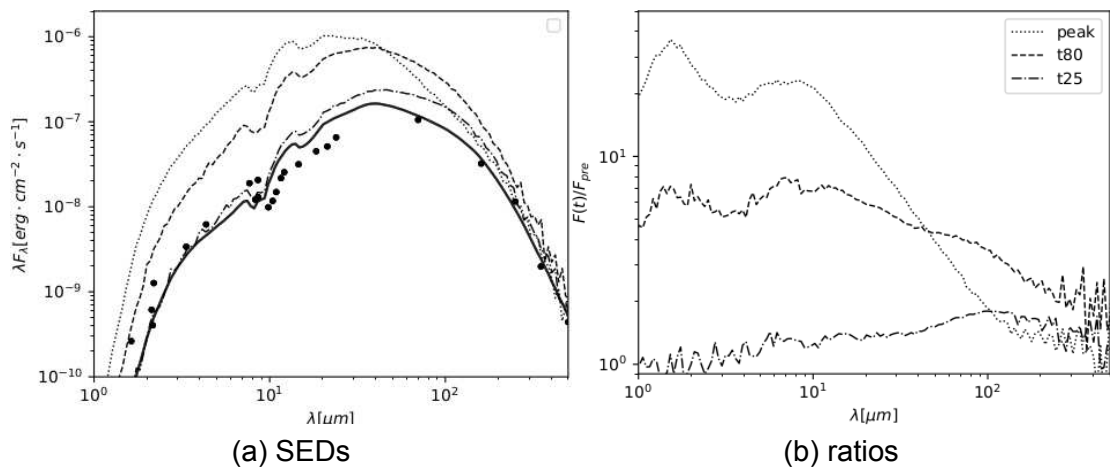

(a) SEDs      (b) ratios

Figure A.18: SEDs (**left**) and ratios (**right**) of G323's mean model at three different times. All times refer to the source luminosity variation. Simulations are performed with TORUS as described in Sect. 3.3. Clearly, the afterglow duration depends on the wavelength (and is much longer towards longer wavelengths). In between peak and t80, the FIR is still rising, while the NIR/MIR already decays. In the FIR, the differences are smaller.

### A.1.5.3  TDRT parameter-study

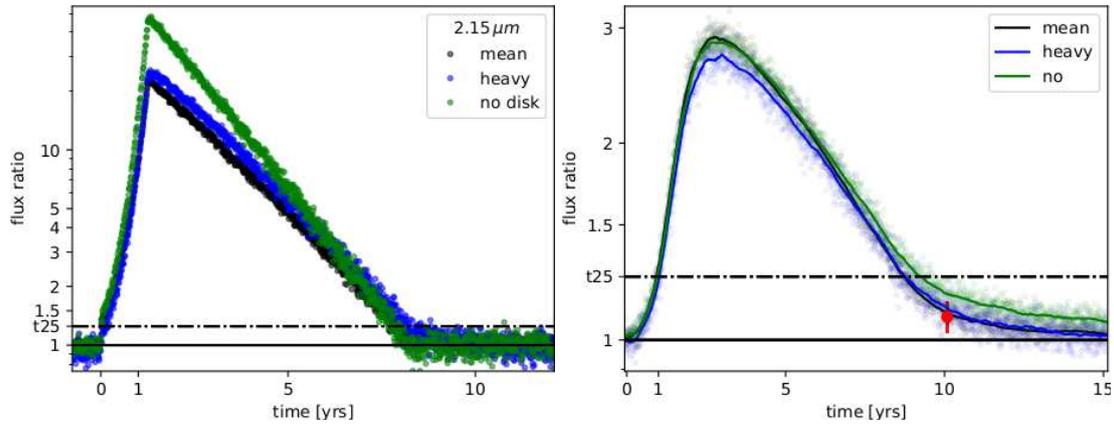

(a) The disk

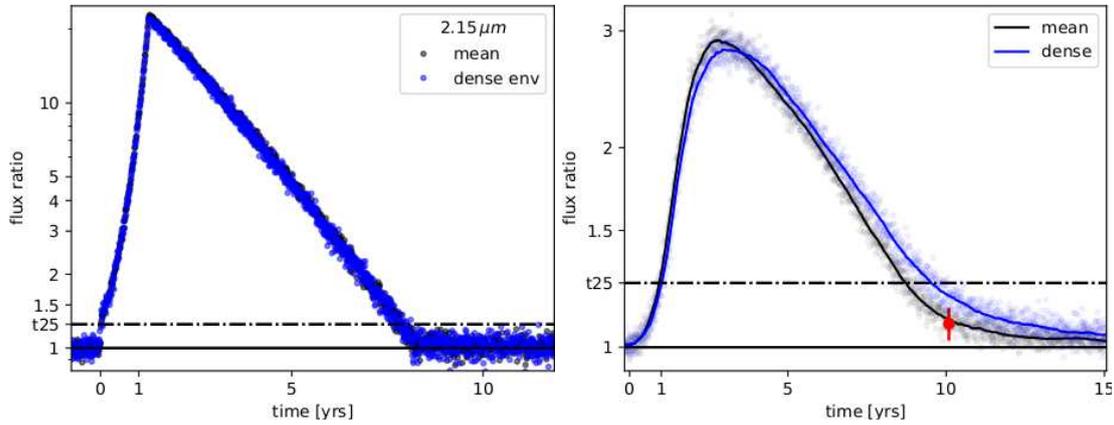

(b) The envelope

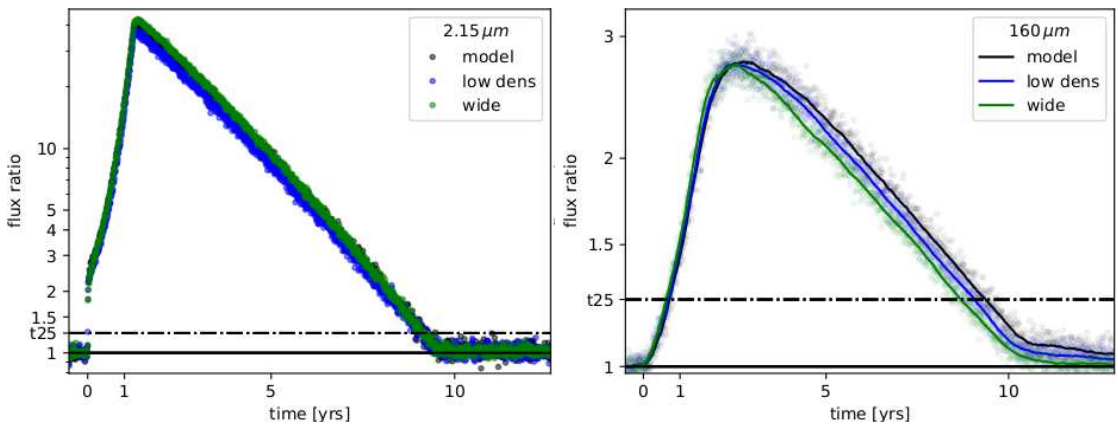

(c) The cavity

Figure A.19: Ks (**left**) and 160 $\mu m$ (**right**) light curve of the models used to test the influence of the disk (**upper**), envelope (**middle**), and cavity (**lower row**) as described in Sect. 3.3.3.3. The Ks peak flux of the no disk model is about 2 times higher for the 'no disk' model, while it stays the same for the other settings. The maximum increase at 160$\mu m$ for the heavy disk is slightly below the other settings (possibly due to shielding), but later on (at the SOFIA observing epoch), both (the mean and the heavy-disk model) equalize. For the dense envelope, the decay is slower. For the cavity models, the curves look similar, but the decay 'starts early' for the wide angle (green) and the low density model (blue).

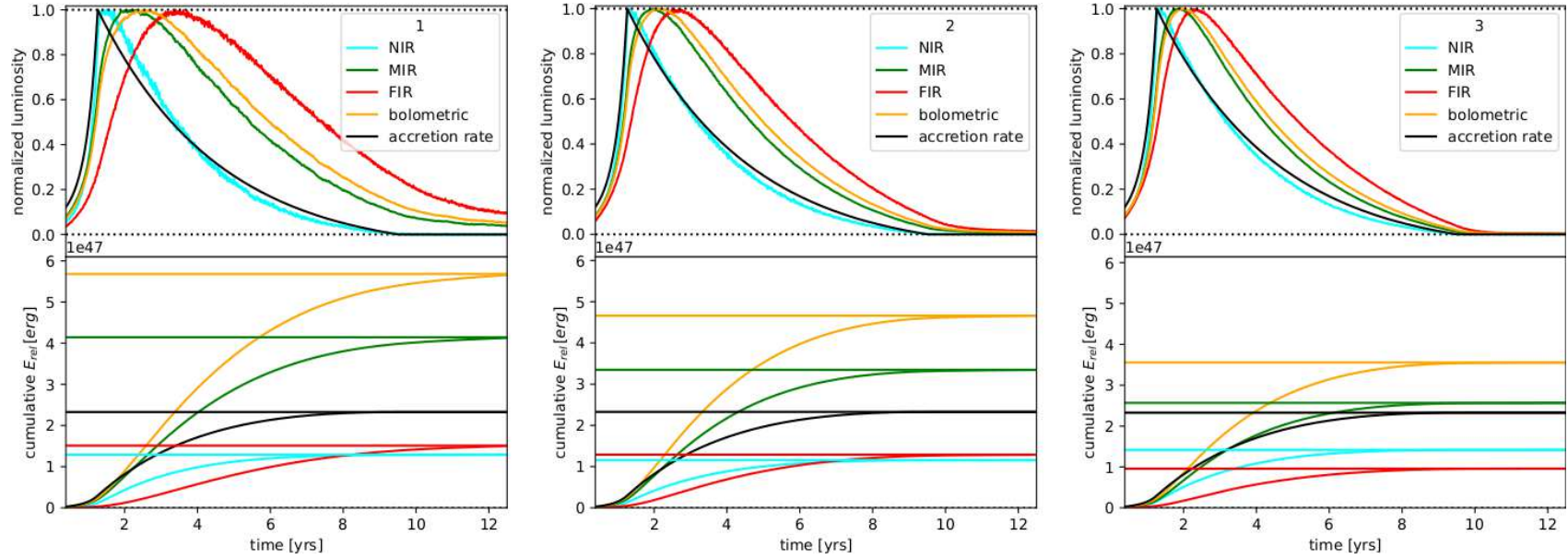

Figure A.20: Same as Fig. A.25 but for a different reference model. The cavity opening angle increases from left to right (the densities are the same). Obviously, the afterglow duration decreases with the cavity opening and more radiation can escape undetected (the apparent total amount of released energy decreases) and at shorter wave-bands (more NIR, less FIR).

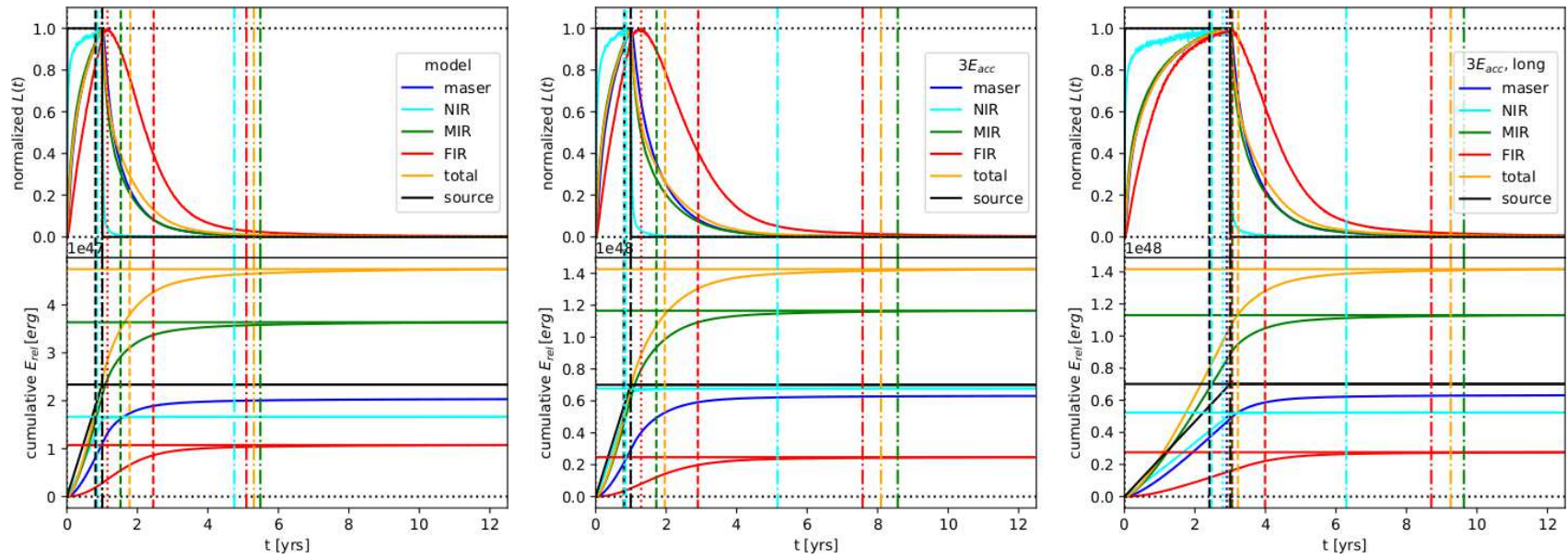

Figure A.21: The afterglow is a record of the burst history. The same as in Fig. 3.11, but for three different rectangular outbursts (see text). Note the different scales. Vertical lines indicate t80 (dashed) and t25 (dashed-dotted).

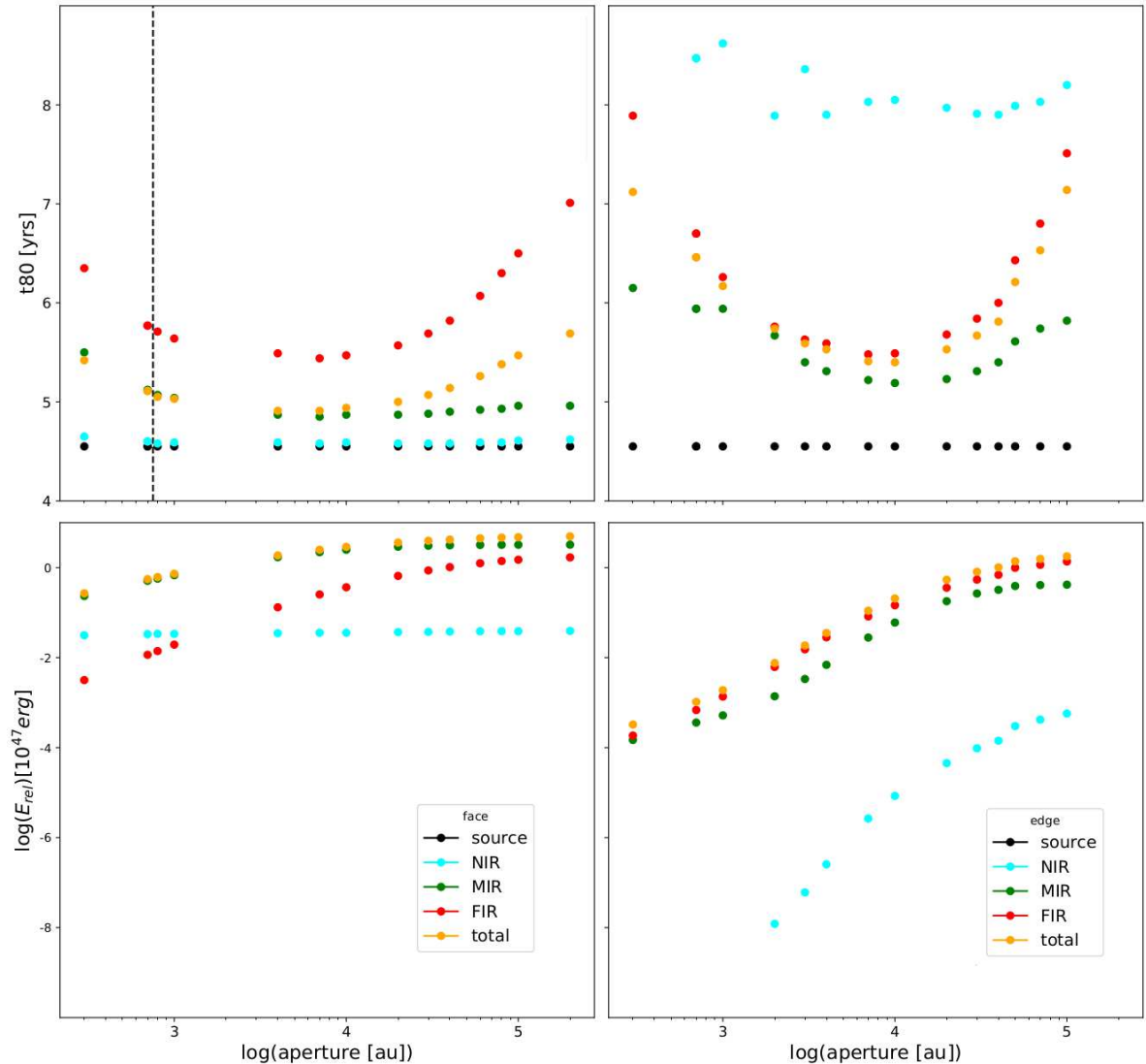

Figure A.22: Timescales (**upper row**) and total released energy (**lower row**) as a function of the aperture for face-on (**left**) and edge-on (**right**) configurations. The extent of the disk is indicated in the upper left plot (vertical line). The total released energy is integrated (in the respective wavelength-regime) over the whole afterglow duration. Surprisingly, the 'edge-on value' is below the corresponding 'face-on value' even in the FIR at all apertures. This is because of the high optical depth (even at FIR wavelengths, see Fig. A.11). In the face-on case, all NIR flux is included in the smallest aperture already, while it steadily increases with the aperture in the edge-on case. In the face-on case the NIR-photons are 'direct' photons, in the edge-on case, there are mostly scattered.

### A.1.5.4   Results (G323 models)

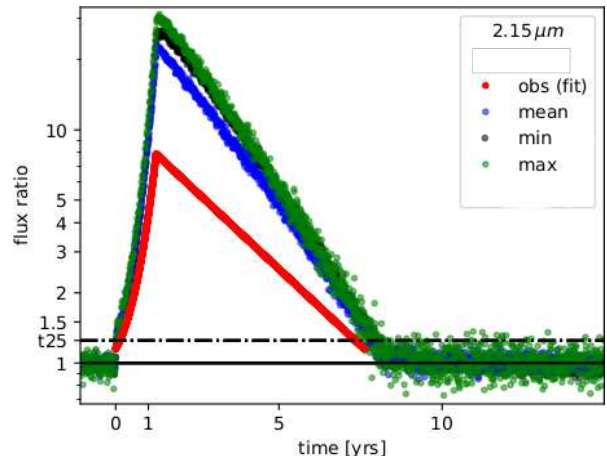

Figure A.23: Ks-light curve of the models described in Sect. 3.3.4.

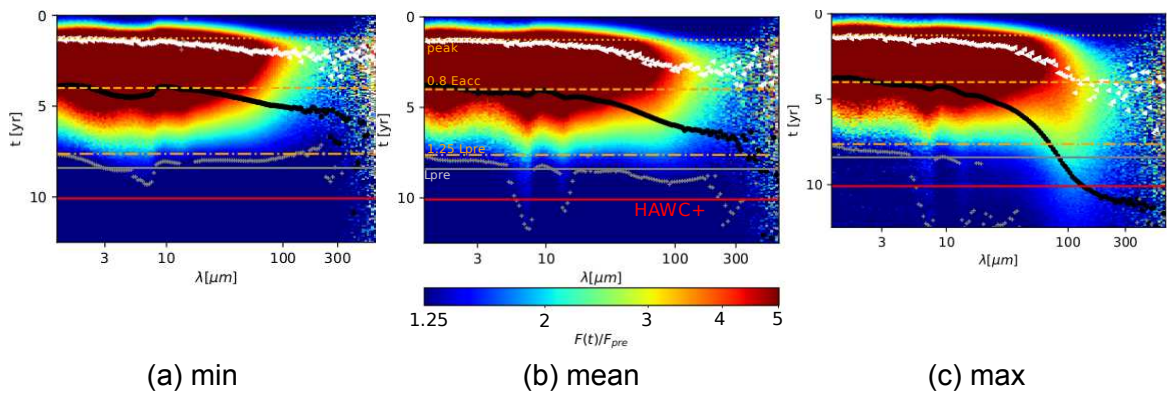

(a) min                    (b) mean                    (c) max

Figure A.24: Dynamic ratios of min (**left**), mean (**middle**) and max (**right**) model showing possible afterglows for G323 (as they depend on the local dust-distribution). The burst is the same for all cases. The differences (due to the local dust configuration) are remarkably large.

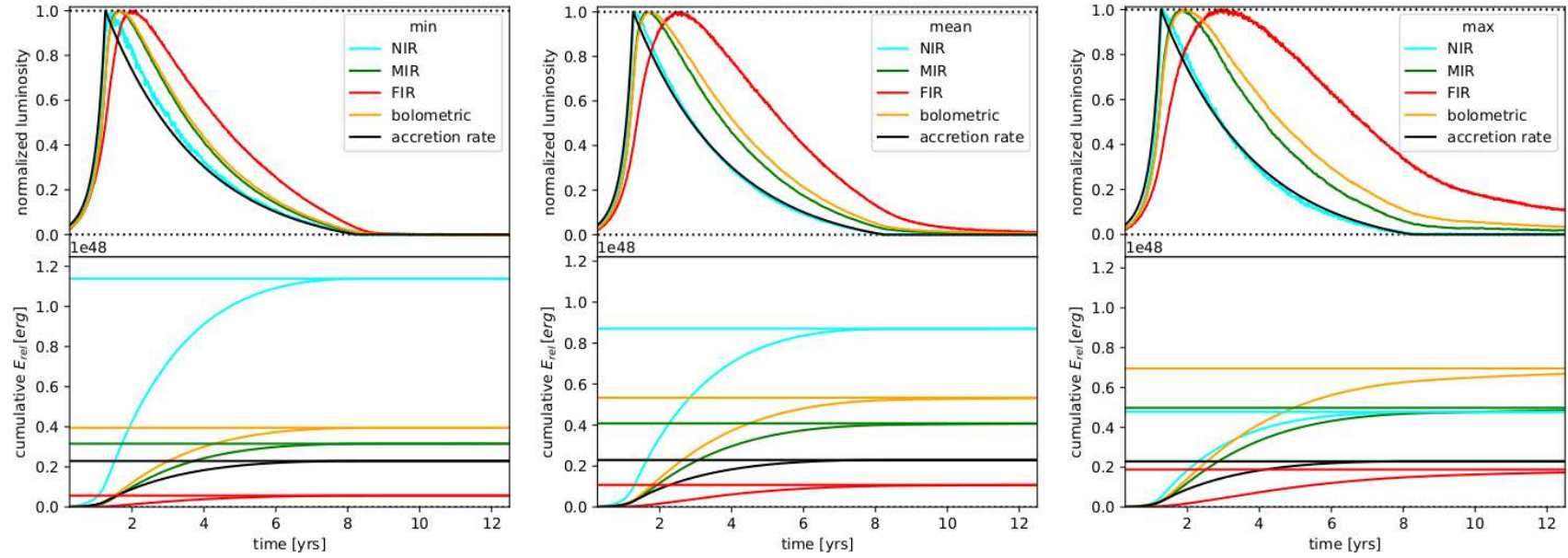

Figure A.25: Normalized luminosity (upper row) and apparent cumulative released energy (lower row) among the wavelength regimes of the models described in Sect. 3.3.4. The NIR is scaled by a factor of 50 for better visibility. The extinction increases from left to right, while the outflow angle decreases. The burst is the same in all cases. Obviously, the heat remains longest in the max-configuration (densest, smallest cavity opening), which powers the FIR afterglow years after it ends in the other configurations (min, mean).

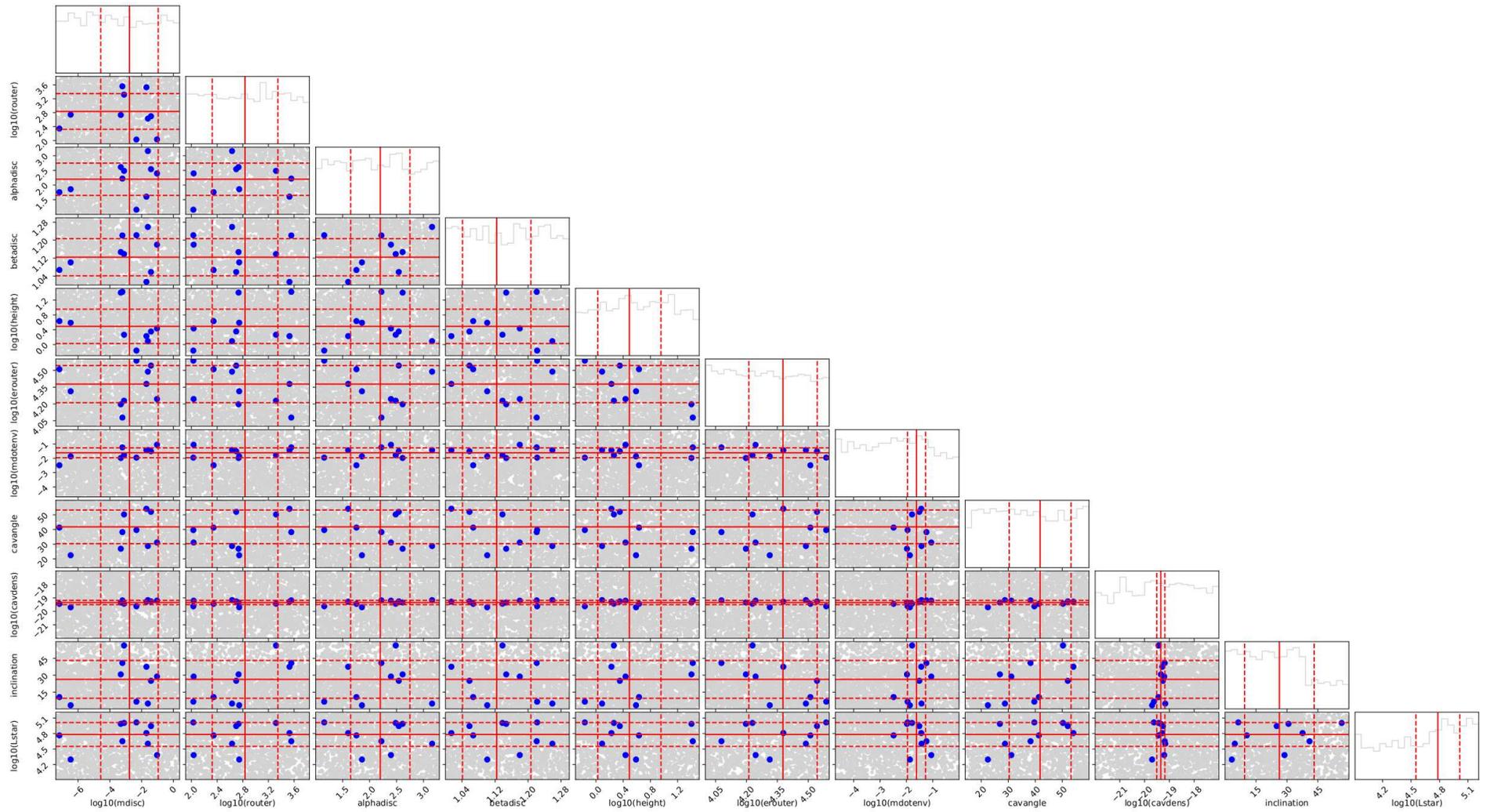

Figure A.26: Corner-plot for the G323 preburst fit. The red lines indicate the mean model (including the 1$\sigma$ confidence intervals).

### A.1.5.5 Burst energy (G323 models)

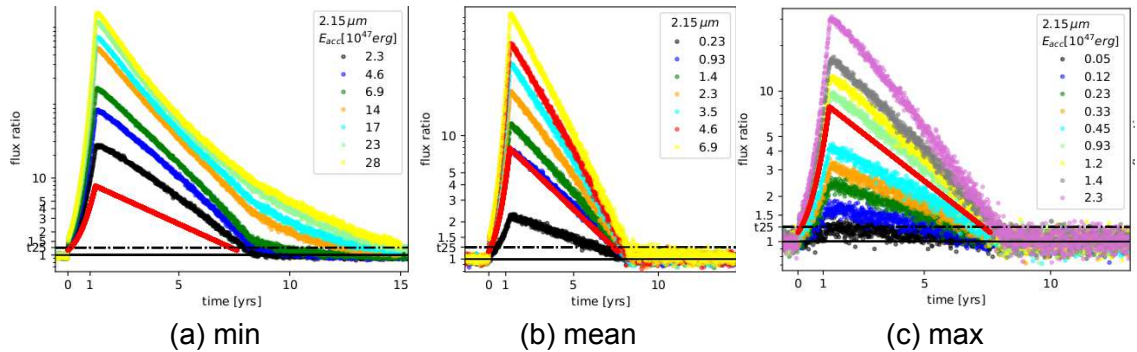

(a) min      (b) mean      (c) max

Figure A.27: Ks-light curve for min (left), mean (middle) and max (right) model featuring different bursts (in $10^{47}\,erg$, color-coded). The fit to the observed Ks-increase is shown in red for comparison. The ratio depends rather on the energy input, then on the setting. The best agreement is reached for $E_{acc} = 0.93 \cdot 10^{47}\,erg$. For the min setting and the highest burst energies, the Ks-curves end later, this might be due to the innermost dust getting unrealistically hot ($r_{min}$ is fixed to 60 $au$ configuration, that is $3 \cdot R_{sub}$ of the mean model).

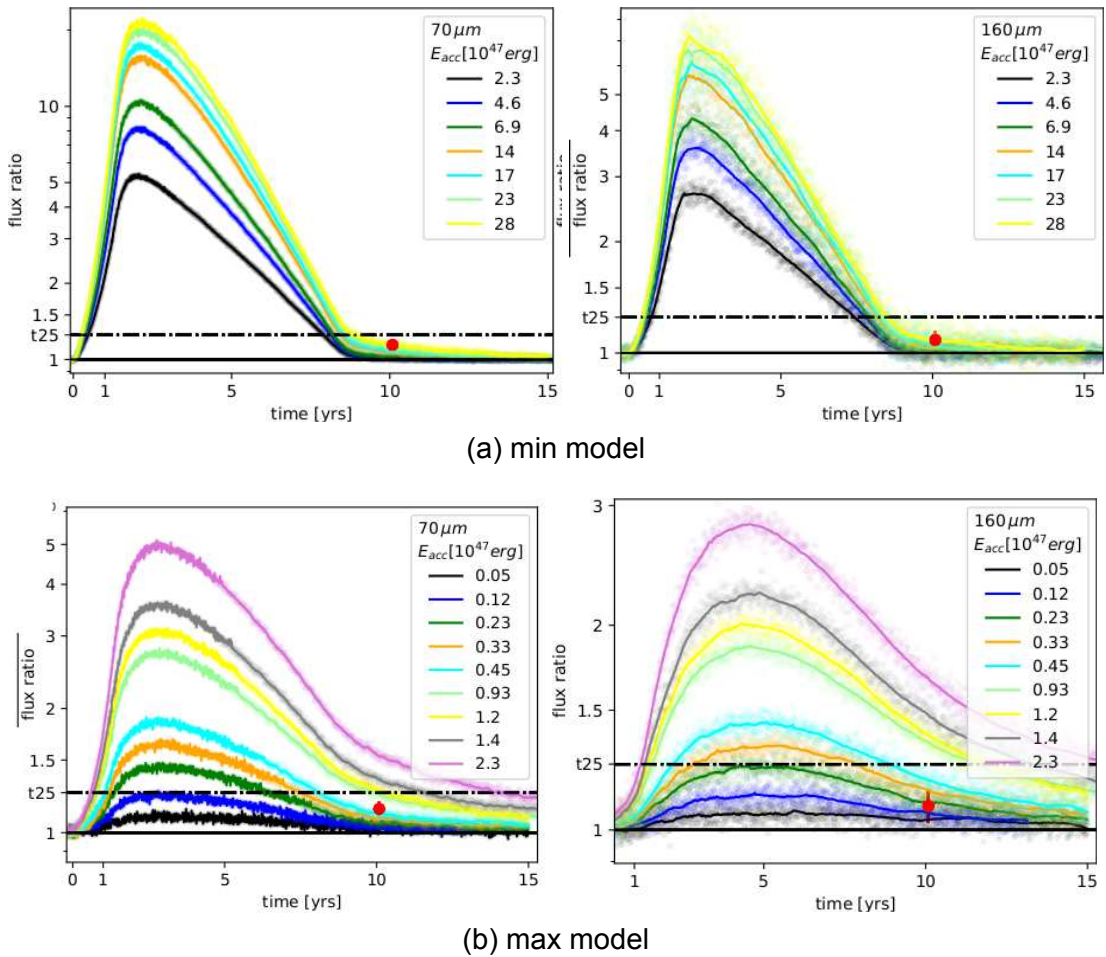

(a) min model

(b) max model

Figure A.28: Same as 3.26 but for min/max model (upper/lower row respectively). If the dust configuration is such, that the afterglow timescale is shortest (i.e., min model), then the burst needed to reproduce the HAWC+ data must be the most energetic.

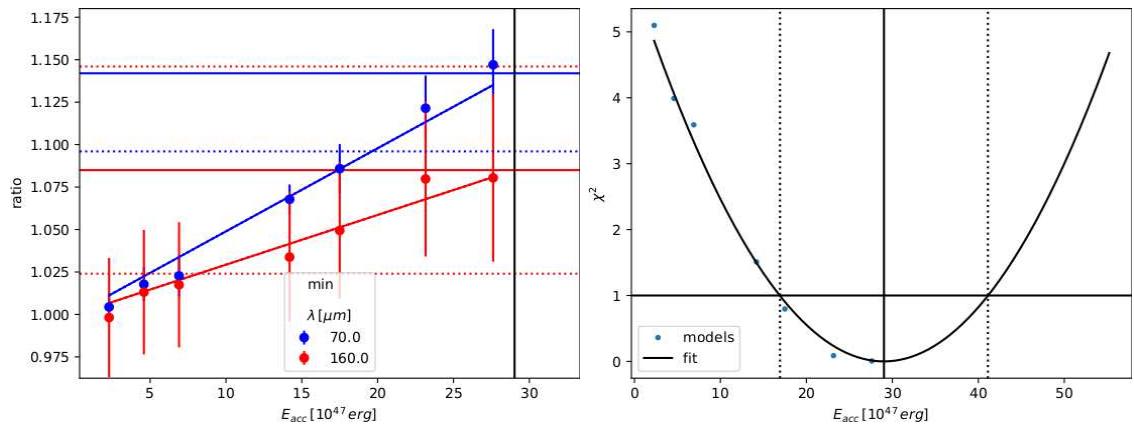

(a) min model

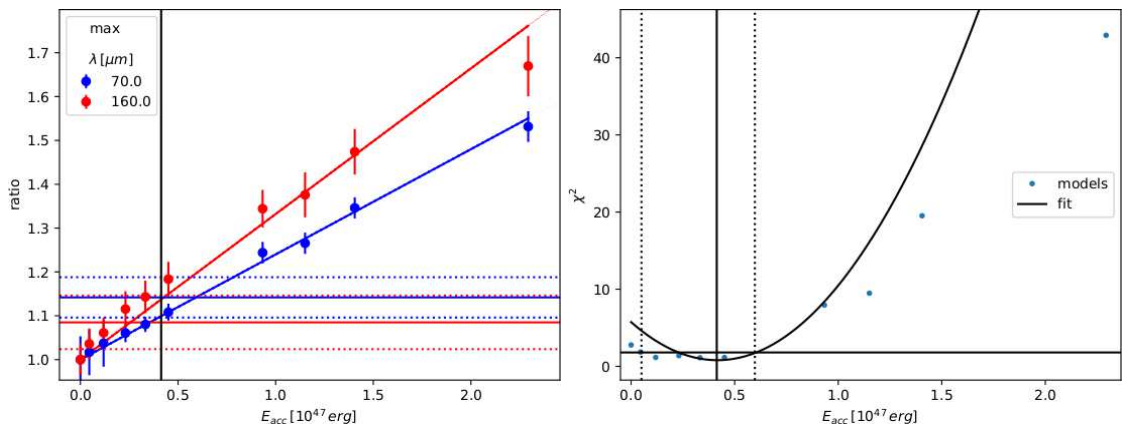

(b) max model

Figure A.29: Same as Fig. 3.27, but for the min/max model (**upper/lower** row). For these settings, the burst energy needed to explain the HAWC+ data is maximized/minimized.

**A.1.5.6   Discussion**

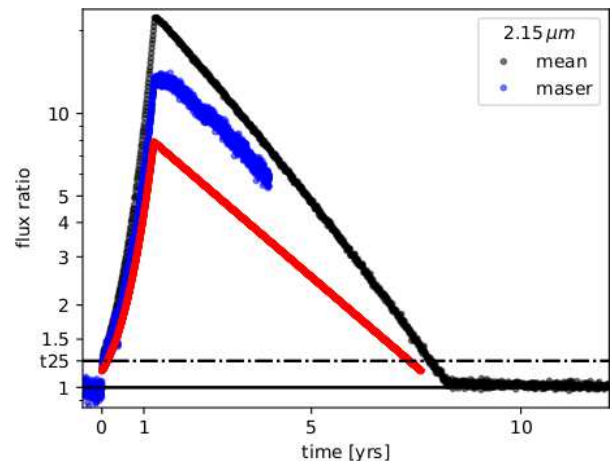

Figure A.30: Predicted Ks-ratio for the mean setting, as compared to the one we used for the maser-curve. As the latter includes dust-sublimation and features a smaller time step, it is the better guess.

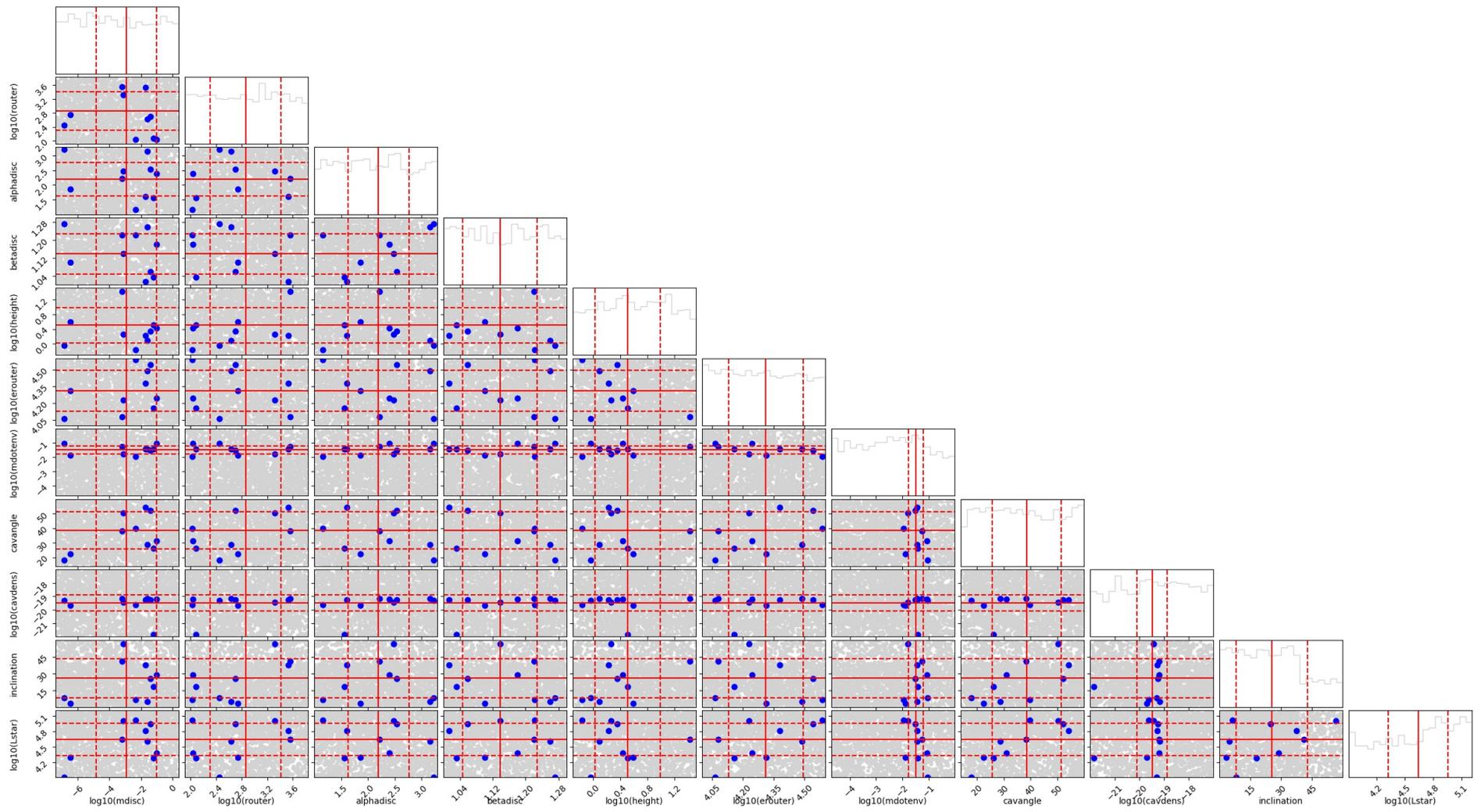

Figure A.31: Same as Fig. A.26 but with all FIR data points excluded between 30 and 350 $\mu$m. In case of G323, the mean values still agree quite well.

## A.1.6   Tfitter

### A.1.6.1   Benchmark

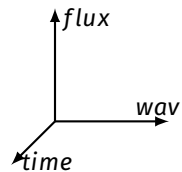

Figure A.32: Sketch of the Tcube-format. It features the dimension [SED (i.e., wav, flux/flux density ± error), time] or in numbers (for the benchmark setting): [(200, 200 ± 200), 100]. A projection is given in Fig. A.34, right in transparent (without errors).

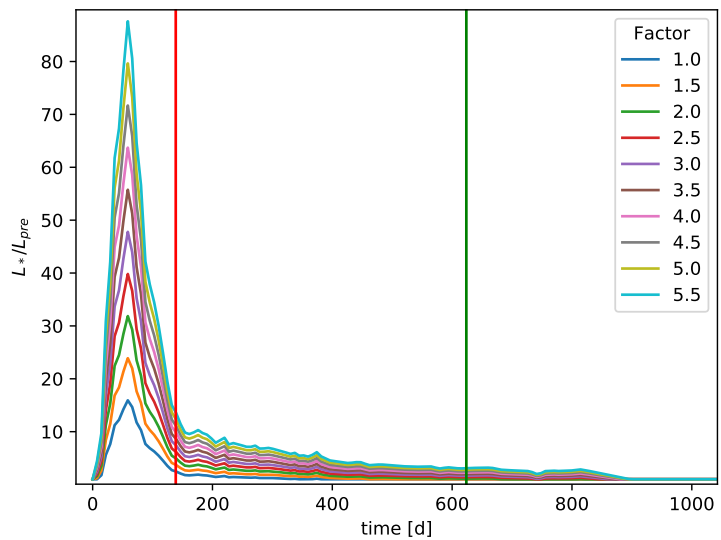

Figure A.33: Templates for the burst-induced luminosity variation, based on G358's maser flux light curve. These are used as input for the TFitter. Vertical lines indicate the dates of the observing epochs.

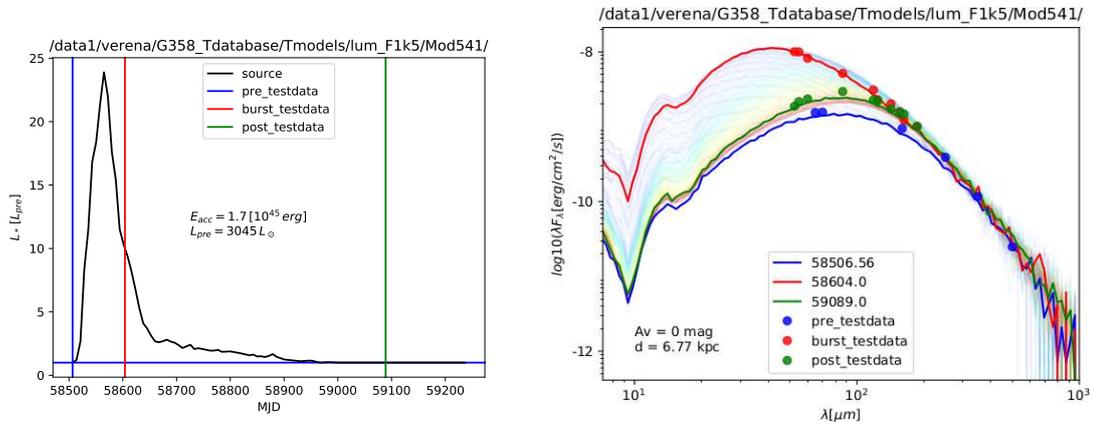

Figure A.34: Examples for the TFitter input: **left:** source luminosity variation (black) and observing epochs (vertical lines). **Right:** epoch-data (dots) with the uncorrected epoch SEDs (interpolated to the respective date) of the first model in the pool (dark colors). The entire Tcube (all SEDs of that model) is shown in transparent (with the time color-coded).

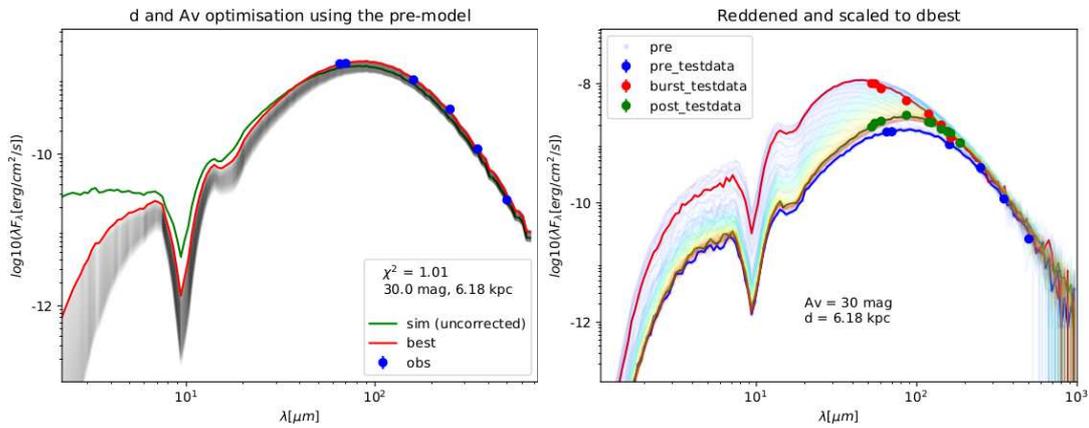

Figure A.35: Examples for the TFitter d, Av correction: **left:** original preburst simulation (green) together with the range of values, spanned by the adapted ranges for d/Av (gray). The best fit is given in red (the corresponding values are in the legend). **Right:** same as Fig. A.34 right, but with d/Av-correction (according to the pre-fit).

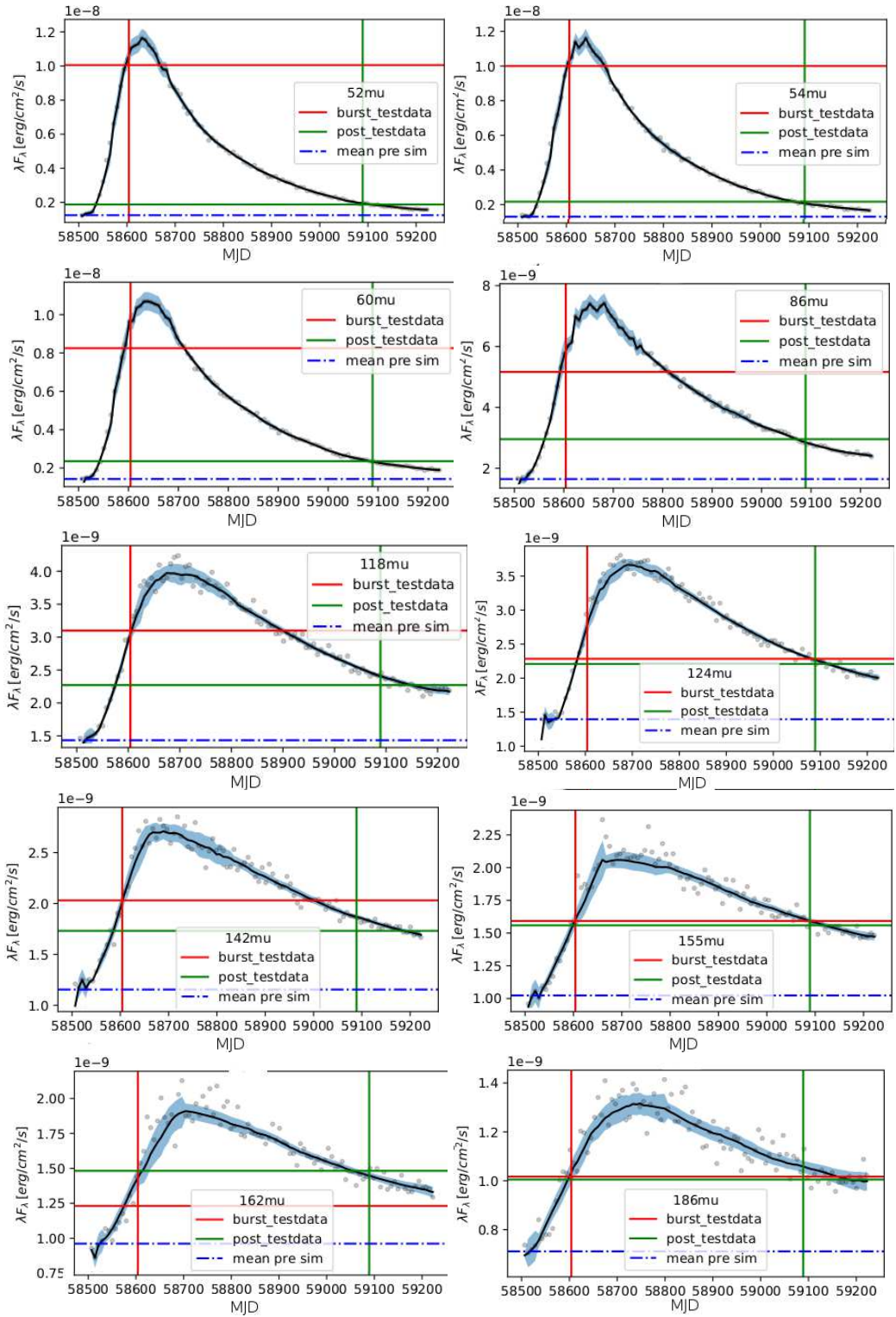

Figure A.36: Simulated (gray dots) and optimized light curves (solid black line with confidence-intervals in light blue) for the test model for all observed burst-/post- wavelengths. Test data (colored lines) is given for comparison. The y-scales are different, x is the same for all plots.

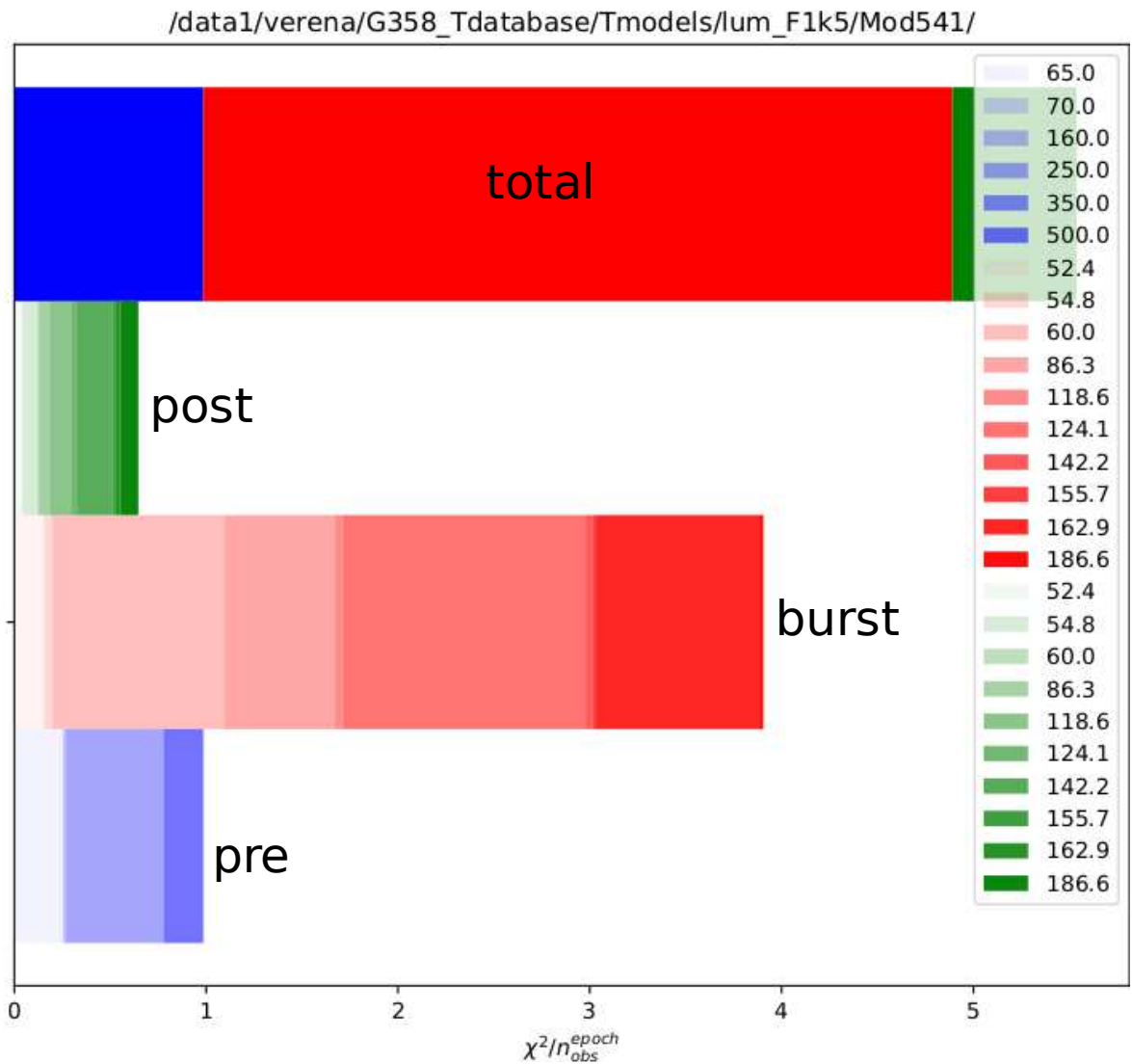

Figure A.37: Example for the fit result for the test model, $X^2$-values split by epoch and wavelength. The values are divided by the numbers of data-points per epoch. The uppermost row gives the weighted sum (i.e., the value used to compare the models). Obviously, the $X^2$-value for this particular model is dominated by the burst. The deviation does not systematically depend on the wavelength (as expected).

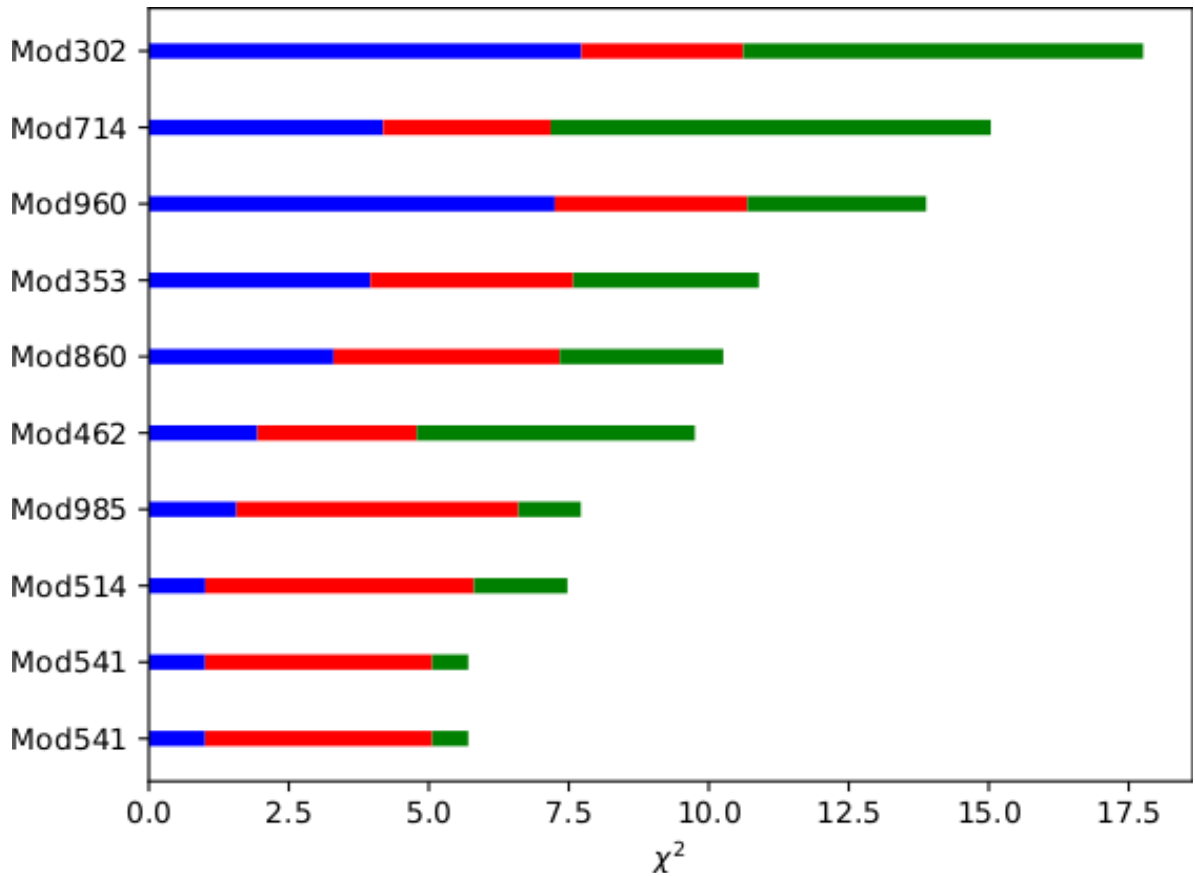

Figure A.38: $\chi^2$-bars for the 10 best models split by epoch (labels are input dirs). The deviation of the best models is dominated by the burst. Note that the test-model appears twice, as it was included twice in the input-list for practical reasons.

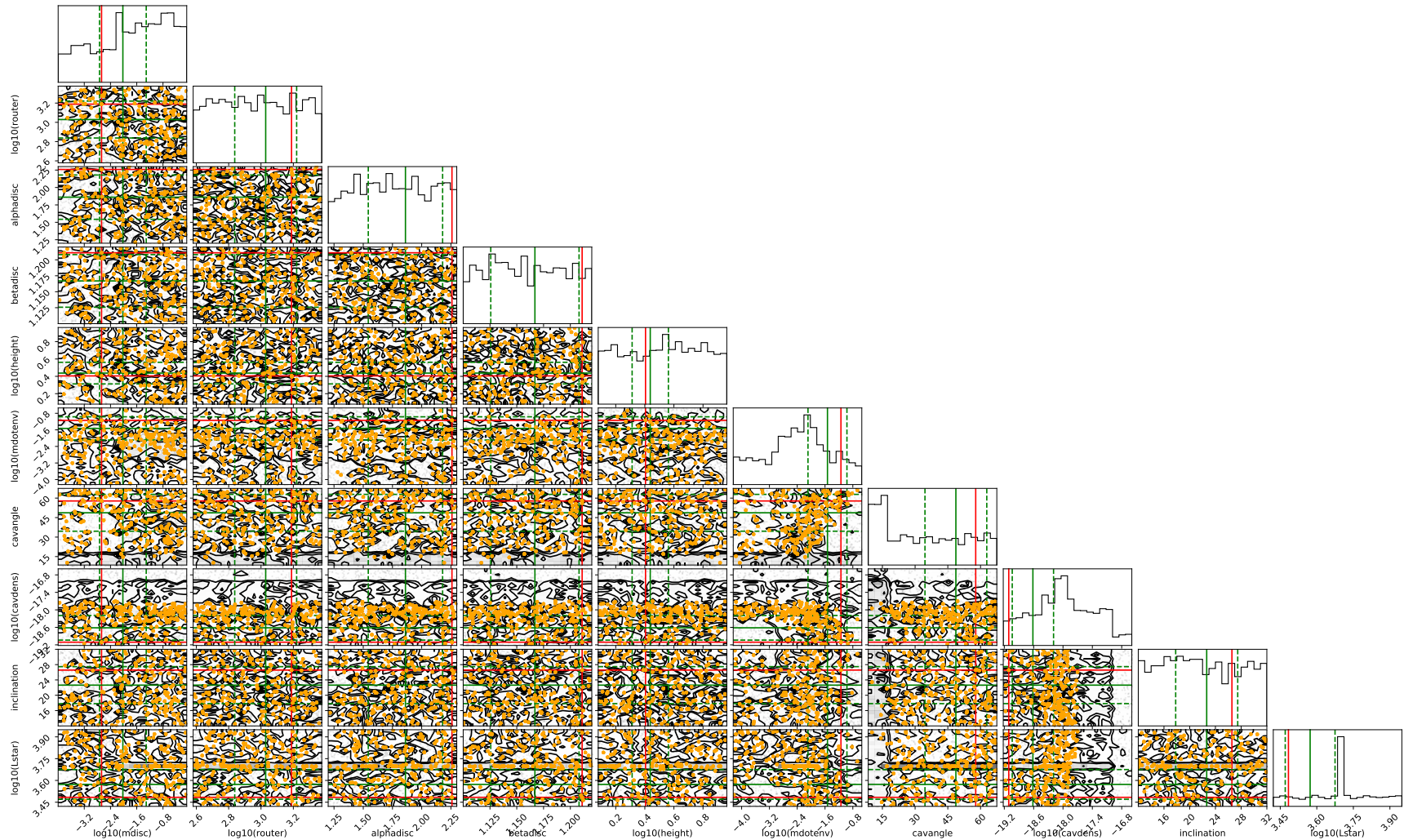

Figure A.39: Corner plot showing all pre-models used for the TFitter benchmark (black). Red indicates the best model, and orange all Tmodels. Green lines are for the mean premodel (and confidence intervals), which obviously differs from the best model (although 6 out of 10 parameters shown are 'the same' within the errors). The protostellar luminosity was only varied for a bunch of models. The total number of models is 3027 (pre) or 837 (Tmodels). A corner plot of the Tmodels including the fit result is given in Fig. A.40.

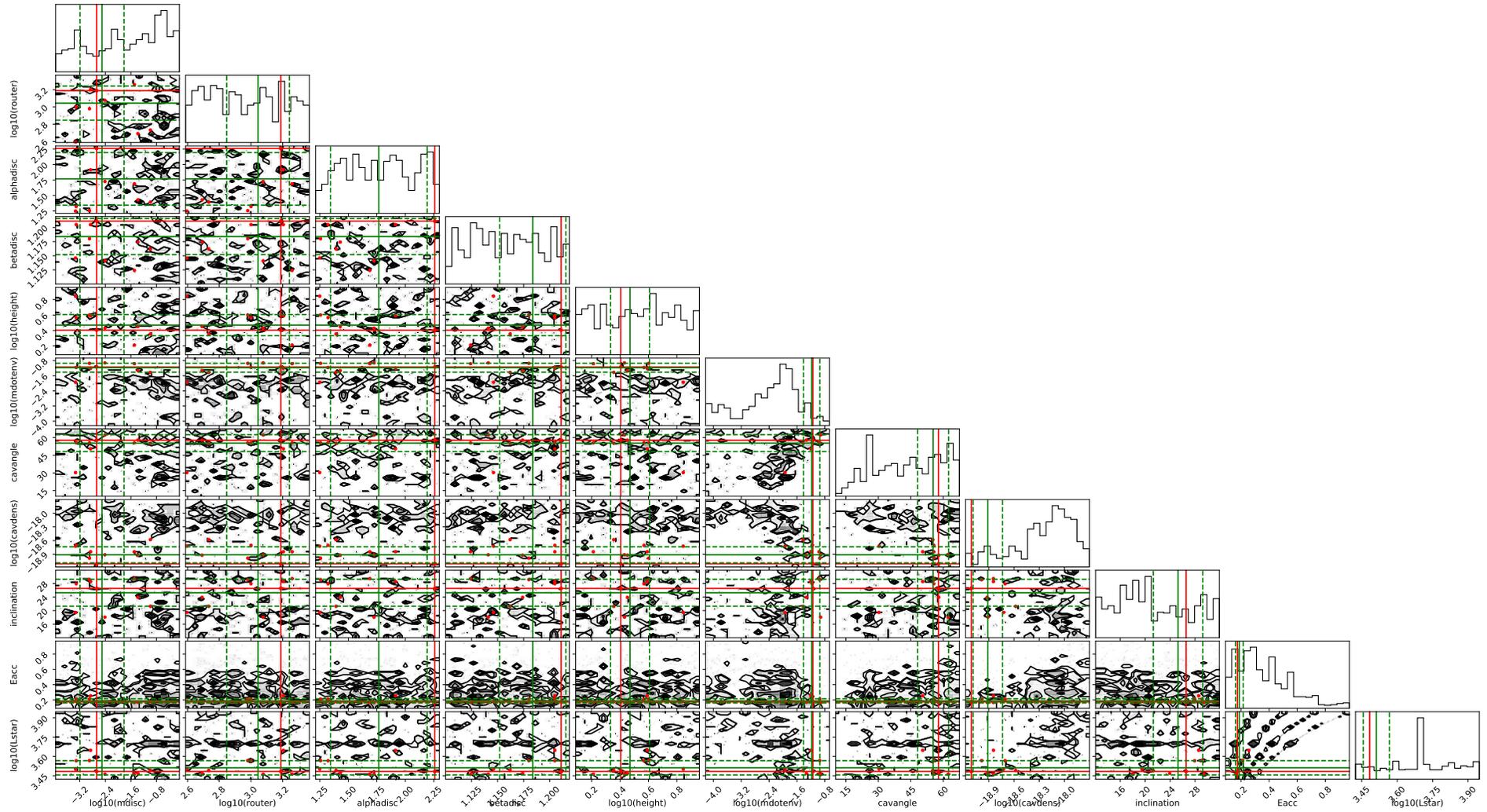

Figure A.40: Corner plot showing all Tmodels used for the TFitter benchmark (black). The best model is indicated by the red lines (same as the test model). The weighted mean parameters and the corresponding 1σ confidence intervals are colored green. All mean parameters despite $\alpha_{disk}$ agree with the test model ('input values'). The mean Tmodel (this plot) has slightly smaller confidence intervals than the mean premodel (Fig. A.39.

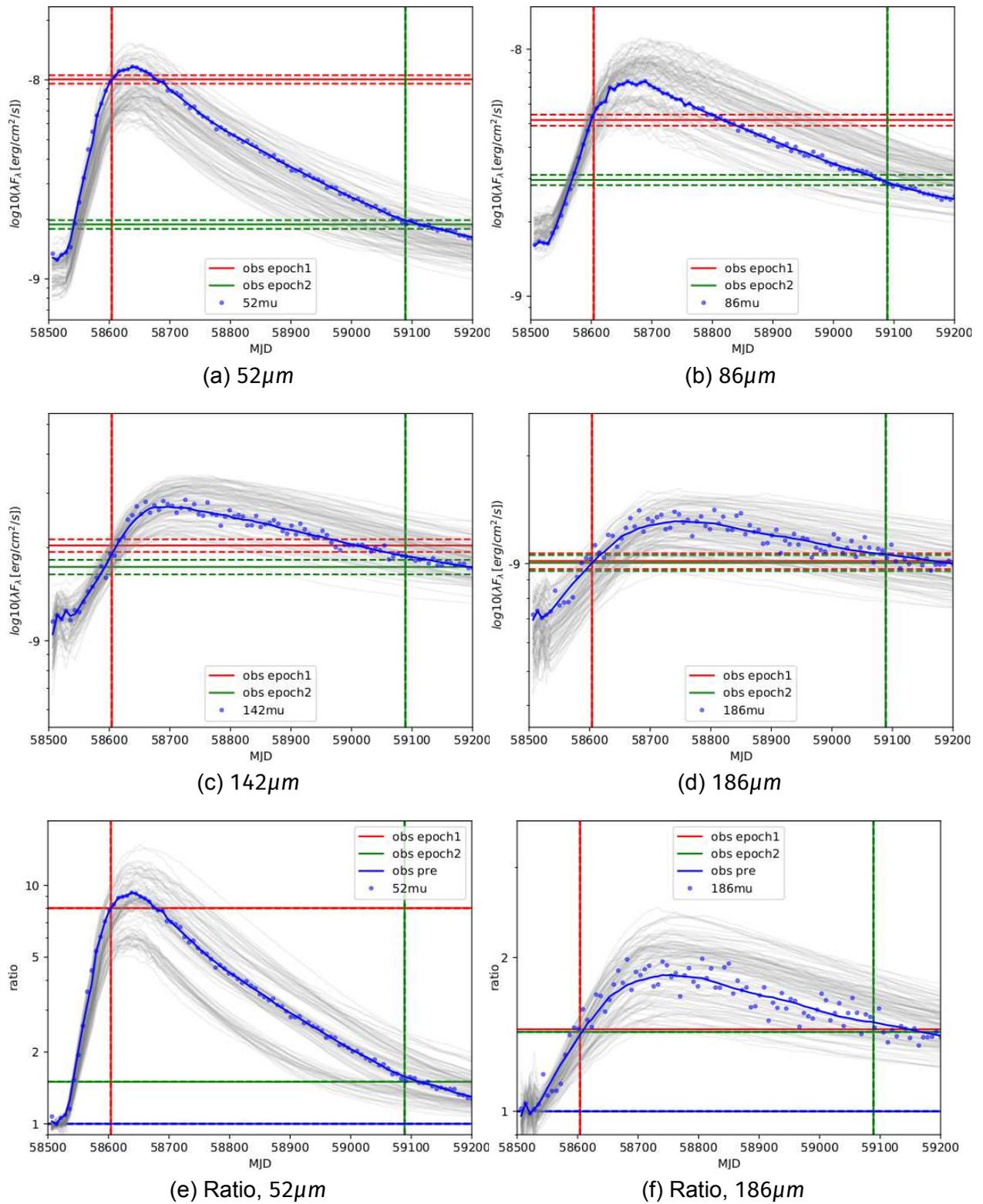

(a) $52\mu m$

(b) $86\mu m$

(c) $142\mu m$

(d) $186\mu m$

(e) Ratio, $52\mu m$

(f) Ratio, $186\mu m$

Figure A.41: Results of the benchmark fit with the TFitter (value-fit): Optimized light-curves (4 examples) for the 100 best models (gray) compared to the test-data (color). For the best model, the scatter/fit is indicated in blue (dots/solid line). The lowermost row gives the ratio-curves for 2 out of the 4 examples for comparison. The synthetic noise is increased (as expected).

## A.1.6.2 TFitter benchmark, ratio

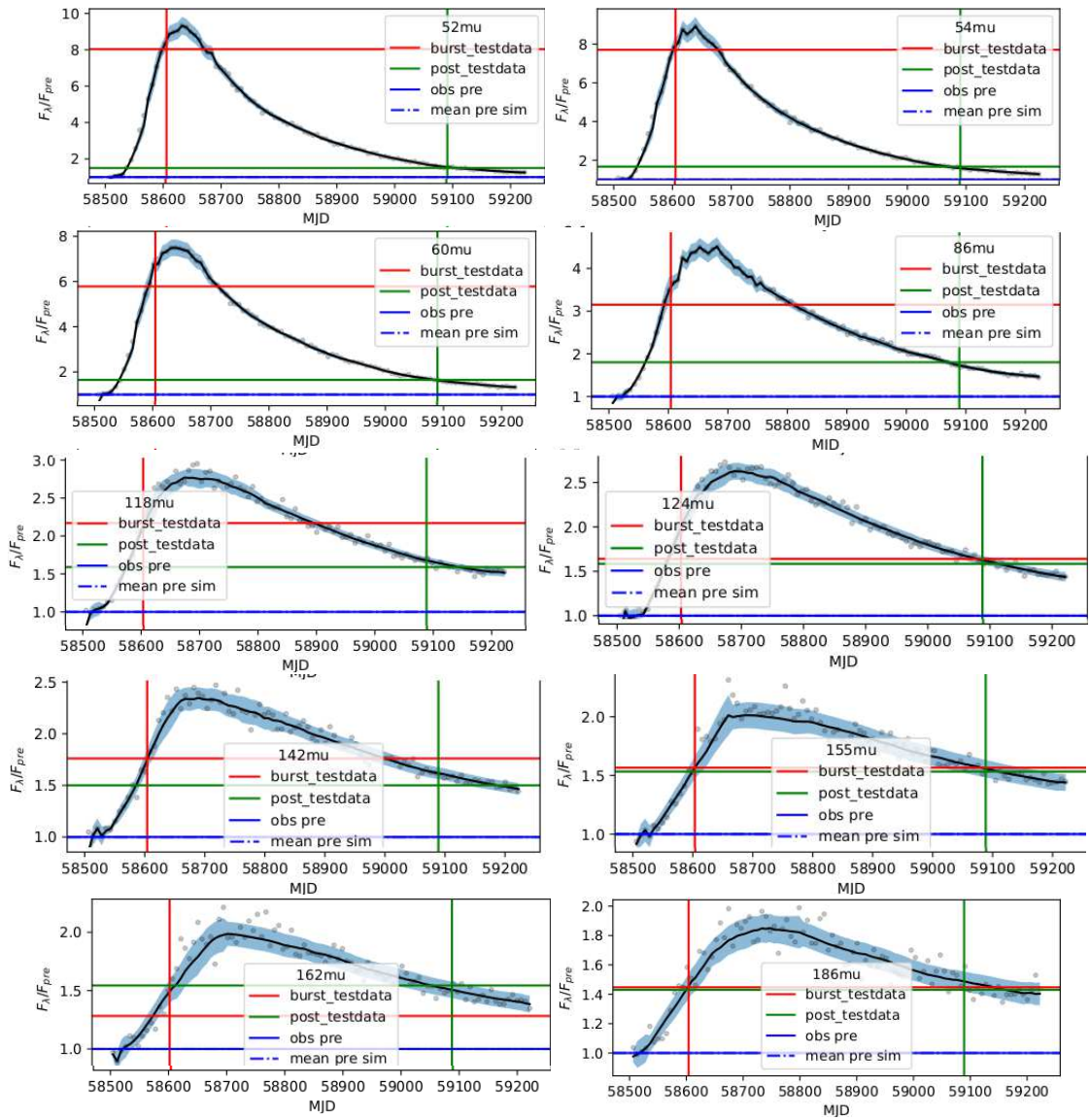

Figure A.42: Same as Fig. A.36 but for the ratio fit.

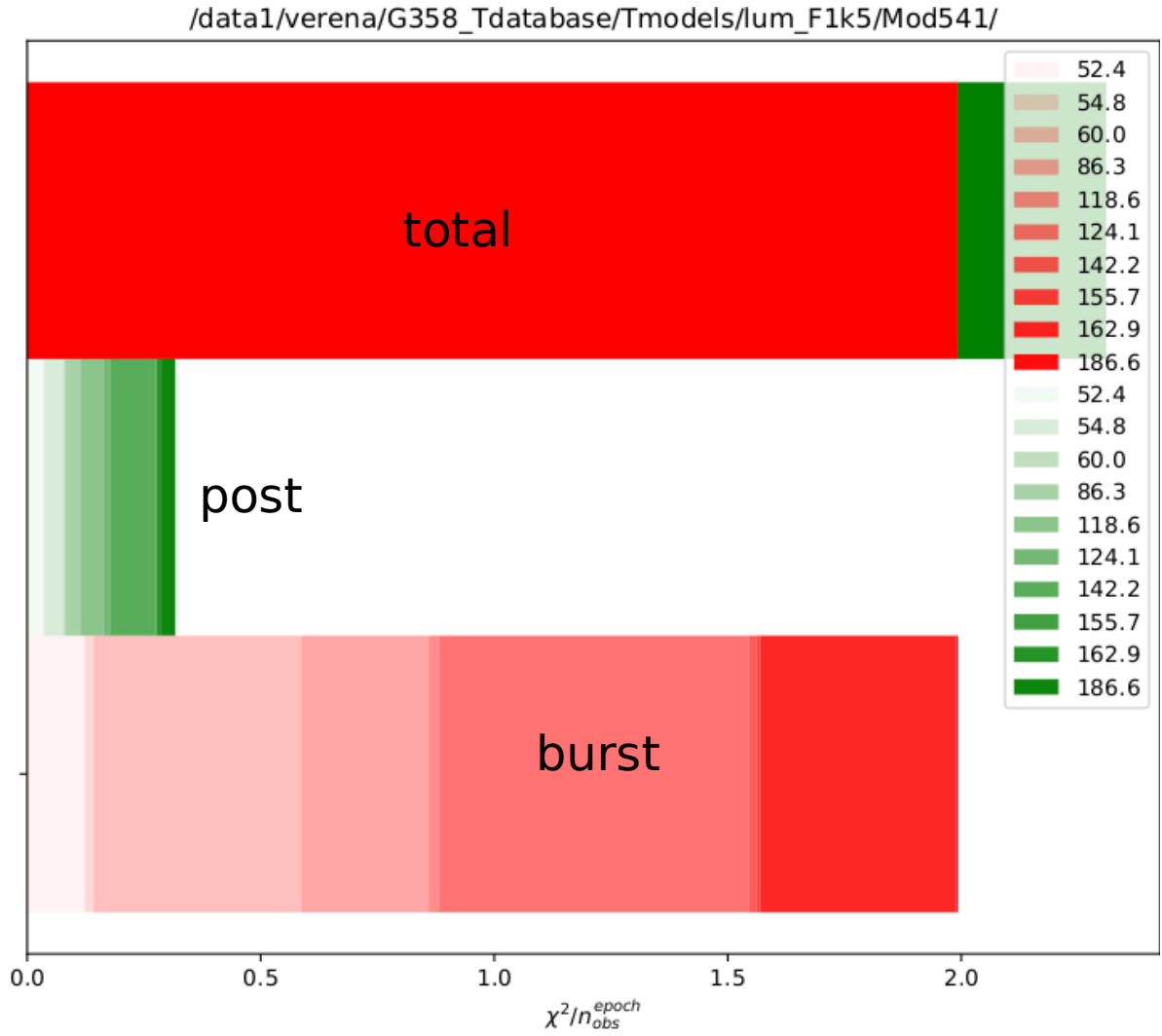

Figure A.43: Same as Fig. A.37 but for the ratio fit (of the test model). Again, the burst dominates the $\chi^2$-value (same model). The $\chi^2$ values are lower than before (for both epochs) due to the larger errors adapted for the simulation ('higher' synthetic noise).

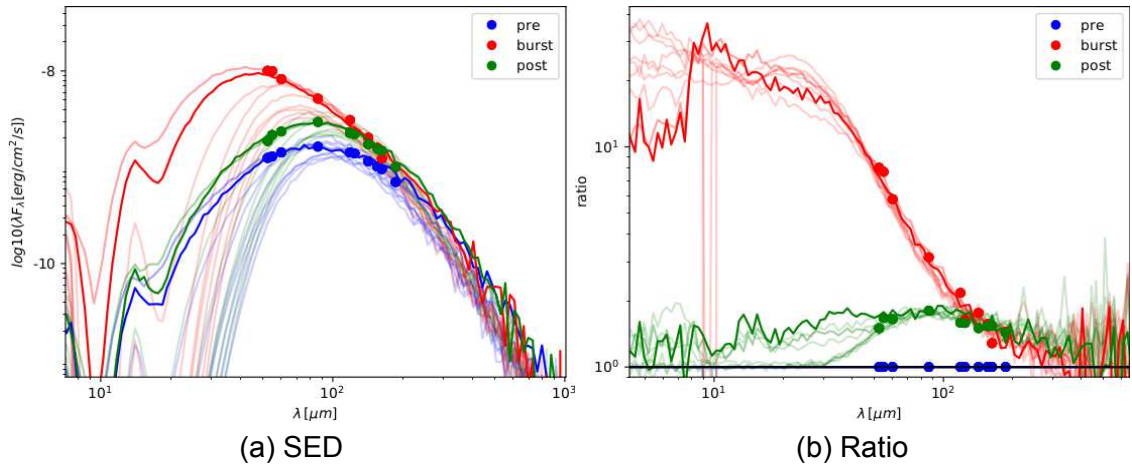

(a) SED                      (b) Ratio

Figure A.44: Same as Fig. 4.2 but for the ratio fit. All ratios match the test-data pretty well ('contrary' to the SEDs) as expected.

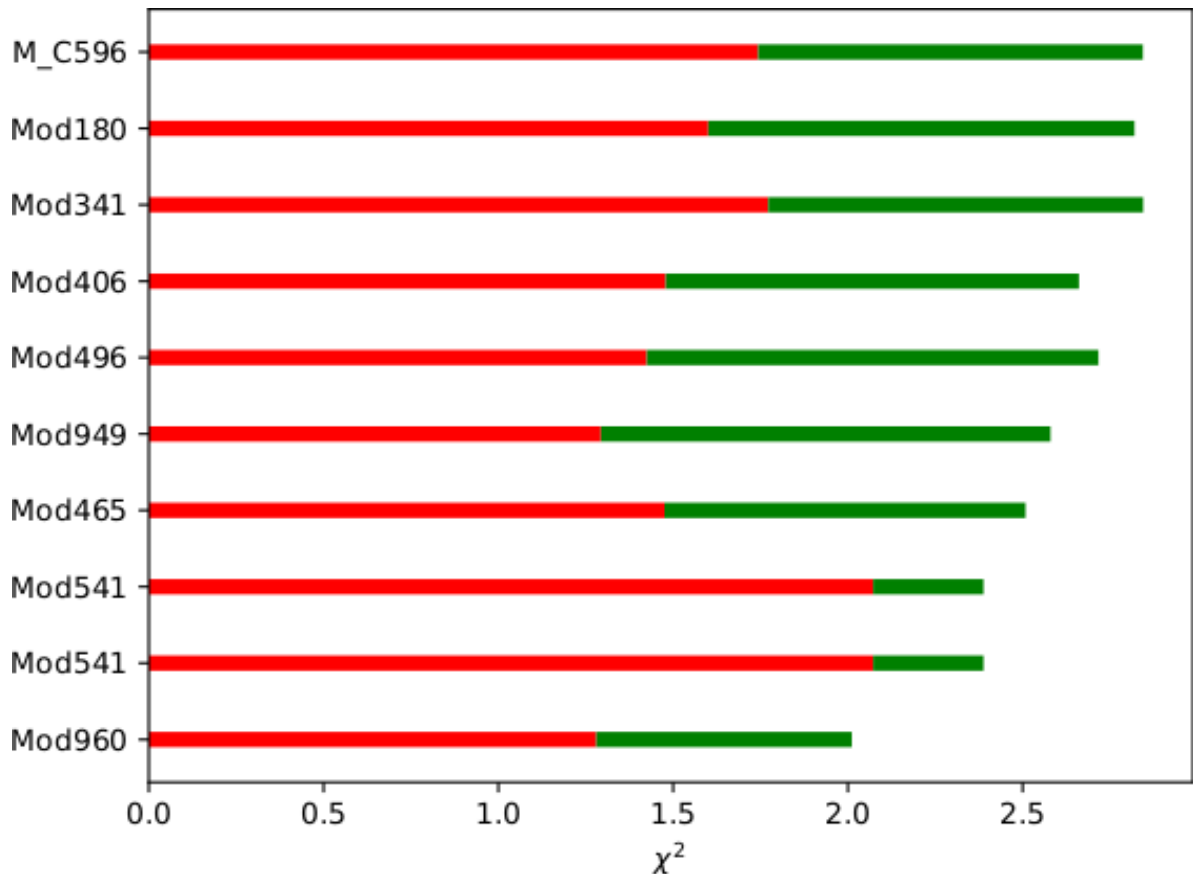

Figure A.45: Same as Fig. A.38 but for ratio-fit.

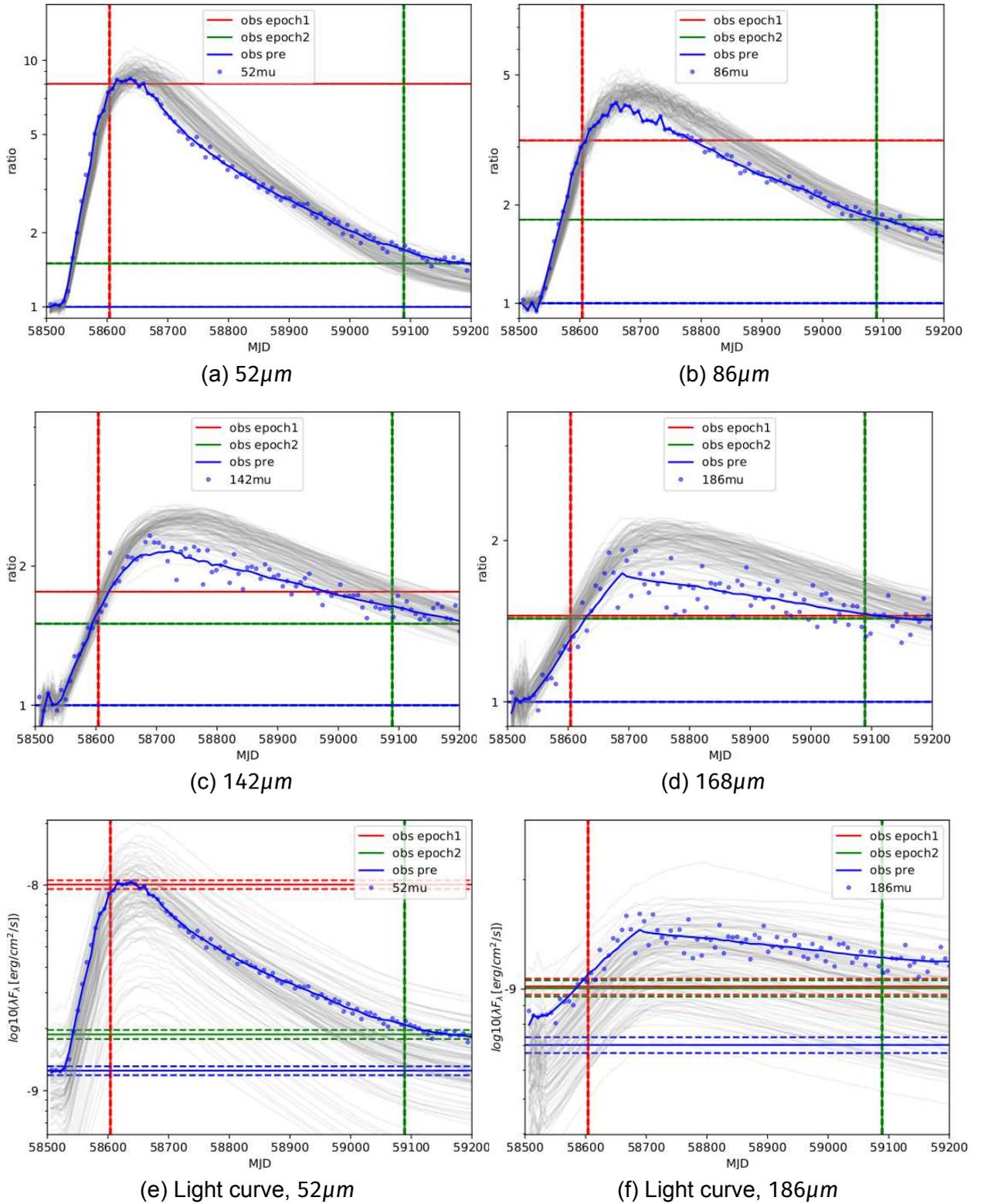

(a) 52μm

(b) 86μm

(c) 142μm

(d) 168μm

(e) Light curve, 52μm

(f) Light curve, 186μm

Figure A.46: Results of the small-signal benchmark fit with the TFitter (ratio-fit): Optimized ratio-curves (4 examples) for the 100 best models (gray) compared to the test-data (color). For the best model, the scatter/fit is indicated in blue (dots/solid line). The lowermost row gives the light curves for 2 out of the 4 examples for comparison.

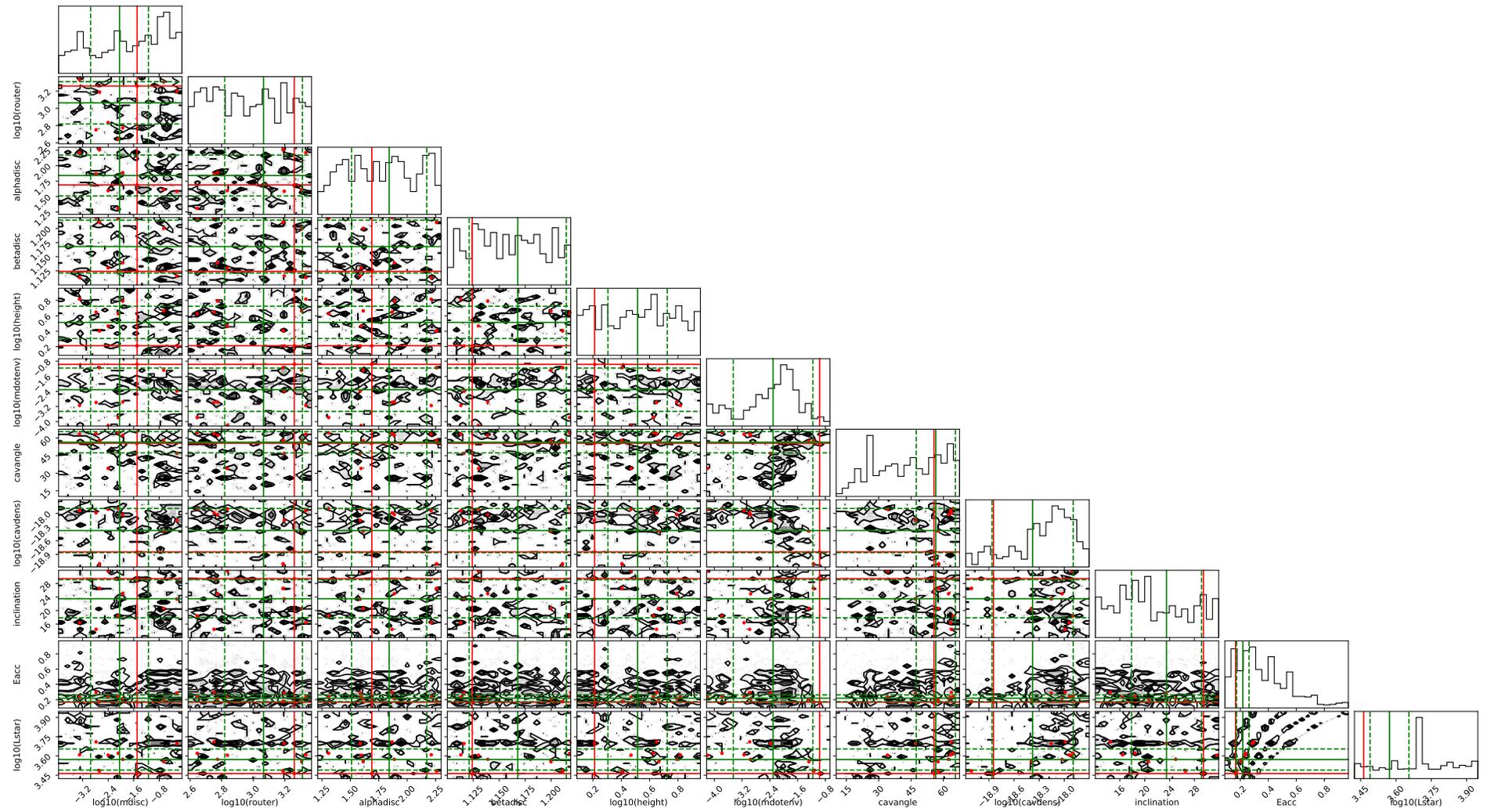

Figure A.47: Corner plot showing all Tmodels used for the TFitter benchmark ratio-fit (black). The best model is indicated by the red lines (different to the test model). The weighted mean parameters and the corresponding 1$\sigma$ confidence intervals are indicated in green. The results agree with the standard fit (value fit shown in Fig. A.40.

### A.1.6.3  TFitter, flexible burst onsets

/data1/verena/G358_Tdatabase/Tmodels/lum_F1k5/Mod541/

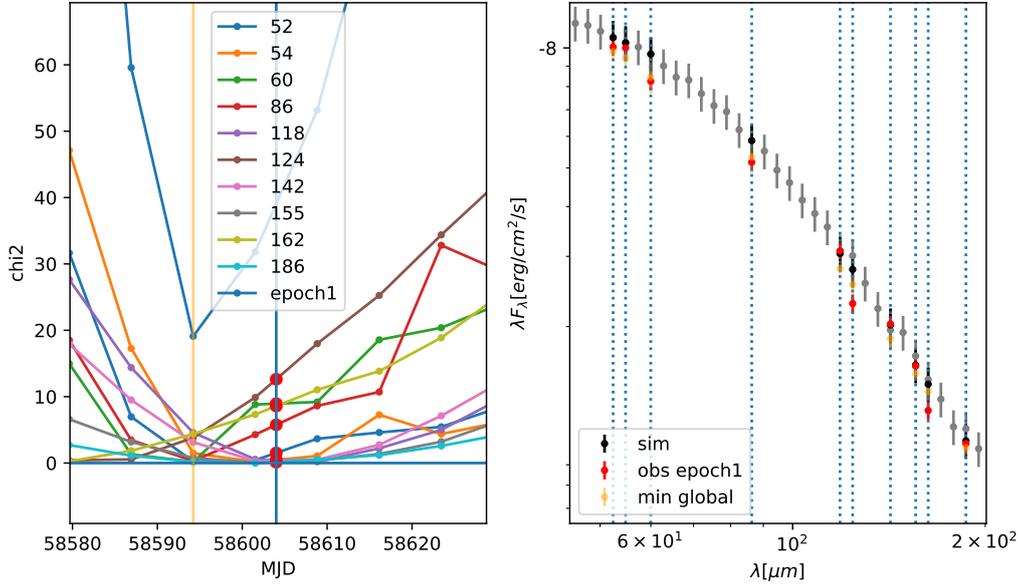

(a) Burst $X^2$ for flexible onsets

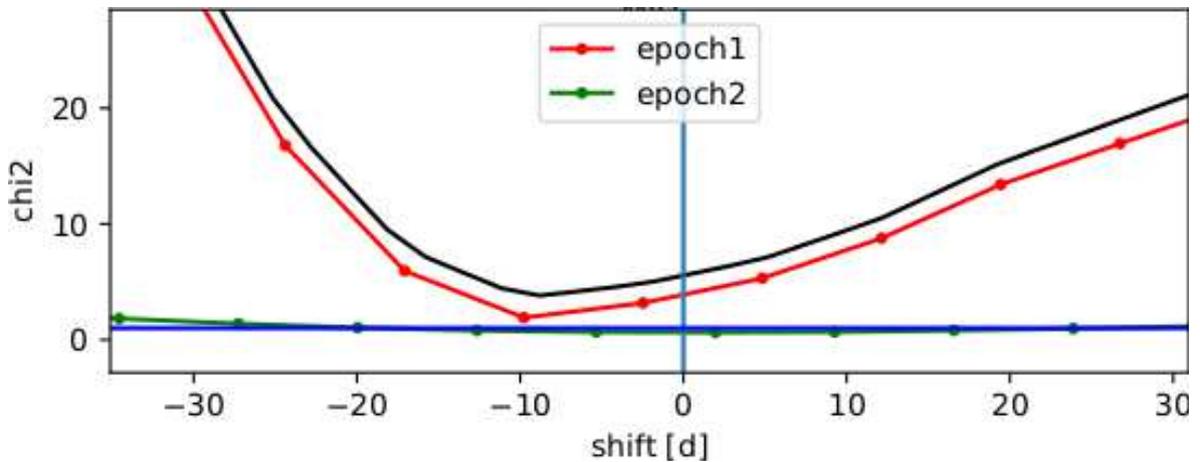

(b) Total and epoch $X^2$ for flexible onsets

Figure A.48: Example for the implementation of a flexible burst onset. The **upper left panel** shows the $X^2$-value of the burst for each band (legend) and total (blue, epoch 1) as a function of the observing date. The vertical blue line indicates the time of the 'real' test-observation, the orange horizontal line marks the date, where the agreement in between simulation and test data is highest (i.e., $X^2_{min}$). The corresponding data points are given in the **upper right panel**. In this case, the improvement is rather a 'stochastic artifact' than a 'real thing'. The **lower row** shows the $X^2$ values of each epoch (color-coded) as function of the time-shift for the test model. The burst-value (red) varies strongly close to the date of the test observation, while the post-value (green) is more stable. This is a general bias for flexible burst onsets. The sum is given in black. The best agreement is achieved, if the burst starts earlier by ≈ 10 days.

### A.1.6.4   TFitter, G358

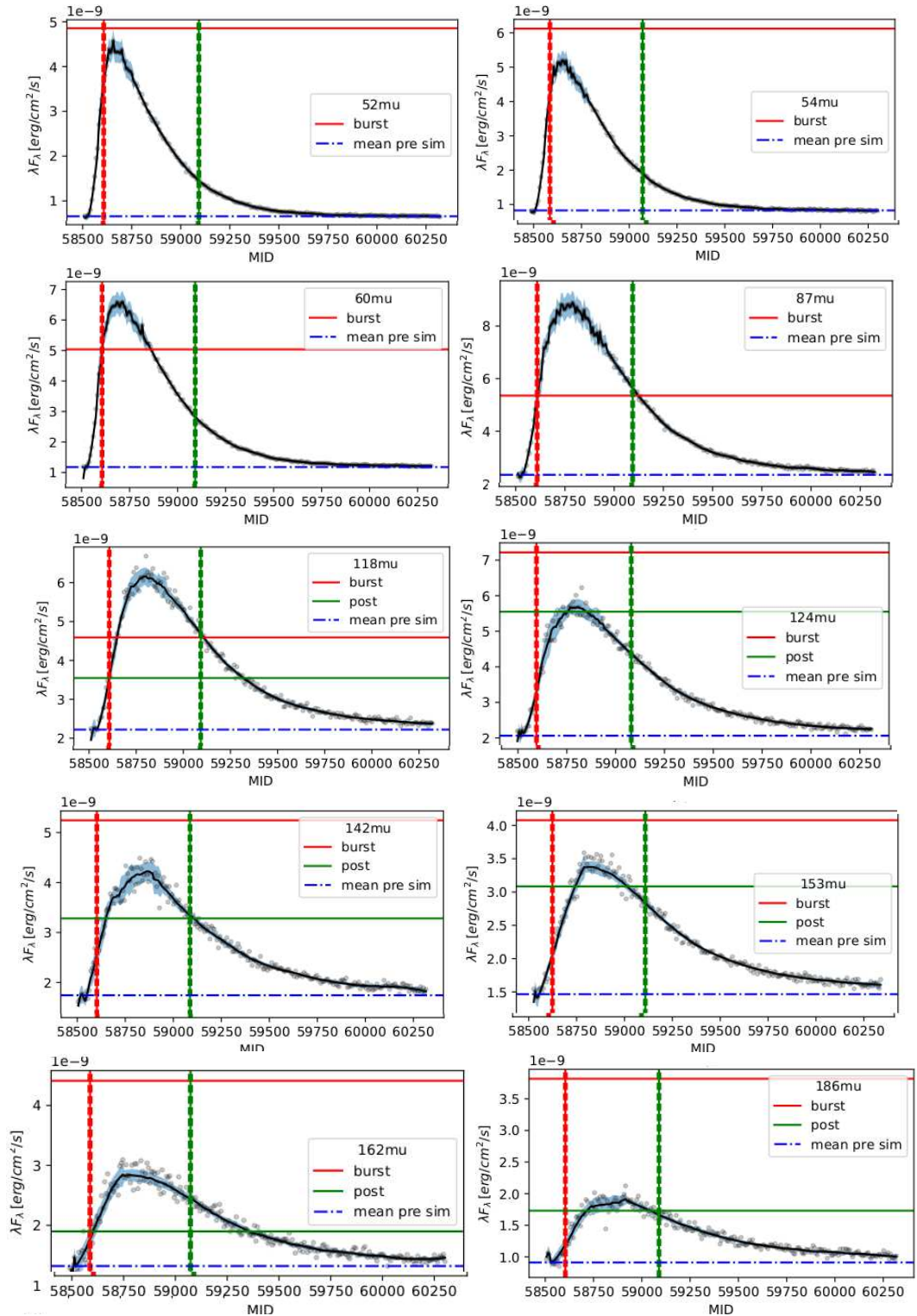

Figure A.49: Same as Fig. A.36, but for the adapted static mean model. Obviously, during the burst, the model flux is below the observation (in most of the bands). Note that there are no postburst observations in the blue channel.

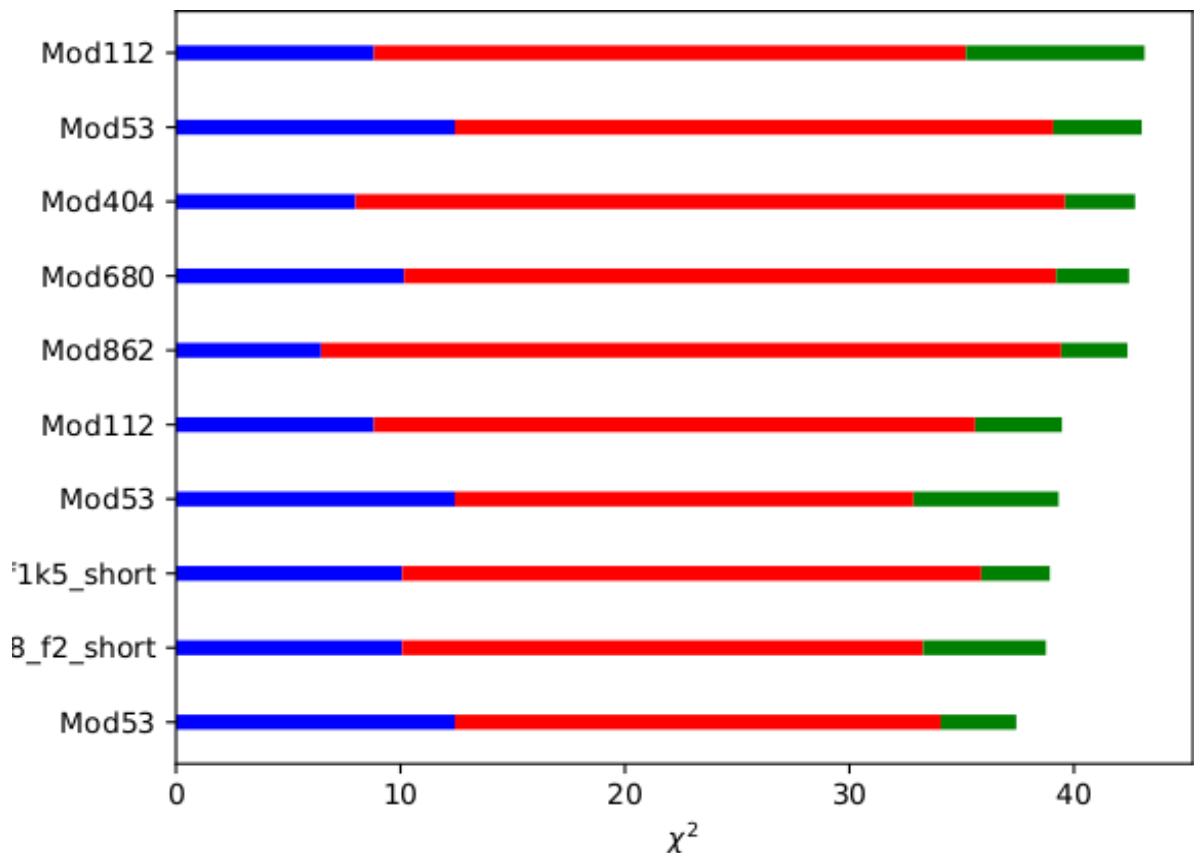

Figure A.50: Same as Fig. A.38 but for the fit with the real data (instead of the test-data). The $\chi^2$-values are much higher, indicating that there are some issues (the models are not appropriate to fit the data and/or the data needs re-calibration).

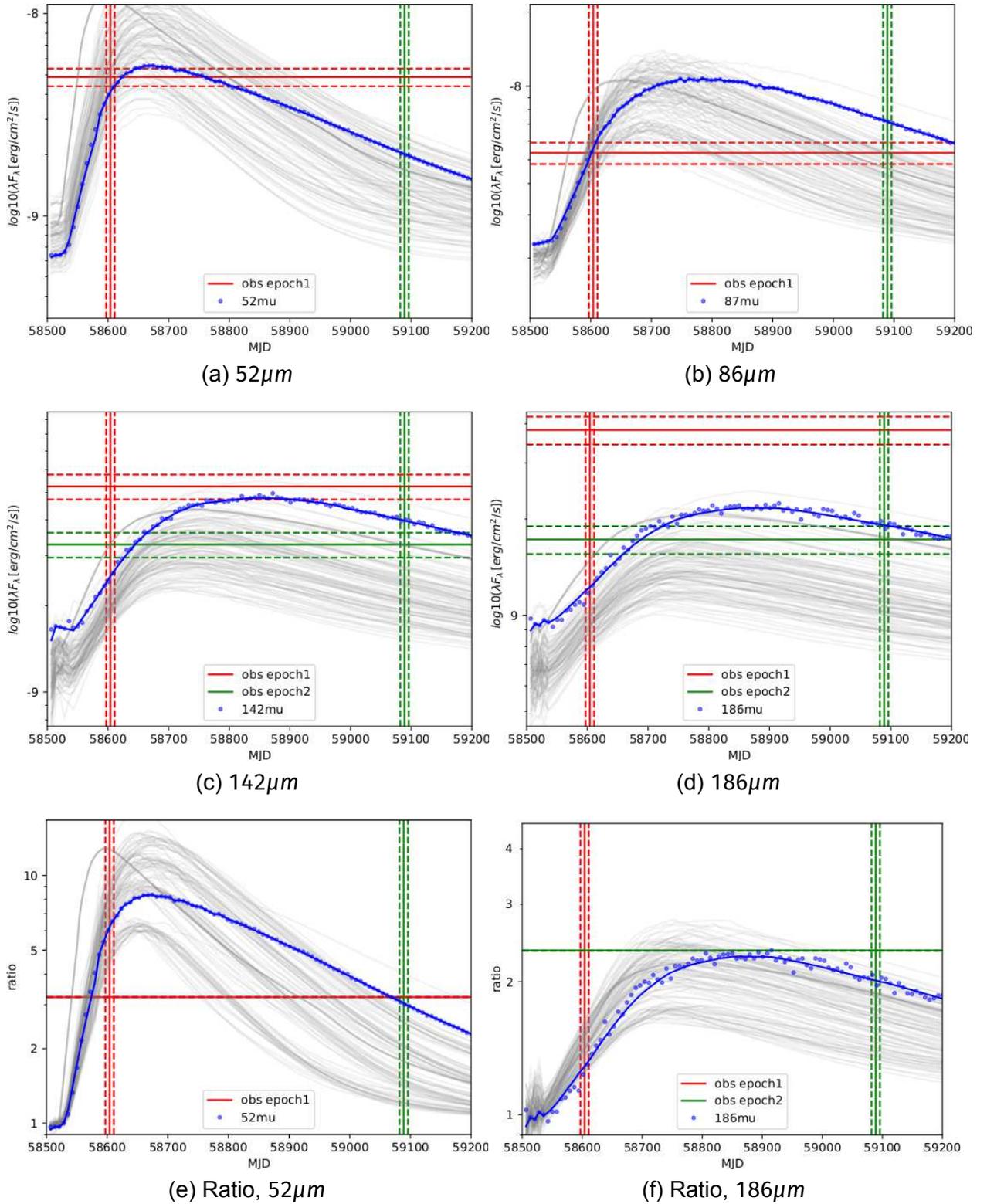

(a) 52$\mu m$

(b) 86$\mu m$

(c) 142$\mu m$

(d) 186$\mu m$

(e) Ratio, 52$\mu m$

(f) Ratio, 186$\mu m$

Figure A.51: Same as A.41, but for the fit of the G358 data (instead of the test data).

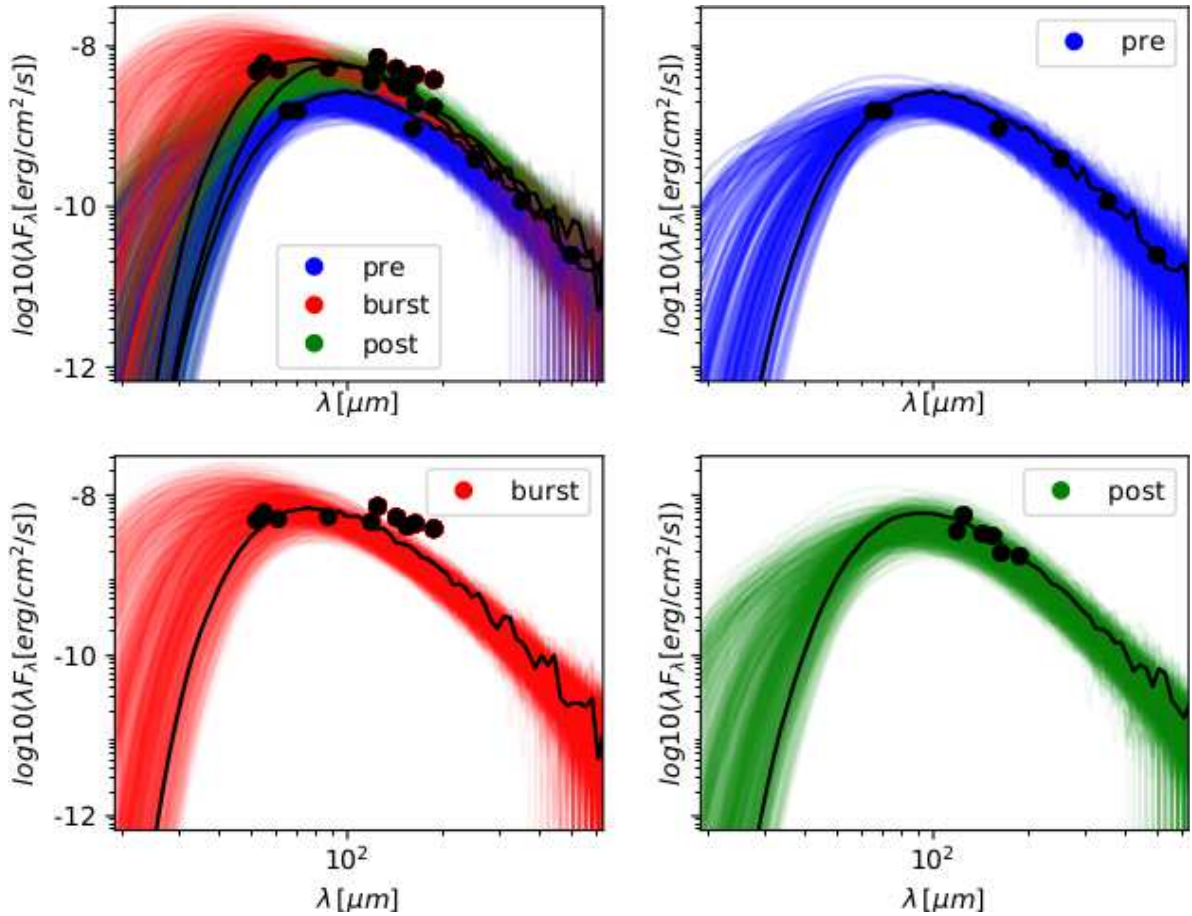

Figure A.52: TORUS-pool: The upper left shows all epochs (similar to the right panel of Fig. 4.1). The other panels show the epochs, as given in the labels. The best model is indicated in black. The pool does not cover the burst SED longward ≈ 120$\mu m$ (all other data points are covered by the pool).

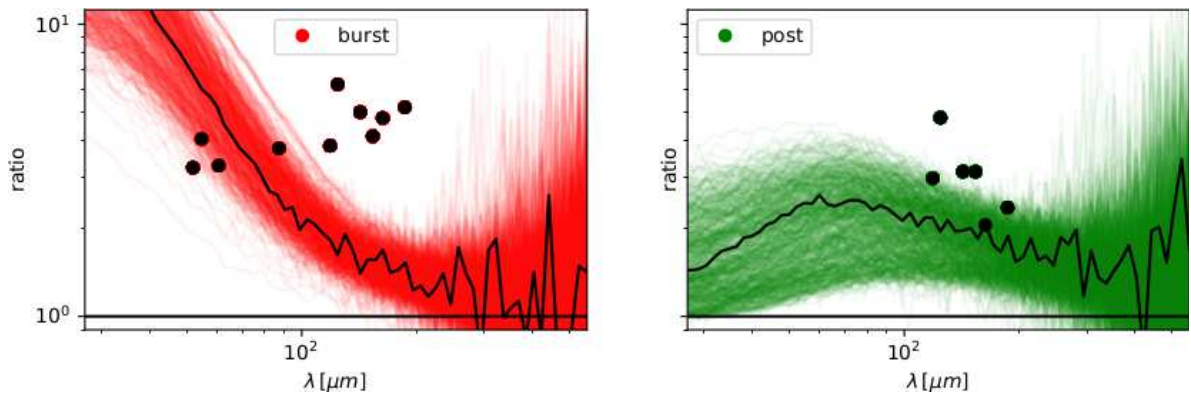

Figure A.53: Same as the lower row of Fig. A.52, but now the ratios are plotted. The misalignment of burst models and observations might point to different burst history (as an earlier onset/less steep profile), whereas the post-pool might indicate a higher burst energy. But these conclusions are pretty speculative and need more investigations.

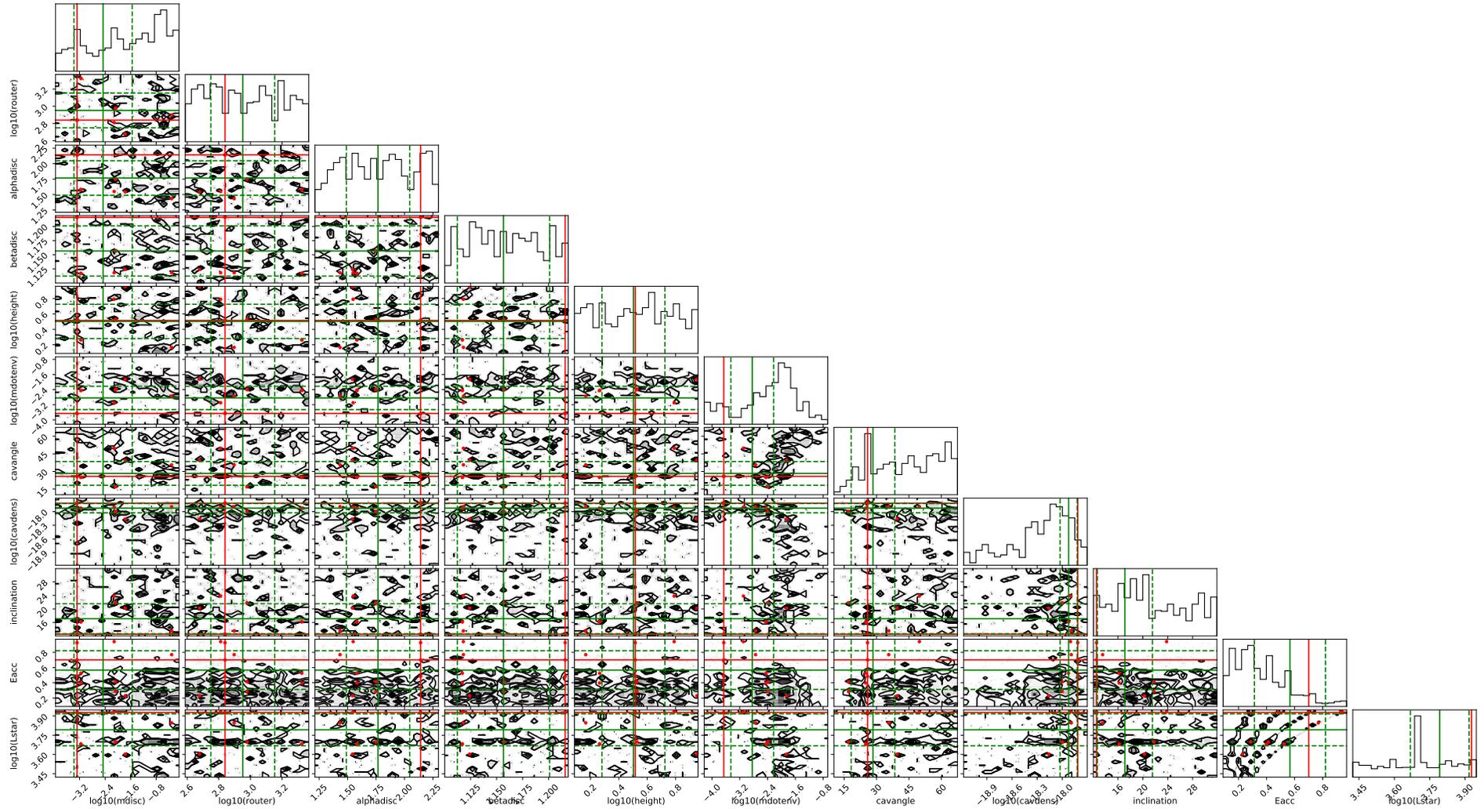

Figure A.54: Same as Fig. A.40 but for fit with real data. Interestingly, the 10 best models do not share the same source template. The fits cannot reproduce the burst; therefore, the results should be taken with care.

## A.2 Tables

### A.2.1 G358

Table A.1: SED of G358.93-0.03 MM3. These are the values, used for the MM3 fit. Note that the fluxes at $\lambda \geq 889\,\mu m$, are MM3-fluxes, whereas the fluxes at $\lambda \leq 24\,\mu m$ are total fluxes (but the other sources are much younger and do 'almost' not emit in the NIR/MIR). Facilities and instruments are given behind corresponding references. Published in [130].

| $\lambda$ [$\mu m$] | $\lambda F_\lambda$ [$erg\,cm^{-2}\,s^{-1}$] | Ref. |
|---|---|---|
| 1.63 | $1.16 \pm 0.07 \times 10^{-12}$ | 1 |
| 2.13 | $1.67 \pm 0.02 \times 10^{-11}$ | 1 |
| 3.55 | $3.64 \pm 0.02 \times 10^{-10}$ | 2 |
| 4.49 | $6.10 \pm 0.03 \times 10^{-10}$ | 2 |
| 5.73 | $9.73 \pm 0.05 \times 10^{-10}$ | 2 |
| 7.0 | $6.90 \pm 0.49 \times 10^{-10}$ | 3 |
| 7.87 | $5.30 \pm 0.04 \times 10^{-10}$ | 2 |
| 11.6 | $3.54 \pm 0.07 \times 10^{-10}$ | 4 |
| 15.0 | $6.46 \pm 0.20 \times 10^{-10}$ | 3 |
| 22.1 | $5.13 \pm 0.13 \times 10^{-10}$ | 4 |
| 23.7 | $4.49 \pm 0.13 \times 10^{-10}$ | 5 |
| 889 | $1.34 \pm 0.04 \times 10^{-13}$ | 6 |
| 1282 | $2.43 \pm 0.10 \times 10^{-14}$ | 6 |
| 1420 | $1.50 \pm 0.19 \times 10^{-14}$ | 6 |
| 1532 | $1.06 \pm 0.08 \times 10^{-14}$ | 6 |

(1) [102] (VISTA/VIRCAM); (2) [115] (*Spitzer*/IRAC)
; (3) [109] (ISO/ISOCAM); (4) [37] (ALLWISE); (5) [52] (*Spitzer*/MIPS); (6) [19] (ALMA, SMA)

Table A.2: Total- and MM1-preburst fluxes of G358.93-0.03. We obtained the MM1 flux densities used for the SED fit by removing the contribution from all other sources in the field (including MM3). Facilities and instruments are given behind corresponding references. Published in [130].

| Wavelength [$\mu m$] | Total flux [$erg\,cm^{-2}\,s^{-1}$] | Ref. | MM1 flux [$erg\,cm^{-2}\,s^{-1}$] |
|---|---|---|---|
| 2.15 | $\leq 5.86 \pm 0.59 \times 10^{-13}$ | 1 | |
| 24 | $\leq 5.25 \pm 0.53 \times 10^{-11}$ | 1 | |
| 65 | $3.10 \pm 0.06 \times 10^{-9}$ | 1 | $1.55 \pm 0.03 \times 10^{-9}$ |
| 70 | $3.13 \pm 0.38 \times 10^{-9}$ | 1 | $1.57 \pm 0.06 \times 10^{-9}$ |
| 160 | $1.90 \pm 0.21 \times 10^{-9}$ | 1 | $9.50 \pm 0.52 \times 10^{-10}$ |
| 250 | $7.82 \pm 1.04 \times 10^{-10}$ | 1 | $3.91 \pm 0.27 \times 10^{-10}$ |
| 350 | $2.34 \pm 0.52 \times 10^{-10}$ | 1 | $1.17 \pm 0.26 \times 10^{-10}$ |
| 500 | $4.98 \pm 2.82 \times 10^{-11}$ | 1 | $2.49 \pm 0.33 \times 10^{-11}$ |
| 850 | $4.73 \pm 0.15 \times 10^{-12}$ | 2 | $2.36 \pm 0.08 \times 10^{-12}$ |
| 870 | $4.03 \pm 0.07 \times 10^{-12}$ | 3 | $2.00 \pm 0.35 \times 10^{-12}$ |

(1) this work (VISTA/VIRCAM, *Spitzer*/MIPS, SOFIA/FIFI-LS); (2) [111] (JCMT/SCUBA-2); (3) [19] (APEX/LABOCA)

Table A.3: Total- and MM1-burst flux densities of G358.93-0.03: Similar to Table A.2 but for the burst epoch. We assume that during the burst, only the luminosity of MM1 increased, while all other sources remained constant. Facilities and instruments are given behind corresponding references. Published in [130].

| wavelength [$\mu m$] | total flux [$erg\,cm^{-2}\,s^{-1}$] | Ref. | MM1 flux [$erg\,cm^{-2}\,s^{-1}$] |
|---|---|---|---|
| 2.15 | $\leq 9.20 \pm 0.92 \times 10^{-13}$ | 1 | |
| 3.4 | $\leq 8.82 \pm 0.89 \times 10^{-11}$ | 1 | |
| 4.6 | $\leq 2.93 \pm 0.30 \times 10^{-10}$ | 1 | |
| 52.0 | $5.51 \pm 0.56 \times 10^{-9}$ | 1 | $4.85 \pm 0.49 \times 10^{-9}$ |
| 54.8 | $6.94 \pm 0.70 \times 10^{-9}$ | 1 | $6.12 \pm 0.62 \times 10^{-9}$ |
| 60.7 | $6.20 \pm 0.62 \times 10^{-9}$ | 1 | $5.04 \pm 0.51 \times 10^{-9}$ |
| 87.2 | $7.48 \pm 0.75 \times 10^{-9}$ | 1 | $5.35 \pm 0.54 \times 10^{-9}$ |
| 118.6 | $6.45 \pm 0.65 \times 10^{-9}$ | 1 | $4.59 \pm 0.46 \times 10^{-9}$ |
| 124.2 | $8.97 \pm 0.90 \times 10^{-9}$ | 1 | $7.22 \pm 0.72 \times 10^{-9}$ |
| 142.2 | $5.26 \pm 0.53 \times 10^{-9}$ | 1 | $5.25 \pm 0.53 \times 10^{-9}$ |
| 153.3 | $5.29 \pm 0.53 \times 10^{-9}$ | 1 | $4.08 \pm 0.41 \times 10^{-9}$ |
| 162.8 | $5.47 \pm 0.55 \times 10^{-9}$ | 1 | $4.41 \pm 0.45 \times 10^{-9}$ |
| 186.4 | $4.58 \pm 0.46 \times 10^{-9}$ | 1 | $3.82 \pm 0.39 \times 10^{-9}$ |
| 889 | $3.81 \pm 0.11 \times 10^{-12}$ | 2 | $1.72 \pm 0.07 \times 10^{-12}$ |
| 1282 | $4.21 \pm 0.03 \times 10^{-13}$ | 2 | |
| 1420 | $2.74 \pm 0.28 \times 10^{-13}$ | 2 | |
| 1532 | $1.88 \pm 0.01 \times 10^{-13}$ | 2 | |

(1) this work; (2.2-m MPG/ESO telescope/GROND, NEOWISE, SOFIA/FIFI-LS); (2) [19] (ALMA, SMA)

Table A.4: Total- and MM1-postburst fluxes: Similar to Table A.2 but for the postburst. We assume that only the luminosity of MM1 has changed, while all other sources remained constant. Facilities and instruments are given behind the corresponding references. Published in [130].

| wavelength [$\mu m$] | total flux [$erg\,cm^{-2}\,s^{-1}$] | Ref. | MM1 flux [$erg\,cm^{-2}\,s^{-1}$] |
|---|---|---|---|
| 118.6 | $5.17 \pm 0.52 \times 10^{-9}$ | 1 | $3.55 \pm 0.36 \times 10^{-9}$ |
| 124.2 | $7.70 \pm 0.77 \times 10^{-9}$ | 1 | $5.55 \pm 0.56 \times 10^{-9}$ |
| 142.2 | $4.74 \pm 0.48 \times 10^{-9}$ | 1 | $3.28 \pm 0.33 \times 10^{-9}$ |
| 153.3 | $4.35 \pm 0.44 \times 10^{-9}$ | 1 | $3.09 \pm 0.31 \times 10^{-9}$ |
| 162.8 | $3.01 \pm 0.31 \times 10^{-9}$ | 1 | $1.90 \pm 0.19 \times 10^{-9}$ |
| 186.4 | $2.53 \pm 0.26 \times 10^{-9}$ | 1 | $1.73 \pm 0.18 \times 10^{-9}$ |
| 889 | $3.81 \pm 0.11 \times 10^{-12}$ | 2 | $1.72 \pm 0.07 \times 10^{-12}$ |

(1) present paper (SOFIA/FIFI-LS); (2) [19] (ALMA)

| Grid spacing | lin | log | log | log | log | log | lin | lin | log | log | lin | lin | log | lin | log |
|---|---|---|---|---|---|---|---|---|---|---|---|---|---|---|---|
| Model | $\chi^2$ | av mag | d kpc | $R_*$ $R_\odot$ | $T_*$ K | $m_{disk}$ $M_\odot$ | $r_{disk}^{max}$ au | $\beta_{disk}$ | $p_{disk}$ | $h_{100}^{disk}$ au | $\rho_0^{env}$ g/cc | $p_{cav}$ | $\theta_0^{cav}$ ° | $\rho_0^{cav}$ g/cc | inc ° | $L_*$ $L_\odot$ |
| **Preburst** | | | | | | | | | | | | | | | | |
| eGYXcOh8_02 | 42.8 | 67.6 | 7.11 | 3.79 | 25590 | 0.000217 | 3110 | 1.18 | -0.131 | 2.34 | $5.25 \times 10^{-19}$ | 1.56 | 31.5 | $1.81 \times 10^{-23}$ | 19.3 | 5437 |
| 90Yt0exl_03 | 46.6 | 65.5 | 6.4 | 5.51 | 19670 | 0.0357 | 1191 | 1.13 | -0.345 | 1.25 | $2.07 \times 10^{-18}$ | 1.39 | 36.4 | $1.1 \times 10^{-21}$ | 24.1 | 4017 |
| PUyhjE8Z_02 | 47.5 | 48.2 | 6.75 | 14.1 | 12130 | $4.01 \times 10^{-8}$ | 278 | 1.28 | -1.46 | 11.4 | $2.07 \times 10^{-17}$ | 1.76 | 23.4 | $2.71 \times 10^{-22}$ | 11.3 | 3797 |
| nTRTrE7X_02 | 51.2 | 59 | 7.11 | 2.71 | 29889 | $8.1 \times 10^{-6}$ | 423 | 1.13 | -0.607 | 1.85 | $9.58 \times 10^{-19}$ | 1.94 | 32.7 | $4.99 \times 10^{-21}$ | 18.1 | 5180 |
| nTRTrE7X_03 | 55 | 47.5 | 6.75 | 2.71 | 29889 | $8.1 \times 10^{-6}$ | 423 | 1.13 | -0.607 | 1.85 | $9.58 \times 10^{-19}$ | 1.94 | 32.7 | $4.99 \times 10^{-21}$ | 23.3 | 5180 |
| TJyUR9bA_02 | 57.8 | 62.6 | 7.11 | 3.91 | 23840 | $3.1 \times 10^{-7}$ | 631 | 1.17 | -0.102 | 12.7 | $7.15 \times 10^{-18}$ | 1.89 | 46 | $9.4 \times 10^{-23}$ | 14.4 | 4366 |
| 4kW1TtMH_04 | 59.1 | 41.8 | 6.4 | 47.3 | 7335 | 0.000299 | 140 | 1.24 | -0.752 | 2.49 | $5.14 \times 10^{-17}$ | 1.35 | 32.9 | $8.17 \times 10^{-22}$ | 30.6 | 5715 |
| IQ0YS3aF_02 | 63.2 | 70 | 6.4 | 6.7 | 17980 | 0.00527 | 3798 | 1.06 | -0.775 | 17.3 | $3.24 \times 10^{-19}$ | 1.06 | 12.5 | $8.66 \times 10^{-22}$ | 13.5 | 4150 |
| qWHuujku_05 | 64.4 | 70 | 6.75 | 38 | 8551 | $2.63 \times 10^{-5}$ | 1015 | 1.23 | -1.62 | 1.52 | $2.33 \times 10^{-18}$ | 1.28 | 47.8 | $9.27 \times 10^{-21}$ | 46.8 | 6832 |
| qWHuujku_06 | 65.2 | 36.2 | 6.75 | 38 | 8551 | $2.63 \times 10^{-5}$ | 1015 | 1.23 | -1.62 | 1.52 | $2.33 \times 10^{-18}$ | 1.28 | 47.8 | $9.27 \times 10^{-21}$ | 50 | 6832 |
| **Burst** | | | | | | | | | | | | | | | | |
| 90Yt0exl_L5.0 | 226 | 65.5 | 6.4 | 12.3 | 19670 | 0.0357 | 1191 | 1.13 | -0.345 | 1.25 | $2.07 \times 10^{-18}$ | 1.39 | 36.4 | $1.1 \times 10^{-21}$ | 24.1 | 20078 |
| 90Yt0exl_L5.5 | 229 | 65.5 | 6.4 | 12.9 | 19670 | 0.0357 | 1191 | 1.13 | -0.345 | 1.25 | $2.07 \times 10^{-18}$ | 1.39 | 36.4 | $1.1 \times 10^{-21}$ | 24.1 | 22081 |
| eGYXcOh8_L4.5 | 234 | 67.6 | 7.11 | 8.04 | 25590 | 0.000217 | 3110 | 1.18 | -0.131 | 2.34 | $5.25 \times 10^{-19}$ | 1.56 | 31.5 | $1.81 \times 10^{-23}$ | 19.3 | 24470 |
| eGYXcOh8_L5.0 | 234 | 67.6 | 7.11 | 8.47 | 25590 | 0.000217 | 3110 | 1.18 | -0.131 | 2.34 | $5.25 \times 10^{-19}$ | 1.56 | 31.5 | $1.81 \times 10^{-23}$ | 19.3 | 27185 |
| 90Yt0exl_L4.5 | 236 | 65.5 | 6.4 | 11.7 | 19670 | 0.0357 | 1191 | 1.13 | -0.345 | 1.25 | $2.07 \times 10^{-18}$ | 1.39 | 36.4 | $1.1 \times 10^{-21}$ | 24.1 | 18077 |
| TJyUR9bA_L5.0 | 236 | 62.6 | 7.11 | 8.74 | 23840 | $3.1 \times 10^{-7}$ | 631 | 1.17 | -0.102 | 12.7 | $7.15 \times 10^{-18}$ | 1.89 | 46 | $9.4 \times 10^{-23}$ | 14.4 | 21828 |
| TJyUR9bA_L4.5 | 243 | 62.6 | 7.11 | 8.3 | 23840 | $3.1 \times 10^{-7}$ | 631 | 1.17 | -0.102 | 12.7 | $7.15 \times 10^{-18}$ | 1.89 | 46 | $9.4 \times 10^{-23}$ | 14.4 | 19644 |
| eGYXcOh8_L5.5 | 244 | 67.6 | 7.11 | 8.88 | 25590 | 0.000217 | 3110 | 1.18 | -0.131 | 2.34 | $5.25 \times 10^{-19}$ | 1.56 | 31.5 | $1.81 \times 10^{-23}$ | 19.3 | 29907 |
| 90Yt0exl_L6.0 | 245 | 65.5 | 6.4 | 13.5 | 19670 | 0.0357 | 1191 | 1.13 | -0.345 | 1.25 | $2.07 \times 10^{-18}$ | 1.39 | 36.4 | $1.1 \times 10^{-21}$ | 24.1 | 24108 |
| qWHuujku_L4.5 | 245 | 36.2 | 6.75 | 80.7 | 8551 | $2.63 \times 10^{-5}$ | 1015 | 1.23 | -1.62 | 1.52 | $2.33 \times 10^{-18}$ | 1.28 | 47.8 | $9.27 \times 10^{-21}$ | 50 | 30745 |
| **Postburst** | | | | | | | | | | | | | | | | |
| IQ0YS3aF_L2.5 | 92.7 | 70 | 6.4 | 10.6 | 17980 | 0.00527 | 3798 | 1.06 | -0.775 | 17.3 | $3.24 \times 10^{-19}$ | 1.06 | 12.5 | $8.66 \times 10^{-22}$ | 13.5 | 10376 |
| nTRTrE7X_L2.5 | 103 | 59 | 7.11 | 4.29 | 29889 | $8.1 \times 10^{-6}$ | 423 | 1.13 | -0.607 | 1.85 | $9.58 \times 10^{-19}$ | 1.94 | 32.7 | $4.99 \times 10^{-21}$ | 18.1 | 12950 |
| IQ0YS3aF_L3.0 | 105 | 70 | 6.4 | 11.6 | 17980 | 0.00527 | 3798 | 1.06 | -0.775 | 17.3 | $3.24 \times 10^{-19}$ | 1.06 | 12.5 | $8.66 \times 10^{-22}$ | 13.5 | 12448 |
| 90Yt0exl_L3.0 | 107 | 65.5 | 6.4 | 9.54 | 19670 | 0.0357 | 1191 | 1.13 | -0.345 | 1.25 | $2.07 \times 10^{-18}$ | 1.39 | 36.4 | $1.1 \times 10^{-21}$ | 24.1 | 12051 |
| TJyUR9bA_L3.0 | 110 | 62.6 | 7.11 | 6.77 | 23840 | $3.1 \times 10^{-7}$ | 631 | 1.17 | -0.102 | 12.7 | $7.15 \times 10^{-18}$ | 1.89 | 46 | $9.4 \times 10^{-23}$ | 14.4 | 13097 |
| 90Yt0exl_L2.5 | 110 | 65.5 | 6.4 | 8.71 | 19670 | 0.0357 | 1191 | 1.13 | -0.345 | 1.25 | $2.07 \times 10^{-18}$ | 1.39 | 36.4 | $1.1 \times 10^{-21}$ | 24.1 | 10044 |
| nTRTrE7X_L3.0 | 114 | 59 | 7.11 | 4.69 | 29889 | $8.1 \times 10^{-6}$ | 423 | 1.13 | -0.607 | 1.85 | $9.58 \times 10^{-19}$ | 1.94 | 32.7 | $4.99 \times 10^{-21}$ | 18.1 | 15540 |
| TJyUR9bA_L3.5 | 114 | 62.6 | 7.11 | 7.32 | 23840 | $3.1 \times 10^{-7}$ | 631 | 1.17 | -0.102 | 12.7 | $7.15 \times 10^{-18}$ | 1.89 | 46 | $9.4 \times 10^{-23}$ | 14.4 | 15281 |
| eGYXcOh8_L2.5 | 116 | 67.6 | 7.11 | 5.69 | 25590 | 0.000217 | 3110 | 1.18 | -0.131 | 2.34 | $5.25 \times 10^{-19}$ | 1.56 | 31.5 | $1.81 \times 10^{-23}$ | 19.3 | 13591 |
| nTRTrE7X_L2.0 | 118 | 59 | 7.11 | 3.83 | 29889 | $8.1 \times 10^{-6}$ | 423 | 1.13 | -0.607 | 1.85 | $9.58 \times 10^{-19}$ | 1.94 | 32.7 | $4.99 \times 10^{-21}$ | 18.1 | 10362 |
| **mean model** | | | | | | | | | | | | | | | | |
| mean | | 60.5 | 6.77 | 8.38* | 16834* | $8.42 \times 10^{-5}$ | 952 | 1.16 | -0.592 | 3.37 | $2.98 \times 10^{-18}$ | 1.57 | 33.7 | $6.37 \times 10^{-22}$ | 21.8 | 4984* |
| sigma | | 9.7 | 1.05 | 2.86 | 1.66 | 72.2 | 2.51 | 0.06 | 0.480 | 2.72 | 4.14 | 0.31 | 10.2 | 7.42 | 10.1 | 1.22 |

* Pre-burst-value

## A.2.2   G323

Table A.5: G323 preburst flux densities. Facilities and instruments are given behind corresponding references. Observations with complementing postburst measurements are bold.

| $\lambda$ [$\mu$m] | $F_\nu$ [Jy] | $\Delta F_\nu$ [Jy] | Ref. |
|---|---|---|---|
| 1.02 | 0.000407 | 0.000005 | 1 |
| 1.24 | 0.00336 | 0.0002 | 2 |
| 1.25 | 0.00297 | 0.00001 | 1 |
| 1.63 | 0.143 | 0.02 | 1 |
| 1.65 | 0.0239 | 0.0018 | 2 |
| 2.13 | 0.437 | 0.05 | 1 |
| 2.19 | 0.926 | 0.141 | 3 |
| 2.16 | 0.291 | 0.03 | 2 |
| 3.35 | 3.79 | 0.21 | 4 |
| 4.35 | 9 | 1.48 | 5 |
| 7.67 | 48.5 | 5 | 10 |
| 8.28 | 33.6 | 1.4 | 5 |
| 8.58 | 59.1 | 6 | 10 |
| 8.61 | 37.3 | 0.3 | 6 |
| 9.83 | 32.4 | 3 | 10 |
| 10.4 | 40.7 | 5 | 10 |
| 10.9 | 54.2 | 6 | 10 |
| 11.6 | 84.1 | 5 | 7 |
| 12.1 | 103 | 5 | 5 |
| 14.6 | 154 | 9 | 5 |
| 18.4 | 276 | 4 | 6 |
| 21.3 | 364 | 22 | 5 |
| 23.9 | 522 | 26 | 7 |
| **70** | **2459** | **23** | 8 |
| **160** | **1721** | **70** | 8 |
| 250 | 962 | 54 | 9 |
| 350 | 232 | 8 | 9 |
| 500 | 73.6 | 1.8 | 9 |
| 870 | 13.3 | 1.33 | 9 |

(1) [102] (VISTA/VIRCAM); (2) [69] 2MASS; (3) Johnson; (4) [37] WISE; (5) [91] (6) [154] AKARI; (7) [71] IRAS; (8) [97] (Herschel/PACS); (9) [126] (ATLASGAL/APEX) The flux has been scaled, as mentioned in [42, Section 3.1].; (10) [51] extracted from the IRAS-LRS Spectrum.

| $\lambda$ [$\mu m$] | $F_v$ [Jy] | $\Delta F_v$ [Jy] | Aperture ["] |
|---|---|---|---|
| 53 | 2602 | 260 | 7.8 |
| 62 | 2669 | 270 | 12.2 |
| **70** | **2809** | **110** | **12.2** |
| 89 | 3067 | 310 | 12.2 |
| 154 | 1928 | 190 | 17.0 |
| **160** | **1867** | **73** | **17.0** |
| 214 | 1295 | 130 | 21.3 |

Table A.6: G323 postburst flux densities as measured with SOFIA/HAWC+. Values in bold are interpolated in wavelength space to match the PACS bands (preburst).

| Component | Parameter | Sampling | Adapted range |
|---|---|---|---|
| Protostar | $L_*[L_\odot]$ | log | $8 \cdot 10^3 - 1.6 \cdot 10^5$ |
| Disk | $M_{disk}[M_\odot]$ | log | $4 \cdot 10^{-8} - 2.5$ |
| | $r_{disk}[au]$ | log | $80 - 7000$ |
| | $\beta_{disk}$ | lin | $1 - 1.3$ |
| | $\alpha_{disk}$ | lin | $1 - 3.3$ |
| | $h_{disk}[au]$ | log | $0.5 - 33$ |
| Envelope | $\dot{M}_{env}[M_\odot/yr]$ | log | $2 \cdot 10^{-5} - 1$ |
| | $r_{env}[au]$ | log | $(1 - 4) \cdot 10^4$ |
| Cavity | $\rho_{cav}[g/cm^3]$ | log | $10^{-22} - 10^{-17}$ |
| | $\Theta_{cav}$ [°] | lin | $10 - 60$ |
| View | $i$ [°] | lin | $0 - 60$ |

Table A.7: Adapted parameter spaces and sampling for the TORUS premodels, all densities/-masses are total values (dust+gas), where we assume a dust:gas-ratio of 100.

| $\lambda$ [$\mu m$] | obs | mean | min | max |
|---|---|---|---|---|
| 2.19 | $1.3 \pm 0.2$ | 0.6 | 0.6 | 0.3 |
| 70 | $110 \pm 1$ | 92 | 63 | 135 |
| 160 | $32 \pm 1$ | 22 | 7 | 50 |

Table A.8: Comparison of the preburst flux densities as observed and modeled (with Av=18mag) for wavelengths with (post-)/burst-measurements. Values are in $10^{-9} \, erg/cm^2/s$.

| Nr | name | $\chi^2$ | d | Av | $m_{disk}$ | $r_{disk}$ | $\alpha_{disk}$ | $\beta_{disk}$ | $h_{100}$ | $r_{env}$ | $\dot{M}_{env}$ | $\Theta_{cav}$ | $\rho_{cav}$ | i | $L_*$ |
|---|---|---|---|---|---|---|---|---|---|---|---|---|---|---|---|
| | | | kpc | mag | $M_\odot$ | au | | | au | au | $1.33 \cdot M_\odot/yr$ | ° | $10^{-20} g/cm^3$ | ° | $L_\odot$ |
| | | | log | lin | log | log | lin | lin | log | log | log | lin | log | lin | log |
| min | | | | | $4 \cdot 10^{-8}$ | 80 | 1.0 | 1.0 | 0.5 | 10000 | $2 \cdot 10^{-5}$ | 14 | 0.01 | 0 | 8000 |
| max | | | | | 2.5 | 7000 | 3.3 | 1.3 | 33 | 40000 | 1 | 60 | 1000 | 60 | 160000 |
| 1 | MCX48 | 140 | 4.48 | 17 | 0.04 | 498 | 2.542 | 1.058 | 2.244 | 34877 | 0.03084 | 52 | 5.5 | 25 | 88059 |
| 2 | MCX24 | 263 | 3.9 | 19 | 0.0201 | 3381 | 1.603 | 1.014 | 1.69 | 23928 | 0.03654 | 54 | 5.2 | 37 | 64540 |
| 3 | MCY437 | 306 | 3.7 | 17 | 0.00062 | 3605 | 2.22 | 1.221 | 26.51 | 11982 | 0.05759 | 38 | 6.5 | 40 | 44781 |
| 4 | MCY916 | 346 | 3.7 | 17 | 0.00079 | 2083 | 2.485 | 1.138 | 1.836 | 16950 | 0.01633 | 50 | 3.7 | 56 | 101215 |
| 5 | MCX831 | 395 | 4.48 | 19 | 0.00473 | 106 | 1.159 | 1.222 | 0.6861 | 38660 | 0.0111 | 39 | 2.4 | 6 | 103690 |
| 6 | MCX152 | 396 | 3.7 | 19 | $3.4 \cdot 10^{-7}$ | 555 | 1.859 | 1.101 | 3.885 | 20568 | 0.01352 | 22 | 2.0 | 3 | 19793 |
| 7 | MCX391 | 475 | 4.48 | 18 | 0.02483 | 429 | 3.16 | 1.259 | 1.231 | 30800 | 0.03746 | 28 | 6.83 | 4 | 40115 |
| 8 | MCZ27 | 489 | 3.7 | 18 | 0.09312 | 108 | 2.397 | 1.18 | 2.72 | 17546 | 0.08764 | 31 | 6.2 | 28 | 24158 |
| 9 | MCY891 | 529 | 4.0 | 17 | 0.00049 | 542 | 2.613 | 1.147 | 25.37 | 15747 | 0.01049 | 26 | 4.6 | 30 | 97363 |
| 10 | MCX92 | 556 | 3.9 | 18 | $6.8 \cdot 10^{-8}$ | 220 | 1.757 | 1.067 | 4.295 | 32383 | 0.003268 | 41 | 3.6 | 10 | 58087 |
| mean | | 90 | 3.9 | 19 | 0.0017 | 684 | 2.2 | 1.12 | 3.1 | 24000 | 0.024 | 42 | 4.5 | 26 | 60587 |
| sigma | | | | | 64 | 3.25 | 0.6 | 0.08 | 2.9 | 1.5 | 2.2 | 11 | 1.5 | 17 | 1.7 |
| min | | | | | $2.7 \cdot 10^{-5}$ | 200 | 1.7 | 1.03 | 1.1 | 16000 | 0.011 | 31 | 3.0 | 9 | 36000 |
| max | | | | | 0.11 | 2100 | 2.8 | 1.20 | 9.0 | 36000 | 0.053 | 53 | 6.8 | 43 | $10^5$ |
| Tmin | | 610 | 3.9 | 19 | $2.7 \cdot 10^{-5}$ | 684 | 2.2 | 1.12 | 3.1 | 16000 | 0.011 | 53 | 3.0 | 26 | 60587 |
| Tmax | | 180 | 3.9 | 19 | 0.11 | 684 | 2.2 | 1.12 | 3.1 | 36000 | 0.053 | 31 | 6.8 | 26 | 60587 |

Table A.9: Parameters of the mean model (orange), the 10 best fits, and the models with minimum/maximum afterglow duration (yellow). The adapted ranges for the preburst models are given in gray for comparison. Log-sampled values imply geometric means (with $x_{min}$ =< $x$ > $\cdot\sigma$). All values are total (gas+dust). Note the factor 1.33 in the unit of $\dot{M}_{env}$.

### A.2.3 TFitter

Table A.10: TFitter benchmark fit result (value fit). The test model (used to create the input test data) is shown in gray. It was recovered by the TFitter.

| dir | $m_{disk}$ $m_\odot$ | $r_{disk}$ $au$ | $\alpha_{disk}$ | $\beta_{disk}$ | $h_{100}$ $au$ | $\dot{M}_{env}$ $m_\odot/yr$ | $\Theta_{cav}$ ° | $\rho_{cav}$ $10^{-19}g/cc$ | i ° | $E_{gcc}$ $10^{46}erg$ | $L_*$ $10^3 L_\odot$ | d kpc | Av mag | $\chi^2$ |
|---|---|---|---|---|---|---|---|---|---|---|---|---|---|---|
| | log | log | lin | lin | log | log | lin | log | lin | | log | log | lin | |
| Mod541/ | 0.0021 | 1500 | 2.26 | 1.21 | 2.5 | 0.072 | 57.8 | 0.8 | 26.6 | 0.17 | 3.1 | 6.18 | 30 | 5.56 |
| Mod541/ | 0.0021 | 1500 | 2.26 | 1.21 | 2.5 | 0.072 | 57.8 | 0.8 | 26.6 | 0.17 | 3.1 | 6.18 | 30 | 5.56 |
| Mod514/ | 0.0005 | 1000 | 1.32 | 1.21 | 3.7 | 0.068 | 55.7 | 2.0 | 27.9 | 0.21 | 3.7 | 6.63 | 38.4 | 7.60 |
| Mod985/ | 0.0039 | 1190 | 1.73 | 1.14 | 2.7 | 0.120 | 62.3 | 1.2 | 18.1 | 0.17 | 2.9 | 6.07 | 30 | 7.61 |
| Mod462/ | 0.0014 | 1600 | 1.91 | 1.21 | 3.9 | 0.062 | 50.5 | 1.4 | 28.6 | 0.26 | 4.5 | 7.12 | 65.8 | 9.79 |
| Mod860/ | 0.1020 | 542 | 1.39 | 1.16 | 2.3 | 0.038 | 55.2 | 2.6 | 21.3 | 0.18 | 3.1 | 6.07 | 30 | 10.2 |
| Mod353/ | 0.0013 | 954 | 1.26 | 1.18 | 4.0 | 0.113 | 55.6 | 1.0 | 29.1 | 0.21 | 3.7 | 6.87 | 30 | 10.9 |
| Mod960/ | 0.0322 | 1800 | 1.69 | 1.12 | 1.6 | 0.108 | 55.6 | 1.4 | 29.4 | 0.17 | 2.9 | 6.07 | 30 | 13.9 |
| Mod714/ | 0.0005 | 394 | 1.25 | 1.15 | 7.0 | 0.011 | 30.3 | 1.9 | 19.5 | 0.18 | 3.2 | 6.87 | 70 | 15.1 |
| Mod302/ | 0.0002 | 450 | 2.0 | 1.19 | 4.2 | 0.011 | 30.8 | 2.0 | 16.8 | 0.20 | 3.5 | 7.12 | 70 | 17.8 |
| mean | 0.0024 | 1100 | 1.8 | 1.2 | 3 | 0.06 | 54 | 1.3 | 25 | 0.19 | 3.3 | | | |
| sigma | 5 | 1.6 | 0.4 | 0.03 | 1.4 | 2 | 9 | 1.5 | 5 | 0.03 | 1.1 | | | |

Table A.11: Same as Tab. A.10 but for the benchmark ratio fit. $\chi^2$ is only burst+post value. The test model (gray) is among the best models. The mean parameters of the ratio-fit match those of the value-fit within the errors.

| dir | $m_{disk}$ $m_\odot$ | $r_{disk}$ $au$ | $\alpha_{disk}$ | $\beta_{disk}$ | $h_{100}$ $au$ | $\dot{M}_{env}$ $m_\odot/yr$ | $\Theta_{cav}$ ° | $\rho_{cav}$ $10^{-19}g/cc$ | $i$ ° | $E_{acc}$ $10^{46}erg$ | $L_*$ $10^3 L_\odot$ | $d$ kpc | $Av$ mag | $\chi^2$ |
|---|---|---|---|---|---|---|---|---|---|---|---|---|---|---|
| | log | log | lin | lin | log | log | lin | log | lin | | log | log | lin | |
| Mod960/ | 0.032 | 1800 | 1.69 | 1.12 | 1.6 | 0.108 | 55.6 | 1.4 | 29.4 | 0.166 | 2.9 | 6.07 | 30 | 2.03 |
| Mod541/ | 0.002 | 1500 | 2.26 | 1.21 | 2.5 | 0.072 | 57.8 | 0.8 | 26.6 | 0.174 | 3.1 | 6.18 | 30 | 2.31 |
| Mod541/ | 0.002 | 1500 | 2.26 | 1.21 | 2.5 | 0.072 | 57.8 | 0.8 | 26.6 | 0.174 | 3.1 | 6.18 | 30 | 2.31 |
| Mod465/ | 0.001 | 2200 | 2.21 | 1.11 | 6.6 | 0.001 | 63.3 | 11.5 | 16.4 | 0.233 | 4.1 | 6.07 | 30 | 2.43 |
| Mod949/ | 0.220 | 2000 | 1.61 | 1.2 | 4.6 | 0.053 | 57.5 | 4.1 | 20.6 | 0.206 | 3.6 | 6.07 | 30 | 2.58 |
| Mod496/ | 0.004 | 685 | 1.59 | 1.13 | 2.9 | $6 \cdot 10^{-5}$ | 34.1 | 12.3 | 31.7 | 0.231 | 4.0 | 6.07 | 30 | 2.71 |
| Mod406/ | 0.008 | 442 | 1.31 | 1.22 | 4.2 | 0.0002 | 64.5 | 12.7 | 18.4 | 0.239 | 4.2 | 6.07 | 30 | 2.72 |
| Mod341/ | 0.011 | 594 | 1.89 | 1.14 | 6.2 | 0.001 | 62.4 | 9.4 | 25 | 0.215 | 3.8 | 6.07 | 30 | 2.84 |
| Mod180/ | 0.002 | 564 | 1.87 | 1.22 | 4.5 | 0.001 | 63.7 | 10.2 | 14.4 | 0.294 | 5.1 | 6.07 | 30 | 2.84 |
| MC596/ | 0.567 | 1600 | 1.59 | 1.12 | 1.6 | 0.004 | 47.0 | 7.0 | 23.4 | 0.285 | 5.0 | 6.18 | 70 | 2.89 |
| mean | 0.009 | 1200 | 1.8 | 1.2 | 3.3 | 0.05 | 56 | 4.2 | 24 | 0.22 | 3.7 | | | |
| sigma | 8 | 1.8 | 0.3 | 0.05 | 1.6 | 14 | 9 | 3.1 | 6 | 0.04 | 1.2 | | | |

Table A.12: Same as Tab. A.10 but for the fit with the real data. The data (especially the burst) cannot be fitted by the models pretty well. Some models appear more than once, but with different bursts (see burst energy $E_{acc}$). The adapted mean model is highlighted in gray.

| dir | $m_{disk}$ $m_\odot$ | $r_{disk}$ au | $\alpha_{disk}$ | $\beta_{disk}$ | $h_{100}$ au | $\dot{M}_{env}$ $m_\odot/yr$ | $\Theta_{cav}$ ° | $\rho_{cav}$ $10^{-19}g/cc$ | i ° | $E_{acc}$ $10^{46} erg$ | $L_*$ $10^3 L_\odot$ | d kpc | Av mag | $\chi^2$ |
|---|---|---|---|---|---|---|---|---|---|---|---|---|---|---|
| | log | log | lin | lin | log | log | lin | log | lin | | log | log | lin | |
| Mod53/ | 0.0005 | 690 | 2.14 | 1.22 | 3.28 | 0.00022 | 25.8 | 15 | 12.5 | 0.70 | 8.5 | 6.07 | 30 | 37.4 |
| f2short/ | 0.0084 | 950 | 1.75 | 1.16 | 3.37 | 0.00403 | 17 | 10 | 21.8 | 0.41 | 5.0 | 6.75 | 30 | 38.8 |
| f1k5short/ | 0.0084 | 950 | 1.75 | 1.16 | 3.37 | 0.00403 | 17 | 10 | 21.8 | 0.29 | 5.0 | 6.75 | 30 | 38.9 |
| Mod53/ | 0.0005 | 690 | 2.14 | 1.22 | 3.28 | 0.00022 | 25.8 | 15 | 12.5 | 0.93 | 8.5 | 6.07 | 30 | 39.3 |
| Mod112/ | 0.0007 | 2100 | 1.58 | 1.12 | 1.84 | 0.00359 | 24.9 | 12 | 16.2 | 0.40 | 4.8 | 6.18 | 30 | 39.5 |
| Mod862/ | 0.4810 | 790 | 1.43 | 1.12 | 1.47 | 0.00180 | 35.6 | 13 | 13.5 | 0.77 | 7.0 | 6.52 | 38.4 | 42.4 |
| Mod680/ | 0.0168 | 480 | 1.54 | 1.12 | 8.93 | 0.01410 | 39.5 | 7 | 20.3 | 0.23 | 4.0 | 6.18 | 30 | 42.5 |
| Mod404/ | 0.0076 | 650 | 1.55 | 1.12 | 6.17 | 0.00080 | 49.8 | 13 | 23.8 | 0.94 | 8.6 | 6.4 | 32.1 | 42.7 |
| Mod53/ | 0.0005 | 690 | 2.14 | 1.22 | 3.28 | 0.00022 | 25.8 | 15 | 12.5 | 0.48 | 8.5 | 6.07 | 30 | 43.0 |
| Mod112/ | 0.0007 | 2100 | 1.58 | 1.12 | 1.84 | 0.00359 | 24.9 | 12 | 16.2 | 0.53 | 4.8 | 6.18 | 30 | 43.1 |
| mean | 0.0034 | 890 | 1.8 | 1.2 | 3.2 | 0.0014 | 28 | 12 | 17 | 0.57 | 6.2 | | | |
| sigma | 8.2 | 1.6 | 0.28 | 0.044 | 1.7 | 4.1 | 10 | 1.3 | 4.4 | 0.25 | 1.3 | | | |

## A.3 Input-files

### A.3.1 TORUS

```
# TORUS, G323 mean model, input file
nsource 1
radius1 27.39 !rsun
teff1 17385 !K
mass1 13.1 !msun
contflux1 blackbody

! grid
readgrid T
writegrid F
inputfile grid.dat
amrgridsize 71356617 ! radial extend 10**10cm
maxdepthamr 25
!mindepthamr 8
amr2d T !2d model, zylindrical
rhofloor 1e-30 ! minimum density

! dust
dustphysics T
ndusttype 3
tsub1 1600
graintype1 silD03_nk.txt !
graindensity1 3.6 ! g/cc
amin1 0.005 ! mu
amax1 0.250 ! mu
qdist1 3.5 ! index of size distribution s^-qdist 0.001
grainfrac1 0.00625 ! dust:gas

tsub2 1600
graintype2 Cper_nk.txt
graindensity2 2.2 ! g/cc
amin2 0.005 !mu
amax2 0.250 !mu
qdist2 3.5 ! index of size distribution s^-qdist 0.001
```

grainfrac2 0.0025

tsub3 1600
graintype3 Cpar_nk.txt
graindensity3 2.2 ! g/cc
amin3 0.005 ! mu
amax3 0.250 ! mu
qdist3 3.5 ! index of size distribution s^-qdist 0.001
grainfrac3 0.00125

geometry shakara
mdisc 0.001727 ! total mass (gas+dust) solar mass
rinner 450 ! Rstellar (not solar) Radius
router 684 ! au
alphadisc 2.199 ! radiale structur
betadisc 1.124 ! verticale structur
height 3.100 ! au scaleheight at 100au
heightsplitfac 1. ! 1 cell per heightunit (increasing with r)

erinner 60 ! au envelope inner radius
erouter 23849 ! au
mdotenv 0.02412 ! total accretion rate (gas+dust) msol/yr
cavangle 41.8 ! fullangle deg
cavdens 4.5e-20 ! total density (dust+gas) in g/cc=g/cm**3

radeq F
lucy_undersampled 1.0
spectrum F
jansky T
lambdaInMicrons T
nphotspec 1000000
distance 4018 ! in pc (if uncommented sed is measured at 100pc)
sednumlam 200
sedlammin 0.12 ! mu
sedlammax 1000. ! mu
inclination 26.32

timedepimage F

npix 100

timedep T
timestart 0.d0 ! years
timeend 25d0 ! years
risestart 0.0d0
varystart  0.0d0 ! years
varyend 25d0 ! years
lumfactor -1 ! if set to -1 read source lv from lum_file.dat
lumdecaytime 1e300 !after 0.1 yr it is at e/10 with L0*e**-((t-t0)/a)
lumrisetime 1e300
ntime 2501

nphotons 1e7

## A.3.2  TFitter

```
# TSEDfitter input file
object_name = G358
# distance
dsim_kpc = 6.77
d_range = [6.07, 7.12]
n_dist = 10
# foreground extinction
Av_range = [30, 70]
n_Av = 20

# times of the observations
nobs = 3
MJD_obs1 = start # 2009
MJD_obs2 = 58604 # 2019 May 1.
MJD_obs3 = 59089 # 2020 August 28.

# epoch names
epo1 = pre
epo2 = burst
epo3 = post

# accretion rate variation
# source input
lum_file = lum_file.dat # contains L*/Lpre for all times
step_source_lv = 7.3 # timestep
# reference date
MJD_peak = 58565 # Maserpeak −9 days (fitted best shift)
# uncertainty of assumed start
start_unc = [−7, 7]

# inputlist
models_in = TSEDfitter_inputfiles.dat
dir = ./
prefitresult = TFitter_result_value/newTSEDfitter.inp
```

## A.4   The TFitter

### A.4.1   Basic Steps

In this section, the basic steps of the TFitter are described. The TFitter delivers **diagnostic plots** automatically for the first model in the input list (see Fig. A.34 to A.38). This is useful for quickly finding errors in the input. We put the best pre-model (used to create the test-data) first in the input-list. Therefore, diagnosis plots serve as part of the benchmark (Sect. 4.2.1) already. The basic steps of the TFitter are as follows.

1. Read in the files (see Fig. A.34, example input file in the Appendix). The main parameters are as follows.

   - Observations: SEDs, times, dirs/filenames, uncertainty of burst onset (see Sect. A.4.2)
   - Simulations: TSEDs, burst profile (length of the pre-period if existent, $E_{acc}$), dedicated parameters (time step, $L_*$), Tdirs/filenames

2. Convert all times to a **common time for sim/obs**: We use the MJD, but setting the burst onset to 0 equally works.

3. Format the data-sets to Tcubes (one for each Tdir, see sketch Fig. A.32)

4. Our simulations show **numerical scatter**. Therefore, it is necessary to optimize them (before fitting). We applied a **stepwise interpolation in time** on the Tcubes. Although this sounds like a pure technical detail, it is actually a crucial step, since it can impact the results and should therefore be handled with care (see Sects. 5.3 and 5.3).

5. Get **d, Av, $X^2_{pre}$ for the preburst** (re-fit or from input file). Note that in principle d, Av could be estimated from a different epoch (or even as mean of all epochs), but the pre-SED is usually the most reliable (best wavelength-coverage, relaxed setting). **Redden and scale all Tcubes** accordingly (see Fig. A.35).

6. Interpolate the Tcubes to the **wavelength space**[1] of the observing epochs.

7. Get the **TSEDs for the observing epochs**, as visualized for an example in Fig. A.36. In our case, this is basically an interpolation in time (to the MJD of the burst and post) on the optimized TCubes.

---

[1] The wavelength-space of each epoch can be different (we use a common wavelength-space, which includes all wavelengths from all epochs despite the preburst). We interpolate the flux values linearly in the log-log space (this is the same as in the sedfitter [122]). If the scatter is huge, fitting (in an appropriate wavelength interval) could be better (not implemented yet).

8. Perform value or ratio **fits of all (burst-/post-) epochs**.

9. **Sort the Tdirs according to their epoch-combined** $\chi^2$-values. The epoch-combined $\chi^2$-values are given as sum of the $\chi^2$-values of the individual epochs, e.g. $\chi^2_{total} = \chi^2_{pre} + \chi^2_{epoch1} + ...$, where $\chi^2_{epoch}$ refers to the reduced $\chi^2$-value, i.e. it is divided by the number of datapoints (in each SED). This means, that all epochs (including the preburst) are counted equally. Different weighting factors could be easily implemented in the future. A visualization of the $\chi^2$-values for each epoch (split by data-points) is shown for one model in Fig. A.37. An option to sort by another value, such as $\chi^2$ of a dedicated epoch (or even a data point) could be implemented in the future for further testing.

   The final step is to **store the results**. We output a list containing the best models, as well as two '.pkl'-files (with a full list including parameters and fit results + an extra list with $\chi^2$-values for each data point/epoch). From this, the usual analysis can be performed. It is possible to modify the TFitter to return additional/different files.

**A remark**   There are possibilities to speed up the program (for example, to reduce the sizes of the Tcubes in Step 3). However, this was not necessary in our case. On the contrary, our concern is the synthetic noise of the Tmodels, rather than the computation time of the TFitter.

## A.4.2   Implementation of a time-shift

Until now, we had assumed that the burst onset was known. In general, this is not the case in reality. The best guess of the burst onset comes from an NIR-lightcurve (e.g. Ks in case of G323) or from the accompanying maser-flare (NIR dark bursters as G358). However, the actual burst probably starts earlier (timescale of ≈ days to weeks). This is visible e.g. in Fig. 3.11, which compares the variation in accretion rate with the expected NIR and maser curves (for the G323 model). Both the NIR and the maser curve peak slightly offset the adapted accretion-rate variation (although the correlation with the NIR curve is pretty good). One idea to get a more realistic estimate of the burst onset is to correlate the burst template and the output light curve for the band used as a template (or for the maser curve). The (time) uncertainty of the (best) correlation then reflects the uncertainty of the 'most realistic' burst onset (this approach is used for G358, see Sect. 4.3). It is not yet the best solution. However, the TFitter has been modified to allow a flexible burst onset (within a given range). In its current realization, this range is the same for all Tmodels. The TFitter now computes the $\chi^2$ values for all possible time shifts (within the length of the simulated time interval)[2]. Then the best value within the given range is returned for each Tmodel. Fig. A.48 shows the $\chi^2$ values for the test model and the burst epoch as a function of the date of the test observation (left panel) for each wavelength. For the test data, the minimum should be reached exactly on the adapted date (vertical blue line). However, there is a shift by 10 days, most likely caused by stochastic errors (see also the SEDs in the right panel). Fig. A.48 shows $X^2$ as a function of the time shift. The best shift is determined by the burst epoch alone (this is probably a bias of the method).

---

[2]For the future only the range of importance should be computed in order to reduce the computation time.

We note that the fit with the time shift is possibly not the best solution and that a more iterative approach should be favored. Nevertheless, it is a tool to 'probe' the burst template, as systematic shifts can hint at 'false' burst onsets or shapes. Clearly, this needs further tests with real data!

# List of Publications:

1. *Infrared observations of the flaring maser source G358.93−0.03 - SOFIA confirms an accretion burst from a massive young stellar object* (**published**)
   B. Stecklum, **V. Wolf**, H. Linz, A. Caratti o Garatti, S. Schmidl, S. Klose, J. Eislöffel, Ch. Fischer, C. Brogan, R. Burns, O. Bayandina, C. Cyganowski, M. Gurwell, T. Hunter, N. Hirano, K.-T. Kim, G. MacLeod, K. M. Menten, M. Olech, G. Orosz, A. Sobolev, T. K. Sridharan, G. Surcis, K. Sugiyama, J. van der Walt, A. Volvach, and Y. Yonekura, A&A, 646:A161, February 2021.

2. *The accretion burst of the massive young stellar object G323.46−0.08 - A showcase on afterglows and echoes of episodic accretion* (**in preparation**)
   V. Wolf, B. Stecklum, A. Caratti o Garatti, Ch. Fischer, T. Harries, J. Eislöffel, H. Linz, and A. Ahmadi

# List of Talks:

**Tautoloquium**

| | |
|---|---|
| 2018 | *SED-Modeling of NIRS3 (an HMYSO with detected accretion burst)* |
| 2020 | *Light echoes with Mol3D - some showcases* |
| 2021 | *SOFIA in action: G358-MM1's accretion burst confirmed* |
| 2022 | *Time-dependent radiative transfer modeling with TORUS on the example of G323's accretion outburst* |

**Kiel**

| | |
|---|---|
| 2018 | *Accretion-bursts and the disks around HMYSOs* |

**Exeter**

| | |
|---|---|
| 2019 | *Light echoes with Mol3D - some showcases* |

**Advisory board meeting in Tautenburg**

| | |
|---|---|
| 2023 | *Time-dependent modeling of thermal dust emission from protostellar outbursts* |

# List of scientific Posters:

**2020**  **Advisory Board Meeting**
*A Breakthrough in Massive Star Formation Research*
*–The Accretion Burst from S255IR NIRS3*
B. Stecklum, J. Eislöffel, V. Wolf, A. Drabent, NIRS3 Collaboration

**2021**  **EAS annual meeting**
*Analysis of the accretion burst in the MYSO G358.9-0.03* [131]
B. Stecklum, V. Wolf, J. Eislöffel, S. Klose, S. Schmidl, H. Linz,
A. Caratti o Garatti, Ch. Fischer
**Star Formation: From Clouds to Discs**
*Accretion bursts in high-mass young stellar objects*
Caratti o Garatti, Stecklum, Wolf

**2022**  **EAS annual meeting**
*IR afterglows from MYSO accretion bursts - A unique science case*
*for SOFIA*
B. Stecklum, V. Wolf, A. Caratti o Garatti, T. Harries, H. Linz

**2023**  **Advisory Board meeting**
*Episodic Accretion in High-Mass Star Formation*
V. Wolf, B. Stecklum, J. Eislöffel, A. Caratti o Garatti, C. Fischer,
T. Harries, H. Linz, M2O collaboration
**PPVII**
*Episodic Accretion in High-Mass Star Formation*
B. Stecklum, V. Wolf, J. Eislöffel, A. Caratti o Garatti, C. Fischer,
T. Harries, H. Linz, M2O collaboration

# Ehrenwörtliche Erklärung

1. dass mir die geltende Promotionsordnung bekannt ist;

2. dass ich die Dissertation selbst angefertigt habe, keine Textabschnitte eines Dritten oder eigener Prüfungsarbeiten ohne Kennzeichnung übernommen und alle von mir benutzten Hilfsmittel, persönlichen Mitteilungen und Quellen in meiner Arbeit angegeben habe;

3. dass bei der Auswahl und Auswertung folgenden Materials mich die nachstehend aufgeführten Personen in der jeweils beschriebenen Weise unterstützt haben:

   - Dr. Bringfried Stecklum – durch seine fachliche Betreuung der vorliegenden Arbeit, sowie die Aufbereitung der Daten (Archivdaten und Fotometrie der SOFIA daten, SED Dekomposition) und die gemeinsame Arbeit an den beiden wissenschaftlichen Publikationen (davon eine in prep.), auf deren Konzept diese Thesis aufbaut

   - Prof. Tim Harries – durch das Bereitstellen und Modifizieren des TORUS Codes

   - Dr. Christian Fischer – durch seine Hilfe bei der Aufnahme und Kalibration der Daten

   - Dr. Jochen Eislöffel, Dr. Alessio Caratti o Garatti, Dr. Hendrik Linz – durch die Diskussionen und Hinweise zur episodischen Akkretion und die gemeinsame Arbeit

   - Dr. Ross Burns – durch die Koordination der M2O (welche essentiell für unsere Beobachtungen war) und den Austausch zur Ausbreitung der Hitzewelle in G358

4. dass die Hilfe einer kommerziellen Promotionsvermittlerin/eines kommerziellen Promotionsvermittlers nicht in Anspruch genommen wurde und dass Dritte weder unmittelbar noch mittelbar geldwerte Leistungen für Arbeiten erhalten haben, die im Zusammenhang mit dem Inhalt der vorgelegten Dissertation stehen;

5. dass die Dissertation noch nicht als Prüfungsarbeit für eine staatliche oder andere wissenschaftliche Prüfung eingereicht wurde;

6. dass eine gleiche, eine in wesentlichen Teilen ähnliche oder eine andere Abhandlung bei einer anderen Hochschule als Dissertation nicht eingereicht wurde.

_________________________                    _________________________

Ort und Datum                                Unterschrift